\documentclass[%
reprint,
superscriptaddress,
amsmath,amssymb,
iop,longbibliography,
jcp,
floatfix,
]{revtex4-2}

\usepackage{graphicx}%
\usepackage{verbatim}
\usepackage[version=4]{mhchem}
\usepackage{xcolor}
\usepackage{physics}
\usepackage[normalem]{ulem}
\usepackage{pdfpages}
\usepackage{pgffor}
\usepackage{listings} %
\usepackage{csquotes}
\usepackage{bm}

\usepackage{hyperref}
\hypersetup{
    colorlinks=true,
    linkcolor=blue,
    filecolor=magenta,      
    urlcolor=cyan,
    pdftitle={Overleaf Example},
    pdfpagemode=FullScreen,
    }
\usepackage{accents}
\newcommand{\dbtilde}[1]{\accentset{\approx}{#1}}
\newcommand{\vxi}{\boldsymbol{\xi}}
 \usepackage{amsmath}

\newcommand{\der}[2]{\ensuremath{\frac{\partial #1}{\partial #2}}\xspace} %

\newcommand{\bec}{\ensuremath{Z^*_{\alpha j}}\xspace}
\newcommand{\ipi}{{i-PI}\xspace}
\newcommand{\rev}[1]{#1}

\usepackage{listings}

\usepackage{color}
\definecolor{gray}{rgb}{0.4,0.4,0.4}
\definecolor{darkblue}{rgb}{0.0,0.0,0.6}
\definecolor{cyan}{rgb}{0.0,0.6,0.6}
\definecolor{lightgray}{rgb}{0.9,0.9,0.9}

\lstdefinelanguage{XML}
{
  basicstyle=\ttfamily,
  morestring=[s]{"}{"},
  morecomment=[s]{?}{?},
  morecomment=[s]{!--}{--},
  commentstyle=\color{darkgreen},
  moredelim=[s][\color{black}]{>}{<},
  moredelim=[s][\color{red}]{\ }{=},
  stringstyle=\color{blue},
  identifierstyle=\color{maroon}
}

\usepackage{xspace} %

\makeatletter
\AtBeginDocument{\let\LS@rot\@undefined}
\makeatother

\begin{document}

\title{i-PI 3.0: a flexible \rev{and} efficient framework for advanced atomistic simulations}

\author{Yair Litman}%
\affiliation{Y. Hamied Department of Chemistry,  University of Cambridge,  Lensfield Road,  Cambridge,  CB2 1EW, UK\looseness=-1}
\newcommand{\yl}[1]{{\color{blue}{#1}}}

\author{Venkat Kapil}
\affiliation{Y. Hamied Department of Chemistry,  University of Cambridge,  Lensfield Road,  Cambridge,  CB2 1EW, UK\looseness=-1}
\affiliation{Dept. of Physics and Astronomy, University College London, 17-19 Gordon St, London WC1H 0AH, UK\looseness=-1}
\affiliation{Thomas Young Centre \&  London Centre for Nanotechnology, 19 Gordon St, London WC1H 0AH, UK\looseness=-1}
\newcommand{\vk}[1]{{\color{pink}#1}}

\author{Yotam M.\ Y.\ Feldman}
\affiliation {School of Chemistry, Tel Aviv University, Tel Aviv 6997801, Israel.}

\author{Davide Tisi}
\affiliation{Laboratory of Computational Science and Modeling, Institut des Mat\'eriaux, \'Ecole Polytechnique F\'ed\'erale de Lausanne, 1015 Lausanne, Switzerland}

\author{Tomislav Begušić}
\affiliation{Div. of Chemistry and Chemical Engineering, California Institute of Technology, Pasadena, California 91125, USA\looseness=-1}

\author{Karen Fidanyan}
\affiliation{MPI for the structure and Dynamics of Matter, Hamburg, Germany}

\author{Guillaume Fraux}
\affiliation{Laboratory of Computational Science and Modeling, Institut des Mat\'eriaux, \'Ecole Polytechnique F\'ed\'erale de Lausanne, 1015 Lausanne, Switzerland}

\author{Jacob Higer}
\affiliation {School of Physics, Tel Aviv University, Tel Aviv 6997801, Israel.}

\author{Matthias Kellner}
\affiliation{Laboratory of Computational Science and Modeling, Institut des Mat\'eriaux, \'Ecole Polytechnique F\'ed\'erale de Lausanne, 1015 Lausanne, Switzerland}

\author{Tao E. Li}
\affiliation{Department of Physics and Astronomy, University of Delaware, Newark, Delaware 19716, USA\looseness=-1}

\author{Eszter S. Pós}
\affiliation{MPI for the structure and Dynamics of Matter, Hamburg, Germany}

\author{Elia Stocco}
\affiliation{MPI for the structure and Dynamics of Matter, Hamburg, Germany}

\author{George Trenins}
\affiliation{MPI for the structure and Dynamics of Matter, Hamburg, Germany}

\author{Barak Hirshberg}
\affiliation {School of Chemistry, Tel Aviv University, Tel Aviv 6997801, Israel.}
\altaffiliation{The Center for Computational Molecular and Materials Science, Tel Aviv University, Tel Aviv 6997801, Israel}

\author{Mariana Rossi}
\affiliation{MPI for the structure and Dynamics of Matter, Hamburg, Germany}

\author{Michele Ceriotti}
\email{michele.ceriotti@epfl.ch}
\affiliation{Laboratory of Computational Science and Modeling, Institut des Mat\'eriaux, \'Ecole Polytechnique F\'ed\'erale de Lausanne, 1015 Lausanne, Switzerland}

\newcommand{\mc}[1]{{\color{blue}#1}}
\newcommand{\dt}[1]{{\color{orange}#1}}

\date{\today}%

\begin{abstract}
Atomic-scale simulations have progressed tremendously over the past decade, 
largely due to the availability of \rev{machine-learning interatomic potentials}. These potentials \rev{combine} the accuracy of electronic structure calculations with \rev{the ability to reach} extensive length and time scales. 
\rev{The \ipi{} package facilitates integrating the latest developments in this field with advanced modeling techniques, thanks to a modular software architecture based on inter-process communication through a socket interface. 
The choice of Python for implementation facilitates rapid prototyping but can add computational overhead. 
In this new release, we carefully benchmarked and optimized \ipi{} for several common simulation scenarios, making such overhead negligible when \ipi{} is used to model systems up to tens of thousands of atoms using widely adopted machine learning \rev{interatomic} potentials, such as Behler-Parinello, DeePMD and MACE neural networks.}
We also present the implementation of several new features, including an efficient algorithm to model bosonic and fermionic exchange, a framework for uncertainty quantification to be used in conjunction with machine-learning potentials, a communication infrastructure that allows deeper integration with electronic-driven simulations, and an approach to simulate coupled photon-nuclear dynamics in optical or plasmonic cavities. 
\end{abstract}

\maketitle

\section{Introduction}

The atomic-scale modeling of molecules and materials is a constantly evolving field driven by the interest of researchers in discovering new fundamental phenomena, interpreting complex experimental results, and, together with an increasing number of commercial enterprises, improving, discovering, and developing new materials.\cite{vonLilienfeld2020,Butler2018,Kulik_2022,Deringer_AdvMat_2019,10.1126/sciadv.1701816,Tao2021} 
Several computational developments in recent years \rev{extended dramatically} the problems that can be addressed by atomistic simulations, far exceeding the steady improvement of the underlying hardware. Arguably, the largest contribution to such fast progress can be attributed to
 the \rev{development of machine learning} interatomic potentials (MLIPs).
 \rev{MLIPs approach the accuracy of the underlying \textit{ab initio} data, but at a small fraction of the computational cost.} They have 
 allowed access to system sizes and time scales which were previously only accessible to empirical interatomic potentials.\cite{Lu_CPC_2021, Behler_Ang_2017,batatia2024foundation,Unke_ChemRev_2021} %
 This has driven a renewed interest in complex simulation algorithms that were thought inapplicable for \rev{studies} that require 
 \rev{a first-principles}
 description of interatomic interactions, such as the investigation of catalytic reactions \cite{Bruix2019,Schlexer_ChemCatCHem_2019,10.1021/acscatal.0c03472} or of the thermodynamic stability of materials.\cite{cheng_ab_2019,PhysRevB.95.094203,10.1021/acs.jctc.2c00343,kapil_complete_2022} However, the lack of easy-to-use implementations of such approaches has become an important factor that hinders \rev{a more widespread adoption}

Since its first release \rev{in 2014}~\cite{IPIv1}, one of \ipi's primary objectives has been \enquote{lowering the implementation barrier to bring state-of-the-art sampling and atomistic modeling techniques to \textit{ab initio} and empirical interatomic potentials programs}. The first \rev{version} of \ipi was \rev{designed for} path integral molecular dynamics (PIMD) simulations, hence the code name. Since then, and thanks to a community effort, it has evolved into a \enquote{universal force engine for advanced molecular simulations}, including \rev{both} standard and advanced simulations algorithms, as described in the following sections.\cite{IPIV2} \\
In the last few years, our efforts have adapted to the changing needs of the field. While continuing with the implementation of new and advanced features, we have improved the scalability of the code backend, as required by newly available and computationally efficient MLIPs. We also adopted automatic testing workflows to \rev{strengthen} code stability and enhance our reproducibility standards.

\ipi's capabilities as a versatile tool for atomistic simulations can be demonstrated by the wide range of applications in which it has been employed. These examples range from the dynamics of charge density wave materials~\cite{schob+24spp}, \rev{interpretation of neutron scatter experiments},\cite{Linker2024}  \rev{and nuclear motion associated to electronic excitations,\cite{jacobs+es2024, Liu2023,Gigli2023}} to the elucidation of the thermodynamic properties of aqueous electrons,\cite{lan_simulating_2021,Lan_Angew_2022, Novelli2023} the calculation of first principles phase diagrams,\cite{Cheng_Nature_2020, Kapil_Nature_2022, Bore2023} and the study of aqueous solutions and interfaces.\cite{Litman_NatChem_2024,Litman_JPCL_2023, shepherd_efficient_2021, fidanyan_jcp_2023, delaPuente2023,Inoue_JPCL_2023}
Moreover, by virtue of its modular structure,  \ipi is already being used as a simulation engine with newly developed MLIP architectures~\cite{Chmiela2023,batatia2024foundation,kovacs_mace-off23_2023} \rev{making it easy to use them in combination with advanced sampling techniques. It can also be used as } a toolbox to combine different simulation \enquote{modes} to design \rev{sophisticated modeling workflows that involve multiple types of interatomic potentials, free-energy-sampling software\cite{PLUMED} as well as post-processing of trajectory data.\cite{ross+20jctc}}

This manuscript reports on the progress made in the last 5 years since the release of version 2.0, by describing the newly adopted maintenance strategies, the improved performance, and the newly-implemented features. The manuscript is organised as follows:
In Sec. II, we start with a short review of the code architecture, followed by a description of the improvements made to the code's robustness and usability and a list of the main features available in this release. 
In Sec. III, we present a detailed discussion of the code efficiency for production and large-scale simulations while in Sec. IV, we describe \rev{some selected} new features since the last code release \rev{discussing their implementation and examples of their application}.
In Sec. V, we showcase how the modularity and flexibility of \ipi{} help the implementation of advanced simulation setups that would otherwise require new codes to be written and Sec. VI concludes the article.

\begin{figure*}
    \centering
    \includegraphics[width=0.7\linewidth]{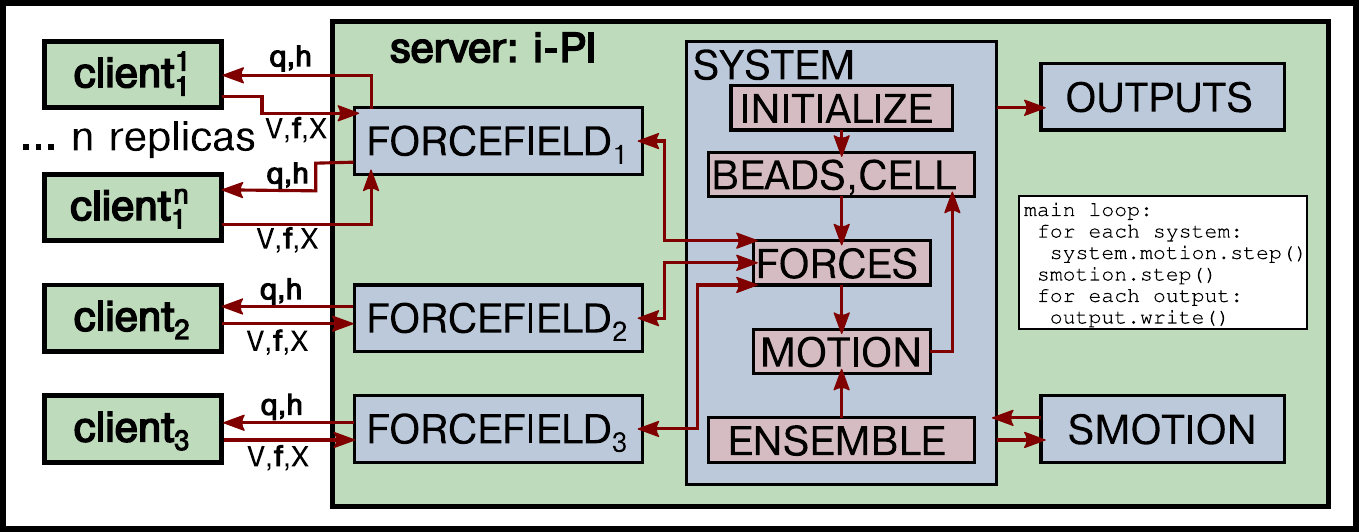}
\caption{Schematic representation of the i-PI code structure and the server-client communication. 
The physical system is defined by one or more replicas (beads), and the sampling conditions (e.g. temperature and pressure) by an \emph{Ensemble} class. 
The forces acting on the atoms are constructed based on a combination of energy contributions evaluated by one or more instances of the \emph{Forcefield} class (three in this example).   
Each \emph{Forcefield} instance communicates with a different client sending positions ($\textbf{q}$) and lattice vectors ($\textbf{h}$), and receiving energy
($V$), forces ($\textbf{f}$), stresses and possible extra properties ($X$) as a JSON formatted string. 
In simulations with multiple replicas, each \emph{Forcefield}  instance can accept connections from several clients, to achieve parallel evaluation.
The \emph{Motion} class determines the evolution of the atoms (e.g. time integration, or geometry optimization), while ``system motion'' classes (\emph{SMotion}) can act on multiple systems, e.g. to carry out replica exchange simulations. The output class handles writing the output and restart files.
}
    \label{fig:structure}
\end{figure*}

\section{Program Overview}

\ipi~ is a force engine that operates with a client-server paradigm. \ipi~ acts as a server in command of the evolution of the nuclear positions while one or more instances of the client code handle the evaluation of \rev{energy, forces, and possibly other properties for individual configurations.} 
Designed to be universal, \ipi's architecture is agnostic to the identity of the force providers, with a general and flexible backend that accommodates a wide range of client codes. 

The code is \rev{distributed under a dual GPL and MIT license, and it is} written in Python 3, a high-level, general-purpose interpreted programming language known for its emphasis on code readability. This choice facilitates rapid prototyping of new ideas and relatively easy code maintenance when compared to compiled languages traditionally used in scientific computing. Additionally, the availability of highly optimised numerical libraries allows Python-based software to deliver competitive performance. Indeed, even computing-intensive electronic structure codes have released Python implementations in recent years.\cite{gpaw2024,pyscf_2020}  A detailed discussion of the code efficiency and improvements since the previous release is presented in Sec.  \ref{sec:Efficieny}.

The communication between \ipi~ and the client codes is carried out through sockets, enabling local connections by shared-memory inter-process communication (UNIX domain sockets) or between two hosts connected to a local network or to the Internet (TCP/IP sockets).
During the initialization, \ipi~creates the sockets and awaits the connection of one or more clients. When the connection is established, the simulation loop starts: i) \ipi communicates the nuclear positions and the cell lattice vectors to the active clients, ii) the clients evaluate the required properties -- typically energies, forces, and stresses -- and send these back, and iii)  \ipi uses this data to displace the atoms and to calculate the new nuclear positions, restarting the loop. 
As \ipi~ is designed to be client-agnostic, 
clients must be initialized independently before they can connect to the sockets. \rev{Importantly, the identity and the number of atoms must be defined consistently in the client code according to its own input mechanism and remain unchanged throughout the simulation.}

A schematic representation of the code structure and the server-client communication is presented in Fig. \ref{fig:structure}. \ipi is structured in a modular way that represents the underlying physics of the target simulation as faithfully as possible.
To achieve this, the code is organized around the \emph{System} class which encodes all the information related to the physical system, such as the number and identity of atoms, the ensemble to be sampled, the number of replicas in the path integral simulations, the initialization procedure, and the algorithm for evolving the nuclear positions. 
Forces are managed through the \emph{Forces} class, which can combine multiple force components, each of which is computed following the strategy specified in a \emph{Forcefield} class. This two-layer approach is particularly advantageous for algorithms where the total forces and energies are obtained as the sum of terms computed at different accuracy levels,\cite{mark-mano08jcp,mark-mano08cpl,Gigli2023,tisi2024thermal,kapi+16jcp,kapil_modeling_2019} or for different portions of the system.\cite{litm+17jcp} 
Simulation techniques that require the evolution of many systems simultaneously make use of the \emph{SMotion} (Systems Motion) class. This class enables the definition of evolution steps that combine the (possibly many) different systems,
facilitating, for example, exchanges between system replicas as done in replica exchange simulations, \rev{or the application of biases, like in metadynamics}.
Finally, estimators can be defined within the code or computed by the provided post-processing tools or user-generated scripts. 
An updated list of code features is reported in Section \ref{sec:list_of_features}.

\subsection{Code philosophy and development model}

\paragraph*{Many available client codes:} 
\ipi~ is connected to a multitude of electronic structure and machine learning codes, either directly or through its ASE~\cite{ASE} or LAMMPS~\cite{LAMMPS} interfaces. This wide-ranging connectivity means that a single implementation of an advanced simulation method is readily available to all the users of these codes, which significantly amplifies the potential impact of each contribution to \ipi. \rev{Furthermore, this capability enables a direct comparison between different client codes within an identical simulation setup.}

\rev{\paragraph*{Ease of implementation of new interfaces:}

Adding a new interface to \ipi{} is simple. For codes that provide a Python interface, this can be implemented by adding a small new file to the Python driver directory (\lstinline{drivers/py/pes/}).
For compiled software, the driver loop can usually be rather self-contained, modelled after the molecular dynamics or geometry optimization loop. 
The FORTRAN driver distributed with \ipi{} provides a good example, as well as utilities to open and communicate over sockets that can be reused in third-party codes, while the LAMMPS\cite{LAMMPS} implementation of \lstinline{fix_ipi} can be used as a template of a C++ interface. 
In all cases, the structure (atom numbers and types) should be initialized \emph{before} communication is established, and that \ipi{} does not guarantee that successive calls provide similar structures (as in a MD trajectory) which might require some care when the force evaluator makes a continuity assumption for neighbor list updates or wavefunction extrapolation. 
}

\paragraph*{Lower development and maintenance effort:}  
In scenarios where a developer aims to use a newly implemented algorithm \rev{for nuclear motion} with different types of codes, such as a density-functional theory (DFT) plane wave code to study periodic systems, or a quantum chemistry code to study gas phase molecules at a higher electronic structure level, two separate implementations would be required. By integrating the algorithm into \ipi, only one implementation is needed and there is only one code to maintain. %

\paragraph*{Automatic (re)evaluation of physical quantities:} The infrastructure of \ipi automatically manages updates of energy, forces, stresses, and other structure-dependent properties by establishing dependencies between physically related variables. This architecture enables developers to concentrate solely on the algorithm's implementation making the development process more focused and efficient.

\paragraph*{Participative development model:}  
\rev{
The modular structure of the code simplifies the contribution of self-contained implementations of new methods. These methods are explicitly listed in a dedicated section of the documentation, giving developers visibility and recognition for their contributions. Many of the applications presented in this paper are examples of this model.}

\paragraph*{Code Maintenance and Long-term Sustainability:}
\rev{
As with most open-source projects, \ipi{} is developed and maintained by a community of volunteer contributors. 
Several of the authors of this manuscript, who are core developers and hold permanent or long-term academic positions, are involved in the governance of the code. These people have administrator rights on the main repository, which reduces the risk that the code base will be abandoned. 
We encourage power users to also help maintain the core infrastructure and the documentation; in preparation for this release, we have dedicated a special focus on improving the development infrastructure, as we discuss in the next Section.}

\subsection{Code robustness and usability \label{sec:regtests}}

The increased number of \ipi~ users and \rev{contributors since the last release} 
required enhancing the maintainability, robustness, and standardization of the code. These improvements involve five distinct aspects:

\paragraph*{Documentation:} \rev{We refactored the documentation of \ipi{}, which is now generated using \texttt{Sphinx}, is deployed automatically when new features are merged to the \lstinline{main} branch,} and is linked to the main code webpage~\footnote{\rev{\ipi official documentation: \url{https://ipi-code.org/i-pi/index.html}}}. 
The documentation has been updated to reflect the current status of the code, and a section about how to contribute new features to the code has been added to reduce the barrier for new developers. 
An automatic parsing of the \rev{Python docstrings} throughout \ipi~ensures a complete listing of all possible keywords and the corresponding explanation, which is automatically updated as soon as new features are implemented and merged in the main branch.

\paragraph*{Examples:} We have completely restructured the provided examples to make them easier to navigate and as intuitive as possible for newcomers. The examples are now organized into three directories, namely \textit{features},  \textit{clients}, and \textit{hpc\_scripts}. The \textit{features} directory displays extensive examples of the different implemented functionalities, all of which can be run locally using the built-in Python or Fortran driver, without requiring additional software installation.
The \textit{clients} directory includes examples tailored to specific driver codes, highlighting the unique syntax and tags needed for an adequate setup when used in connection to \ipi. 
At the moment, we provide examples for
    ASE,\cite{ASE}
    CASTEP,\cite{CASTEP}
    CP2K,\cite{CP2K}
    LAMMPS,\cite{LAMMPS}
    DFTB+,\cite{DFTB+} 
    ffsGDML,\cite{ffsGDML}
    FHI-aims,\cite{FHI-AIMS}
    librascal,\cite{librascal}
    Quantum ESPRESSO,\cite{Qespresso,QE-2017,QE-2020}
    VASP,\cite{VASP}
    elphmod,\cite{elphmod}
    and YAFF.\cite{YAFF}
We \rev{reiterate} that many other codes can also be connected to \ipi\ through their ASE and LAMMPS interfaces.  Finally, motivated by the number of user queries on this topic, the \textit{hpc\_scripts} directory provides examples with the corresponding submission scripts to run the simulations on high-performance computing platforms.

\paragraph*{Demonstrations and tutorials:} We have added  \rev{a \textit{demos} directory that contains examples of more complicated setups,}
to illustrate the application of \ipi~ in realistic simulation scenarios, combining several of the features showcased in the \textit{examples} directory. 
Some of these demonstration examples also cover some of the use of the pre- and post-processing scripts accompanying the code. Users specifically interested in learning the theory of path integral simulations and obtaining hands-on practice using \ipi~ are referred to the online courses developed with this goal.\footnote{\rev{The  \enquote{Path Integral Methods in Atomistic Modelling} is hosted at \url{https://app.courseware.epfl.ch/learning/course/course-v1:EPFL+X+2022/home} and code tutorials at \url{https://github.com/i-pi/piqm2023-tutorial/}}}

\paragraph*{Continuous integration:} 
We have implemented continuous integration workflows to \rev{automatically test for the code's functionality}. One automatic workflow reviews all the simulations in the \textit{examples} directory by running them for a few steps and quickly verifying if they are functioning.
A second workflow performs regression tests. In these cases, the simulations are performed for the required amount of steps and the simulation outputs are compared against pre-computed reference values. Together, these workflows guarantee that any new modification to the code neither alters the results of simulations \rev{that are tested} nor breaks the provided examples. Additional workflows, less critical for the functioning of the code, focus on checking the formatting syntax using Flake8 
\footnote{ \rev{See 
\url{https://flake8.pycqa.org/}} for more information about Flake8} and Black \footnote{ \rev{See \url{https://github.com/psf/black}
for more information about Black} }, or managing the updates of PyPI releases.

\rev{
\paragraph*{Developers' documentation and workflows}
We have included and updated the development guidelines, that are discussed in the main documentation, as well as in the \lstinline{CONTRIBUTING.md} file in the main repository. 
Contributors are encouraged to use the standard \emph{git} workflows, such as creating a fork and opening a pull request, and expect code review that will involve both the correctness of the algorithm and the consistency with \ipi{}'s coding style and quality. 
}

\subsection{Main Features\label{sec:list_of_features}}

We conclude this Section by listing the main features currently available in \ipi{}.
\rev{A comprehensive list of features, indicating the main contributors, is maintained as part of the documentation.}

\rev{\subsubsection{Features available in the 2.0 release:}}
\paragraph*{Classical and path integral molecular dynamics:} in the {\em NVE}, {\em NVT}, {\em NPT}
and constant stress ({\em NsT}) ensembles.

\paragraph*{Thermostats:} stochastic velocity rescaling,\cite{buss+07jcp}
path-integral Langevin equation thermostats,\cite{ceri+10jcp}
generalized Langevin equation (GLE) thermostats,
including the optimal sampling,\cite{ceri+09prl,ceri+10jctc} quantum, 
\cite{ceri+09prl2}, $\delta$,\cite{ceri-parr10pcs}
hot-spot,~\cite{dett+17jctc} and PIGLET thermostats,\cite{ceri-mano12prl} GLE thermostats for use with path-integral correlation functions,\cite{ross+18jcp} fast forward Langevin thermostat,\cite{hija+18jcp} Langevin sampling for noisy and/or dissipative forces.\cite{PhysRevB.73.041105,PhysRevLett.98.066401}

\paragraph*{Barostats:} isotropic barostats,\cite{Bussi_JCP_2009,IPIv1} anisotropic barostats,\cite{Raiteri_JPhys_2011}
 and its path integral adaptations.\cite{Martyna_JCP_1999, Kapil_JCTC_2019b}
 
\paragraph*{Geometry optimizers:} steepest descent, \rev{conjugate gradient}, L-BFGS, trust radius BFGS, nudged elastic band \cite{Henkelman_JCP_2000_I,Henkelman_JCP_2000_II,fidanyan_jcp_2023} and mode following saddle point search.\cite{Nichols_JCP_1990}

\paragraph*{Enhanced sampling and free energy calculations:} thermodynamic integration,\cite{ross+16prl} metadynamics, replica exchange MD,\cite{petr+15jcc} and quantum alchemical transformations.\cite{liu+13jpcc,chen+16jpcl}

\paragraph*{Multiple time-stepping (MTS):} MTS in real time and imaginary time, also known as ring polymer contraction.\cite{mark-mano08jcp,mark-mano08cpl,kapi+16jcp,litm+17jcp}

\paragraph*{Approximate quantum dynamics calculations:} ring polymer molecular dynamics (RPMD),\cite{crai-mano04jcp,habe+13arpc} centroid molecular dynamics (CMD),\cite{cao-voth93jcp, cao_JCP_94} thermostatted RPMD,\cite{ross+14jcp, ross+18jcp}, ring polymer instanton rates and tunnelling splittings.\cite{Litman_JACS_2019} 

\paragraph*{Advanced path integral simulations and estimators:} 
fourth-order path integrals,\cite{kapi+16jcp2}
perturbed path integrals,\cite{polt-tkat16cs}
the scaled-coordinate heat capacity estimator,\cite{yama05jcp}
isotope fractionation estimations,\cite{ceri-mark13jcp}  particle momentum distribution,\cite{lin+10prl} reweighted fourth-order path integral MD.\cite{ceri+12prsa,jang-voth01jcp}

\rev{\subsubsection{Features added after the 2.0 release:}}

\rev{
\paragraph*{Efficient integrators:} BAOAB splitting~\cite{kapil_modeling_2019}, and Cayley integrator~\cite{Korol2019} for larger PIMD timesteps.

\paragraph*{Barostats:} MTTK fully flexible barostat~\cite{Martyna1996}, MTS implementation, and barostat implementation for the Suzuki-Chin~\cite{kapil_modeling_2019} PIMD.

\paragraph*{Advanced path integral simulations and estimators:} 
Suzuki-Chin double virial operator estimator for calculating heat capacity ~\cite{kapil_modeling_2019} and Open chain path integrals~\cite{kapi+18jpcb} for estimating momentum dependent operators

\paragraph*{Vibrational analysis:} through the harmonic Hessian,
independent mode framework, the vibrational self-consistent field, and self-consistent phonons.\cite{Kapil_JCTC_2019}

\paragraph*{Particle exchange Monte Carlo:} for efficient sampling of poorly ergodic mixtures ~\cite{Imbalzano_JCP_2021}}

\paragraph*{Path Integral Molecular Dynamics (PIMD) for indistinguishable particles:} Methods for simulating bosons,\cite{hirshberg2019path} with improved quadratic scaling,\cite{10.1063/5.0173749} and fermions, through a reweighting procedure.\cite{10.1063/5.0008720}

\paragraph*{Cavity MD (CavMD):} Simulation of systems confined in optical or plasmonic cavities.\cite{Li2020Water, Li2022RPMDCav}

\paragraph*{Nuclear propagation by additional dynamical variables:} non-adiabatic tunnelling rates in metallic systems,\cite{Litman_JCP_2022_I,Litman_JCP_2022_II} molecular dynamics with time-dependent external fields.

\paragraph*{Ensemble models:} committee models, baseline energy model and uncertainty estimation.\cite{Imbalzano_JCP_2021,kellner2024uncertainty}

\rev{\paragraph*{Extracting dynamical information from thermostatted simulations:} for vibrational spectra and diffusion coefficients from  PIGLET thermostatted simulations~\cite{kapil_inexpensive_2020}}

\paragraph*{Non-linear spectroscopy:} Evaluation of non-linear spectroscopic observables by equilibrium non-equilibrium simulations for classical and quantum nuclei.\cite{Hasegawa_JCP_2006,Begusic_JCP_2022,Begusic_NatComm_2023}

\rev{\paragraph*{T$_{\text{e}}$ path integral coarse-graining simulations (PIGS) simulations:} for approximate quantum dynamics using molecular dynamics on an MLIP representing the centroid potential of mean force~\cite{musil_quantum_2022, kapil_first-principles_2023}. }

\begin{figure}[tbp]
\centering
\includegraphics[width=1.0\linewidth]{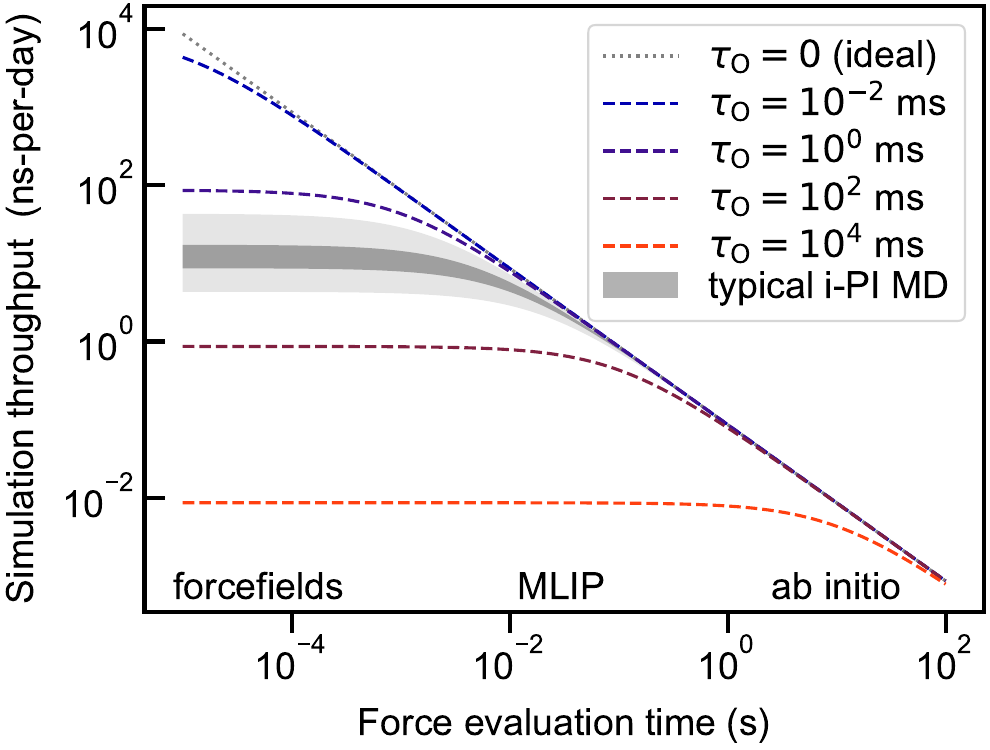}
\caption{Simplified picture of the impact of a given overhead $\tau_{\text{O}}$ on the simulation throughput (expressed in ns/day, assuming a time step of 1~fs) as a function of the force evaluation cost. \rev{The larger the force evaluation time, the lower the impact a given $\tau_{\text{O}}$ on the simulation throughput. }
An overhead of about 10 ms (typical for a moderately complicated simulation with \ipi{} 2.6) is negligible when used with \emph{ab initio} force evaluations, but becomes significant for the most efficient (or heavily parallelised) \rev{MLIPs}. 
}
    \label{fig:simple-overhead-scheme}
\end{figure}

\begin{figure*}[tbhp]
    \centering
     \includegraphics[width=\textwidth]{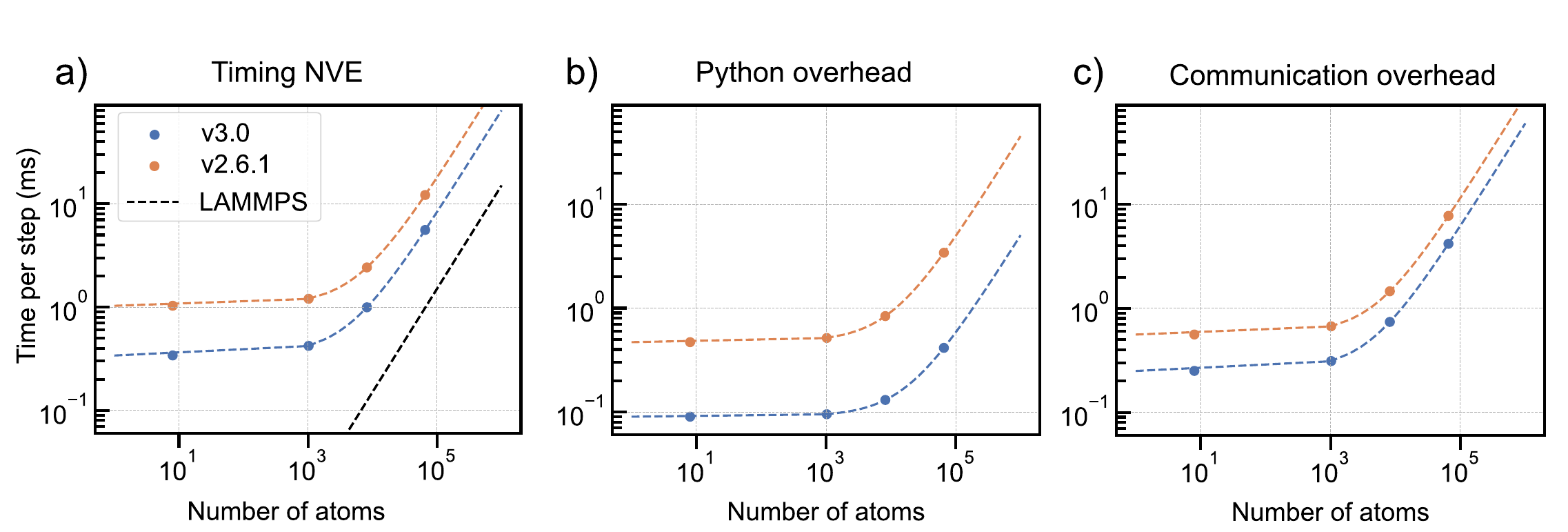}
\caption{
(a) System size dependence of the timing per step when performing a NVE simulation of an ideal gas using LAMMPS, and when using \ipi{} with a dummy driver that returns zero energy and forces, communicating through UNIX sockets. 
The \rev{cost of a simulations run with \ipi{} is} broken down into (b) a base Python overhead (obtained by performing the \ipi{} simulation with no forcefield) and the communication overhead (c), \rev{that is obtained by subtracting the Python overhead from the cost of a simulation that does communicate with the dummy driver.}
Markers correspond to actual simulations while the dashed lines are fits to $\tau_{\text{O}} = A+B N_\text{a}$, where $N_\text{a}$ is the number of atoms. \rev{All simulations were run on nodes with Intel Xeon E5-2690 v3 @ 2.60GHz processors using a single core.} 
} \label{fig:ipi_efficiency} 
\end{figure*}

\section{Efficiency of large-scale simulations\label{sec:Efficieny}}
\ipi's socket-based communication and Python architecture offer convenience and flexibility to implement new and advanced simulation algorithms, facilitating integration with other codes.     This flexibility, however, comes at the price of a computational overhead associated with UNIX- and TCP-socket communications, and Python's inherent inefficiency compared to lower-level languages such as Fortran, C, and C++. %

To analyse \ipi's overhead on the performance of the simulations, it is instructive to separate the intrinsic wall-clock time needed to evaluate energy and forces, $\tau_\text{F}$, and \rev{the time overhead associated only to}  \ipi, $\tau_\text{O}$, such that the length of simulation that can be obtained for a certain computational effort is proportional to $1/(\tau_\text{F}+\tau_{O})$. 
The initial focus of \ipi{} was on \textit{ab initio} force providers, where $\tau_\text{F}$ is of the order of \rev{several} seconds, and therefore an overhead of $\tau_\text{O}\sim10$~ms was inconsequential ( see Fig.~\ref{fig:simple-overhead-scheme}).
However, with MLIPs becoming the primary tool for large-scale, reactive calculations, an overhead of several ms can become the performance bottleneck. Therefore, we have \rev{improved the efficiency of} \ipi{} for this release. 

\begin{table*}[thp]
    \centering
\begin{tabular}{ccccccccc}
        \hline\hline
&    &  &      &  &  \multicolumn{2}{c}{$A$ (ms) }& \multicolumn{2}{c}{$B$ ($\mu$s/$N_{\text{a}}$) }
\\
 type & replicas & socket &  ensemble    & thermostat&\ipi 2.6 & \ipi 3.0& \ipi 2.6 & \ipi 3.0\\
    \hline
    MD & 1  & none   & NVE   & none     &  0.47 &   0.09 &  0.06 &   0.02 \\ %
    MD & 1  & UNIX   & NVE   & none     &  1.03 &   0.34 &  0.17 &   0.08 \\ %
\multicolumn{5}{c}{UNIX socket communication overhead} &
0.56  & 0.25   & 0.11 & 0.06 \\
    MD & 1  & TCP/IP & NVE   & none     & 21.54 &   1.26 &  0.64 &   0.10 \\ %
\multicolumn{5}{c}{TCP/IP socket communication overhead} &
21.07  & 1.20   & 0.56 & 0.08 \\
\hline
    PIMD        & 2  & none   & NPT   & pile\_l  &  1.39 &   0.47 &  0.94 &   0.60 \\ %
    PIMD        & 2  & UNIX   & NPT   & pile\_l  &  2.62 &   1.27 &  1.32 &   0.76 \\ %
    \multicolumn{5}{c}{UNIX socket communication overhead} & 1.23  & 0.80   & 0.38 & 0.16 \\
    PIMD        & 2  & TCP/IP & NPT   & pile\_l  & 32.64 &   2.45 &  1.68 &   0.77 \\
    \multicolumn{5}{c}{TCP/IP socket communication overhead} & 31.25  & 1.98   & 0.74 & 0.17 \\
        \hline\hline\end{tabular}
    \caption{Comparison between the timings for the tests in the \texttt{ ipi\_tests/profiling} directory with \rev{\ipi{}  version 3.0.0-beta} or 2.6.3. The coefficients $A$ and  $B$ can be used to estimate the overhead for a given number of atoms $N_\text{a}$, as $\tau_{\text{O}} = A+B N_\text{a}$. In the table, (PI)MD indicate simulations performed with (path integral) MD; ``socket'' indicates the type of communication, \emph{none} corresponding to simulations that avoid entirely force evaluations. ``pile\_l" stands for the path-integral Langevin thermostat.\cite{ceri+10jcp}. \rev{The purpose of this table is to make the communication overhead of MD and PIMD runs evident. The numbers are obtained by subtracting the UNIX or TCP/IP socket values from the respective line where socket is \emph{none}. }\rev{All the simulations were run on nodes with 2 Intel(R) Xeon(R) Platinum 8360Y processors using a single core.}} 
    \label{tab:test_profiling}
\end{table*}

\begin{table*}[thp]
    \centering
\begin{tabular}{ccccccccc}
        \hline\hline
&    &  &      &  &  \multicolumn{2}{c}{$A$ (ms) }& \multicolumn{2}{c}{$B$ ($\mu$s/$N_{\text{a}}$) }
\\
 type & replicas & socket &  ensemble    & thermostat&\ipi 2.6 & \ipi 3.0& \ipi 2.6 & \ipi 3.0\\
    \hline
    MD & 1  & none   & NVT   & svr      &  0.56 &   0.14 &  0.07 &   0.03 \\ %
    MD & 1  & none   & NVT   & langevin &  0.86 &   0.22 &  0.24 &   0.21 \\ %
    MD & 1  & none   & NPT   & langevin &  1.45 &   0.47 &  0.37 &   0.32 \\ %
    MD & 1  & none   & NPT   & svr      &  1.13 &   0.38 &  0.22 &   0.13 \\ %
    PIMD        & 32 & none   & NPT   & langevin &  2.73 &   0.67 & 17.21 &  12.97 \\ %
    REMD        & $4\times 6$  & none   & SC-NPT & langevin & 27.80 &  16.86 & 11.29 &  12.50 \\ %
        \hline\hline\end{tabular}
    \caption{Comparison between the timings for additional tests in the \texttt{ ipi\_tests/profiling} directory with \rev{\ipi{}  version 3.0.0-beta} or 2.6.3, presented in the same way as in Table~\ref{tab:test_profiling}. In the table, (PI)MD indicate simulations performed with (path integral) MD; Thermostats are chosen between stochastic velocity rescaling, \emph{svr} and \cite{bussi_canonical_2007} plain Langevin. The last line corresponds to a complicated setup combining path integral and replica-exchange molecular dynamics (REMD) with 4 beads and 6 ensemble replicas and a finite-difference evaluation of Suzuki-Chin (SC) high-order path integrals.\cite{suzu95pla,kapi+16jcp2} \rev{All the simulations were run avoiding force evaluations entirely (socket = \emph{none}), on nodes with 2 Intel(R) Xeon(R) Platinum 8360Y processors using a single core. The communication timings reported in Table~\ref{tab:test_profiling} should be added to the timings reported here for a realistic estimation, keeping into account, when relevant, the number of replicas.}} 
    \label{tab:test_profiling_2}
\end{table*}

\subsection{Estimating the overhead of \ipi{}}

\rev{The cost per MD step of \ipi{} simulations can be roughly modelled as a constant term plus a per-atom contribution. This overhead can be estimated as $\tau_\text{O}\approx A + B N_\text{a}$. Note that, for such a wide range of system sizes, it is not advisable to determine the coefficients by linear regression because the intercept would be affected by large relative errors. Therefore, we determined $A$ as the time per step of the smaller simulation (8 atoms) and $B$ as the time per step per atom of the largest one ($2^{16}$ atoms).}

\rev{We start by analysing the simplest possible case of a constant-energy simulation of an ideal gas, with no forces acting between atoms and therefore $\tau_\text{F}\approx0$. (Fig.~\ref{fig:ipi_efficiency}). In this case, the advantage of a C++ implementation such as that in LAMMPS is evident. 
For the largest system sizes, LAMMPS is about 10 times faster than \ipi{}, but for a few atoms the difference is much larger: LAMMPS can perform millions of steps per second, whereas \ipi{} 2.6 saturates at about 1000 steps per second (Fig.~\ref{fig:ipi_efficiency}a).}

It is also instructive to partition $\tau_\text{O}$ into a contribution associated with the Python overhead and the one associated with the socket communication. 
We estimated the former by disabling force evaluation altogether, which can be achieved by setting \lstinline{weight="0"} as an attribute of the \lstinline{<force>} components. \rev{We estimated the latter by subtracting the Python overhead from the overall cost of a simulation communicating with a dummy driver that simply returned zero energy and forces (Fig.~\ref{fig:ipi_efficiency}b-c)}.

Fig.~\ref{fig:ipi_efficiency} also compares the \rev{computational efficiency of the} stable 2.6 version with \rev{that of the} current 3.0 release. In the latter, we performed several optimizations to reduce the overhead associated with accessing the \ipi{} \emph{depend} objects,\cite{IPIv1} substituted explicit loops with array operations, and removed other obvious bottlenecks from the implementation. This reduced the overhead of the NVE integrator by a factor of 3-5, and the communication overhead by a factor of 2.

\rev{In Table~\ref{tab:test_profiling}, we report the values of the $A$ and $B$ coefficients for different simulation setups enabling the calculation of the communication overhead, while in Table~\ref{tab:test_profiling_2}, we report the values of the $A$ and $B$ associated with the Python overhead for different nuclear motion algorithms. Thus, the total overall \ipi overhead, $\tau_\text{O}$,  can be obtained by adding the TCP/IP or UNIX socket timings from Table~\ref{tab:test_profiling} to the algorithm-specific times from Table~\ref{tab:test_profiling_2}}. In all cases, the new release dramatically reduces both constants so that for most classical MD simulations and up to a few 1000s of atoms, one can now expect an overhead of a few ms per step. \rev{It is thus possible to perform tens of millions of simulation steps per day when disregarding the force evaluation costs}.

The \rev{Python} overhead is larger when considering simulations with multiple replicas and/or complicated setups that involve, e.g. high-order path integration or replica exchange\rev{, as evidenced in Table~\ref{tab:test_profiling_2}}. Often, however, these advanced simulation techniques \rev{decrease the} overall cost of a calculation by reducing the number of force evaluations \rev{necessary} to reach the desired \rev{statistical} accuracy. \rev{This} is a key advantage of using \ipi{} over faster, but less feature-rich, alternatives. 
An example, also relevant to classical MD trajectories, is the use of BAOAB integrators,\cite{leim+13jcp} which allow doubling the time step in simulations of liquid water. 
A technical detail we also want to highlight from Table~\ref{tab:test_profiling} is the $>20\times$ reduction of the constant overhead for TCP/IP communication. This \rev{was achieved by} disabling Nagle's algorithm,\cite{Nagle1984} which was found to interfere negatively with the transfer of small messages in the \ipi{} communication protocol. This change has to be enacted also on the client side, \rev{as currently done in the driver codes we distribute with \ipi{}}, so we urge developers of packages that contain an \ipi{} interface to apply similar changes.

\begin{figure}
\centering
\includegraphics[width=0.8\columnwidth]{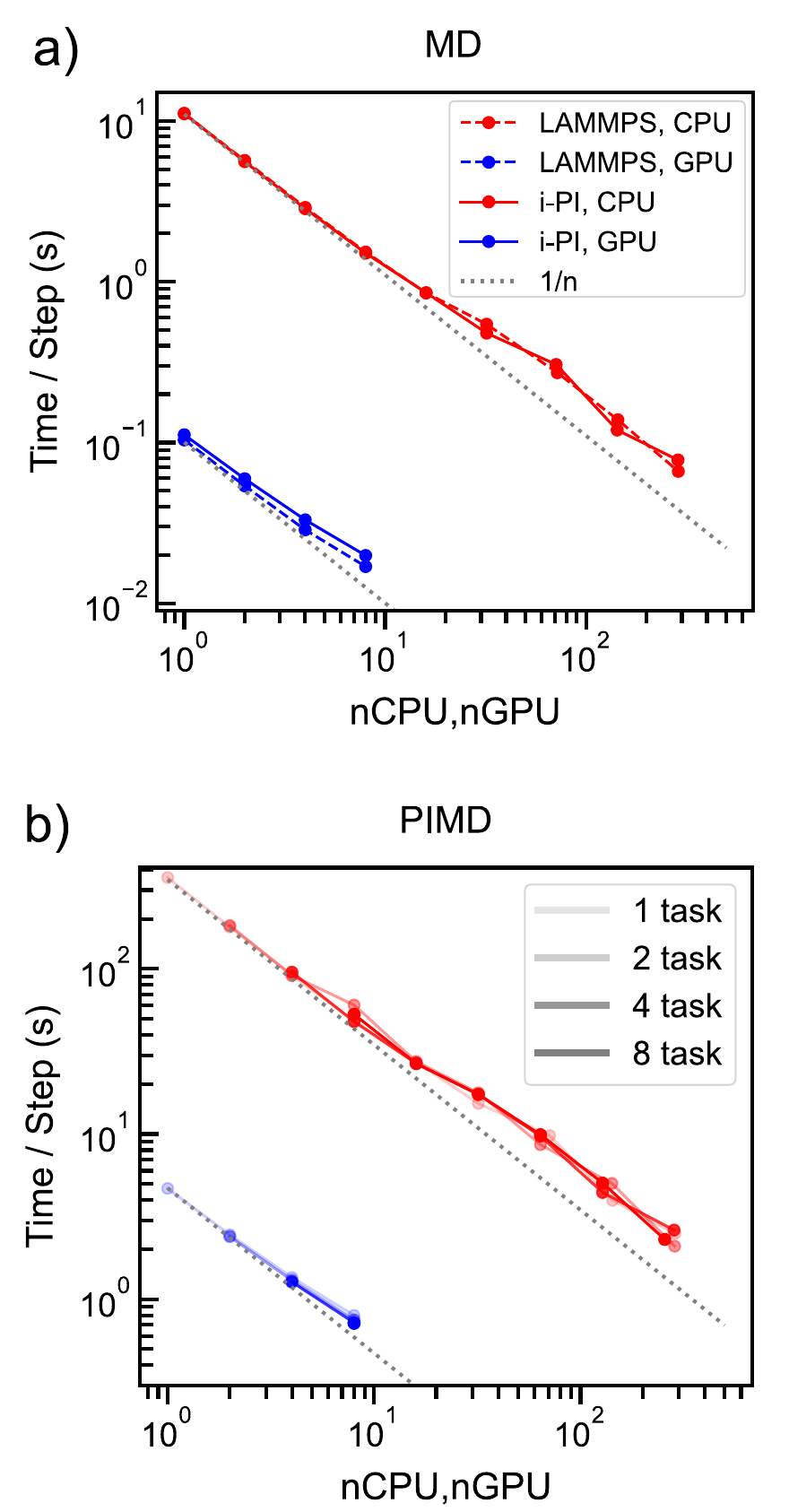}
\caption{Timings for 1 step of NVT dynamics with Langevin thermostatting, for a box containing 4096 water molecules at ambient temperature and density, using a DeePMD model,\cite{deepmd} using the LAMMPS implementation.\cite{DeePMD-kitv1,DeePMD-kitv2} 
Simulations were run on nodes with 2 Intel(R) Xeon(R) Platinum 8360Y, with 36 cores each (CPU) and nodes with 2 NVIDIA V100 cards (GPU).
Parallelization is based on MPI, leveraging the LAMMPS domain decomposition.
\rev{Dotted lines are guides for the eye corresponding to an ideal $1/n$ scaling.}
(a) Timings for MD simulations. Dashed and full lines correspond to simulations performed with LAMMPS, and i-PI using LAMMPS/DeePMD as a driver, respectively.
(b) Timings for PIMD simulations with 32 replicas. 
The shading of the curves indicates the number of independent LAMMPS tasks that are connected to i-PI, and the number of processing units indicates the sum across all tasks.}
    \label{fig:ipi-deepmd}
\end{figure}

\subsection{Efficient performance of \ipi{} on realistic parallelised simulations}

Even though this release significantly cuts the \ipi{} overhead $\tau_O$, and we are committed to further reducing it in the future, it is important to determine whether and when it matters in practical use cases. 
To this end, we consider the task of simulating 4096 water molecules at room temperature, in the canonical \emph{NVT} ensemble, using a stochastic velocity rescaling thermostat.
\rev{Based on Table~\ref{tab:test_profiling}, we expect that using \ipi{} 3.0 should entail an overhead of 1.3~ms when using a UNIX socket, and 2.5~ms when using a TCP/IP socket, as necessary for simulations running across many nodes.
}

We start by considering the LAMMPS implementation of a DeePMD neural network, that is used both to perform MD directly using its LAMMPS implementation,\cite{DeePMD-kitv1,DeePMD-kitv2,Lu_CPC_2021} and as a driver for running classical MD and path integral molecular dynamics with \ipi{}. 
On the hardware we use for this test, 
\rev{detailed in the caption of Fig.~\ref{fig:ipi-deepmd}},
\rev{the evaluation of energies and forces on one CPU requires approximately 10 s/step}, which is several orders of magnitude \rev{larger} than the estimated \rev{$\tau_O$}.

In practical scenarios, however, the MLIP calculation will be heavily parallelized, either using a multi-CPU implementation or using one or more GPUs. As shown in Fig.~\ref{fig:ipi-deepmd},  \ipi{} and LAMMPS having essentially equivalent performance up to the highest parallelization level for CPUs (up to 4 nodes, 288 cores), consistent with the small predicted $\tau_O$.
The GPU implementation of DeePMD is much faster and can be parallelized efficiently up to about 8 GPUS (4 nodes), for which (with a time step of 0.5~fs) LAMMPS clocks almost 2.5 ns of simulation per day. 
\rev{In our initial benchmarks we observed the empirical $\tau_O$ with LAMMPS DeePMD to be up to 10 times larger than when using the dummy driver, which we could trace to unnecessary reconstructions of the neighbor list at the LAMMPS side of the interface. 
We could optimize the LAMMPS interface of the driver, minimizing the number of neighbor list updates, and reducing the observed overhead to about 3~ms for the 8-GPU run.
This is still almost 2 times larger than predicted, because of the additional bookkeeping that is needed to interface \ipi{}  with a feature-rich code such as LAMMPS, in comparison to the minimalistic dummy driver. 
Nevertheless, these optimizations have reduced $\tau_O$ to less than 20\%{} of the overall simulation step even for the highly efficient multi-GPU setup, allowing to perform more than 2~ns of simulation per day with a relatively large number of atoms. }
For a PIMD setup, \ipi{} allows exploiting an additional level of parallelism by running multiple clients in parallel. Given that this is a trivial type of parallelization, it can improve the scaling whenever the client code shows less-than-ideal scaling, as observed for the GPU implementation of DeePMD \rev{in Fig.~\ref{fig:ipi-deepmd}b, even though the speedup is minimal in this case. }

\begin{table}[htbp]
    \centering
    \begin{tabular}{c c c c c| c c }
    &  &  &   &    &  \multicolumn{2}{c}{throughput  } \\ 
    &  &  &   &    &  \multicolumn{2}{c}{(ns/day)  } \\ 
    \hline
      model & architecture & $n_\text{CPU}$/$n_\text{GPU}$ & $n_{\text{H}_2\text{O}}$ &  $\tau
_\text{F}$/s & direct   &   w/\ipi{}   \\
    \hline
  DeePMD  &  CPU (Xeon)  & 288  & 4096  & 0.066       & 0.65      & \rev{0.56} \\    
  DeePMD  &  GPU (V100)  & 8    & 4096  & 0.017       & 2.54      & \rev{2.16} \\
  MACE    &  GPU (A100)  & 1    & 4096  & 1.47        & 0.03  & 0.03 \\  
  MACE    &  GPU (A100)  & 1    & 64    & 0.047       & 0.94      & 0.92 \\  
  BPNN    &  CPU (Xeon)  & 768  & 4096  & 0.015       & \rev{2.79} & \rev{2.10} \\  
\hline\hline
    \end{tabular}
    \caption{An overview of the expected throughput for $NVT$ MD simulations of liquid water \rev{(with a time step of 0.5~fs)} for different system sizes, hardware, \rev{number of processing units ($n_\text{CPU}$/$n_\text{GPU}$)} and MLIPs. DeePMD simulation timings correspond to the highest level of parallelization in Fig.~\ref{fig:ipi-deepmd}\rev{. The Simulations were run on nodes with 2 Intel(R) Xeon(R) Platinum 8360Y, with 36 cores each (CPU) and nodes with 2 NVIDIA V100 cards (GPU)}; MACE simulations use the ASE driver with the MACE-MP-0 model(S),\cite{batatia2024foundation} single precision and were run on one A100 GPU card.
Behler-Parrinello neural network (BPNN) simulations use the model from Ref.~\cite{ravindra_quasi-one-dimensional_2023}, and were run using the n2p2 interface in LAMMPS, on a Intel Xeon E5-2690 v3 @ 2.60GHz.
 }
    \label{tab:practical-examples}
\end{table}

\subsection{Near-ideal performance of \ipi{} with popular MLIPs}
These experiments show that, in practical use cases, the overhead introduced by \ipi{} is negligible or small up to high levels of parallelization of the MLIP code -- although one should reassess these conclusions depending on the hardware, the specific MLIP implementation, and the type of simulation performed. 
As shown in Fig.~\ref{fig:simple-overhead-scheme}, a practical recipe to determine what throughput one can expect out of the combination of \ipi{} with your favourite \rev{machine learning} code is to determine the overhead $\tau_\text{O}$, either from Table~\ref{tab:test_profiling} or by running a short simulation with a dummy driver.
A dummy calculation is also suitable to tune the simulation parameters (and potentially to help remove bottlenecks from other simulation pathways\rev{-- some examples of profiling runs can be found in the \textit{test} directory}). %
The cost of evaluating energy and forces, $\tau_\text{F}$, also depends on the type of MLIP, system size and the hardware of choice. Thus, by comparing $\tau_\text{O}$ and $\tau_\text{F}$, it is possible to determine whether \rev{$\tau_O$} is significant.

As shown in Table~\ref{tab:practical-examples}, for both small and medium-sized supercells, and for DeePMD, n2p2 and MACE models, a standard constant-temperature MD simulation is limited by $\tau_\text{F}$. 
\rev{Even though, as shown by the case of LAMMPS, the overhead of a specific client might be larger than that of the minimalistic driver distributed with \ipi{}, this type of analysis helps determine whether \ipi{} is a suitable choice for the modeling task, or whether one should look for (or implement) a more performant, compiled-language implementation.} 
In summary, using \ipi{} as a driver does not prevent reaching a few million energy evaluations per day, corresponding to several ns for typical MD time steps. \rev{In addition, it} allows access to advanced simulation methods, such as path integral molecular dynamics, as well as facile interfacing with new MLIPs implementations, \rev{which might make \ipi{} the preferred choice even in cases in which it entails a small performance penalty}.

\section{Description of Selected Features}

\subsection{Bosonic and Fermionic Path Integral Molecular Dynamics}

Exchange effects, arising from bosonic and fermionic statistics, are important in simulations of phenomena such as superfluidity,\cite{cepe95rmp} supersolidity,\cite{PhysRevLett.128.045301} and superconductivity.
However, ordinary PIMD simulations assume that the atoms are distinguishable particles and obey Boltzmann statistics.
The biggest challenge in going beyond this assumption and performing PIMD simulations of indistinguishable particles is including all the possible identical particle permutations in the partition function. These are taken into account by connecting rings of permuted particles to form larger polymers.\cite{chan-woly81jcp}
Since there are $N!$ permutations of $N$ particles, the simulation, in principle, should sum over the spring forces from $N!$ configurations, which quickly becomes intractable. 

\begin{figure}[tbp]
    \centering
    \includegraphics[width=0.9\columnwidth]{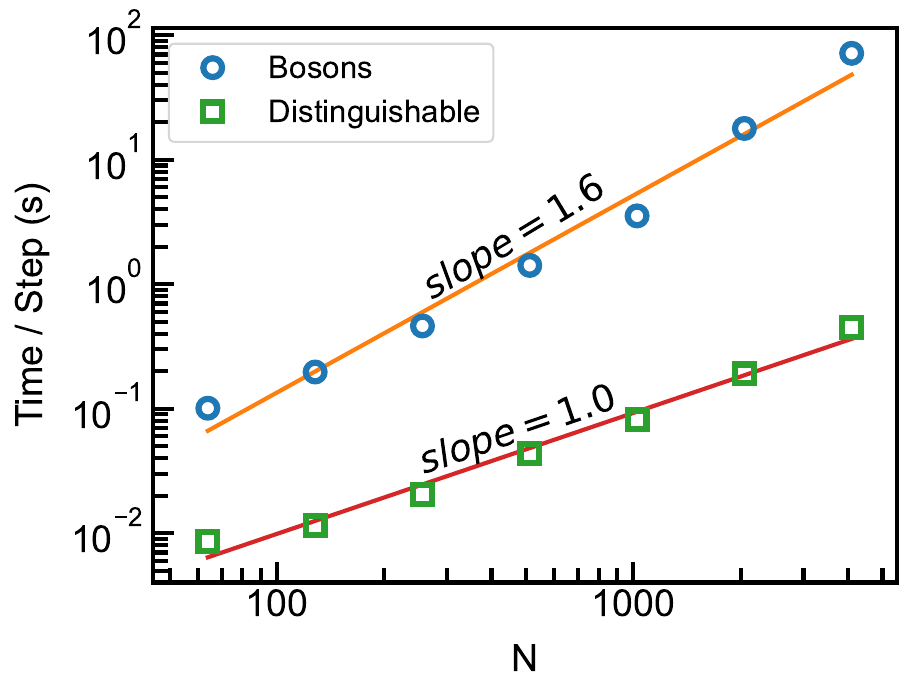}
        \caption{
        Scaling with the number of particles of the propagation of the spring forces in Cartesian coordinates, for distinguishable and bosonic \rev{trapped, non-interacting} systems.
        Both simulations were run with the \texttt{bab} propagator with \texttt{nmts=10}.
        \rev{Simulations were run on a node with two AMD EPYC 7763 Processors with 64 cores each, blocking the entire node for every run and using a single i-Pi force-field client.}
        }
    \label{fig:exchange_scaling}
\end{figure}

Version 3.0 of \ipi~ supports PIMD simulations that include exchange effects, using an algorithm introduced by Hirshberg \textit{et al.}\cite{hirshberg2019path, 10.1063/5.0173749} In this approach, PIMD simulations of bosons scale quadratically,\cite{10.1063/5.0173749} rather than exponentially, with $N$. 
This is achieved by replacing the standard ring polymer potential
with $V_B^{(N)}$, which is defined by the recurrence relation,
\begin{equation}\label{eq:recurrence}
    e^{-\beta V_B^{(N)}} = \frac{1}{N} \sum_{k=1}^{N}{e^{-\beta \left(V_B^{(N-k)} + E_N^{(k)}\right)}}.
\end{equation}
Here, $E_N^{(k)}$ is the spring potential of a ring polymer connecting the particles $N-k+1,\ldots,N$ in a cycle, and $V_B^{(0)} = 0$.
This potential \rev{and the forces it induces} are computed in quadratic time with $N$ and in linear time with the number of beads.\cite{10.1063/5.0173749}

The \ipi implementation allows the efficient simulation of systems composed of $\mathord{\sim} 1000$ particles. A comparison of the scalability of bosonic and standard PIMD simulations in \ipi is presented in Fig.~\ref{fig:exchange_scaling}, using the same integrator (\lstinline{bab}, see below) in both cases\rev{, for a system of trapped non-interacting particles}. The bosonic simulation has quadratic scaling while the distinguishable one scales linearly. For $N=64$ particles, the cost of bosonic exchange effects is approximately $12$ times that of standard PIMD, and for $N=1024$ it is only $44$ times slower with the same integrator.

\begin{figure}[tbp]
    \centering
    \includegraphics[width=0.9\columnwidth]{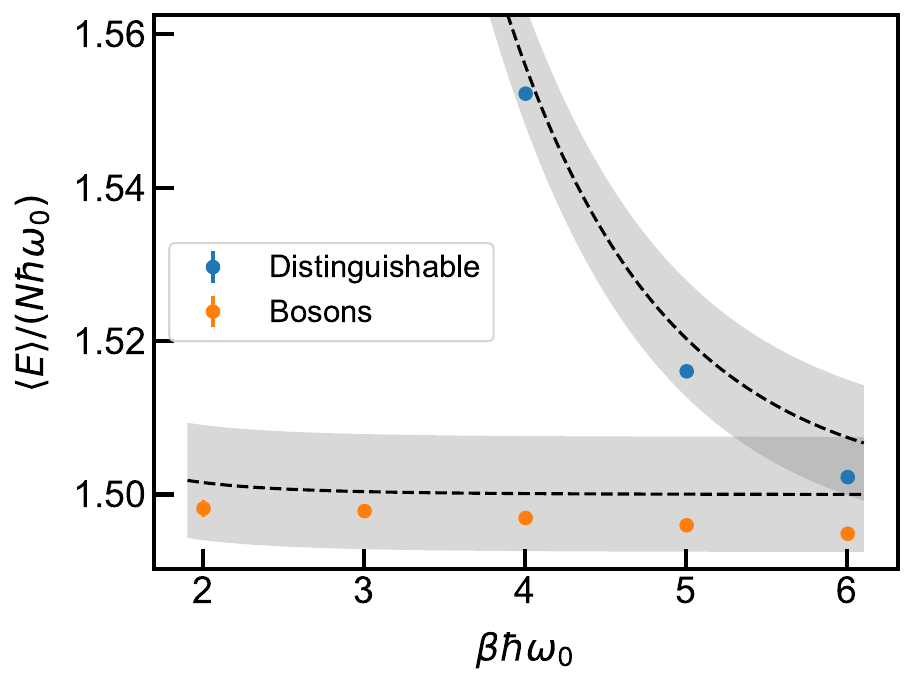}
        \caption{
        The energy per particle of $N=512$ trapped non-interacting bosons as a function of temperature (indicated by the dimensionless number $\beta \hbar \omega_0$, where $\omega_0$ is the harmonic trap constant), with $P=42$ beads. The \rev{dashed lines indicate the analytical results and the} shaded areas indicate an error region of $\pm 0.5\%$.
        }
    \label{fig:exchange_noninteracting_energy}
\end{figure}

\rev{Performing} bosonic PIMD in \ipi is straightforward and involves adding the following section to the xml input file:
\begin{minipage}{\linewidth}
\begin{lstlisting}
<normal_modes propagator='bab'>
    <bosons>...</bosons>
    <nmts>M</nmts>
</normal_modes>
\end{lstlisting}
\end{minipage}
The bosonic atoms are listed %
using the \texttt{bosons} tag. %
This can be done as a list of atoms (zero-based), such as \lstinline{<bosons id='index'> [0, 1, 2] </bosons>}, 
or by a list of labels, such as \lstinline{<bosons id='label'> ['He'] </bosons>}.
At the moment, the integration should be done in Cartesian coordinates (\texttt{normal\_modes propagator='bab'}), because an analytic expression for the normal-modes propagator of the free ring polymer part of the Hamiltonian is not known for bosons.
Because the frequencies of the springs are typically higher than the physical frequencies, the use of multiple-time stepping in the Cartesian integrator is recommended.~\cite{Tuckerman1993} %
\rev{This is done by setting the \texttt{nmts} tag (standing for the number of iterations for each multiple-time stepping level) to \texttt{M}, where \texttt{M} is the ratio between the time step used for the physical forces and that used for the spring term.\cite{IPIV2}} 
In Fig.~\ref{fig:exchange_noninteracting_energy}, we present the average energy of 512 trapped, non-interacting cold bosons. By comparing with distinguishable particle simulations, the relevance of exchange effects becomes evident.

\begin{figure}[tbph]
    \centering
    \includegraphics[width=0.9\columnwidth]{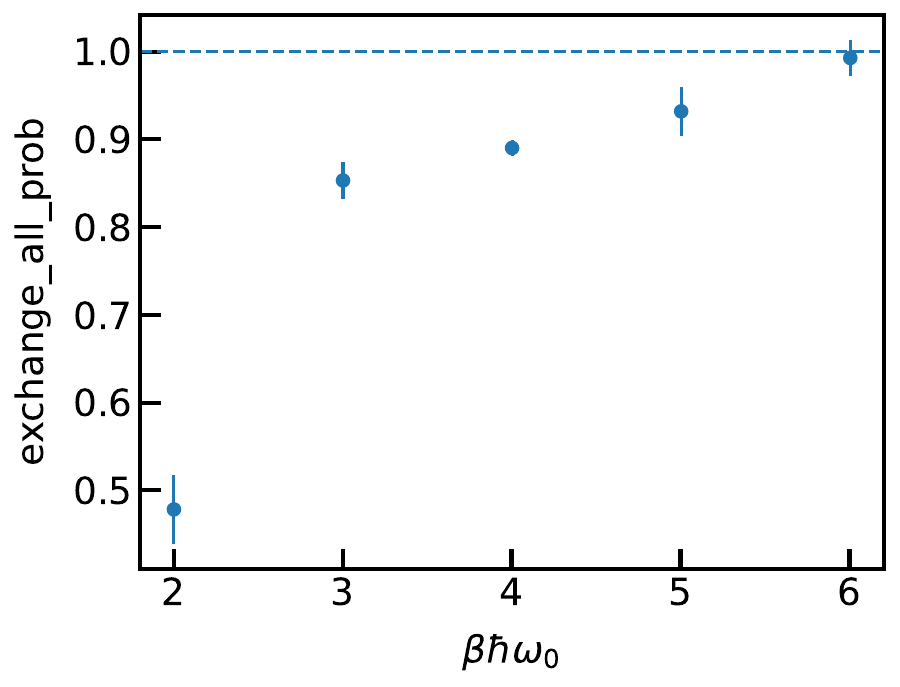}
        \caption{
        The probability of the ring polymer configuration that connects all $N=16$ non-interacting trapped particles at different temperatures (indicated by the dimensionless number $\beta \hbar \omega_0$, where $\omega_0$ is the harmonic trap constant).
        }
    \label{fig:exchange_all_prob}
\end{figure}
A way to assess the extent of exchange effects in bosonic simulations is to analyse the following probabilities, which can be accessed through the properties tag in \ipi:
\rev{(i)} \verb|exchange_distinct_prob|, which is the probability of the ring-polymer configuration corresponding to distinguishable particles: 
\begin{equation}
    \frac{1}{N! \cdot e^{-\beta V_B^{(N)}}}e^{-\beta\left(\sum_{i=1}^{N}{E_i^{(1)}}\right)},
\end{equation}
where $E_i^{(1)}$ is the spring potential of the ring polymer connecting the beads of particle $i$ in a separate ring.
This quantity tends to $1$ in the regimes where the exchange is negligible. 
\rev{(ii)} \verb|exchange_all_prob|, which is the probability of the ring-polymer configuration connecting \emph{all} the bosons, scaled by $1/N$:
\begin{equation}
    \frac{1}{e^{-\beta V_B^{(N)}}}e^{-\beta \cdot E_N^{(N)}},
\end{equation}
where $E_N^{(N)}$ is the spring potential of the ring polymer connecting all the particles $1 \rightarrow 2 \ldots \rightarrow \ldots \rightarrow N \rightarrow 1$ in one large ring.
This quantity is high in regimes of effective exchange between all the particles. In Fig.~\ref{fig:exchange_all_prob}, we show the average of \verb|exchange_all_prob| in converged simulations of trapped cold noninteracting bosons as a function of temperature. This quantity tends to $1$ in the limit $T \to 0$, $\beta \to \infty$, in which all permutations are equally probable.~\cite{hirshberg2019path}

\begin{figure}[btp]
    \centering
        \centering
        \includegraphics[width=.85\linewidth]{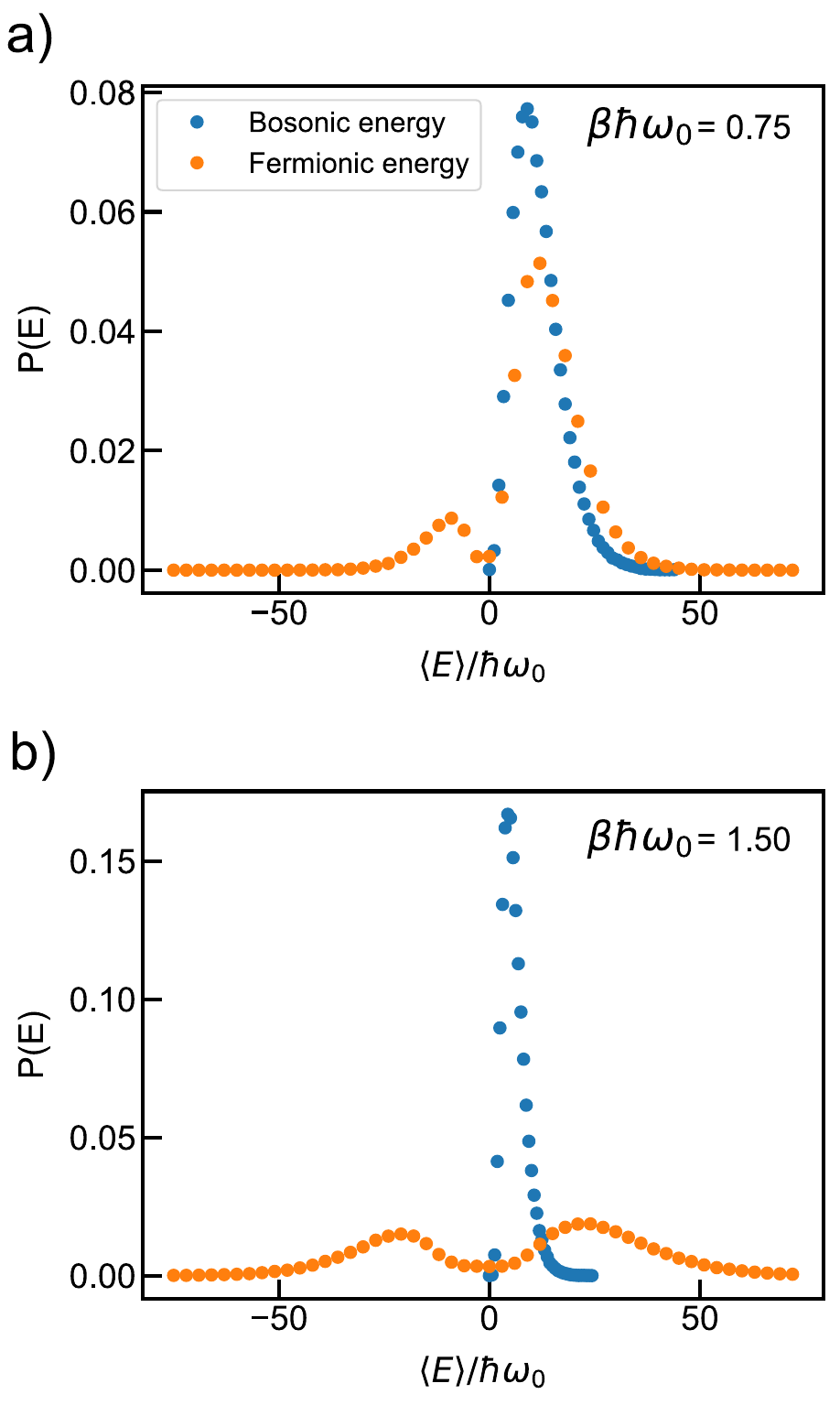}
        \caption{
        The distribution of bosonic and fermionic energy estimator, $P(E)$, for $N=3$ trapped non-interacting particles at two representative temperatures indicated by the dimensionless number $\beta \hbar \omega_0$, where $\omega_0$ is the harmonic trap constant\rev{: (a) high temperature exhibiting mild cancellations between the two peaks of the bimodal distribution of the fermionic energy estimator, and (b) low temperature where the cancellation is near-complete and the sign problem is more severe.}
        } \label{fig:exchange_fermion_energy_hist}
\end{figure}

Several estimators require modifications when performing bosonic PIMD simulations. For example, the primitive estimator for the kinetic energy (\texttt{kinetic\_td}) is modified according to Eq.~\ref{eq:recurrence} (as shown in Eq.~2 in the SI of ~\citet{hirshberg2019path}) while the standard virial kinetic energy estimator can be used for bosons in confined systems. On the other hand, the centroid-virial kinetic estimator is not applicable due to the lack of an appropriate centroid definition.

Lastly, fermionic statistics can be obtained by first performing bosonic simulations and then reweighting the results.\cite{10.1063/5.0008720,10.1063/5.0030760,10.1063/5.0171930} This is done by using the fermionic sign estimator (the \texttt{fermionic\_sign} property of \ipi). For every position-dependent operator, the corresponding fermionic estimator is obtained from the bosonic estimator by multiplying with the fermionic sign, and dividing by the average fermionic sign in the simulation (see Eq.~(9) of Ref.~\cite{10.1063/5.0008720}). The average sign is always positive but for simulations at low temperatures or with large $N$, where the sign problem is severe, it gets close to zero. Fig.~\ref{fig:exchange_fermion_energy_hist} illustrates the behaviour of the fermionic energy estimator for $N=3$ non-interacting, trapped fermions at several temperatures. While the distribution of the bosonic energy estimator throughout the simulation is always positive, the fermionic energy estimator changes sign, its distribution is bi-modal around $E=0$, and more symmetric at lower temperatures. The near-complete cancellations of positive and negative contributions make the mean fermionic energy harder to converge in comparison to the bosonic case.

\subsection{Nuclear propagation by additional dynamical variables \label{sec:jason}}

\ipi 3.0 can employ numerical quantities provided by the driver (beyond the usual forces and stress arrays) to run advanced types of dynamics. This supplementary communication is achieved through an additional string that can be passed in the \texttt{extras} field of \ipi within the same socket communication. When such string is JSON-formatted \rev{it is automatically parsed, and the additional quantities returned by the driver are made} available in the code and accessible as a dictionary. More importantly, when the data is numeric, it is added as a property of the force object, guaranteeing a seamless integration with other \ipi features, and becoming available to be used in any type of algorithm.
In the next section, we present two examples of advanced simulations that employ electronic structure quantities as \enquote{additional} variables to modify the atomic forces and therefore alter the nuclear dynamics.

\subsubsection{Non-adiabatic tunnelling rates in metallic systems}

The vanishing band gap in metals allows nuclear movements to induce electronic excitations, even at relatively low velocities.\cite{Wodtke_ChemSocRev_2016,Kavokine2022} These effects are not captured by the Born-Oppenheimer approximation and, therefore, are collectively referred to as non-adiabatic effects (NAEs).
NAEs in metals are qualitatively different from the ones found in molecules due to the continuous manifold of available empty electronic states present in the former, and thus, they require a different set of theoretical approaches. 
A frequently used technique to simulate these types of NAEs is to decorate the otherwise Born-Oppenheimer dynamics with an electronic frictional kernel and a stochastic force.\cite{HeadGordon_JCP_1995,Dou_PRL_2017,Dou_JCP_2018} The electronic friction formalism has been derived from first principles starting from the exact factorization of the wave function, initially at 0 K,\cite{Martinazzo_PRA_2022,Martinazzo_PRL_2022} and more recently at finite temperatures.\cite{martinazzo2023quantum} Moreover, in the case of a theory of non-interacting electrons, such as DFT, the friction kernel can be computed explicitly from \textit{ab initio} calculations.\cite{Maurer_PRB_2016}

When studying the kinetics of light atom diffusing in metals,
\rev{nuclear quantum effects (NQEs)} can become as important as NAEs. To address this complex problem some of the authors have recently derived a theoretical framework that can capture both types of effects.\cite{Litman_JCP_2022_I} The method called 
ring polymer instanton with explicit friction (RPI-EF) was derived 
by combining the semiclassical instanton rate theory in its path integral discretized form,\cite{Richardson_Review_2018} with the \textit{ab initio} electronic friction formalism. The framework is general enough to incorporate the spatial and frequency dependence of the friction tensor and, owing to the computational efficiency of the parent methods, \cite{Litman_PRL_2020,Box_IOP_2023} it is suitable for first-principles calculations of reaction rates in complex, high-dimensional systems.\cite{Litman_JCP_2022_II}

The RPI-EF method is algorithmically similar to classical Eyring transition rate theory and requires the search of a first-order saddle point. In this case, however, the saddle point is not the reaction transition state but rather an imaginary time trajectory that makes the Euclidean action stationary. This special trajectory is known as instanton and can be interpreted as the main tunnelling pathway.\cite{Richardson_JCP_2009} In the limiting case of a position-independent friction kernel (see a more general expression in\rev{cluding the position dependence of the friction kernel in} Ref. \cite{Litman_JCP_2022_I}) the discretized Euclidean action, $S_P$ can be expressed as
\begin{equation}
\begin{split}
S_P = S_P^\text{sys} + \hbar \frac{\beta}{P}
\sum_{l=-P/2+1}^{P/2}  \sum_{i=1}^{3N}  \frac{\tilde{\eta}(\omega_l){\omega_l}}{2}(Q_i^{(l)})^2,
\end{split}
\end{equation}
\noindent where $P$ is the number of discretization points or beads, $N$ is the number of atoms, $\beta$ is the inverse temperature, $\bm{Q}^{(l)}$ represents the free ring-polymer normal mode coordinates, $\omega_l$ are the free \rev{ring-polymer (RP)} normal mode frequencies, and $\tilde{\eta}$ is the Laplace transform of the (electronic) friction kernel. 
\rev{$S_P$ is related to the potential energy of the ring polymer $U_P$ by $S_P/\hbar = \beta_PU_P$.}
The RPI-EF calculation consists of a series of geometry optimizations \rev{on $U_P$} using progressively more discretization points until the converged instanton pathway is achieved.
A detailed description of the individual steps involved in RPI and RPI-EF simulations, as well as practical guidelines, can be found elsewhere.\cite{Beyer_PCL_2016,Litman_thesis,Litman_JCP_2022_II}

\begin{figure}[btp]
  \centering
  \includegraphics[width=.95\linewidth]{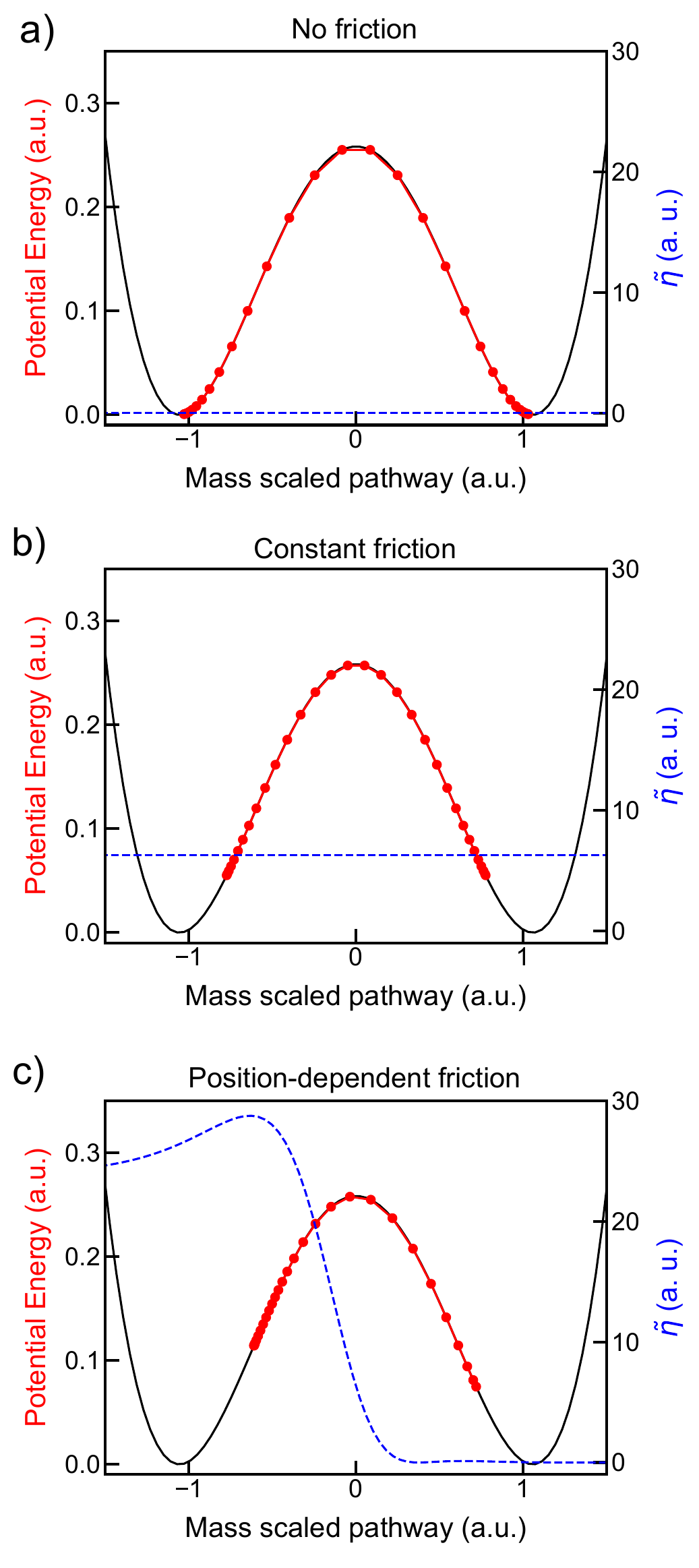}
          \caption{ Instanton pathways obtained in double-well  potential with a) no friction, b) constant friction and c) position-dependent friction. The instanton pathways, potential energy surface and friction profiles, $\tilde{\eta}$, are depicted as red dotted lines, black solid lines and blue dashed lines, respectively. The analytical expression of the potential and friction profiles can be found in Ref. \cite{Litman_JCP_2022_II}.}
    \label{fig:InstantonFriction}
\end{figure}

The \ipi~ implementation of RPI-EF builds upon the RPI implementation from the previous \ipi~ release. The primary distinction regards the additional terms that contain the friction kernel. In the case of position-dependent friction, these terms require updates at each step of the geometry optimization, which has been made possible by the automatic parsing of JSON-formatted data returned by the client. 
The friction terms are enabled by setting the friction flag inside the instanton block, \verb|frictionSD|. Since the theory assumes that the frequency dependence of the friction kernel is \rev{the same for all} positions, the user needs to provide the normalized spectral density, $\tilde{\eta} (\omega)$, that can be specified with the flag \verb|fric_spec_dens| and the energy at which it has been normalized to one by using the flag \verb|fric_spec_dens_ener|. A minimal example of the section \textbf{motion} to run RPI-EF simulations with \ipi~ is

\begin{minipage}{\linewidth}
\begin{lstlisting}
<motion mode='instanton'>
  <instanton mode='rate'>
    <friction>true</friction>
    <fric_spec_dens mode='file'> 
         fric_spec_dens.dat 
    </fric_spec_dens>
    <fric_spec_dens_ener> 0.1 
    </<fric_spec_dens_ener>
    <hessian_update>powell</hessian_update>
    </instanton>
</motion>
\end{lstlisting}
\end{minipage}

In Fig. \ref{fig:InstantonFriction}, we present the instanton geometry obtained for a \rev{1D} double well potential without \rev{including any} friction, \rev{including a} constant friction \rev{profile}, and \rev{including an analytical} position-dependent friction \rev{profile}. The inclusion of constant friction causes the particle to behave more classically, resulting in the shrinkage of the instanton geometry when compared to the calculation without friction. The \rev{result with} position-dependent friction shows that the instanton geometry avoids regions with large friction. \rev{As discussed in Refs.~\cite{Litman_JCP_2022_II,bridge2024}, this indicates} that non-adiabatic effects can steer the main tunnelling pathway and, consequently, steer quantum dynamics.

\subsubsection{Dynamics with time-dependent external field}

Recent experiments have shown that the excitation of infrared-active phonons through THz laser pulses leads to a vast phenomenology in different kinds of materials.\cite{SALEN20191}
These phenomena are non-thermally driven by microscopic processes 
rather than originating from ultrafast heating effects.\cite{RevModPhys.93.041002}
The possibility for driving materials, ranging from traditional solids to liquids and interfaces, across phase transitions and into metastable states that are not accessible at thermal equilibrium, has opened the possibility of dynamical materials design.\cite{Disa2021,PhysRevLett.118.054101}

\begin{figure}[htb]
    \centering
     \includegraphics[width=0.95\linewidth]{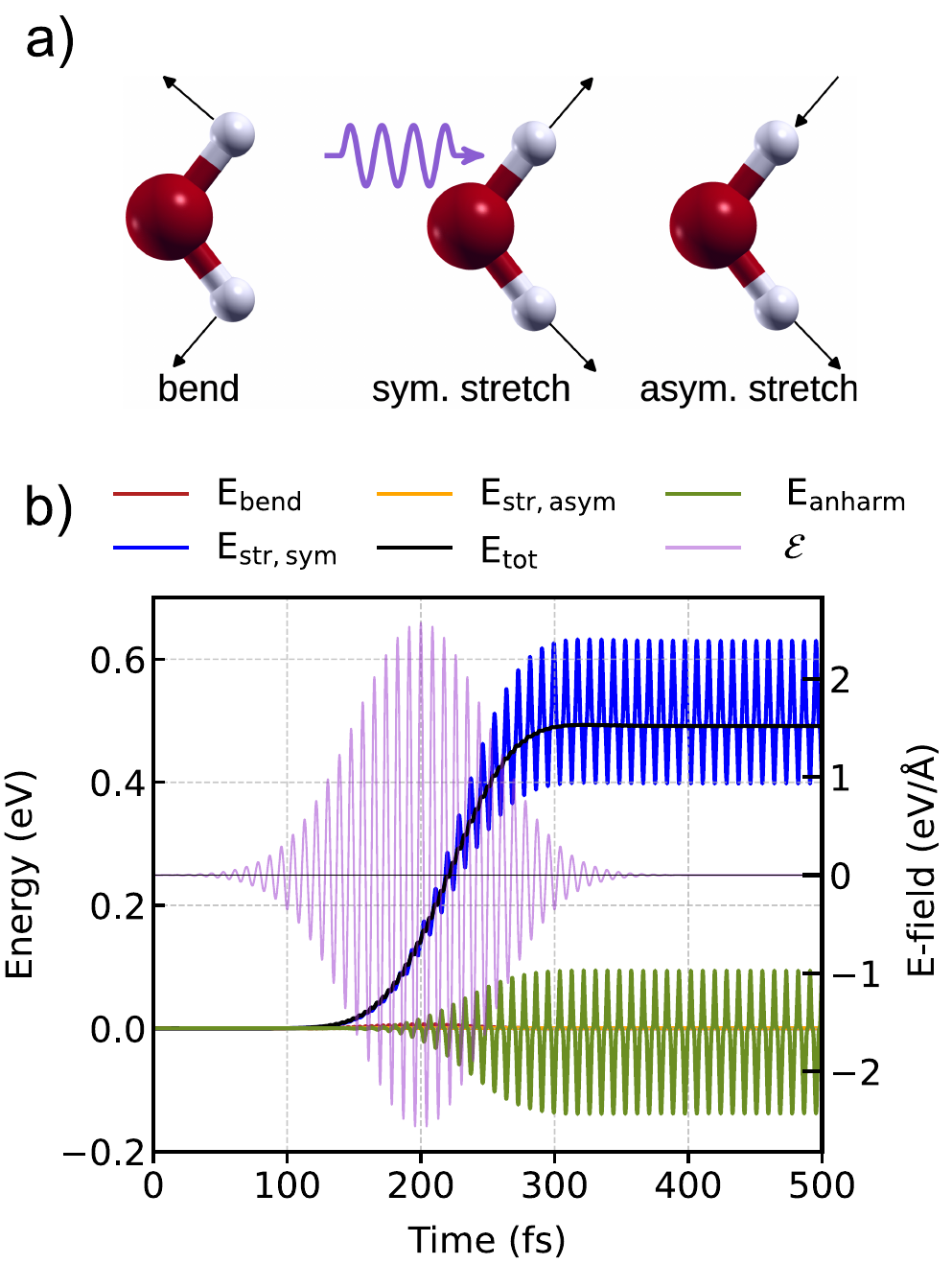} 
    \caption{(a) Vibrational modes of a water monomer: bending $\rm E_{\rm bend}$, symmetric-stretch $\rm E_{\rm str,~sym}$, and asymmetric-stretch $E_{\rm str,~asym}$. (b) Contributions to the total energy along a driven trajectory.  Time series of the harmonic contribution to the energy of different vibrational modes, the total energy $\rm E_{\rm tot}$, and the anharmonic contribution $\rm E_{\rm anharm}$ of a water monomer along a driven trajectory. The $\rm E_{\rm str,~sym}$ mode \rev{(blue)} is resonant with the applied electric field $\boldsymbol{\mathcal{E}}$ (shown in purple). \rev{The contributions from $\rm E_{\rm str,~asym}$ (yellow) and $\rm E_{\rm bend}$ (red) remain close to zero throughout.}}\label{fig:water-driven-contributions}
\end{figure}

Many of these phenomena rely on the existence of anharmonic coupling between different vibrational modes that lead to the so-called nonlinear effects in vibrational dynamics.\cite{PhysRevB.89.220301,Forst2011}
A popular approach to simulate these experiments and capture the associated nonlinear dynamics is based on the electric dipole approximation (EDA). 
Within EDA, one can assume that the changes induced by the electric field $\boldsymbol{\mathcal{E}}$ on the electronic structure are not significant, and expand the electric enthalpy function of a system around the vanishing-field condition up to first order.
This leads to the following expression for the (external field dependent) potential energy
\begin{align}\label{eq:EDA}
    \mathcal{U}(\boldsymbol{\mathcal{E}}, \bm{q}) \simeq V(\bm{q})|_{\boldsymbol{\mathcal{E}}=0} - \boldsymbol{\mathcal{E}} \cdot \bm{\mu}(\bm{q}).
\end{align}
where $\bm{\mu}$ is the dipole moment of the system in the absence of the electric field and $\bm{q}$ represents the atomic positions. Eq.~\ref{eq:EDA} can be evaluated with a variety of theories. For example, when employing DFT, evaluating the expression above would require a full self-consistent convergence of the Kohn-Sham (KS) energy and the dipole corresponding to this converged electronic density. In periodic systems, the dipole would have to be expressed in terms of the system polarization, calculated with the Berry-phase formalism.\cite{SPALDIN20122}

Assuming a time-dependent electric field, $\boldsymbol{\mathcal{E}}(t)$, the effect of the external field manifests itself on the nuclear dynamics through the nuclear forces, $\bm{F}$,

\begin{align}
    \bm{F}
    = - \der{\mathcal{U}(\boldsymbol{\mathcal{E}}, \bm{q})}{\bm{q}}
    = - \der{V(\bm{q})}{\bm{q}}\bigg\vert_{\boldsymbol{\mathcal{E}}=0} + \boldsymbol{\mathcal{E}}(t) \cdot\der{\bm{\mu}}{\bm{q}} ,\label{eq:forces}
\end{align}
where the first term corresponds to the usual forces in the absence of an external field. The second term can be conveniently expressed in terms of the \textit{dynamical charges}, or Born effective charges (BEC) \bec, defined as,
\begin{align}\label{eq:bec-definition}
    \bec 
    = \, \frac{1}{e}\der{\mu_{\alpha}}{q_j},
\end{align}
where $e$ is the elementary charge (1 in atomic units) $\alpha=x,y,z$ runs through the Cartesian components of the total dipole and $j=1,\dots,3N$ runs through the degrees of freedom of the system.
With this definition, the last term in Eq.~\ref{eq:forces}, henceforth $\bm{F}^{\text{ext}}$, reads
\begin{align}
    F^{\text{ext}}_j(t) = \sum_{\alpha=x,y,z} \mathcal{E}_{\alpha}(t) \, \bec.
\end{align}

For a fixed atomic configuration, the BEC can be computed by finite differences of the dipole values with respect to the atomic positions. In the context of DFT, it is also possible to exploit density functional perturbation theory (DFPT).\cite{baroni2001phonons, Shang_2018}

The \ipi implementation for driven vibrational motion is based on the definition of a time-dependent integrator for the equations of motion, called \texttt{EDAIntegrator}, built on top of the existing integrators. The role of this integrator is simply to add $\bm{F}^{\text{ext}}(t)$ to the momenta within the Verlet algorithm. This was implemented in a new \emph{motion} class of i-PI, called \texttt{DrivenDynamics}, which inherits all the attributes and methods of the \emph{Dynamics} class.

We implemented a plane-wave linearly-polarized electric field with a Gaussian envelope function, of the following form,
\begin{align}
    \boldsymbol{\mathcal{E}}\left(t\right) = \boldsymbol{\mathcal{E}}_{\text{amp}}\cos \left( \omega t \right) \frac{1}{{\sigma \sqrt{2\pi}}} e^{-\frac{{(x - m)^2}}{{2\sigma^2}}} \label{eq:efield}
\end{align}
where $\omega$ is the frequency of the plane-wave, $\boldsymbol{\mathcal{E}}_{\text{amp}}$ is the polarization vector of the electric field, and $m$ and $\sigma^2$ are the mean and variance of the Gaussian envelope function, respectively. Other types of field profiles can be added straightforwardly.

The BECs and dipoles are communicated by the client at every new nuclear configuration through the \ipi~ \texttt{extras} field as a JSON formatted string.
An example of the communication of such a client is provided in the Python driver code distributed with \ipi~ (see \texttt{driverdipole}). Alternatively, BECs can be provided as a \texttt{txt} file and kept fixed during the dynamics.

An example of the section \texttt{motion} to run dynamics driven by a Gaussian-enveloped electric field with fixed BECs reads

\begin{minipage}{\linewidth}
{\small
\begin{lstlisting}
<motion mode='driven_dynamics'> 
 <driven_dynamics mode='eda-nve'>
  <timestep units='femtosecond`> 0.3 </timestep>
   <efield>
    <amp   units='atomic_unit'> [0,5e-2,0]  </amp>   
    <freq  units='thz'>         115     </freq>            
    <peak  units='picosecond'>  0.2        </peak>     
    <sigma units='picosecond'>  0.05       </sigma>
   </efield>
  <bec mode='file'> bec.txt  </bec> 
 </driven_dynamics>
</motion>
\end{lstlisting}
}
\end{minipage}
In this example, the
\verb|amp|,
\verb|freq|,
\verb|peak|, and 
\verb|sigma| tags refer to 
$\boldsymbol{\mathcal{E}}_{\text{amp}}$,
$\omega$, $m$
and $\sigma$ as defined in Eq. \ref{eq:efield}, respectively.
The file \texttt{bec.txt} contains a 2D array of shape $\left(3N,3\right)$. The columns correspond to the Cartesian components of the dipole and the rows to the degrees of freedom $j$ in Eq.~\ref{eq:bec-definition}. The current implementation only supports the NVE ensemble, but \rev{can be easily combined with} other ensembles.

As a benchmark example, a single water monomer has been driven using fixed BECs and the  Partridge-Schwenke water model.\cite{10.1063/1.473987} The parameters adopted for the electric field are the ones shown in the input example above. The resulting trajectory, initialized with vanishing velocities has then been projected along the 3 vibrational modes of the system, which have been computed using the harmonic vibrational analysis provided by \ipi. The vibrational modes are shown in Fig.~\ref{fig:water-driven-contributions}a. The oscillating frequency of the field at around 3836 cm$^{-1}$ (115 THz) was chosen to be resonant with the symmetric-stretch mode.

Fig.~\ref{fig:water-driven-contributions}b shows the energy given by the sum of the three harmonic vibrational mode energies ($\rm E_{\rm bend}, E_{\rm sym.stretch}, E_{\rm asym.stretch}$). $\rm E_{\rm anharm}$ is computed as the difference between the total energy and the sum of the harmonic energies. The profile of the applied field is also shown in Fig.~\ref{fig:water-driven-contributions}.
As expected, the resonant vibrational mode is excited and the energy of the system increases while the driving field assumes non-negligible values. Afterwards, the field decays to zero but the system remains excited due to the lack of a dissipation mechanism. By comparing the different components of the energy, it is evident that the harmonic contribution is dominated by the symmetric-stretch mode and that in this simple example the anharmonicity plays a relatively minor role.\subsection{Ensemble models and error estimation}

This version of \ipi{} implements error estimation for MLIPs based on ensembles. The idea is that, for each configuration $A$, multiple potentials $V^{(k)}(A)$ are computed, together with the associated forces. 
The simulation is then evolved based on the mean of the ensemble $\bar{V}(A)=\frac{1}{n_\text{ens}} \sum_k V^{(k)}(A)$, where 
$n_\text{ens}$ is the number of potentials in the ensemble,
and the error is estimated as the standard deviation $\sigma^2(A)$. 
While, in general, ensemble approaches entail a large computational overhead because $n_\text{ens}$ potentials have to be evaluated at each time step, there are cases in which this overhead can be largely avoided, as e.g. when employing linear or kernel methods,\cite{musi+19jctc} or when applying a high-level of weight sharing between different neural networks e.g. using a ``shallow-ensemble" on the last-layer.\cite{Lakshminarayanan2017,kunapuli2023ensemble,ashukha2021pitfalls,Rahaman2021,kellner2024uncertainty,bigi2024prediction}
\ipi{} provides two alternative modes to simulate nuclear (thermo)dynamics including ensemble-based uncertainty estimates, depending on the characteristics of the underlying framework. 

\begin{figure}
\includegraphics[width=0.80\linewidth]{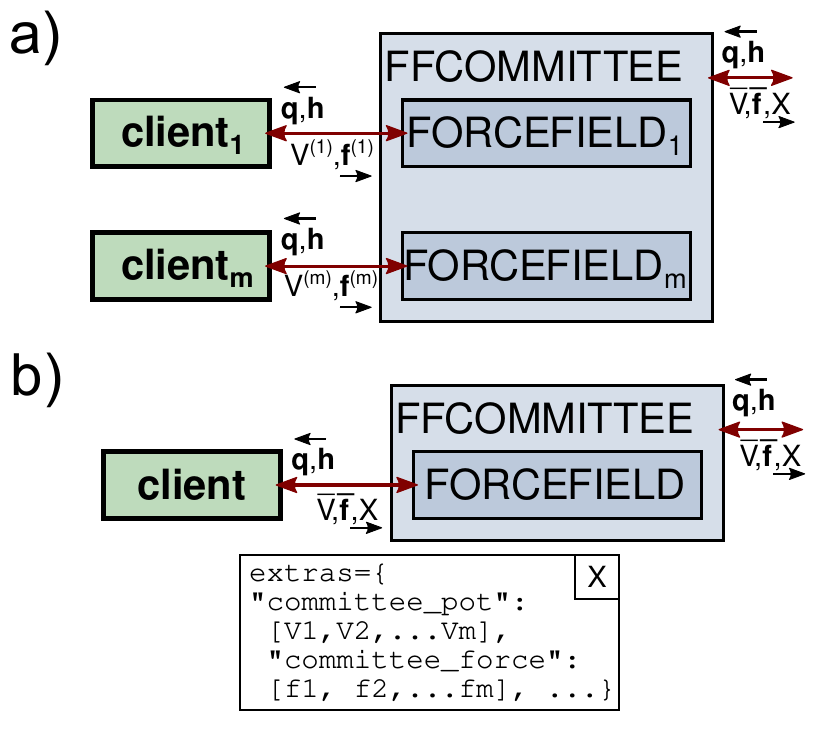}
\caption{Schematic representation of the implementation of the \emph{FFCommittee} class, that combines the members of an ensemble model to evaluate the mean potential and force, that are used to propagate the dynamics. 
The statistics of the ensemble are stored in a JSON-formatted \emph{extras} string, depicted in the figure with the $X$ label. 
This class can handle two types of models. The first one, depicted in panel a, works by collecting data from one forcefield per committee member, which is suitable for fully independent models. Alternatively, a single forcefield can compute all ensemble members, returning them to i-PI as a JSON string with the same syntax (panel b), which is then used to compute the error. This second mode is particularly suitable for shallow models with a high degree of weight sharing. 
}
\end{figure}

The first entails independent models that are run as separate clients, by specifying multiple socket connectors as part of a \lstinline{<ffcommittee>} tag. 
For instance,
\begin{minipage}{\linewidth}
\begin{lstlisting}
<ffcommittee name='my_committee'>
   <ffsocket name='driver-1'>
   [...]
   </ffsocket>
   <ffsocket name='driver-2'>
   [...]
   </ffsocket>
   [...]
</ffcommittee>
\end{lstlisting}
\end{minipage}
The associated class takes care of collecting potentials, forces and stresses from the various sockets and averaging them. 
Even though a \lstinline{ffcommittee} section contains multiple sockets, one should use a single \lstinline{<force>} block  
referring to the name of the committee (\texttt{my\_committee} in this example) in the definition of the system.  
A linear committee calibration scaling \cite{musi+19jctc} can also be applied with the keyword \lstinline{<alpha>}. 

In the second mode, which is suitable when using a shallow ensemble model with weight sharing, a single client computes all the ensemble members. In this case, the client passes the average energies, forces and virials to \ipi~ in the standard way and also passes the ensemble information as a JSON string into the \texttt{extras} field. 
The string contains the list of potentials, forces, and virials into fields named \lstinline{committee_pot}, \lstinline{committee_force}, \lstinline{committee_virial}. 
We provide a reference implementation of such a client in the FORTRAN driver distributed with \ipi{}.
It is then sufficient to specify a single client in the committee section 
\begin{minipage}{\linewidth}
\begin{lstlisting}
<ffcommittee name='my_committee'>
 <ffsocket name='driver'>
 [...]
 </ffsocket>
 <parse_json> True </parse_json>
</ffcommittee>
\end{lstlisting}
\end{minipage}
and to activate the flag \lstinline{parse_json} that instructs the committee class to parse the ensemble data out of the string. Also in this case, the \lstinline{forces} section should refer to \lstinline{"my_committee"} and not to the name of the inner forcefield.

For both modes, committee potentials, forces, as well as the computed uncertainty, are collected into an \emph{extras} dictionary corresponding to the same schema used for the JSON string. 
The different diagnostics can then be printed out using the generic output features, e.g. specifying \lstinline{extra_type='committee_pot'} to print the energies of all members of the ensemble. 

The committee forcefield implements a rudimentary active-learning mechanism. 
By specifying an output filename and an error threshold, every structure with a predicted uncertainty above the threshold encountered at any point of a calculation will be stored so it can be used to improve the MLIP model in an iterative manner. 
It is also possible to build a \emph{weighted baseline} model,\cite{Imbalzano_JCP_2021} in which one forcefield, whose name is specified in the \lstinline{<baseline_name>} flag, is interpreted as a reference potential, to which the model described by a committee is added as a correction. 
The baseline is assumed to have a constant uncertainty $\sigma_\text{base}$, that can be set with the option \lstinline{<baseline_uncertainty>}. 
Rather than simply computing the overall potential as $V=V_\text{base}+\bar{V}$, the weighted baseline potential is computed as 
\begin{equation}
V(A) = V_\text{base}(A) + \frac{\sigma^2_\text{base}}{\sigma^2_\text{base} + \sigma^2(A)} \bar{V}(A)
\end{equation}
which uses the correction when the predicted error is smaller than the baseline error, and switches to the more reliable (although less accurate for structures that are well-predicted by the committee) baseline potential. 

\begin{figure}
\includegraphics[width=1.0\linewidth]{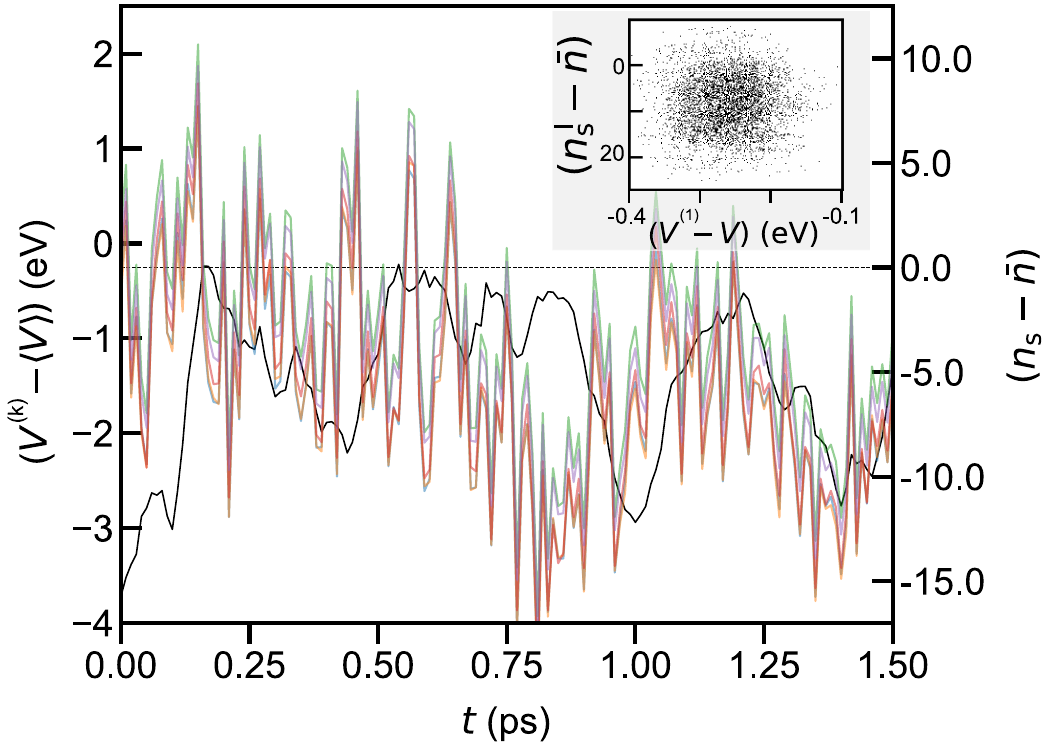}
\caption{
Trajectories of the various energy predictions of the DPOSE \cite{kellner2024uncertainty} committee model (plotted in constrasting colors) and of the order parameter that determines the pinning potential, Eq.~\ref{eq:pinning} (black). 
The inset shows the near-Gaussian correlation plot between the order parameter and the deviation of one of the ensemble members from the mean potential. 
\label{fig:ensemblemu-trajectory}
}
\end{figure}

\subsubsection{Computing the solid-liquid chemical potential for water}

We discuss a representative example of the use of \lstinline{ffcommittee} to compute errors of the solid-liquid chemical potential $\Delta\mu$  for an ice/water system, that showcases several of the advanced features in \ipi{}. 
This type of simulation can be used to estimate the melting point of water $T_\text{m}$, by repeating the procedure at different temperatures and finding the temperature at which $\Delta\mu=0$. 
As shown in Ref.~\citenum{Imbalzano_JCP_2021}, once a collection of $\Delta\mu$ values associated with the different ensemble members has been obtained, it is easy to determine the error on  $T_\text{m}$ by ensemble propagation. 

For this example, we use a Behler-Parrinello-style neural network,\cite{behl-parr07prl} based on smooth-overlap of atomic positions (SOAP)\cite{bart+13prb} features and trained on the dataset of Ref.~\citenum{cheng_ab_2019}. 
The model generates an ensemble of 5 energy predictions, differing only by the last-layer weights and trained by optimizing the negative log-likelihood following the direct propagation of shallow ensembles (DPOSE) protocol.\cite{kellner2024uncertainty}
We run a simulation with 336 water molecules, arranged in an elongated simulation cell containing coexisting water and ice. 
A restraining \emph{pinning potential} to an anchor point, $\bar{n}$,
\begin{equation}
V_\text{pin}(A) = \frac{1}{2} k (n_\text{S}(A)-\bar{n})^2
\label{eq:pinning}
\end{equation}
is applied using a suitable collective variable that estimates the number of molecules in a solid-like environment, $n_\text{S}(A)$,\cite{pede+15jcp} as implemented in PLUMED\cite{PLUMED} \rev{and here used through its interface with \ipi.}
At the melting point, the potential restraints the solid fraction to the target value $\bar{n}$, whereas away from equilibrium the mean deviation is proportional to the solid-liquid molar free-energy difference, $\Delta\mu\propto \left<n_\text{S}(A)-\bar{n}\right>$.
Further details on the setup can be inferred from the example files included in the \lstinline{demos/ensemble-deltamu} directory. It should be noted that both size and duration of the simulations are insufficient to obtain a converged estimate of the melting point, as this is only intended as a demonstrative example. 

\begin{figure}
\includegraphics[width=1.0\linewidth]{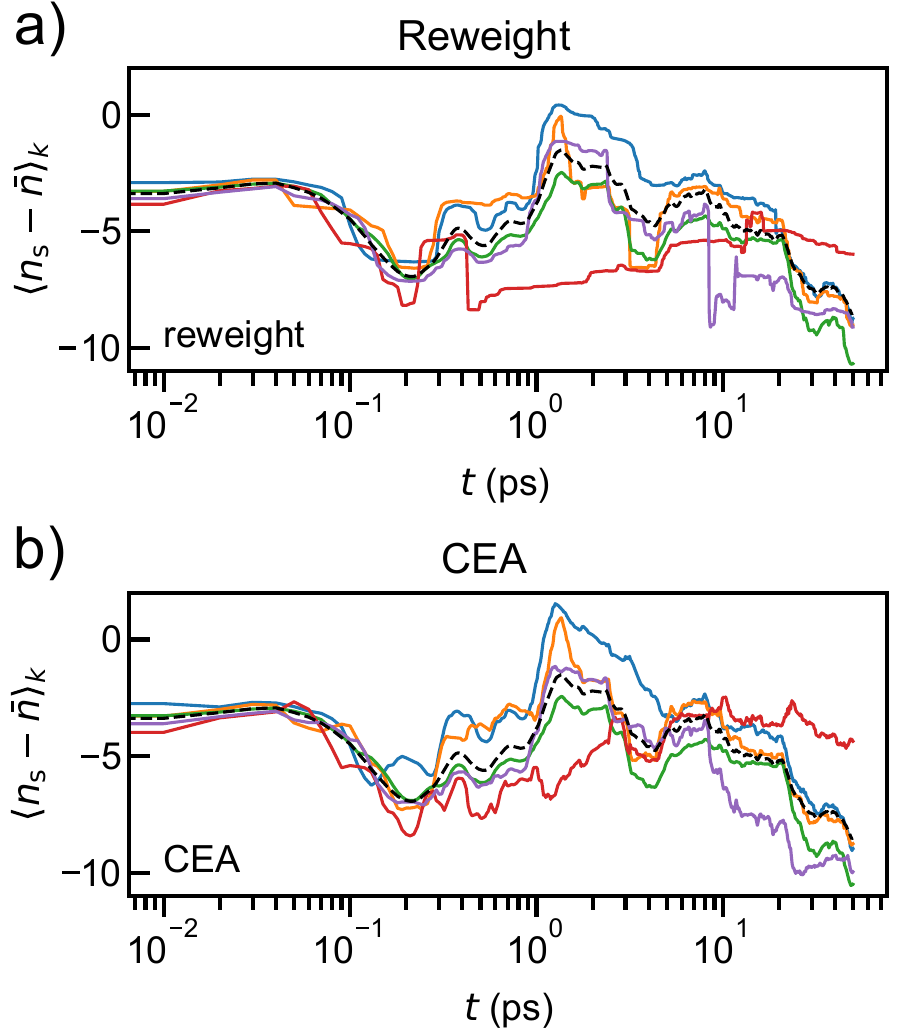}
\caption{
Cumulative sums of the reweighted mean $\langle n_\text{S}-\bar{n}\rangle_k$ computed using the $k$-th ensemble member (plotted in contrasting colors). The dashed black line indicates the cumulative average of $n_{\text{S}}$ computed without reweighting.
Panel a) shows the average computed with Boltzmann reweighting while panel b)  uses a statistically-better-behaved cumulant expansion approximation (CEA). 
\label{fig:ensemblemu-cumsum}
}
\end{figure}

Fig.~\ref{fig:ensemblemu-trajectory} shows a small segment of a trajectory obtained with the DPOSE model\cite{kellner2024uncertainty}. The mean potential undergoes thermal fluctuations, and the ensemble of predictions has a varying spread along the trajectory. Similar to what was observed in Ref.~\citenum{kellner2024uncertainty}, the members of the ensemble differ mostly by a rigid shift, reflecting a systematic bias in the prediction of the potential. 
Meanwhile, the order parameter fluctuates about a value that is slightly below the target, indicating that the simulation temperature (290~K) is above the melting point for this water model -- at least for this under-converged supercell size.
There is no obvious pattern in the correlation between the total potential energy and the order parameter: at this temperature, and for such a comparatively small system, random fluctuations dominate over the enthalpy change associated with the changes in the solid fraction. 
Importantly, the cross-correlations between the solid fraction and the ensemble potential shift $(V^{(k)}(A)-\bar{V}(A))$ show approximately Gaussian behaviour (see inset in Fig.~\ref{fig:ensemblemu-trajectory}). 
This simplifies the estimation of the average deviation $\left<(n_\text{S}(A)-\bar{n})\right>$ from a single trajectory driven by $\bar{V}$.
The basic idea is to use Torrie-Valleau reweighting,\cite{torr-vall99jcp} i.e. weighting each snapshot by $e^{-\beta(V^{(k)}(A)-\bar{V}(A))}$. 

As shown in Fig.~\ref{fig:ensemblemu-cumsum}, the cumulative averages computed with the different weighting factors converge to different values -- providing an ensemble of predictions for $\langle n_\text{s}-\bar{n}\rangle_k$.
The sharp, sudden changes in the cumulative average reflect the presence of very large weighting factors, that occur whenever the deviations of individual predictions from the mean potential are large relative to $k_\text{B}T$.
The resulting statistical inefficiency\cite{ceri+12prsa} would become worse as the system size is increased, eventually making the procedure completely ineffective. Following  Ref.~\citenum{Imbalzano_JCP_2021}, one can use instead a cumulant expansion approximation (CEA) that estimates reweighted average under the assumption that the target observable and the potential are joint Gaussians, and yields 
\begin{equation}
\langle n_\text{S} \rangle_k \approx 
\langle n_\text{S} \rangle - \beta  [
\langle n_\text{S}(V^{(k)}-\bar{V}) \rangle -
\langle n_\text{S}\rangle \langle (V^{(k)}-\bar{V}) \rangle].
\end{equation}
This approximation introduces a systematic error because the distribution is \emph{not} exactly Gaussian, but as can be seen in Fig.~\ref{fig:ensemblemu-trajectory}, the qualitative behaviour reflects the one seen for the direct reweighting. The statistical error is well-behaved also for large simulations, making the CEA a more generally applicable strategy.
We provide a commented example to perform this kind of simulation in the \lstinline{demos/ensemble-deltamu}; the post-processing needed to evaluate the reweighted averages is implemented in the \lstinline{i-pi-committee-reweight} script. 

By repeating the simulation at different temperatures one can determine the temperature dependence of $\Delta \mu$ for each ensemble member separately, and use the separately-computed values of $T_m$ to estimate the error in the determination of the melting point. 
It is worth noting that in Ref.~\citenum{Imbalzano_JCP_2021} the melting point was determined to be $290\pm 5$~K, with a setup similar to the one discussed here, but using a deep ensemble of models, fitted on the same dataset. 
This estimate is consistent with our observation that $\langle n_\text{s}-\bar{n}\rangle$ at 290~K is very close to zero. The discrepancy with Ref.~\citenum{cheng_ab_2019}, where  the classical melting point for the reference DFT 
was determined by free-energy perturbation to be at $\approx 275$~K can be attributed to finite-size effects, that tend to destabilise the liquid and increase the melting point. 
\subsection{Cavity Molecular Dynamics for Polaritonics}

For molecules confined in optical cavities, molecular transitions may form polaritons, hybrid light-matter states with cavity photon modes.\cite{Ribeiro2018,Li2022Review,Fregoni2022,Mandal2023ChemRev,Ruggenthaler2023} Of particular interest is the vibrational strong coupling regime,\cite{Shalabney2015,Long2015} in which vibrational polaritons are formed when a molecular vibrational mode is near resonant with an IR cavity photon mode.  Under vibrational strong coupling, many molecular properties, including chemical reaction rates and energy transfer rates, can be significantly modified in the liquid phase.\cite{Thomas2016,Thomas2019_science,Xiang2020Science}
For a better understanding of these exciting experiments, both the molecules and the cavity photons need to be propagated self-consistently with realistic parameters. This requirement is beyond the scope of standard molecular dynamics schemes, whereby the time-dependent dynamics of the photon field are not explicitly propagated. 
    
The 3.0 version of the \ipi package provides an efficient method for simulating vibrational strong coupling: the CavMD approach.\cite{Li2020Water,Li2022RPMDCav} Within the framework of CavMD, both nuclei and cavity photons in the \rev{IR} domain are treated classically, propagated by the following  light-matter Hamiltonian under the ground electronic state: $H_{\rm QED} = H_{\rm M} + H_{\rm F}$,\cite{Flick2017,Li2020Water} where $H_{\rm M}$ denotes the standard molecular Hamiltonian composed of the kinetic and potential terms, and 
\begin{equation}\label{eq:cavmd_H}
        H_{\rm F}  = \sum_{k,\lambda} \frac{\dbtilde{p}_{k,\lambda}^2}{2m_{k,\lambda}} + \frac{1}{2}m_{k,\lambda}\omega_{k,\lambda}^2\left (
        \dbtilde{q}_{k,\lambda} + \frac{\widetilde{\varepsilon}_{k,\lambda} d_{\lambda}}{m_{k,\lambda}\omega_{k,\lambda}^2}
        \right )^2 .
\end{equation}
Here, $\dbtilde{p}_{k, \lambda}$, $\dbtilde{q}_{k, \lambda}$, $\omega_{k,\lambda}$, and $m_{k, \lambda}$ denote the momentum, position, frequency, and auxiliary mass for the cavity photon mode defined by a wave vector $\mathbf{k}$ and polarization direction $\vxi_\lambda$ with $\mathbf{k}\cdot \vxi_\lambda = 0$. For example, when the cavity is placed along the $z$ direction, $\lambda = x, y$. $d_{\lambda}$ is the electronic ground-state dipole moment for the molecular system projected along the direction of $\vxi_\lambda$, and $\widetilde{\varepsilon}_{k,\lambda}$ represents the effective light-matter coupling strength. 

\begin{figure}[tbp]
  \centering
  \includegraphics[width=.85\linewidth]{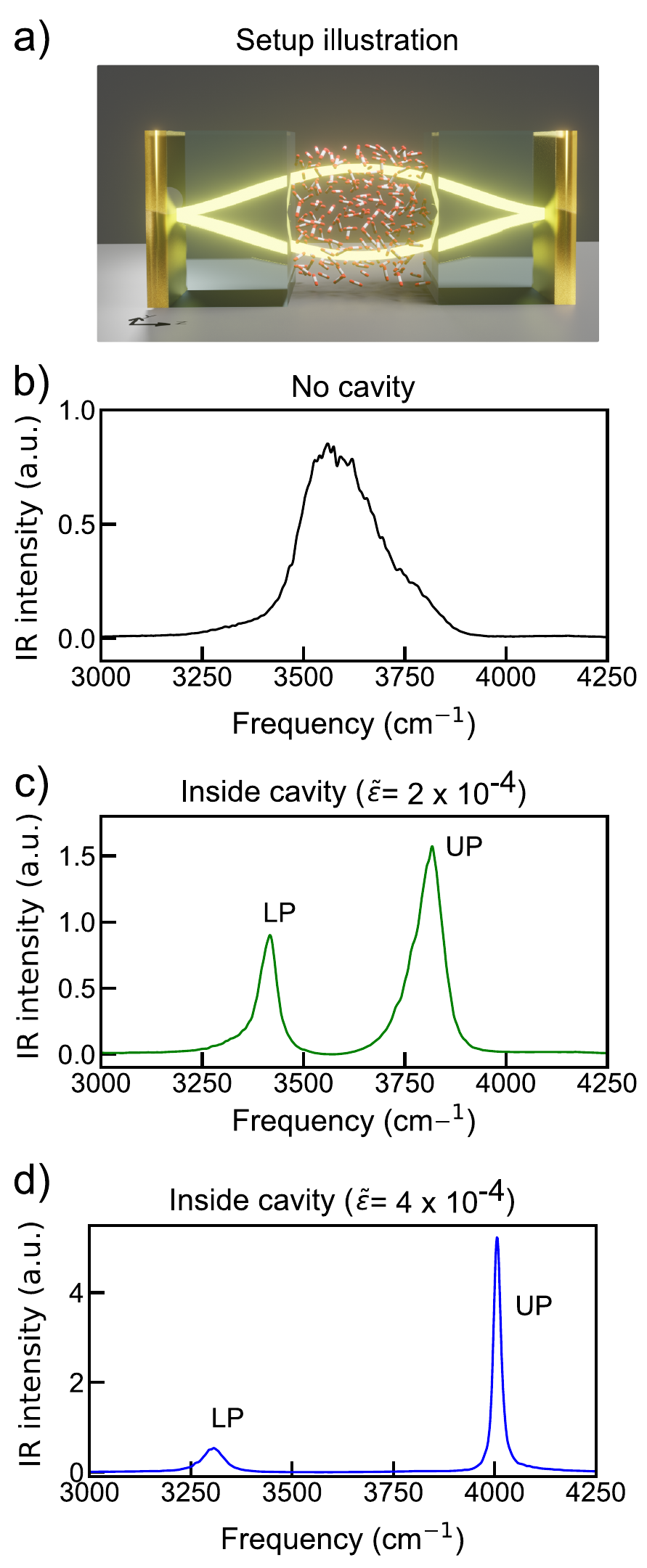}
 \caption{CavMD simulations: Panel (a) illustrates the simulation setup of CavMD, whereby a large ensemble of molecules is confined in an optical cavity under vibrational strong coupling.  
    The rest of the panels are adapted from Ref. \citenum{Li2020Water}, \rev{with permission from the authors}, and show the IR spectra of the OH stretch mode of liquid water \rev{without a cavity} and inside a cavity with different light-matter coupling strength, $\widetilde{\varepsilon}$. LP and UP stand for lower and upper polariton, respectively.}
		\label{fig:cavmd_demo}
    \end{figure}

The basic implementation strategy of CavMD is as follows. The cavity photon modes are stored in \ipi as \textbf{L} (light) atoms. The coordinates and velocities of both the nuclei and the cavity photon modes are propagated in \ipi. A new \emph{FFCavPhSocket} class, which inherits from the original \emph{FFSocket} class, evaluates the forces of both the nuclei and the cavity photon modes. Similar to \emph{FFSocket}, \emph{FFCavPhSocket}  calls external client codes to evaluate the nuclear forces associated with $H_{\rm M}$, via the socket interface. After obtaining this component of the total forces, \emph{FFCavPhSocket} performs additional calculations to obtain the photonic forces \rev{and cavity modifications on the nuclear forces} as prescribed by Eq. \ref{eq:cavmd_H}. Because this implementation encapsulates the cavity-related calculations within the \emph{FFCavPhSocket} class, many advanced features in \ipi, including RPMD, \cite{Li2020Water} and ring-polymer instantons, can be directly applied to study vibrational strong coupling without additional modifications of the \ipi code. 

The CavMD implementation was used to simulate the IR spectra of liquid water under vibrational strong coupling.\cite{Li2020Water} The key \ipi input section to simulate vibrational strong coupling is as follows: 
\begin{minipage}{\linewidth}    
\begin{lstlisting}
<ffcavphsocket name='lammps' mode='unix'>
    <address>h2o-cl-cavmd</address>
    <apply_photon> True </apply_photon>
    <E0> 4e-4 </E0>
    <omega_c units='inversecm'> 
       3550.0 
    </omega_c>
    <charge_array> [...] </charge_array>
</ffcavphsocket>
\end{lstlisting}
\end{minipage}
In the \emph{FFCavPhSocket} section above, a cavity mode 
with frequency $\omega_{\rm c} = 3550$ cm$^{-1}$ was coupled to liquid water with an effective light-matter coupling strength  $\widetilde{\varepsilon} = 4\times 10^{-4}$ a.u. Under the default setting, the cavity mode is polarised along both the $x$ and $y$ directions. The partial charges of all molecular atoms are given in the \texttt{charge\_array} section of \emph{FFCavPhSocket}, and are used to evaluate the molecular dipole moment in Eq. \eqref{eq:cavmd_H}. 
In Fig. \ref{fig:cavmd_demo} we show a schematic picture of the setup along with the IR spectra of water inside and outside the cavity.     As shown in Fig. \ref{fig:cavmd_demo}c, via classical CavMD simulations, the wide OH band of liquid water splits into a pair of polariton states, the upper polariton (UP) and the lower polariton (LP). Beyond this study, CavMD was also used to explore novel energy transfer and chemical reaction pathways under vibrational strong coupling. \cite{Li2020Nonlinear,Li2021Solute,Li2023QMMM}

\section{Showcase of Advanced Simulation Setups}

\subsection{Path Integral Coarse-Grained Simulations}
\begin{figure}
    \centering
    \includegraphics[width=0.95\columnwidth]{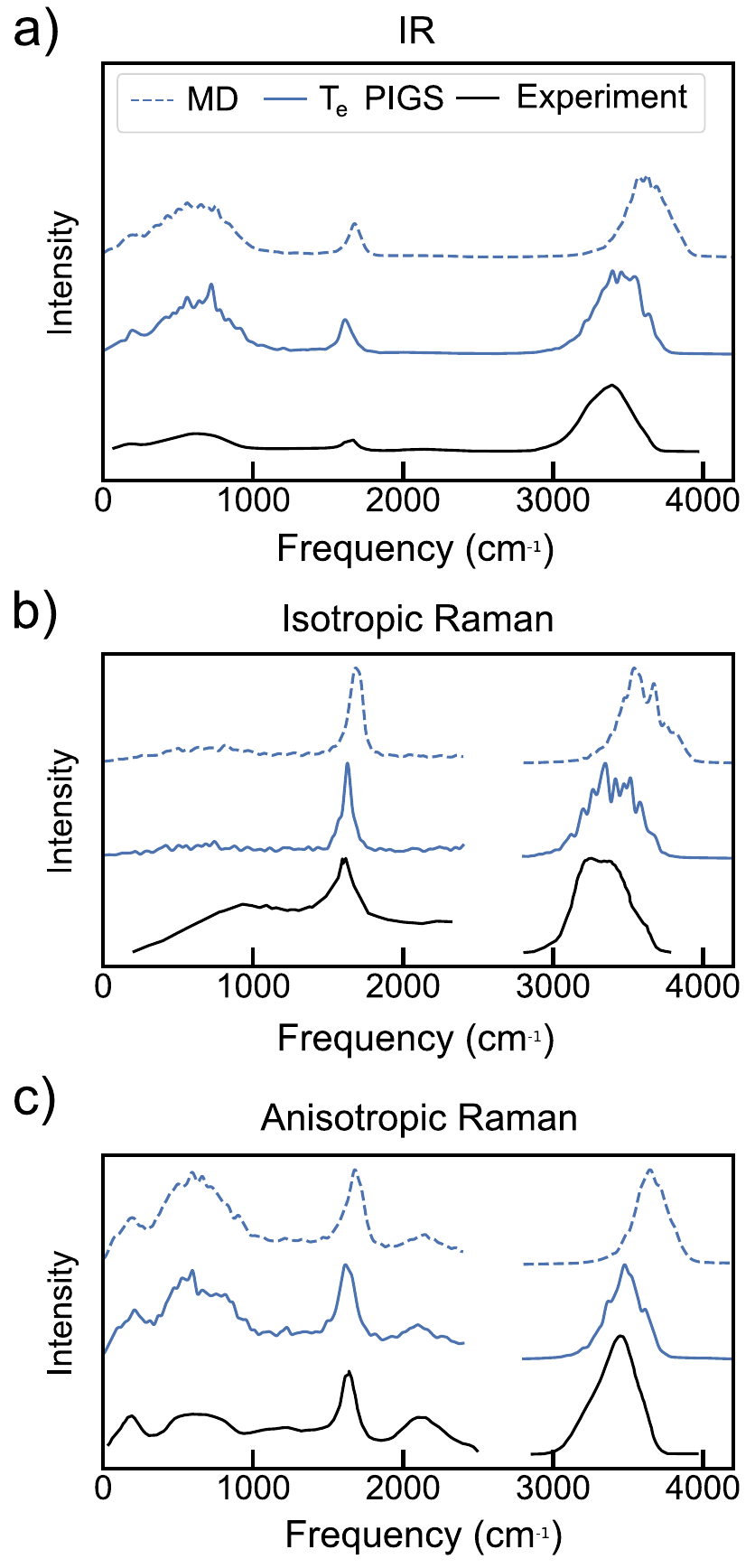}
    \caption{IR, isotropic Raman, and anisotropic Raman vibrational spectrum of water at 300 K using the T$_{\text{e}}$ PIGS method. Adapted from Ref.~\citenum{kovacs_mace-off23_2023} \rev{with permission (CC BY 4.0 Licence)}.} 
    \label{fig:tepigs_water}
\end{figure}

\rev{While PIMD has demonstrated success in incorporating quantum nuclear motion for equilibrium properties, extracting accurate dynamical information from path-integral simulations is a longstanding challenge~\cite{althorpe_path-integral_2021}.} 
\rev{For instance, in estimating the vibrational spectra, established path-integral approximations to quantum dynamics such as with RPMD~\cite{crai-mano04jcp}, CMD~\cite{cao-voth93jcp,cao_JCP_94} and variants~\cite{ross+14jcp, ross+18jcp, willatt_approximating_2018, trenins_path-integral_2019, haggard_testing_2021}, present artefacts such as unphysical shifted or broadened peaks.\cite{witt+09jcp, ross+14jcp}}
In addition, the computational overhead of path-integral simulations and the impact of the artefacts become more pronounced with lowering temperature. 

~\citet{musil_quantum_2022} recently introduced a machine learning framework to significantly enhance the accuracy and efficiency of approximating quantum dynamics based on CMD. 
\rev{Their approach, referred to as T$_\text{e}$ PIGS, has two components. 
The PIGS method is introduced to estimate an effective potential with an MLIP, which represents the centroid potential of the mean force, to drive the dynamics of the centroid.
Second, the  T$_\text{e}$ \textit{ansatz} estimates the potential of mean force at an elevated temperature to effectively address the ``curvature problem" of standard CMD~\cite{witt+09jcp}.
This \textit{ansatz} attenuates the spurious redshift in the centroid-based vibrational spectra of high-frequency quantized modes and results in a temperature-transferable effective potential for approximate quantum dynamics.}
Combining these components, the \rev{T$_\text{e}$ PIGS method yields accurate modeling of vibrational spectra of large systems via inexpensive classical MD on an effective centroid MLIP.} 
Together with an accurate potential energy surface, we observe quantitative agreement with experiments for IR, Raman and sum-frequency generation vibrational spectra in bulk and interfacial water~\cite{kapil_first-principles_2023}.

This approach is easily implemented using the flexible functionality of \ipi{} as demonstrated for \rev{the calculation of accurate IR and Raman spectra of liquid water} (see Fig.~\ref{fig:tepigs_water}) using the \texttt{MACE-OFF23(S)} general organic \rev{MLIP~\cite{kovacs_mace-off23_2023} and an ML model for the polarization and polarizability of bulk water.\cite{kapil_first-principles_2023}}
\rev{A fully documented example with scripts to generate the spectra is provided in the \ipi\ \textit{demos} directory.}
For a given potential energy surface $V(\mathbf{q})$, \rev{with $\mathbf{q}$ the Cartesian positions,  the T$_{\text{e}}$ PIGS ML centroid potential of mean force can be trained using a single high-temperature PIMD simulation.}
\rev{The training data includes} the centroid positions $\bar{\mathbf{q}} = \sum_{j=1}^{P} \mathbf{q^{(j)}}/P$ \rev{with $\mathbf{q}^{(j)}$ the position of the $j$-th replica}, the \rev{instantaneous estimator of the mean centroid force} $\bar{\mathbf{f}}_{\text{c}} = \sum_{j=1}^{P} \mathbf{f^{(j)}}/P$, \rev{where $\mathbf{f}^{(j)}$ is the force acting on the $j$-th replica}, and the physical force acting on the centroid $\bar{\mathbf{f}}_{\text{p}} = - \partial V(\bar{\mathbf{q}})/\partial \bar{\mathbf{q}}$. 
The elevated temperature was set to $T_e$ = 500\,K, and the centroid potential of mean force was fit from a 10 ps long path integral simulation with $P=8$, with the centroid positions and forces sampled every 10 fs (20 timesteps). The centroid positions and \rev{instantaneous} forces (\rev{$\bar{\mathbf{q}}$ and $\bar{\mathbf{f}}_\mathrm{c}$}) can be printed out as follows,
\begin{minipage}{\linewidth}
\begin{lstlisting}
    <trajectory filename='xc' stride='20' 
       format='xyz' cell_units='ase'> 
       x_centroid{ase} 
    </trajectory>
    <trajectory filename='centroid_force' 
       stride='20' format='ase'> 
       f_centroid  
    </trajectory>
\end{lstlisting}
\end{minipage}

\rev{The quantity  $\bar{\mathbf{f}}_{\text{p}}$ can be calculated in a post-processing step with a ``replay" simulation, which allows the calculation and output of properties on a list of centroid positions. Alternatively, the physical force on the centroid can be printed on the fly by defining a ``dummy" force component with a zero weight that receives centroid positions using,}
\begin{minipage}{\linewidth}
\begin{lstlisting}
    <forces>
      [...]
      <force forcefield='maceoff23' weight='0'
      nbeads='1'> 
      </force>
    </forces>
\end{lstlisting}
\end{minipage}
\rev{Printing out this force component requires a single line:}
\begin{minipage}{\linewidth}
\begin{lstlisting}
    <trajectory filename='physical_force' 
       stride='20'  format='ase'> 
       forces_component_raw(1) 
    </trajectory>
\end{lstlisting}
\end{minipage}

\rev{With this data, a force matching method can be used to obtain the T$_{\text{e}}$ PIGS MLIP by regressing the centroid force against the centroid position.
This can be done transparently using standard MLIP packages like \texttt{n2p2}~\cite{singraber_library-based_2019}, \texttt{nequip}~\cite{batzner_e3-equivariant_2022}, \texttt{mace}~\cite{batatia_mace_2022}, by replacing physical positions and forces, as is typically done while training MLIPs, with positions and forces associated with the centroid.
To simplify the training problem, we advise to learn the difference between $\bar{\mathbf{f}}_{\text{c}}$ and $\bar{\mathbf{f}}_{\text{p}}$. 
The scripts necessary for dataset generation, curation and training with \texttt{mace} are provided in the \ipi{} \textit{demos} directory. } 

\rev{After the MLIP is trained in this fashion, performing a T$_{\text{e}}$ PIGS simulation is as simple as running} classical molecular dynamics on the sum of the potential energy surface $V(\mathbf{q})$ and the T$_{\text{e}}$ PIGS MLIP. This can also done in \ipi{} by
exploiting again the modularity of forces. \rev{We only need to define two forcefield sockets for the potential energy surface and the T$_{\text{e}}$ PIGS MLIP}:
\begin{minipage}{\linewidth}
\begin{lstlisting}
  <ffsocket name='maceoff23' mode='unix'
  pbc='false'>
    <address> driver </address>
  </ffsocket>
  <ffsocket name='maceoff23-pigs' mode='unix'
  pbc='false'>
    <address> driver-pigs </address>
  </ffsocket>
\end{lstlisting}
\end{minipage}
The addition of the potentials, along with the calculation of all derivatives like the force and stress, is done under the hood by defining the forces of the system class comprising two force components that communicate with the forcefield sockets:
\begin{minipage}{\linewidth}
\begin{lstlisting}
    [...]
    <forces>
      <force forcefield='maceoff23'> </force>
      <force forcefield='maceoff23-pigs'>
      </force>
    </forces>
    [...]
\end{lstlisting}

\rev{The remaining steps for estimating IR and Raman spectra, as shown in Fig.~\ref{fig:tepigs_water}, include predicting the total polarization and polarizability tensors and estimating their time correlation functions as detailed in Ref.~\citenum{kapil_first-principles_2023}. 
The scripts for these steps are in the \textit{demos} directory.} 

\end{minipage}

\begin{figure}
  \centering
  \includegraphics[width=.95\linewidth]{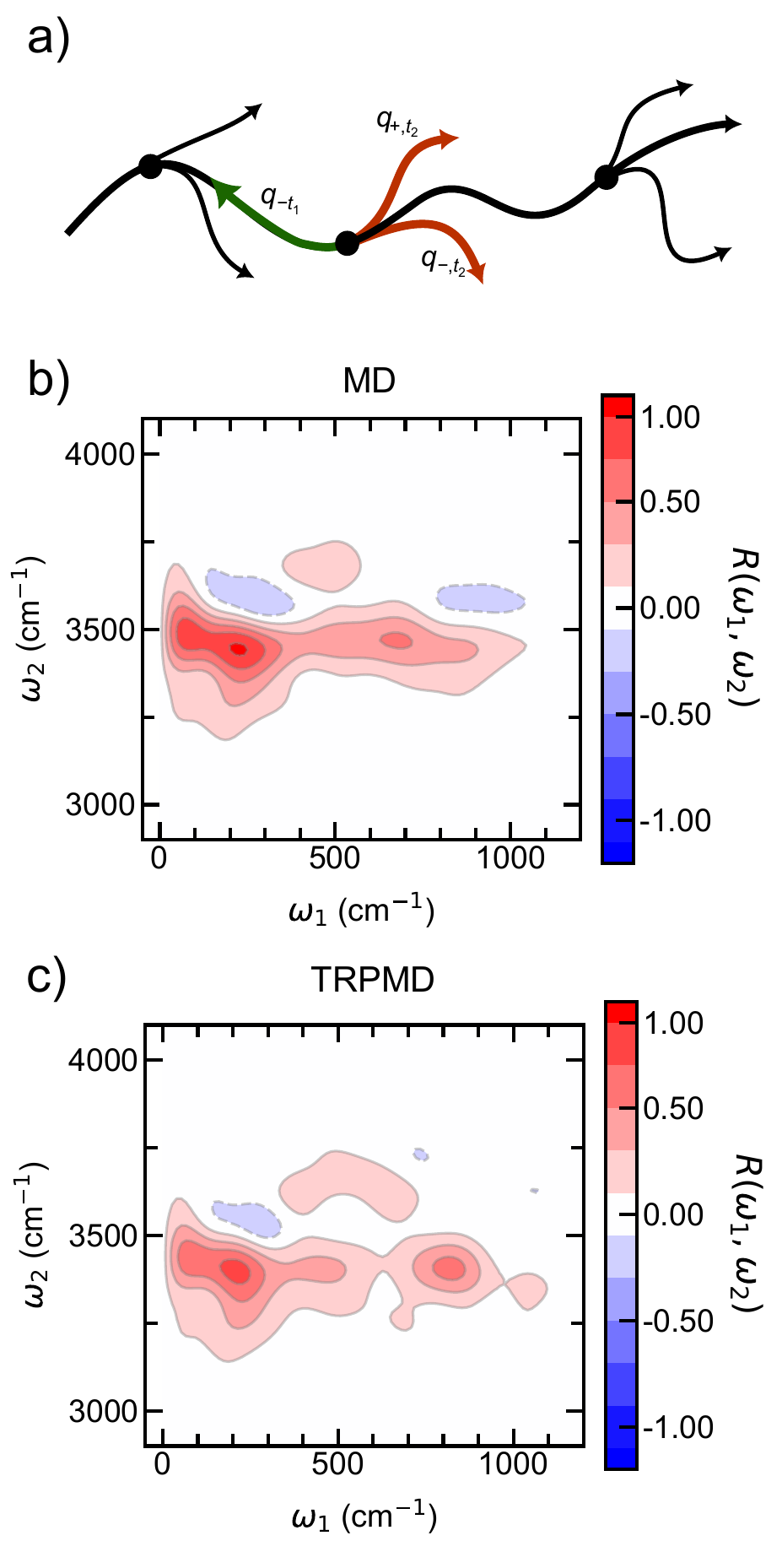}
    \caption{Two-dimensional IR-Raman spectroscopy: Panel (a) shows the schematic of the equilibrium-nonequilibrium (RP)MD approach (see text). Panel (b) and (c) depict the frequency-dependent response function corresponding to the two-dimensional THz-IR-Raman (also called THz-IR-visible \cite{Grechko_NatComm_2018}) spectrum of liquid water, computed using classical MD and TRPMD, respectively. Figure adapted from Ref.~\citenum{Begusic_NatComm_2023}, with permission (CC-BY Licence).}
    \label{fig:2D_IR_Raman}
\end{figure}

\subsection{2D IR-Raman spectra of liquid water}

Nonlinear, multi-time response functions are associated with different flavours of time-resolved and multidimensional spectroscopic techniques.\cite{Mukamel} Here, we show an example provided in the \ipi{}~\textit{demos} directory, that simulates the two-dimensional IR-Raman spectrum of liquid water in which the sample interacts first with two mid- or far-IR pulses through a dipole interaction and then with a near-IR or visible pulse through a Raman process.\cite{Grechko_NatComm_2018,Seliya_JCP_2024} The spectrum is obtained as a two-dimensional Fourier sine transform of the two-time response function

\begin{equation}
R(t_1, t_2) = -\frac{1}{\hbar^2} \langle[[\hat{\alpha}(t_1 + t_2),\hat{\mu}(t_1)],\hat{\mu}(0)]\rangle \label{eq:R_t}
\end{equation}
where $\hat{\alpha}$ is the polarizability operator and $\hat{\mu}$ the dipole moment. The example presented in \texttt{demos/2D-IR-Raman} implements the equilibrium-nonequilibrium RPMD method of Refs.~\citenum{Begusic_JCP_2022} and~\citenum{Begusic_NatComm_2023}, which approximates the quantum-mechanical response function \rev{of Eq.}~\ref{eq:R_t}. As a special case when the number of ring-polymer beads $P=1$, the method corresponds to the MD approach of Hasegawa and Tanimura.\cite{Hasegawa_JCP_2006} In passing, we note that other methods based on imaginary-time path integrals have been reported to calculate non-linear response functions using approximations to fully-symmetrized multi-time Kubo correlation functions.\cite{Jung_JCP_2020,Jung_JCP_2018,Tong_JCP_2020}

The equilibrium-nonequilibrium method requires sampling the initial distribution as in a typical PIMD simulation and running three trajectories for each initial phase-space point ($q_0$, $p_0$)---one backward equilibrium trajectory along which we evaluate the dipole moment and two nonequilibrium trajectories that are initialized with a momentum kick, i.e., starting from ($q_{\pm, 0} = q_0$, $p_{\pm, 0}= p_0 \pm \varepsilon \mu^{\prime}(q_0) / 2$) (see schematics in Fig.~\ref{fig:2D_IR_Raman}a). The two-time response function is then computed as 
\begin{equation}
R^{(2)}(t_1, t_2) = \frac{\beta}{\varepsilon} \langle [ \alpha(q_{+, t_2}) - \alpha(q_{-, t_2})] \dot{\mu}(q_{-t_1}) \rangle, \label{eq:R2_t_RPMD}
\end{equation}
where $\langle \cdot \rangle$ now represents a classical average over the ring-polymer equilibrium distribution at inverse temperature $\beta$. $\varepsilon$ is a parameter that controls the magnitude of the perturbation (``kick'') applied to the nonequilibrium trajectories. Position-dependent operators $\alpha(q)$ and $\mu(q)$ in Eq.~\ref{eq:R2_t_RPMD} are evaluated as averages over $P$ beads of the ring polymer, like in the case of linear RPMD correlation functions.\cite{crai-mano04jcp} 

The evaluation of nonequilibrium trajectories, two-time response function, and two-dimensional spectrum is implemented in the scripts provided in \texttt{demos/2D-IR-Raman}. The \texttt{noneqm-traj.py} script uses checkpoint files and $z$-dipole nuclear gradients stored along a longer RPMD trajectory to run multiple shorter nonequilibrium trajectories (see Fig.~\ref{fig:2D_IR_Raman}a). In this way, the long RPMD trajectory serves both for sampling initial conditions and computing the response function. To properly sample the canonical ensemble, the whole calculation is typically repeated with independent RPMD trajectories. The \texttt{noneqm-response.py} script then computes the response function by reading in the dipole moments and polarizabilities along the long equilibrium trajectory and nonequilibrium trajectories. Finally, two-dimensional IR-Raman spectroscopy strongly depends on the nonlinearity of the polarizabilities and dipole moments (electrical anharmonicity). To study these and other phenomena that are sensitive to such nonlinearities, we implemented a simple truncated dipole-induced-dipole model for water in \ipi's internal Fortran driver (option \texttt{-m water\_dip\_pol}). Classical and TRPMD two-dimensional IR-Raman spectra of liquid water at 300 K are displayed in Fig.~\ref{fig:2D_IR_Raman}b showing that NQEs play a minor role in this observable. Overall, this example illustrates how minor scripting and modifications outside of the core \ipi~code can be used to implement new methods and study various types of dynamical properties.

\begin{figure}[htp]
    \centering
        \includegraphics[width=0.9\columnwidth]{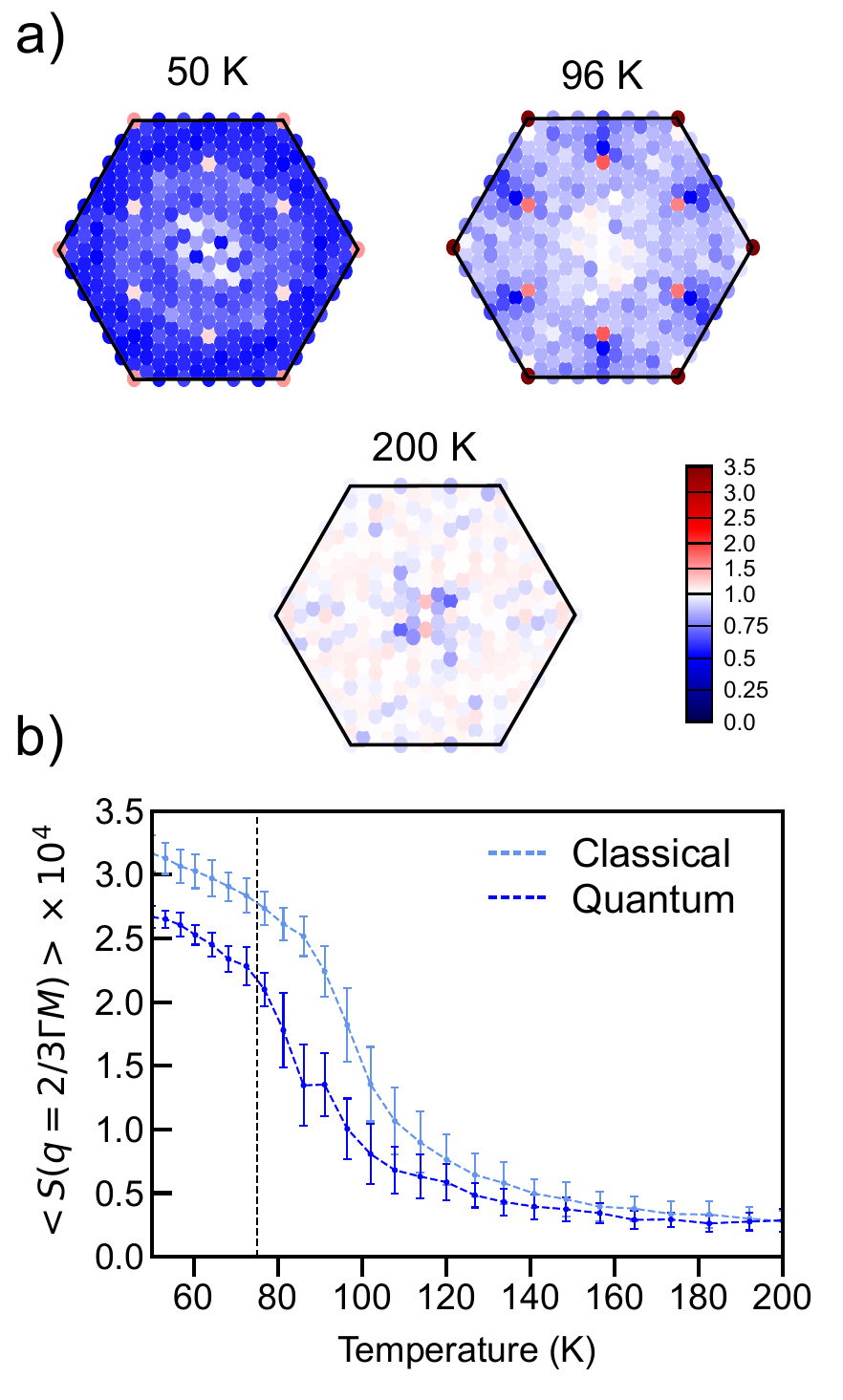}
    \caption{(a) Ratio of the structure factors \rev{obtained from classical ($S_{\mathrm{CL}}$) and quantum path-integral ($S_{\mathrm{PI}}$) simulations} $S_\text{CL}/S_\text{PI}$ performed on a $18 \times 18$ supercell of monolayer 1H-TaS$_2$. A value close
to 1 (indicated in white) corresponds to minimal differences between classical and
quantum simulations. The peaks at $\bm{q} = 2/3 \Gamma M$ and $\bm{q} = K$ for $T = 50$ K are characteristic of the $3 \times 3$
CDW phase. \rev{(b)} Temperature-dependent structure factor $\langle S(\bm{q})\rangle_T$ at the characteristic CDW wavevector
$\bm{q}=2/3 \Gamma M$ obtained from classical MD replica-exchange MD (light blue) and path integral replica-exchange MD (\rev{dark} blue). The experimental value (for the bulk) is represented by the vertical black dashed line. The shift of the inflection point of the PIMD curve toward the experimental value is due to
nuclear quantum effects. Figure and caption adapted from Ref.~\citenum{schob+24spp} (CC-BY Licence).}
     \label{fig:elphmod-cdw}
\end{figure}

\subsection{Melting of Charge Density Waves in 2D Materials}

Downfolded lattice models aim to reduce the complexity of the full electronic problem by treating only a few low-energy electronic states (around the Fermi energy) explicitly and defining a simplified Hamiltonian where all other high-energy states are integrated out.\cite{aryas+book} The parameters required for this dimensionality reduction can be obtained directly from \textit{ab initio} calculations. There are several ways to include the effect of screening of the high-energy electrons on the interaction terms related to the explicit degrees of freedom of the downfolded Hamiltonian,\cite{berge+23prx} and making appropriate choices in this respect is necessary to reproduce the fully \textit{ab initio} potential energy surface of certain materials.\cite{schob+24spp}

\ipi{} has recently been employed in this context, through its new interface with the \lstinline{elphmod} code.\cite{elphmod} With \lstinline{elphmod}, it is possible to set up downfolded models for the electronic interactions and for the electron-nuclear interactions, the latter being far less common to find in literature. Having access to the nuclear degrees of freedom in the Hamiltonian makes it possible to calculate nuclear forces efficiently, in the order of milliseconds per MD step per atom.\cite{schob+24spp}

In connection with \ipi{}, this capability allows the simulation of different types of molecular dynamics with such downfolded models. While \rev{MLIPs} could play a similar role, there is an  added value of physical interpretability when employing these models. Moreover, even if downfolded models cannot be applied to all types of materials, for suitable systems they are straightforward to parametrize.

In Ref.~\citenum{schob+24spp}, large-scale replica-exchange and path-integral replica exchange simulations were performed with the \ipi{}/\lstinline{elphmod} combination in order to study the charge-density-wave (CDW) phase transition in monolayer 1H-TaS$_2$. The downfolded model was obtained from PBE density-functional theory calculations. In particular, the phase transition between the 3$\times$3 CDW and the symmetric structure was studied. We note that capturing the nuclear dynamics involved in such phase transitions calls for techniques that include non-perturbative anharmonic effects such as (PI)MD. In addition, as explicitly shown in Ref.~\citenum{schob+24spp}, finite-size effects on the phase transition are significant for small unit cells of these systems, calling for potentials that are efficient \rev{and accurate} enough to allow the simulation of larger unit cells containing thousands of atoms.  

Experimentally, the ``melting'' temperature of the 3 $\times$ 3 CDW in bulk TaS$_2$ is measured at around 75 K.\cite{zhao+20prb} While the phase-transition temperature is not known for the monolayer, it is accepted that it should not strongly deviate from the bulk temperature. 
In Fig.~\ref{fig:elphmod-cdw}, we show the ratio between classical and quantum $\mathbf{q}$-resolved structure factors \rev{in panel (a)} and the intensity of the CDW $3\times3$ diffraction peaks with varying temperature \rev{in panel (b)}. The inflection point of these curves is shown to correlate with a peak in heat capacity calculated from the simulations. Nuclear quantum effects decrease the transition temperature by around 20 K, bringing it in closer agreement with the experimental estimate. 

Such simulations open the path for a closer analysis of particular phonon modes that contribute to the CDW phase transition, an analysis of the co-existence of different CDW phases at metastable regions, and the mapping of more complex CDW phase diagrams. Given the close connection between CDW phases and superconductivity, this is a promising field of study. An example of how to set up and run the \lstinline{elphmod} code in connection with \ipi{} is included in the \rev{\textit{examples/clients} directory}.

\section{Conclusions}

\rev{
 In this paper, we described the current capabilities of the \ipi code.
The fundamental reasoning behind the design decisions made during its early development, namely an architecture that allows rapid prototyping in Python, and utilities to facilitate implementing complicated sampling algorithms, still prove to be very useful. However, some of the underlying assumptions made in the past are no longer valid.
For example, the widespread adoption of MLIPs has dramatically reduced the cost of energy evaluation relative to explicit electronic-structure calculations, making the overhead associated with the evolution of the atomic coordinates a more pressing concern. 
In this paper, we showed that relatively superficial changes in the implementation of some of the core classes in \ipi{} substantially accelerated the most severe bottlenecks, ensuring that \ipi{} can now be used in combination with state-of-the-art MLIPs without adding significant overhead, even when pushing the energy evaluation to high levels of parallelism.

At the same time, the modular design of \ipi\ makes it easy to implement radically new modeling ideas, bringing algorithmic speed-ups and new useful features to the codebase. 
A bosonic PIMD implementation with quadratic scaling, the implementation of the ring polymer instanton with explicit friction method, a framework for dynamics driven by an electric field, a framework for simulations of light-matter coupling, and a better interface with ensembles of MLIPs including uncertainty estimation, are the five examples we highlight. 
Many useful, and sophisticated, simulation workflows can also be realized without major implementation efforts, by simply combining existing methods with external scripts and suitable post-processing. We demonstrate this with three applications: path integral coarse-grained simulations for vibrational spectra, calculations of the 2D-IR spectrum of water, and the melting of 2D charge-density waves. 

As MLIPs become faster and enable simulations with larger lengths and time scales, it might be necessary to revisit some of the deeper architectural decisions of \ipi to further reduce the overhead inherent to the serial Python-based implementation.
The work that was made in preparation for this release demonstrates that these kinds of optimizations can coexist with a modular design, that supports the fast implementation of new and exciting methods for advanced molecular simulations.
}

\begin{acknowledgments}

Y.L. has been funded by the Deutsche Forschungsgemeinschaft (DFG, German Research Foundation) project number 467724959. 
V.K. acknowledges support from the Ernest Oppenheimer Early Career Fellowship and the Sydney Harvey Junior Research Fellowship, Churchill College, University of Cambridge. V.K is grateful for computational support from the Swiss National Supercomputing Centre under project s1209, the UK national high-performance computing service, ARCHER2, for which access was obtained via the UKCP consortium and the EPSRC grant ref EP/P022561/1, and the Cambridge Service for Data Driven Discovery (CSD3). T.B. acknowledges financial support from the Swiss National Science Foundation through the Early Postdoc Mobility Fellowship (grant number P2ELP2-199757). B.H. acknowledges support by the USA-Israel Binational Science Foundation (grant No. 2020083) and the Israel Science Foundation (grants No.\ 1037/22 and 1312/22). Y.F. was supported by Schmidt Science Fellows, in partnership with the Rhodes Trust. T.E.L. is supported by start-up funds from the University of Delaware Department of Physics and Astronomy. M.R. and E.S. acknowledge computer time from the Max Planck Computing and Data Facility (MPCDF), funding from the IMPRS-UFAST program and the Lise-Meitner Excellence program. M.R. thanks Jan Berges for a careful read of section V C.
M.C., M.K. and D.T. acknowledge funding from the Swiss National Science Foundation (SNSF) under the projects CRSII5\_202296 and 200020\_214879. M.C. and D.T. also acknowledge the support from the MARVEL National Centre of Competence in Research (NCCR) M.C.~acknowledges funding from the European Research Council (ERC) under the European Union’s Horizon 2020 research and innovation programme Grant No.~101001890-FIAMMA. \rev{We thank Eli Fields for helpful comments.}
\end{acknowledgments}

\section{Data Availability}
\rev{\ipi{} can be obtained downloaded from \url{https://github.com/i-pi/i-pi}. The tag \lstinline{v3.0.0-beta} was
used to generate the examples and benchmarks reported in this manuscript}. 
The data required to reproduce the figures and benchmark simulations are available at \href{https://github.com/i-pi/ipiv3_data}{\texttt{https://github.com/i-pi/ipiv3\_data}}.


\begin{thebibliography}{202}%
\makeatletter
\providecommand \@ifxundefined [1]{%
 \@ifx{#1\undefined}
}%
\providecommand \@ifnum [1]{%
 \ifnum #1\expandafter \@firstoftwo
 \else \expandafter \@secondoftwo
 \fi
}%
\providecommand \@ifx [1]{%
 \ifx #1\expandafter \@firstoftwo
 \else \expandafter \@secondoftwo
 \fi
}%
\providecommand \natexlab [1]{#1}%
\providecommand \enquote  [1]{``#1''}%
\providecommand \bibnamefont  [1]{#1}%
\providecommand \bibfnamefont [1]{#1}%
\providecommand \citenamefont [1]{#1}%
\providecommand \href@noop [0]{\@secondoftwo}%
\providecommand \href [0]{\begingroup \@sanitize@url \@href}%
\providecommand \@href[1]{\@@startlink{#1}\@@href}%
\providecommand \@@href[1]{\endgroup#1\@@endlink}%
\providecommand \@sanitize@url [0]{\catcode `\\12\catcode `\$12\catcode
  `\&12\catcode `\#12\catcode `\^12\catcode `\_12\catcode `\%12\relax}%
\providecommand \@@startlink[1]{}%
\providecommand \@@endlink[0]{}%
\providecommand \url  [0]{\begingroup\@sanitize@url \@url }%
\providecommand \@url [1]{\endgroup\@href {#1}{\urlprefix }}%
\providecommand \urlprefix  [0]{URL }%
\providecommand \Eprint [0]{\href }%
\providecommand \doibase [0]{https://doi.org/}%
\providecommand \selectlanguage [0]{\@gobble}%
\providecommand \bibinfo  [0]{\@secondoftwo}%
\providecommand \bibfield  [0]{\@secondoftwo}%
\providecommand \translation [1]{[#1]}%
\providecommand \BibitemOpen [0]{}%
\providecommand \bibitemStop [0]{}%
\providecommand \bibitemNoStop [0]{.\EOS\space}%
\providecommand \EOS [0]{\spacefactor3000\relax}%
\providecommand \BibitemShut  [1]{\csname bibitem#1\endcsname}%
\let\auto@bib@innerbib\@empty
\bibitem [{\citenamefont {von Lilienfeld}\ and\ \citenamefont
  {Burke}(2020)}]{vonLilienfeld2020}%
  \BibitemOpen
  \bibfield  {author} {\bibinfo {author} {\bibfnamefont {O.~A.}\ \bibnamefont
  {von Lilienfeld}}\ and\ \bibinfo {author} {\bibfnamefont {K.}~\bibnamefont
  {Burke}},\ }\bibfield  {title} {\bibinfo {title} {Retrospective on a decade
  of machine learning for chemical discovery},\ }\href
  {https://doi.org/10.1038/s41467-020-18556-9} {\bibfield  {journal} {\bibinfo
  {journal} {Nature Communications}\ }\textbf {\bibinfo {volume} {11}},\
  \bibinfo {pages} {4895} (\bibinfo {year} {2020})}\BibitemShut {NoStop}%
\bibitem [{\citenamefont {Butler}\ \emph {et~al.}(2018)\citenamefont {Butler},
  \citenamefont {Davies}, \citenamefont {Cartwright}, \citenamefont {Isayev},\
  and\ \citenamefont {Walsh}}]{Butler2018}%
  \BibitemOpen
  \bibfield  {author} {\bibinfo {author} {\bibfnamefont {K.~T.}\ \bibnamefont
  {Butler}}, \bibinfo {author} {\bibfnamefont {D.~W.}\ \bibnamefont {Davies}},
  \bibinfo {author} {\bibfnamefont {H.}~\bibnamefont {Cartwright}}, \bibinfo
  {author} {\bibfnamefont {O.}~\bibnamefont {Isayev}},\ and\ \bibinfo {author}
  {\bibfnamefont {A.}~\bibnamefont {Walsh}},\ }\bibfield  {title} {\bibinfo
  {title} {Machine learning for molecular and materials science},\ }\href
  {https://doi.org/10.1038/s41586-018-0337-2} {\bibfield  {journal} {\bibinfo
  {journal} {Nature}\ }\textbf {\bibinfo {volume} {559}},\ \bibinfo {pages}
  {547} (\bibinfo {year} {2018})}\BibitemShut {NoStop}%
\bibitem [{\citenamefont {Kulik}\ \emph {et~al.}(2022)\citenamefont {Kulik},
  \citenamefont {Hammerschmidt}, \citenamefont {Schmidt}, \citenamefont
  {Botti}, \citenamefont {Marques}, \citenamefont {Boley}, \citenamefont
  {Scheffler}, \citenamefont {Todorović}, \citenamefont {Rinke}, \citenamefont
  {Oses}, \citenamefont {Smolyanyuk}, \citenamefont {Curtarolo}, \citenamefont
  {Tkatchenko}, \citenamefont {Bartók}, \citenamefont {Manzhos}, \citenamefont
  {Ihara}, \citenamefont {Carrington}, \citenamefont {Behler}, \citenamefont
  {Isayev}, \citenamefont {Veit}, \citenamefont {Grisafi}, \citenamefont
  {Nigam}, \citenamefont {Ceriotti}, \citenamefont {Schütt}, \citenamefont
  {Westermayr}, \citenamefont {Gastegger}, \citenamefont {Maurer},
  \citenamefont {Kalita}, \citenamefont {Burke}, \citenamefont {Nagai},
  \citenamefont {Akashi}, \citenamefont {Sugino}, \citenamefont {Hermann},
  \citenamefont {Noé}, \citenamefont {Pilati}, \citenamefont {Draxl},
  \citenamefont {Kuban}, \citenamefont {Rigamonti}, \citenamefont {Scheidgen},
  \citenamefont {Esters}, \citenamefont {Hicks}, \citenamefont {Toher},
  \citenamefont {Balachandran}, \citenamefont {Tamblyn}, \citenamefont
  {Whitelam}, \citenamefont {Bellinger},\ and\ \citenamefont
  {Ghiringhelli}}]{Kulik_2022}%
  \BibitemOpen
  \bibfield  {author} {\bibinfo {author} {\bibfnamefont {H.~J.}\ \bibnamefont
  {Kulik}}, \bibinfo {author} {\bibfnamefont {T.}~\bibnamefont
  {Hammerschmidt}}, \bibinfo {author} {\bibfnamefont {J.}~\bibnamefont
  {Schmidt}}, \bibinfo {author} {\bibfnamefont {S.}~\bibnamefont {Botti}},
  \bibinfo {author} {\bibfnamefont {M.~A.~L.}\ \bibnamefont {Marques}},
  \bibinfo {author} {\bibfnamefont {M.}~\bibnamefont {Boley}}, \bibinfo
  {author} {\bibfnamefont {M.}~\bibnamefont {Scheffler}}, \bibinfo {author}
  {\bibfnamefont {M.}~\bibnamefont {Todorović}}, \bibinfo {author}
  {\bibfnamefont {P.}~\bibnamefont {Rinke}}, \bibinfo {author} {\bibfnamefont
  {C.}~\bibnamefont {Oses}}, \bibinfo {author} {\bibfnamefont {A.}~\bibnamefont
  {Smolyanyuk}}, \bibinfo {author} {\bibfnamefont {S.}~\bibnamefont
  {Curtarolo}}, \bibinfo {author} {\bibfnamefont {A.}~\bibnamefont
  {Tkatchenko}}, \bibinfo {author} {\bibfnamefont {A.~P.}\ \bibnamefont
  {Bartók}}, \bibinfo {author} {\bibfnamefont {S.}~\bibnamefont {Manzhos}},
  \bibinfo {author} {\bibfnamefont {M.}~\bibnamefont {Ihara}}, \bibinfo
  {author} {\bibfnamefont {T.}~\bibnamefont {Carrington}}, \bibinfo {author}
  {\bibfnamefont {J.}~\bibnamefont {Behler}}, \bibinfo {author} {\bibfnamefont
  {O.}~\bibnamefont {Isayev}}, \bibinfo {author} {\bibfnamefont
  {M.}~\bibnamefont {Veit}}, \bibinfo {author} {\bibfnamefont {A.}~\bibnamefont
  {Grisafi}}, \bibinfo {author} {\bibfnamefont {J.}~\bibnamefont {Nigam}},
  \bibinfo {author} {\bibfnamefont {M.}~\bibnamefont {Ceriotti}}, \bibinfo
  {author} {\bibfnamefont {K.~T.}\ \bibnamefont {Schütt}}, \bibinfo {author}
  {\bibfnamefont {J.}~\bibnamefont {Westermayr}}, \bibinfo {author}
  {\bibfnamefont {M.}~\bibnamefont {Gastegger}}, \bibinfo {author}
  {\bibfnamefont {R.~J.}\ \bibnamefont {Maurer}}, \bibinfo {author}
  {\bibfnamefont {B.}~\bibnamefont {Kalita}}, \bibinfo {author} {\bibfnamefont
  {K.}~\bibnamefont {Burke}}, \bibinfo {author} {\bibfnamefont
  {R.}~\bibnamefont {Nagai}}, \bibinfo {author} {\bibfnamefont
  {R.}~\bibnamefont {Akashi}}, \bibinfo {author} {\bibfnamefont
  {O.}~\bibnamefont {Sugino}}, \bibinfo {author} {\bibfnamefont
  {J.}~\bibnamefont {Hermann}}, \bibinfo {author} {\bibfnamefont
  {F.}~\bibnamefont {Noé}}, \bibinfo {author} {\bibfnamefont {S.}~\bibnamefont
  {Pilati}}, \bibinfo {author} {\bibfnamefont {C.}~\bibnamefont {Draxl}},
  \bibinfo {author} {\bibfnamefont {M.}~\bibnamefont {Kuban}}, \bibinfo
  {author} {\bibfnamefont {S.}~\bibnamefont {Rigamonti}}, \bibinfo {author}
  {\bibfnamefont {M.}~\bibnamefont {Scheidgen}}, \bibinfo {author}
  {\bibfnamefont {M.}~\bibnamefont {Esters}}, \bibinfo {author} {\bibfnamefont
  {D.}~\bibnamefont {Hicks}}, \bibinfo {author} {\bibfnamefont
  {C.}~\bibnamefont {Toher}}, \bibinfo {author} {\bibfnamefont {P.~V.}\
  \bibnamefont {Balachandran}}, \bibinfo {author} {\bibfnamefont
  {I.}~\bibnamefont {Tamblyn}}, \bibinfo {author} {\bibfnamefont
  {S.}~\bibnamefont {Whitelam}}, \bibinfo {author} {\bibfnamefont
  {C.}~\bibnamefont {Bellinger}},\ and\ \bibinfo {author} {\bibfnamefont
  {L.~M.}\ \bibnamefont {Ghiringhelli}},\ }\bibfield  {title} {\bibinfo {title}
  {Roadmap on machine learning in electronic structure},\ }\href
  {https://doi.org/10.1088/2516-1075/ac572f} {\bibfield  {journal} {\bibinfo
  {journal} {Electronic Structure}\ }\textbf {\bibinfo {volume} {4}},\ \bibinfo
  {pages} {023004} (\bibinfo {year} {2022})}\BibitemShut {NoStop}%
\bibitem [{\citenamefont {Deringer}\ \emph {et~al.}(2019)\citenamefont
  {Deringer}, \citenamefont {Caro},\ and\ \citenamefont
  {Csányi}}]{Deringer_AdvMat_2019}%
  \BibitemOpen
  \bibfield  {author} {\bibinfo {author} {\bibfnamefont {V.~L.}\ \bibnamefont
  {Deringer}}, \bibinfo {author} {\bibfnamefont {M.~A.}\ \bibnamefont {Caro}},\
  and\ \bibinfo {author} {\bibfnamefont {G.}~\bibnamefont {Csányi}},\
  }\bibfield  {title} {\bibinfo {title} {Machine learning interatomic
  potentials as emerging tools for materials science},\ }\href
  {https://doi.org/https://doi.org/10.1002/adma.201902765} {\bibfield
  {journal} {\bibinfo  {journal} {Advanced Materials}\ }\textbf {\bibinfo
  {volume} {31}},\ \bibinfo {pages} {1902765} (\bibinfo {year}
  {2019})}\BibitemShut {NoStop}%
\bibitem [{\citenamefont {Bartók}\ \emph {et~al.}(2017)\citenamefont
  {Bartók}, \citenamefont {De}, \citenamefont {Poelking}, \citenamefont
  {Bernstein}, \citenamefont {Kermode}, \citenamefont {Csányi},\ and\
  \citenamefont {Ceriotti}}]{10.1126/sciadv.1701816}%
  \BibitemOpen
  \bibfield  {author} {\bibinfo {author} {\bibfnamefont {A.~P.}\ \bibnamefont
  {Bartók}}, \bibinfo {author} {\bibfnamefont {S.}~\bibnamefont {De}},
  \bibinfo {author} {\bibfnamefont {C.}~\bibnamefont {Poelking}}, \bibinfo
  {author} {\bibfnamefont {N.}~\bibnamefont {Bernstein}}, \bibinfo {author}
  {\bibfnamefont {J.~R.}\ \bibnamefont {Kermode}}, \bibinfo {author}
  {\bibfnamefont {G.}~\bibnamefont {Csányi}},\ and\ \bibinfo {author}
  {\bibfnamefont {M.}~\bibnamefont {Ceriotti}},\ }\bibfield  {title} {\bibinfo
  {title} {Machine learning unifies the modeling of materials and molecules},\
  }\href {https://doi.org/10.1126/sciadv.1701816} {\bibfield  {journal}
  {\bibinfo  {journal} {Science Advances}\ }\textbf {\bibinfo {volume} {3}},\
  \bibinfo {pages} {e1701816} (\bibinfo {year} {2017})}\BibitemShut {NoStop}%
\bibitem [{\citenamefont {Tao}\ \emph {et~al.}(2021)\citenamefont {Tao},
  \citenamefont {Xu}, \citenamefont {Li},\ and\ \citenamefont {Lu}}]{Tao2021}%
  \BibitemOpen
  \bibfield  {author} {\bibinfo {author} {\bibfnamefont {Q.}~\bibnamefont
  {Tao}}, \bibinfo {author} {\bibfnamefont {P.}~\bibnamefont {Xu}}, \bibinfo
  {author} {\bibfnamefont {M.}~\bibnamefont {Li}},\ and\ \bibinfo {author}
  {\bibfnamefont {W.}~\bibnamefont {Lu}},\ }\bibfield  {title} {\bibinfo
  {title} {Machine learning for perovskite materials design and discovery},\
  }\href {https://doi.org/10.1038/s41524-021-00495-8} {\bibfield  {journal}
  {\bibinfo  {journal} {npj Computational Materials}\ }\textbf {\bibinfo
  {volume} {7}},\ \bibinfo {pages} {23} (\bibinfo {year} {2021})}\BibitemShut
  {NoStop}%
\bibitem [{\citenamefont {Lu}\ \emph {et~al.}(2021)\citenamefont {Lu},
  \citenamefont {Wang}, \citenamefont {Chen}, \citenamefont {Lin},
  \citenamefont {Car}, \citenamefont {E}, \citenamefont {Jia},\ and\
  \citenamefont {Zhang}}]{Lu_CPC_2021}%
  \BibitemOpen
  \bibfield  {author} {\bibinfo {author} {\bibfnamefont {D.}~\bibnamefont
  {Lu}}, \bibinfo {author} {\bibfnamefont {H.}~\bibnamefont {Wang}}, \bibinfo
  {author} {\bibfnamefont {M.}~\bibnamefont {Chen}}, \bibinfo {author}
  {\bibfnamefont {L.}~\bibnamefont {Lin}}, \bibinfo {author} {\bibfnamefont
  {R.}~\bibnamefont {Car}}, \bibinfo {author} {\bibfnamefont {W.}~\bibnamefont
  {E}}, \bibinfo {author} {\bibfnamefont {W.}~\bibnamefont {Jia}},\ and\
  \bibinfo {author} {\bibfnamefont {L.}~\bibnamefont {Zhang}},\ }\bibfield
  {title} {\bibinfo {title} {86 pflops deep potential molecular dynamics
  simulation of 100 million atoms with ab initio accuracy},\ }\href
  {https://doi.org/https://doi.org/10.1016/j.cpc.2020.107624} {\bibfield
  {journal} {\bibinfo  {journal} {Computer Physics Communications}\ }\textbf
  {\bibinfo {volume} {259}},\ \bibinfo {pages} {107624} (\bibinfo {year}
  {2021})}\BibitemShut {NoStop}%
\bibitem [{\citenamefont {Behler}(2017)}]{Behler_Ang_2017}%
  \BibitemOpen
  \bibfield  {author} {\bibinfo {author} {\bibfnamefont {J.}~\bibnamefont
  {Behler}},\ }\bibfield  {title} {\bibinfo {title} {First principles neural
  network potentials for reactive simulations of large molecular and condensed
  systems},\ }\href {https://doi.org/https://doi.org/10.1002/anie.201703114}
  {\bibfield  {journal} {\bibinfo  {journal} {Angewandte Chemie International
  Edition}\ }\textbf {\bibinfo {volume} {56}},\ \bibinfo {pages} {12828}
  (\bibinfo {year} {2017})}\BibitemShut {NoStop}%
\bibitem [{\citenamefont {Batatia}\ \emph {et~al.}(2024)\citenamefont
  {Batatia}, \citenamefont {Benner}, \citenamefont {Chiang}, \citenamefont
  {Elena}, \citenamefont {Kovács}, \citenamefont {Riebesell}, \citenamefont
  {Advincula}, \citenamefont {Asta}, \citenamefont {Avaylon}, \citenamefont
  {Baldwin}, \citenamefont {Berger}, \citenamefont {Bernstein}, \citenamefont
  {Bhowmik}, \citenamefont {Blau}, \citenamefont {Carare}, \citenamefont
  {Darby}, \citenamefont {De}, \citenamefont {Pia}, \citenamefont {Deringer},
  \citenamefont {Elijošius}, \citenamefont {El-Machachi}, \citenamefont
  {Falcioni}, \citenamefont {Fako}, \citenamefont {Ferrari}, \citenamefont
  {Genreith-Schriever}, \citenamefont {George}, \citenamefont {Goodall},
  \citenamefont {Grey}, \citenamefont {Grigorev}, \citenamefont {Han},
  \citenamefont {Handley}, \citenamefont {Heenen}, \citenamefont {Hermansson},
  \citenamefont {Holm}, \citenamefont {Jaafar}, \citenamefont {Hofmann},
  \citenamefont {Jakob}, \citenamefont {Jung}, \citenamefont {Kapil},
  \citenamefont {Kaplan}, \citenamefont {Karimitari}, \citenamefont {Kermode},
  \citenamefont {Kroupa}, \citenamefont {Kullgren}, \citenamefont {Kuner},
  \citenamefont {Kuryla}, \citenamefont {Liepuoniute}, \citenamefont {Margraf},
  \citenamefont {Magdău}, \citenamefont {Michaelides}, \citenamefont {Moore},
  \citenamefont {Naik}, \citenamefont {Niblett}, \citenamefont {Norwood},
  \citenamefont {O'Neill}, \citenamefont {Ortner}, \citenamefont {Persson},
  \citenamefont {Reuter}, \citenamefont {Rosen}, \citenamefont {Schaaf},
  \citenamefont {Schran}, \citenamefont {Shi}, \citenamefont {Sivonxay},
  \citenamefont {Stenczel}, \citenamefont {Svahn}, \citenamefont {Sutton},
  \citenamefont {Swinburne}, \citenamefont {Tilly}, \citenamefont {van~der
  Oord}, \citenamefont {Varga-Umbrich}, \citenamefont {Vegge}, \citenamefont
  {Vondrák}, \citenamefont {Wang}, \citenamefont {Witt}, \citenamefont
  {Zills},\ and\ \citenamefont {Csányi}}]{batatia2024foundation}%
  \BibitemOpen
  \bibfield  {author} {\bibinfo {author} {\bibfnamefont {I.}~\bibnamefont
  {Batatia}}, \bibinfo {author} {\bibfnamefont {P.}~\bibnamefont {Benner}},
  \bibinfo {author} {\bibfnamefont {Y.}~\bibnamefont {Chiang}}, \bibinfo
  {author} {\bibfnamefont {A.~M.}\ \bibnamefont {Elena}}, \bibinfo {author}
  {\bibfnamefont {D.~P.}\ \bibnamefont {Kovács}}, \bibinfo {author}
  {\bibfnamefont {J.}~\bibnamefont {Riebesell}}, \bibinfo {author}
  {\bibfnamefont {X.~R.}\ \bibnamefont {Advincula}}, \bibinfo {author}
  {\bibfnamefont {M.}~\bibnamefont {Asta}}, \bibinfo {author} {\bibfnamefont
  {M.}~\bibnamefont {Avaylon}}, \bibinfo {author} {\bibfnamefont {W.~J.}\
  \bibnamefont {Baldwin}}, \bibinfo {author} {\bibfnamefont {F.}~\bibnamefont
  {Berger}}, \bibinfo {author} {\bibfnamefont {N.}~\bibnamefont {Bernstein}},
  \bibinfo {author} {\bibfnamefont {A.}~\bibnamefont {Bhowmik}}, \bibinfo
  {author} {\bibfnamefont {S.~M.}\ \bibnamefont {Blau}}, \bibinfo {author}
  {\bibfnamefont {V.}~\bibnamefont {Carare}}, \bibinfo {author} {\bibfnamefont
  {J.~P.}\ \bibnamefont {Darby}}, \bibinfo {author} {\bibfnamefont
  {S.}~\bibnamefont {De}}, \bibinfo {author} {\bibfnamefont {F.~D.}\
  \bibnamefont {Pia}}, \bibinfo {author} {\bibfnamefont {V.~L.}\ \bibnamefont
  {Deringer}}, \bibinfo {author} {\bibfnamefont {R.}~\bibnamefont
  {Elijošius}}, \bibinfo {author} {\bibfnamefont {Z.}~\bibnamefont
  {El-Machachi}}, \bibinfo {author} {\bibfnamefont {F.}~\bibnamefont
  {Falcioni}}, \bibinfo {author} {\bibfnamefont {E.}~\bibnamefont {Fako}},
  \bibinfo {author} {\bibfnamefont {A.~C.}\ \bibnamefont {Ferrari}}, \bibinfo
  {author} {\bibfnamefont {A.}~\bibnamefont {Genreith-Schriever}}, \bibinfo
  {author} {\bibfnamefont {J.}~\bibnamefont {George}}, \bibinfo {author}
  {\bibfnamefont {R.~E.~A.}\ \bibnamefont {Goodall}}, \bibinfo {author}
  {\bibfnamefont {C.~P.}\ \bibnamefont {Grey}}, \bibinfo {author}
  {\bibfnamefont {P.}~\bibnamefont {Grigorev}}, \bibinfo {author}
  {\bibfnamefont {S.}~\bibnamefont {Han}}, \bibinfo {author} {\bibfnamefont
  {W.}~\bibnamefont {Handley}}, \bibinfo {author} {\bibfnamefont {H.~H.}\
  \bibnamefont {Heenen}}, \bibinfo {author} {\bibfnamefont {K.}~\bibnamefont
  {Hermansson}}, \bibinfo {author} {\bibfnamefont {C.}~\bibnamefont {Holm}},
  \bibinfo {author} {\bibfnamefont {J.}~\bibnamefont {Jaafar}}, \bibinfo
  {author} {\bibfnamefont {S.}~\bibnamefont {Hofmann}}, \bibinfo {author}
  {\bibfnamefont {K.~S.}\ \bibnamefont {Jakob}}, \bibinfo {author}
  {\bibfnamefont {H.}~\bibnamefont {Jung}}, \bibinfo {author} {\bibfnamefont
  {V.}~\bibnamefont {Kapil}}, \bibinfo {author} {\bibfnamefont {A.~D.}\
  \bibnamefont {Kaplan}}, \bibinfo {author} {\bibfnamefont {N.}~\bibnamefont
  {Karimitari}}, \bibinfo {author} {\bibfnamefont {J.~R.}\ \bibnamefont
  {Kermode}}, \bibinfo {author} {\bibfnamefont {N.}~\bibnamefont {Kroupa}},
  \bibinfo {author} {\bibfnamefont {J.}~\bibnamefont {Kullgren}}, \bibinfo
  {author} {\bibfnamefont {M.~C.}\ \bibnamefont {Kuner}}, \bibinfo {author}
  {\bibfnamefont {D.}~\bibnamefont {Kuryla}}, \bibinfo {author} {\bibfnamefont
  {G.}~\bibnamefont {Liepuoniute}}, \bibinfo {author} {\bibfnamefont {J.~T.}\
  \bibnamefont {Margraf}}, \bibinfo {author} {\bibfnamefont {I.-B.}\
  \bibnamefont {Magdău}}, \bibinfo {author} {\bibfnamefont {A.}~\bibnamefont
  {Michaelides}}, \bibinfo {author} {\bibfnamefont {J.~H.}\ \bibnamefont
  {Moore}}, \bibinfo {author} {\bibfnamefont {A.~A.}\ \bibnamefont {Naik}},
  \bibinfo {author} {\bibfnamefont {S.~P.}\ \bibnamefont {Niblett}}, \bibinfo
  {author} {\bibfnamefont {S.~W.}\ \bibnamefont {Norwood}}, \bibinfo {author}
  {\bibfnamefont {N.}~\bibnamefont {O'Neill}}, \bibinfo {author} {\bibfnamefont
  {C.}~\bibnamefont {Ortner}}, \bibinfo {author} {\bibfnamefont {K.~A.}\
  \bibnamefont {Persson}}, \bibinfo {author} {\bibfnamefont {K.}~\bibnamefont
  {Reuter}}, \bibinfo {author} {\bibfnamefont {A.~S.}\ \bibnamefont {Rosen}},
  \bibinfo {author} {\bibfnamefont {L.~L.}\ \bibnamefont {Schaaf}}, \bibinfo
  {author} {\bibfnamefont {C.}~\bibnamefont {Schran}}, \bibinfo {author}
  {\bibfnamefont {B.~X.}\ \bibnamefont {Shi}}, \bibinfo {author} {\bibfnamefont
  {E.}~\bibnamefont {Sivonxay}}, \bibinfo {author} {\bibfnamefont {T.~K.}\
  \bibnamefont {Stenczel}}, \bibinfo {author} {\bibfnamefont {V.}~\bibnamefont
  {Svahn}}, \bibinfo {author} {\bibfnamefont {C.}~\bibnamefont {Sutton}},
  \bibinfo {author} {\bibfnamefont {T.~D.}\ \bibnamefont {Swinburne}}, \bibinfo
  {author} {\bibfnamefont {J.}~\bibnamefont {Tilly}}, \bibinfo {author}
  {\bibfnamefont {C.}~\bibnamefont {van~der Oord}}, \bibinfo {author}
  {\bibfnamefont {E.}~\bibnamefont {Varga-Umbrich}}, \bibinfo {author}
  {\bibfnamefont {T.}~\bibnamefont {Vegge}}, \bibinfo {author} {\bibfnamefont
  {M.}~\bibnamefont {Vondrák}}, \bibinfo {author} {\bibfnamefont
  {Y.}~\bibnamefont {Wang}}, \bibinfo {author} {\bibfnamefont {W.~C.}\
  \bibnamefont {Witt}}, \bibinfo {author} {\bibfnamefont {F.}~\bibnamefont
  {Zills}},\ and\ \bibinfo {author} {\bibfnamefont {G.}~\bibnamefont
  {Csányi}},\ }\href@noop {} {\bibinfo {title} {A foundation model for
  atomistic materials chemistry}} (\bibinfo {year} {2024}),\ \Eprint
  {https://arxiv.org/abs/2401.00096} {arXiv:2401.00096} \BibitemShut {NoStop}%
\bibitem [{\citenamefont {Unke}\ \emph {et~al.}(2021)\citenamefont {Unke},
  \citenamefont {Chmiela}, \citenamefont {Sauceda}, \citenamefont {Gastegger},
  \citenamefont {Poltavsky}, \citenamefont {Schütt}, \citenamefont
  {Tkatchenko},\ and\ \citenamefont {Müller}}]{Unke_ChemRev_2021}%
  \BibitemOpen
  \bibfield  {author} {\bibinfo {author} {\bibfnamefont {O.~T.}\ \bibnamefont
  {Unke}}, \bibinfo {author} {\bibfnamefont {S.}~\bibnamefont {Chmiela}},
  \bibinfo {author} {\bibfnamefont {H.~E.}\ \bibnamefont {Sauceda}}, \bibinfo
  {author} {\bibfnamefont {M.}~\bibnamefont {Gastegger}}, \bibinfo {author}
  {\bibfnamefont {I.}~\bibnamefont {Poltavsky}}, \bibinfo {author}
  {\bibfnamefont {K.~T.}\ \bibnamefont {Schütt}}, \bibinfo {author}
  {\bibfnamefont {A.}~\bibnamefont {Tkatchenko}},\ and\ \bibinfo {author}
  {\bibfnamefont {K.-R.}\ \bibnamefont {Müller}},\ }\bibfield  {title}
  {\bibinfo {title} {Machine learning force fields},\ }\href
  {https://doi.org/10.1021/acs.chemrev.0c01111} {\bibfield  {journal} {\bibinfo
   {journal} {Chemical Reviews}\ }\textbf {\bibinfo {volume} {121}},\ \bibinfo
  {pages} {10142} (\bibinfo {year} {2021})}\BibitemShut {NoStop}%
\bibitem [{\citenamefont {Bruix}\ \emph {et~al.}(2019)\citenamefont {Bruix},
  \citenamefont {Margraf}, \citenamefont {Andersen},\ and\ \citenamefont
  {Reuter}}]{Bruix2019}%
  \BibitemOpen
  \bibfield  {author} {\bibinfo {author} {\bibfnamefont {A.}~\bibnamefont
  {Bruix}}, \bibinfo {author} {\bibfnamefont {J.~T.}\ \bibnamefont {Margraf}},
  \bibinfo {author} {\bibfnamefont {M.}~\bibnamefont {Andersen}},\ and\
  \bibinfo {author} {\bibfnamefont {K.}~\bibnamefont {Reuter}},\ }\bibfield
  {title} {\bibinfo {title} {First-principles-based multiscale modelling of
  heterogeneous catalysis},\ }\href {https://doi.org/10.1038/s41929-019-0298-3}
  {\bibfield  {journal} {\bibinfo  {journal} {Nature Catalysis}\ }\textbf
  {\bibinfo {volume} {2}},\ \bibinfo {pages} {659} (\bibinfo {year}
  {2019})}\BibitemShut {NoStop}%
\bibitem [{\citenamefont {Schlexer-Lamoureux}\ \emph
  {et~al.}(2019)\citenamefont {Schlexer-Lamoureux}, \citenamefont {Winther},
  \citenamefont {Garrido-Torres}, \citenamefont {Streibel}, \citenamefont
  {Zhao}, \citenamefont {Bajdich}, \citenamefont {Abild-Pedersen},\ and\
  \citenamefont {Bligaard}}]{Schlexer_ChemCatCHem_2019}%
  \BibitemOpen
  \bibfield  {author} {\bibinfo {author} {\bibfnamefont {P.}~\bibnamefont
  {Schlexer-Lamoureux}}, \bibinfo {author} {\bibfnamefont {K.~T.}\ \bibnamefont
  {Winther}}, \bibinfo {author} {\bibfnamefont {J.~A.}\ \bibnamefont
  {Garrido-Torres}}, \bibinfo {author} {\bibfnamefont {V.}~\bibnamefont
  {Streibel}}, \bibinfo {author} {\bibfnamefont {M.}~\bibnamefont {Zhao}},
  \bibinfo {author} {\bibfnamefont {M.}~\bibnamefont {Bajdich}}, \bibinfo
  {author} {\bibfnamefont {F.}~\bibnamefont {Abild-Pedersen}},\ and\ \bibinfo
  {author} {\bibfnamefont {T.}~\bibnamefont {Bligaard}},\ }\bibfield  {title}
  {\bibinfo {title} {Machine learning for computational heterogeneous
  catalysis},\ }\href {https://doi.org/https://doi.org/10.1002/cctc.201900595}
  {\bibfield  {journal} {\bibinfo  {journal} {ChemCatChem}\ }\textbf {\bibinfo
  {volume} {11}},\ \bibinfo {pages} {3581} (\bibinfo {year}
  {2019})}\BibitemShut {NoStop}%
\bibitem [{\citenamefont {Ma}\ and\ \citenamefont
  {Liu}(2020)}]{10.1021/acscatal.0c03472}%
  \BibitemOpen
  \bibfield  {author} {\bibinfo {author} {\bibfnamefont {S.}~\bibnamefont
  {Ma}}\ and\ \bibinfo {author} {\bibfnamefont {Z.-P.}\ \bibnamefont {Liu}},\
  }\bibfield  {title} {\bibinfo {title} {Machine learning for atomic simulation
  and activity prediction in heterogeneous catalysis: Current status and
  future},\ }\href {https://doi.org/10.1021/acscatal.0c03472} {\bibfield
  {journal} {\bibinfo  {journal} {ACS Catalysis}\ }\textbf {\bibinfo {volume}
  {10}},\ \bibinfo {pages} {13213} (\bibinfo {year} {2020})}\BibitemShut
  {NoStop}%
\bibitem [{\citenamefont {Cheng}\ \emph {et~al.}(2019)\citenamefont {Cheng},
  \citenamefont {Engel}, \citenamefont {Behler}, \citenamefont {Dellago},\ and\
  \citenamefont {Ceriotti}}]{cheng_ab_2019}%
  \BibitemOpen
  \bibfield  {author} {\bibinfo {author} {\bibfnamefont {B.}~\bibnamefont
  {Cheng}}, \bibinfo {author} {\bibfnamefont {E.~A.}\ \bibnamefont {Engel}},
  \bibinfo {author} {\bibfnamefont {J.}~\bibnamefont {Behler}}, \bibinfo
  {author} {\bibfnamefont {C.}~\bibnamefont {Dellago}},\ and\ \bibinfo {author}
  {\bibfnamefont {M.}~\bibnamefont {Ceriotti}},\ }\bibfield  {title} {\bibinfo
  {title} {Ab initio thermodynamics of liquid and solid water},\ }\href
  {https://doi.org/10.1073/pnas.1815117116} {\bibfield  {journal} {\bibinfo
  {journal} {Proceedings of the National Academy of Sciences}\ }\textbf
  {\bibinfo {volume} {116}},\ \bibinfo {pages} {1110} (\bibinfo {year}
  {2019})}\BibitemShut {NoStop}%
\bibitem [{\citenamefont {Deringer}\ and\ \citenamefont
  {Cs\'anyi}(2017)}]{PhysRevB.95.094203}%
  \BibitemOpen
  \bibfield  {author} {\bibinfo {author} {\bibfnamefont {V.~L.}\ \bibnamefont
  {Deringer}}\ and\ \bibinfo {author} {\bibfnamefont {G.}~\bibnamefont
  {Cs\'anyi}},\ }\bibfield  {title} {\bibinfo {title} {Machine learning based
  interatomic potential for amorphous carbon},\ }\href
  {https://doi.org/10.1103/PhysRevB.95.094203} {\bibfield  {journal} {\bibinfo
  {journal} {Physical Review B}\ }\textbf {\bibinfo {volume} {95}},\ \bibinfo
  {pages} {094203} (\bibinfo {year} {2017})}\BibitemShut {NoStop}%
\bibitem [{\citenamefont {Wengert}\ \emph {et~al.}(2022)\citenamefont
  {Wengert}, \citenamefont {Csányi}, \citenamefont {Reuter},\ and\
  \citenamefont {Margraf}}]{10.1021/acs.jctc.2c00343}%
  \BibitemOpen
  \bibfield  {author} {\bibinfo {author} {\bibfnamefont {S.}~\bibnamefont
  {Wengert}}, \bibinfo {author} {\bibfnamefont {G.}~\bibnamefont {Csányi}},
  \bibinfo {author} {\bibfnamefont {K.}~\bibnamefont {Reuter}},\ and\ \bibinfo
  {author} {\bibfnamefont {J.~T.}\ \bibnamefont {Margraf}},\ }\bibfield
  {title} {\bibinfo {title} {A hybrid machine learning approach for structure
  stability prediction in molecular co-crystal screenings},\ }\href
  {https://doi.org/10.1021/acs.jctc.2c00343} {\bibfield  {journal} {\bibinfo
  {journal} {Journal of Chemical Theory and Computation}\ }\textbf {\bibinfo
  {volume} {18}},\ \bibinfo {pages} {4586} (\bibinfo {year}
  {2022})}\BibitemShut {NoStop}%
\bibitem [{\citenamefont {Kapil}\ and\ \citenamefont
  {Engel}(2022)}]{kapil_complete_2022}%
  \BibitemOpen
  \bibfield  {author} {\bibinfo {author} {\bibfnamefont {V.}~\bibnamefont
  {Kapil}}\ and\ \bibinfo {author} {\bibfnamefont {E.~A.}\ \bibnamefont
  {Engel}},\ }\bibfield  {title} {\bibinfo {title} {A complete description of
  thermodynamic stabilities of molecular crystals},\ }\bibfield  {journal}
  {\bibinfo  {journal} {Proceedings of the National Academy of Sciences}\
  }\textbf {\bibinfo {volume} {119}},\ \href
  {https://doi.org/10.1073/pnas.2111769119} {10.1073/pnas.2111769119} (\bibinfo
  {year} {2022})\BibitemShut {NoStop}%
\bibitem [{\citenamefont {Ceriotti}\ \emph {et~al.}(2014)\citenamefont
  {Ceriotti}, \citenamefont {More},\ and\ \citenamefont
  {Manolopoulos}}]{IPIv1}%
  \BibitemOpen
  \bibfield  {author} {\bibinfo {author} {\bibfnamefont {M.}~\bibnamefont
  {Ceriotti}}, \bibinfo {author} {\bibfnamefont {J.}~\bibnamefont {More}},\
  and\ \bibinfo {author} {\bibfnamefont {D.~E.}\ \bibnamefont {Manolopoulos}},\
  }\bibfield  {title} {\bibinfo {title} {i-pi: A python interface for ab initio
  path integral molecular dynamics simulations},\ }\href
  {https://doi.org/https://doi.org/10.1016/j.cpc.2013.10.027} {\bibfield
  {journal} {\bibinfo  {journal} {Computer Physics Communications}\ }\textbf
  {\bibinfo {volume} {185}},\ \bibinfo {pages} {1019} (\bibinfo {year}
  {2014})}\BibitemShut {NoStop}%
\bibitem [{\citenamefont {Kapil}\ \emph
  {et~al.}(2019{\natexlab{a}})\citenamefont {Kapil}, \citenamefont {Rossi},
  \citenamefont {Marsalek}, \citenamefont {Petraglia}, \citenamefont {Litman},
  \citenamefont {Spura}, \citenamefont {Cheng}, \citenamefont {Cuzzocrea},
  \citenamefont {Meißner}, \citenamefont {Wilkins}, \citenamefont {Helfrecht},
  \citenamefont {Juda}, \citenamefont {Bienvenue}, \citenamefont {Fang},
  \citenamefont {Kessler}, \citenamefont {Poltavsky}, \citenamefont
  {Vandenbrande}, \citenamefont {Wieme}, \citenamefont {Corminboeuf},
  \citenamefont {Kühne}, \citenamefont {Manolopoulos}, \citenamefont
  {Markland}, \citenamefont {Richardson}, \citenamefont {Tkatchenko},
  \citenamefont {Tribello}, \citenamefont {{Van Speybroeck}},\ and\
  \citenamefont {Ceriotti}}]{IPIV2}%
  \BibitemOpen
  \bibfield  {author} {\bibinfo {author} {\bibfnamefont {V.}~\bibnamefont
  {Kapil}}, \bibinfo {author} {\bibfnamefont {M.}~\bibnamefont {Rossi}},
  \bibinfo {author} {\bibfnamefont {O.}~\bibnamefont {Marsalek}}, \bibinfo
  {author} {\bibfnamefont {R.}~\bibnamefont {Petraglia}}, \bibinfo {author}
  {\bibfnamefont {Y.}~\bibnamefont {Litman}}, \bibinfo {author} {\bibfnamefont
  {T.}~\bibnamefont {Spura}}, \bibinfo {author} {\bibfnamefont
  {B.}~\bibnamefont {Cheng}}, \bibinfo {author} {\bibfnamefont
  {A.}~\bibnamefont {Cuzzocrea}}, \bibinfo {author} {\bibfnamefont {R.~H.}\
  \bibnamefont {Meißner}}, \bibinfo {author} {\bibfnamefont {D.~M.}\
  \bibnamefont {Wilkins}}, \bibinfo {author} {\bibfnamefont {B.~A.}\
  \bibnamefont {Helfrecht}}, \bibinfo {author} {\bibfnamefont {P.}~\bibnamefont
  {Juda}}, \bibinfo {author} {\bibfnamefont {S.~P.}\ \bibnamefont {Bienvenue}},
  \bibinfo {author} {\bibfnamefont {W.}~\bibnamefont {Fang}}, \bibinfo {author}
  {\bibfnamefont {J.}~\bibnamefont {Kessler}}, \bibinfo {author} {\bibfnamefont
  {I.}~\bibnamefont {Poltavsky}}, \bibinfo {author} {\bibfnamefont
  {S.}~\bibnamefont {Vandenbrande}}, \bibinfo {author} {\bibfnamefont
  {J.}~\bibnamefont {Wieme}}, \bibinfo {author} {\bibfnamefont
  {C.}~\bibnamefont {Corminboeuf}}, \bibinfo {author} {\bibfnamefont {T.~D.}\
  \bibnamefont {Kühne}}, \bibinfo {author} {\bibfnamefont {D.~E.}\
  \bibnamefont {Manolopoulos}}, \bibinfo {author} {\bibfnamefont {T.~E.}\
  \bibnamefont {Markland}}, \bibinfo {author} {\bibfnamefont {J.~O.}\
  \bibnamefont {Richardson}}, \bibinfo {author} {\bibfnamefont
  {A.}~\bibnamefont {Tkatchenko}}, \bibinfo {author} {\bibfnamefont {G.~A.}\
  \bibnamefont {Tribello}}, \bibinfo {author} {\bibfnamefont {V.}~\bibnamefont
  {{Van Speybroeck}}},\ and\ \bibinfo {author} {\bibfnamefont {M.}~\bibnamefont
  {Ceriotti}},\ }\bibfield  {title} {\bibinfo {title} {i-pi 2.0: A universal
  force engine for advanced molecular simulations},\ }\href
  {https://doi.org/https://doi.org/10.1016/j.cpc.2018.09.020} {\bibfield
  {journal} {\bibinfo  {journal} {Computer Physics Communications}\ }\textbf
  {\bibinfo {volume} {236}},\ \bibinfo {pages} {214} (\bibinfo {year}
  {2019}{\natexlab{a}})}\BibitemShut {NoStop}%
\bibitem [{\citenamefont {Schobert}\ \emph {et~al.}(2024)\citenamefont
  {Schobert}, \citenamefont {Berges}, \citenamefont {van Loon}, \citenamefont
  {Sentef}, \citenamefont {Brener}, \citenamefont {Rossi},\ and\ \citenamefont
  {Wehling}}]{schob+24spp}%
  \BibitemOpen
  \bibfield  {author} {\bibinfo {author} {\bibfnamefont {A.}~\bibnamefont
  {Schobert}}, \bibinfo {author} {\bibfnamefont {J.}~\bibnamefont {Berges}},
  \bibinfo {author} {\bibfnamefont {E.~G. C.~P.}\ \bibnamefont {van Loon}},
  \bibinfo {author} {\bibfnamefont {M.~A.}\ \bibnamefont {Sentef}}, \bibinfo
  {author} {\bibfnamefont {S.}~\bibnamefont {Brener}}, \bibinfo {author}
  {\bibfnamefont {M.}~\bibnamefont {Rossi}},\ and\ \bibinfo {author}
  {\bibfnamefont {T.~O.}\ \bibnamefont {Wehling}},\ }\bibfield  {title}
  {\bibinfo {title} {{Ab initio electron-lattice downfolding: Potential energy
  landscapes, anharmonicity, and molecular dynamics in charge density wave
  materials}},\ }\href {https://doi.org/10.21468/SciPostPhys.16.2.046}
  {\bibfield  {journal} {\bibinfo  {journal} {SciPost Phys.}\ }\textbf
  {\bibinfo {volume} {16}},\ \bibinfo {pages} {046} (\bibinfo {year} {2024})},\
  \Eprint {https://arxiv.org/abs/2303.07261} {arXiv:2303.07261} \BibitemShut
  {NoStop}%
\bibitem [{\citenamefont {Linker}\ \emph {et~al.}(2024)\citenamefont {Linker},
  \citenamefont {Krishnamoorthy}, \citenamefont {Daemen}, \citenamefont
  {Ramirez-Cuesta}, \citenamefont {Nomura}, \citenamefont {Nakano},
  \citenamefont {Cheng}, \citenamefont {Hicks}, \citenamefont {Kolesnikov},\
  and\ \citenamefont {Vashishta}}]{Linker2024}%
  \BibitemOpen
  \bibfield  {author} {\bibinfo {author} {\bibfnamefont {T.~M.}\ \bibnamefont
  {Linker}}, \bibinfo {author} {\bibfnamefont {A.}~\bibnamefont
  {Krishnamoorthy}}, \bibinfo {author} {\bibfnamefont {L.~L.}\ \bibnamefont
  {Daemen}}, \bibinfo {author} {\bibfnamefont {A.~J.}\ \bibnamefont
  {Ramirez-Cuesta}}, \bibinfo {author} {\bibfnamefont {K.}~\bibnamefont
  {Nomura}}, \bibinfo {author} {\bibfnamefont {A.}~\bibnamefont {Nakano}},
  \bibinfo {author} {\bibfnamefont {Y.~Q.}\ \bibnamefont {Cheng}}, \bibinfo
  {author} {\bibfnamefont {W.~R.}\ \bibnamefont {Hicks}}, \bibinfo {author}
  {\bibfnamefont {A.~I.}\ \bibnamefont {Kolesnikov}},\ and\ \bibinfo {author}
  {\bibfnamefont {P.~D.}\ \bibnamefont {Vashishta}},\ }\bibfield  {title}
  {\bibinfo {title} {Neutron scattering and neural-network quantum molecular
  dynamics investigation of the vibrations of ammonia along the solid-to-liquid
  transition},\ }\bibfield  {journal} {\bibinfo  {journal} {Nature
  Communications}\ }\textbf {\bibinfo {volume} {15}},\ \href
  {https://doi.org/10.1038/s41467-024-48246-9} {10.1038/s41467-024-48246-9}
  (\bibinfo {year} {2024})\BibitemShut {NoStop}%
\bibitem [{\citenamefont {Jacobs}\ \emph {et~al.}(2024)\citenamefont {Jacobs},
  \citenamefont {Fidanyan}, \citenamefont {Rossi},\ and\ \citenamefont
  {Cocchi}}]{jacobs+es2024}%
  \BibitemOpen
  \bibfield  {author} {\bibinfo {author} {\bibfnamefont {M.}~\bibnamefont
  {Jacobs}}, \bibinfo {author} {\bibfnamefont {K.}~\bibnamefont {Fidanyan}},
  \bibinfo {author} {\bibfnamefont {M.}~\bibnamefont {Rossi}},\ and\ \bibinfo
  {author} {\bibfnamefont {C.}~\bibnamefont {Cocchi}},\ }\bibfield  {title}
  {\bibinfo {title} {Impact of nuclear effects on the ultrafast dynamics of an
  organic/inorganic mixed-dimensional interface},\ }\href
  {http://iopscience.iop.org/article/10.1088/2516-1075/ad4d46} {\bibfield
  {journal} {\bibinfo  {journal} {Electronic Structure}\ } (\bibinfo {year}
  {2024})}\BibitemShut {NoStop}%
\bibitem [{\citenamefont {Liu}\ \emph {et~al.}(2023)\citenamefont {Liu},
  \citenamefont {Long}, \citenamefont {Fang},\ and\ \citenamefont
  {Prezhdo}}]{Liu2023}%
  \BibitemOpen
  \bibfield  {author} {\bibinfo {author} {\bibfnamefont {Y.}~\bibnamefont
  {Liu}}, \bibinfo {author} {\bibfnamefont {R.}~\bibnamefont {Long}}, \bibinfo
  {author} {\bibfnamefont {W.-H.}\ \bibnamefont {Fang}},\ and\ \bibinfo
  {author} {\bibfnamefont {O.~V.}\ \bibnamefont {Prezhdo}},\ }\bibfield
  {title} {\bibinfo {title} {Nuclear quantum effects prolong charge carrier
  lifetimes in hybrid organic–inorganic perovskites},\ }\href
  {https://doi.org/10.1021/jacs.3c04412} {\bibfield  {journal} {\bibinfo
  {journal} {Journal of the American Chemical Society}\ }\textbf {\bibinfo
  {volume} {145}},\ \bibinfo {pages} {14112–14123} (\bibinfo {year}
  {2023})}\BibitemShut {NoStop}%
\bibitem [{\citenamefont {Gigli}\ \emph {et~al.}(2024)\citenamefont {Gigli},
  \citenamefont {Tisi}, \citenamefont {Grasselli},\ and\ \citenamefont
  {Ceriotti}}]{Gigli2023}%
  \BibitemOpen
  \bibfield  {author} {\bibinfo {author} {\bibfnamefont {L.}~\bibnamefont
  {Gigli}}, \bibinfo {author} {\bibfnamefont {D.}~\bibnamefont {Tisi}},
  \bibinfo {author} {\bibfnamefont {F.}~\bibnamefont {Grasselli}},\ and\
  \bibinfo {author} {\bibfnamefont {M.}~\bibnamefont {Ceriotti}},\ }\bibfield
  {title} {\bibinfo {title} {Mechanism of charge transport in lithium
  thiophosphate},\ }\href {https://doi.org/10.1021/acs.chemmater.3c02726}
  {\bibfield  {journal} {\bibinfo  {journal} {Chemistry of Materials}\ }\textbf
  {\bibinfo {volume} {36}},\ \bibinfo {pages} {1482} (\bibinfo {year}
  {2024})}\BibitemShut {NoStop}%
\bibitem [{\citenamefont {Lan}\ \emph {et~al.}(2021)\citenamefont {Lan},
  \citenamefont {Kapil}, \citenamefont {Gasparotto}, \citenamefont {Ceriotti},
  \citenamefont {Iannuzzi},\ and\ \citenamefont
  {Rybkin}}]{lan_simulating_2021}%
  \BibitemOpen
  \bibfield  {author} {\bibinfo {author} {\bibfnamefont {J.}~\bibnamefont
  {Lan}}, \bibinfo {author} {\bibfnamefont {V.}~\bibnamefont {Kapil}}, \bibinfo
  {author} {\bibfnamefont {P.}~\bibnamefont {Gasparotto}}, \bibinfo {author}
  {\bibfnamefont {M.}~\bibnamefont {Ceriotti}}, \bibinfo {author}
  {\bibfnamefont {M.}~\bibnamefont {Iannuzzi}},\ and\ \bibinfo {author}
  {\bibfnamefont {V.~V.}\ \bibnamefont {Rybkin}},\ }\bibfield  {title}
  {\bibinfo {title} {Simulating the ghost: quantum dynamics of the solvated
  electron},\ }\href {https://doi.org/10.1038/s41467-021-20914-0} {\bibfield
  {journal} {\bibinfo  {journal} {Nature Communications}\ }\textbf {\bibinfo
  {volume} {12}},\ \bibinfo {pages} {766} (\bibinfo {year} {2021})}\BibitemShut
  {NoStop}%
\bibitem [{\citenamefont {Lan}\ \emph {et~al.}(2022)\citenamefont {Lan},
  \citenamefont {Rybkin},\ and\ \citenamefont {Pasquarello}}]{Lan_Angew_2022}%
  \BibitemOpen
  \bibfield  {author} {\bibinfo {author} {\bibfnamefont {J.}~\bibnamefont
  {Lan}}, \bibinfo {author} {\bibfnamefont {V.~V.}\ \bibnamefont {Rybkin}},\
  and\ \bibinfo {author} {\bibfnamefont {A.}~\bibnamefont {Pasquarello}},\
  }\bibfield  {title} {\bibinfo {title} {Temperature dependent properties of
  the aqueous electron},\ }\href
  {https://doi.org/https://doi.org/10.1002/anie.202209398} {\bibfield
  {journal} {\bibinfo  {journal} {Angewandte Chemie International Edition}\
  }\textbf {\bibinfo {volume} {61}},\ \bibinfo {pages} {e202209398} (\bibinfo
  {year} {2022})}\BibitemShut {NoStop}%
\bibitem [{\citenamefont {Novelli}\ \emph {et~al.}(2023)\citenamefont
  {Novelli}, \citenamefont {Chen}, \citenamefont {Buchmann}, \citenamefont
  {Ockelmann}, \citenamefont {Hoberg}, \citenamefont {Head-Gordon},\ and\
  \citenamefont {Havenith}}]{Novelli2023}%
  \BibitemOpen
  \bibfield  {author} {\bibinfo {author} {\bibfnamefont {F.}~\bibnamefont
  {Novelli}}, \bibinfo {author} {\bibfnamefont {K.}~\bibnamefont {Chen}},
  \bibinfo {author} {\bibfnamefont {A.}~\bibnamefont {Buchmann}}, \bibinfo
  {author} {\bibfnamefont {T.}~\bibnamefont {Ockelmann}}, \bibinfo {author}
  {\bibfnamefont {C.}~\bibnamefont {Hoberg}}, \bibinfo {author} {\bibfnamefont
  {T.}~\bibnamefont {Head-Gordon}},\ and\ \bibinfo {author} {\bibfnamefont
  {M.}~\bibnamefont {Havenith}},\ }\bibfield  {title} {\bibinfo {title} {The
  birth and evolution of solvated electrons in the water},\ }\bibfield
  {journal} {\bibinfo  {journal} {Proceedings of the National Academy of
  Sciences}\ }\textbf {\bibinfo {volume} {120}},\ \href
  {https://doi.org/10.1073/pnas.2216480120} {10.1073/pnas.2216480120} (\bibinfo
  {year} {2023})\BibitemShut {NoStop}%
\bibitem [{\citenamefont {Cheng}\ \emph {et~al.}(2020)\citenamefont {Cheng},
  \citenamefont {Mazzola}, \citenamefont {Pickard},\ and\ \citenamefont
  {Ceriotti}}]{Cheng_Nature_2020}%
  \BibitemOpen
  \bibfield  {author} {\bibinfo {author} {\bibfnamefont {B.}~\bibnamefont
  {Cheng}}, \bibinfo {author} {\bibfnamefont {G.}~\bibnamefont {Mazzola}},
  \bibinfo {author} {\bibfnamefont {C.~J.}\ \bibnamefont {Pickard}},\ and\
  \bibinfo {author} {\bibfnamefont {M.}~\bibnamefont {Ceriotti}},\ }\bibfield
  {title} {\bibinfo {title} {Evidence for supercritical behaviour of
  high-pressure liquid hydrogen},\ }\href
  {https://doi.org/10.1038/s41586-020-2677-y} {\bibfield  {journal} {\bibinfo
  {journal} {Nature}\ }\textbf {\bibinfo {volume} {585}},\ \bibinfo {pages}
  {217} (\bibinfo {year} {2020})}\BibitemShut {NoStop}%
\bibitem [{\citenamefont {Kapil}\ \emph {et~al.}(2022)\citenamefont {Kapil},
  \citenamefont {Schran}, \citenamefont {Zen}, \citenamefont {Chen},
  \citenamefont {Pickard},\ and\ \citenamefont
  {Michaelides}}]{Kapil_Nature_2022}%
  \BibitemOpen
  \bibfield  {author} {\bibinfo {author} {\bibfnamefont {V.}~\bibnamefont
  {Kapil}}, \bibinfo {author} {\bibfnamefont {C.}~\bibnamefont {Schran}},
  \bibinfo {author} {\bibfnamefont {A.}~\bibnamefont {Zen}}, \bibinfo {author}
  {\bibfnamefont {J.}~\bibnamefont {Chen}}, \bibinfo {author} {\bibfnamefont
  {C.~J.}\ \bibnamefont {Pickard}},\ and\ \bibinfo {author} {\bibfnamefont
  {A.}~\bibnamefont {Michaelides}},\ }\bibfield  {title} {\bibinfo {title} {The
  first-principles phase diagram of monolayer nanoconfined water},\ }\href
  {https://doi.org/10.1038/s41586-022-05036-x} {\bibfield  {journal} {\bibinfo
  {journal} {Nature}\ }\textbf {\bibinfo {volume} {609}},\ \bibinfo {pages}
  {512} (\bibinfo {year} {2022})}\BibitemShut {NoStop}%
\bibitem [{\citenamefont {Bore}\ and\ \citenamefont
  {Paesani}(2023)}]{Bore2023}%
  \BibitemOpen
  \bibfield  {author} {\bibinfo {author} {\bibfnamefont {S.~L.}\ \bibnamefont
  {Bore}}\ and\ \bibinfo {author} {\bibfnamefont {F.}~\bibnamefont {Paesani}},\
  }\bibfield  {title} {\bibinfo {title} {Realistic phase diagram of water from
  “first principles” data-driven quantum simulations},\ }\bibfield
  {journal} {\bibinfo  {journal} {Nature Communications}\ }\textbf {\bibinfo
  {volume} {14}},\ \href {https://doi.org/10.1038/s41467-023-38855-1}
  {10.1038/s41467-023-38855-1} (\bibinfo {year} {2023})\BibitemShut {NoStop}%
\bibitem [{\citenamefont {Litman}\ \emph {et~al.}(2024)\citenamefont {Litman},
  \citenamefont {Chiang}, \citenamefont {Seki}, \citenamefont {Nagata},\ and\
  \citenamefont {Bonn}}]{Litman_NatChem_2024}%
  \BibitemOpen
  \bibfield  {author} {\bibinfo {author} {\bibfnamefont {Y.}~\bibnamefont
  {Litman}}, \bibinfo {author} {\bibfnamefont {K.-Y.}\ \bibnamefont {Chiang}},
  \bibinfo {author} {\bibfnamefont {T.}~\bibnamefont {Seki}}, \bibinfo {author}
  {\bibfnamefont {Y.}~\bibnamefont {Nagata}},\ and\ \bibinfo {author}
  {\bibfnamefont {M.}~\bibnamefont {Bonn}},\ }\bibfield  {title} {\bibinfo
  {title} {Surface stratification determines the interfacial water structure of
  simple electrolyte solutions},\ }\href
  {https://doi.org/10.1038/s41557-023-01416-6} {\bibfield  {journal} {\bibinfo
  {journal} {Nature Chemistry}\ }\textbf {\bibinfo {volume} {16}},\ \bibinfo
  {pages} {644} (\bibinfo {year} {2024})}\BibitemShut {NoStop}%
\bibitem [{\citenamefont {Litman}\ \emph {et~al.}(2023)\citenamefont {Litman},
  \citenamefont {Lan}, \citenamefont {Nagata},\ and\ \citenamefont
  {Wilkins}}]{Litman_JPCL_2023}%
  \BibitemOpen
  \bibfield  {author} {\bibinfo {author} {\bibfnamefont {Y.}~\bibnamefont
  {Litman}}, \bibinfo {author} {\bibfnamefont {J.}~\bibnamefont {Lan}},
  \bibinfo {author} {\bibfnamefont {Y.}~\bibnamefont {Nagata}},\ and\ \bibinfo
  {author} {\bibfnamefont {D.~M.}\ \bibnamefont {Wilkins}},\ }\bibfield
  {title} {\bibinfo {title} {Fully first-principles surface spectroscopy with
  machine learning},\ }\href {https://doi.org/10.1021/acs.jpclett.3c01989}
  {\bibfield  {journal} {\bibinfo  {journal} {The Journal of Physical Chemistry
  Letters}\ }\textbf {\bibinfo {volume} {14}},\ \bibinfo {pages} {8175}
  (\bibinfo {year} {2023})}\BibitemShut {NoStop}%
\bibitem [{\citenamefont {Shepherd}\ \emph {et~al.}(2021)\citenamefont
  {Shepherd}, \citenamefont {Lan}, \citenamefont {Wilkins},\ and\ \citenamefont
  {Kapil}}]{shepherd_efficient_2021}%
  \BibitemOpen
  \bibfield  {author} {\bibinfo {author} {\bibfnamefont {S.}~\bibnamefont
  {Shepherd}}, \bibinfo {author} {\bibfnamefont {J.}~\bibnamefont {Lan}},
  \bibinfo {author} {\bibfnamefont {D.~M.}\ \bibnamefont {Wilkins}},\ and\
  \bibinfo {author} {\bibfnamefont {V.}~\bibnamefont {Kapil}},\ }\bibfield
  {title} {\bibinfo {title} {Efficient {Quantum} {Vibrational} {Spectroscopy}
  of {Water} with {High}-{Order} {Path} {Integrals}: {From} {Bulk} to
  {Interfaces}},\ }\href {https://doi.org/10.1021/acs.jpclett.1c02574}
  {\bibfield  {journal} {\bibinfo  {journal} {The Journal of Physical Chemistry
  Letters}\ ,\ \bibinfo {pages} {9108}} (\bibinfo {year} {2021})}\BibitemShut
  {NoStop}%
\bibitem [{\citenamefont {Fidanyan}\ \emph {et~al.}(2023)\citenamefont
  {Fidanyan}, \citenamefont {Liu},\ and\ \citenamefont
  {Rossi}}]{fidanyan_jcp_2023}%
  \BibitemOpen
  \bibfield  {author} {\bibinfo {author} {\bibfnamefont {K.}~\bibnamefont
  {Fidanyan}}, \bibinfo {author} {\bibfnamefont {G.}~\bibnamefont {Liu}},\ and\
  \bibinfo {author} {\bibfnamefont {M.}~\bibnamefont {Rossi}},\ }\bibfield
  {title} {\bibinfo {title} {{Ab initio study of water dissociation on a
  charged Pd(111) surface}},\ }\href {https://doi.org/10.1063/5.0139082}
  {\bibfield  {journal} {\bibinfo  {journal} {The Journal of Chemical Physics}\
  }\textbf {\bibinfo {volume} {158}},\ \bibinfo {pages} {094707} (\bibinfo
  {year} {2023})},\ \Eprint {https://arxiv.org/abs/2212.08855} {2212.08855}
  \BibitemShut {NoStop}%
\bibitem [{\citenamefont {de~la Puente}\ and\ \citenamefont
  {Laage}(2023)}]{delaPuente2023}%
  \BibitemOpen
  \bibfield  {author} {\bibinfo {author} {\bibfnamefont {M.}~\bibnamefont
  {de~la Puente}}\ and\ \bibinfo {author} {\bibfnamefont {D.}~\bibnamefont
  {Laage}},\ }\bibfield  {title} {\bibinfo {title} {How the acidity of water
  droplets and films is controlled by the air–water interface},\ }\href
  {https://doi.org/10.1021/jacs.3c07506} {\bibfield  {journal} {\bibinfo
  {journal} {Journal of the American Chemical Society}\ }\textbf {\bibinfo
  {volume} {145}},\ \bibinfo {pages} {25186–25194} (\bibinfo {year}
  {2023})}\BibitemShut {NoStop}%
\bibitem [{\citenamefont {Inoue}\ \emph {et~al.}(2023)\citenamefont {Inoue},
  \citenamefont {Litman}, \citenamefont {Wilkins}, \citenamefont {Nagata},\
  and\ \citenamefont {Okuno}}]{Inoue_JPCL_2023}%
  \BibitemOpen
  \bibfield  {author} {\bibinfo {author} {\bibfnamefont {K.}~\bibnamefont
  {Inoue}}, \bibinfo {author} {\bibfnamefont {Y.}~\bibnamefont {Litman}},
  \bibinfo {author} {\bibfnamefont {D.~M.}\ \bibnamefont {Wilkins}}, \bibinfo
  {author} {\bibfnamefont {Y.}~\bibnamefont {Nagata}},\ and\ \bibinfo {author}
  {\bibfnamefont {M.}~\bibnamefont {Okuno}},\ }\bibfield  {title} {\bibinfo
  {title} {Is unified understanding of vibrational coupling of water possible?
  hyper-raman measurement and machine learning spectra},\ }\href
  {https://doi.org/10.1021/acs.jpclett.3c00398} {\bibfield  {journal} {\bibinfo
   {journal} {The Journal of Physical Chemistry Letters}\ }\textbf {\bibinfo
  {volume} {14}},\ \bibinfo {pages} {3063} (\bibinfo {year}
  {2023})}\BibitemShut {NoStop}%
\bibitem [{\citenamefont {Chmiela}\ \emph {et~al.}(2023)\citenamefont
  {Chmiela}, \citenamefont {Vassilev-Galindo}, \citenamefont {Unke},
  \citenamefont {Kabylda}, \citenamefont {Sauceda}, \citenamefont
  {Tkatchenko},\ and\ \citenamefont {M\"{u}ller}}]{Chmiela2023}%
  \BibitemOpen
  \bibfield  {author} {\bibinfo {author} {\bibfnamefont {S.}~\bibnamefont
  {Chmiela}}, \bibinfo {author} {\bibfnamefont {V.}~\bibnamefont
  {Vassilev-Galindo}}, \bibinfo {author} {\bibfnamefont {O.~T.}\ \bibnamefont
  {Unke}}, \bibinfo {author} {\bibfnamefont {A.}~\bibnamefont {Kabylda}},
  \bibinfo {author} {\bibfnamefont {H.~E.}\ \bibnamefont {Sauceda}}, \bibinfo
  {author} {\bibfnamefont {A.}~\bibnamefont {Tkatchenko}},\ and\ \bibinfo
  {author} {\bibfnamefont {K.-R.}\ \bibnamefont {M\"{u}ller}},\ }\bibfield
  {title} {\bibinfo {title} {Accurate global machine learning force fields for
  molecules with hundreds of atoms},\ }\bibfield  {journal} {\bibinfo
  {journal} {Science Advances}\ }\textbf {\bibinfo {volume} {9}},\ \href
  {https://doi.org/10.1126/sciadv.adf0873} {10.1126/sciadv.adf0873} (\bibinfo
  {year} {2023})\BibitemShut {NoStop}%
\bibitem [{\citenamefont {Kovács}\ \emph {et~al.}(2023)\citenamefont
  {Kovács}, \citenamefont {Moore}, \citenamefont {Browning}, \citenamefont
  {Batatia}, \citenamefont {Horton}, \citenamefont {Kapil}, \citenamefont
  {Witt}, \citenamefont {Magdău}, \citenamefont {Cole},\ and\ \citenamefont
  {Csányi}}]{kovacs_mace-off23_2023}%
  \BibitemOpen
  \bibfield  {author} {\bibinfo {author} {\bibfnamefont {D.~P.}\ \bibnamefont
  {Kovács}}, \bibinfo {author} {\bibfnamefont {J.~H.}\ \bibnamefont {Moore}},
  \bibinfo {author} {\bibfnamefont {N.~J.}\ \bibnamefont {Browning}}, \bibinfo
  {author} {\bibfnamefont {I.}~\bibnamefont {Batatia}}, \bibinfo {author}
  {\bibfnamefont {J.~T.}\ \bibnamefont {Horton}}, \bibinfo {author}
  {\bibfnamefont {V.}~\bibnamefont {Kapil}}, \bibinfo {author} {\bibfnamefont
  {W.~C.}\ \bibnamefont {Witt}}, \bibinfo {author} {\bibfnamefont {I.-B.}\
  \bibnamefont {Magdău}}, \bibinfo {author} {\bibfnamefont {D.~J.}\
  \bibnamefont {Cole}},\ and\ \bibinfo {author} {\bibfnamefont
  {G.}~\bibnamefont {Csányi}},\ }\href
  {https://doi.org/10.48550/arXiv.2312.15211} {\bibinfo {title}
  {{MACE}-{OFF23}: {Transferable} {Machine} {Learning} {Force} {Fields} for
  {Organic} {Molecules}}} (\bibinfo {year} {2023}),\ \bibinfo {note}
  {arXiv:2312.15211}\BibitemShut {NoStop}%
\bibitem [{\citenamefont {Bonomi}\ \emph {et~al.}(2009)\citenamefont {Bonomi},
  \citenamefont {Branduardi}, \citenamefont {Bussi}, \citenamefont {Camilloni},
  \citenamefont {Provasi}, \citenamefont {Raiteri}, \citenamefont {Donadio},
  \citenamefont {Marinelli}, \citenamefont {Pietrucci}, \citenamefont
  {Broglia},\ and\ \citenamefont {Parrinello}}]{PLUMED}%
  \BibitemOpen
  \bibfield  {author} {\bibinfo {author} {\bibfnamefont {M.}~\bibnamefont
  {Bonomi}}, \bibinfo {author} {\bibfnamefont {D.}~\bibnamefont {Branduardi}},
  \bibinfo {author} {\bibfnamefont {G.}~\bibnamefont {Bussi}}, \bibinfo
  {author} {\bibfnamefont {C.}~\bibnamefont {Camilloni}}, \bibinfo {author}
  {\bibfnamefont {D.}~\bibnamefont {Provasi}}, \bibinfo {author} {\bibfnamefont
  {P.}~\bibnamefont {Raiteri}}, \bibinfo {author} {\bibfnamefont
  {D.}~\bibnamefont {Donadio}}, \bibinfo {author} {\bibfnamefont
  {F.}~\bibnamefont {Marinelli}}, \bibinfo {author} {\bibfnamefont
  {F.}~\bibnamefont {Pietrucci}}, \bibinfo {author} {\bibfnamefont {R.~A.}\
  \bibnamefont {Broglia}},\ and\ \bibinfo {author} {\bibfnamefont
  {M.}~\bibnamefont {Parrinello}},\ }\bibfield  {title} {\bibinfo {title}
  {{PLUMED: A portable plugin for free-energy calculations with molecular
  dynamics}},\ }\href@noop {} {\bibfield  {journal} {\bibinfo  {journal}
  {Computer Physics Communications.}\ }\textbf {\bibinfo {volume} {180}},\
  \bibinfo {pages} {1961} (\bibinfo {year} {2009})}\BibitemShut {NoStop}%
\bibitem [{\citenamefont {Rossi}\ \emph {et~al.}(2020)\citenamefont {Rossi},
  \citenamefont {Jur{\'a}skov{\'a}}, \citenamefont {Wischert}, \citenamefont
  {Garel}, \citenamefont {Corminboeuf},\ and\ \citenamefont
  {Ceriotti}}]{ross+20jctc}%
  \BibitemOpen
  \bibfield  {author} {\bibinfo {author} {\bibfnamefont {K.}~\bibnamefont
  {Rossi}}, \bibinfo {author} {\bibfnamefont {V.}~\bibnamefont
  {Jur{\'a}skov{\'a}}}, \bibinfo {author} {\bibfnamefont {R.}~\bibnamefont
  {Wischert}}, \bibinfo {author} {\bibfnamefont {L.}~\bibnamefont {Garel}},
  \bibinfo {author} {\bibfnamefont {C.}~\bibnamefont {Corminboeuf}},\ and\
  \bibinfo {author} {\bibfnamefont {M.}~\bibnamefont {Ceriotti}},\ }\bibfield
  {title} {\bibinfo {title} {Simulating {{Solvation}} and {{Acidity}} in
  {{Complex Mixtures}} with {{First-Principles Accuracy}}: {{The Case}} of
  {{CH}} {\textsubscript{3}} {{SO}} {\textsubscript{3}} {{H}} and {{H}}
  {\textsubscript{2}} {{O}} {\textsubscript{2}} in {{Phenol}}},\ }\href
  {https://doi.org/10.1021/acs.jctc.0c00362} {\bibfield  {journal} {\bibinfo
  {journal} {Journal of Chemical Theory and Computation}\ }\textbf {\bibinfo
  {volume} {16}},\ \bibinfo {pages} {5139} (\bibinfo {year}
  {2020})}\BibitemShut {NoStop}%
\bibitem [{\citenamefont {Mortensen}(2024)}]{gpaw2024}%
  \BibitemOpen
  \bibfield  {author} {\bibinfo {author} {\bibfnamefont {J.~J.}\ \bibnamefont
  {Mortensen}},\ }\bibfield  {title} {\bibinfo {title} {{GPAW: An open Python
  package for electronic structure calculations}},\ }\href
  {https://doi.org/10.1063/5.0182685} {\bibfield  {journal} {\bibinfo
  {journal} {The Journal of Chemical Physics}\ }\textbf {\bibinfo {volume}
  {160}},\ \bibinfo {pages} {092503} (\bibinfo {year} {2024})}\BibitemShut
  {NoStop}%
\bibitem [{\citenamefont {Sun}\ \emph {et~al.}(2020)\citenamefont {Sun},
  \citenamefont {Zhang}, \citenamefont {Banerjee}, \citenamefont {Bao},
  \citenamefont {Barbry}, \citenamefont {Blunt}, \citenamefont {Bogdanov},
  \citenamefont {Booth}, \citenamefont {Chen}, \citenamefont {Cui},
  \citenamefont {Eriksen}, \citenamefont {Gao}, \citenamefont {Guo},
  \citenamefont {Hermann}, \citenamefont {Hermes}, \citenamefont {Koh},
  \citenamefont {Koval}, \citenamefont {Lehtola}, \citenamefont {Li},
  \citenamefont {Liu}, \citenamefont {Mardirossian}, \citenamefont {McClain},
  \citenamefont {Motta}, \citenamefont {Mussard}, \citenamefont {Pham},
  \citenamefont {Pulkin}, \citenamefont {Purwanto}, \citenamefont {Robinson},
  \citenamefont {Ronca}, \citenamefont {Sayfutyarova}, \citenamefont
  {Scheurer}, \citenamefont {Schurkus}, \citenamefont {Smith}, \citenamefont
  {Sun}, \citenamefont {Sun}, \citenamefont {Upadhyay}, \citenamefont {Wagner},
  \citenamefont {Wang}, \citenamefont {White}, \citenamefont {Whitfield},
  \citenamefont {Williamson}, \citenamefont {Wouters}, \citenamefont {Yang},
  \citenamefont {Yu}, \citenamefont {Zhu}, \citenamefont {Berkelbach},
  \citenamefont {Sharma}, \citenamefont {Sokolov},\ and\ \citenamefont
  {Chan}}]{pyscf_2020}%
  \BibitemOpen
  \bibfield  {author} {\bibinfo {author} {\bibfnamefont {Q.}~\bibnamefont
  {Sun}}, \bibinfo {author} {\bibfnamefont {X.}~\bibnamefont {Zhang}}, \bibinfo
  {author} {\bibfnamefont {S.}~\bibnamefont {Banerjee}}, \bibinfo {author}
  {\bibfnamefont {P.}~\bibnamefont {Bao}}, \bibinfo {author} {\bibfnamefont
  {M.}~\bibnamefont {Barbry}}, \bibinfo {author} {\bibfnamefont {N.~S.}\
  \bibnamefont {Blunt}}, \bibinfo {author} {\bibfnamefont {N.~A.}\ \bibnamefont
  {Bogdanov}}, \bibinfo {author} {\bibfnamefont {G.~H.}\ \bibnamefont {Booth}},
  \bibinfo {author} {\bibfnamefont {J.}~\bibnamefont {Chen}}, \bibinfo {author}
  {\bibfnamefont {Z.-H.}\ \bibnamefont {Cui}}, \bibinfo {author} {\bibfnamefont
  {J.~J.}\ \bibnamefont {Eriksen}}, \bibinfo {author} {\bibfnamefont
  {Y.}~\bibnamefont {Gao}}, \bibinfo {author} {\bibfnamefont {S.}~\bibnamefont
  {Guo}}, \bibinfo {author} {\bibfnamefont {J.}~\bibnamefont {Hermann}},
  \bibinfo {author} {\bibfnamefont {M.~R.}\ \bibnamefont {Hermes}}, \bibinfo
  {author} {\bibfnamefont {K.}~\bibnamefont {Koh}}, \bibinfo {author}
  {\bibfnamefont {P.}~\bibnamefont {Koval}}, \bibinfo {author} {\bibfnamefont
  {S.}~\bibnamefont {Lehtola}}, \bibinfo {author} {\bibfnamefont
  {Z.}~\bibnamefont {Li}}, \bibinfo {author} {\bibfnamefont {J.}~\bibnamefont
  {Liu}}, \bibinfo {author} {\bibfnamefont {N.}~\bibnamefont {Mardirossian}},
  \bibinfo {author} {\bibfnamefont {J.~D.}\ \bibnamefont {McClain}}, \bibinfo
  {author} {\bibfnamefont {M.}~\bibnamefont {Motta}}, \bibinfo {author}
  {\bibfnamefont {B.}~\bibnamefont {Mussard}}, \bibinfo {author} {\bibfnamefont
  {H.~Q.}\ \bibnamefont {Pham}}, \bibinfo {author} {\bibfnamefont
  {A.}~\bibnamefont {Pulkin}}, \bibinfo {author} {\bibfnamefont
  {W.}~\bibnamefont {Purwanto}}, \bibinfo {author} {\bibfnamefont {P.~J.}\
  \bibnamefont {Robinson}}, \bibinfo {author} {\bibfnamefont {E.}~\bibnamefont
  {Ronca}}, \bibinfo {author} {\bibfnamefont {E.~R.}\ \bibnamefont
  {Sayfutyarova}}, \bibinfo {author} {\bibfnamefont {M.}~\bibnamefont
  {Scheurer}}, \bibinfo {author} {\bibfnamefont {H.~F.}\ \bibnamefont
  {Schurkus}}, \bibinfo {author} {\bibfnamefont {J.~E.~T.}\ \bibnamefont
  {Smith}}, \bibinfo {author} {\bibfnamefont {C.}~\bibnamefont {Sun}}, \bibinfo
  {author} {\bibfnamefont {S.-N.}\ \bibnamefont {Sun}}, \bibinfo {author}
  {\bibfnamefont {S.}~\bibnamefont {Upadhyay}}, \bibinfo {author}
  {\bibfnamefont {L.~K.}\ \bibnamefont {Wagner}}, \bibinfo {author}
  {\bibfnamefont {X.}~\bibnamefont {Wang}}, \bibinfo {author} {\bibfnamefont
  {A.}~\bibnamefont {White}}, \bibinfo {author} {\bibfnamefont {J.~D.}\
  \bibnamefont {Whitfield}}, \bibinfo {author} {\bibfnamefont {M.~J.}\
  \bibnamefont {Williamson}}, \bibinfo {author} {\bibfnamefont
  {S.}~\bibnamefont {Wouters}}, \bibinfo {author} {\bibfnamefont
  {J.}~\bibnamefont {Yang}}, \bibinfo {author} {\bibfnamefont {J.~M.}\
  \bibnamefont {Yu}}, \bibinfo {author} {\bibfnamefont {T.}~\bibnamefont
  {Zhu}}, \bibinfo {author} {\bibfnamefont {T.~C.}\ \bibnamefont {Berkelbach}},
  \bibinfo {author} {\bibfnamefont {S.}~\bibnamefont {Sharma}}, \bibinfo
  {author} {\bibfnamefont {A.~Y.}\ \bibnamefont {Sokolov}},\ and\ \bibinfo
  {author} {\bibfnamefont {G.~K.-L.}\ \bibnamefont {Chan}},\ }\bibfield
  {title} {\bibinfo {title} {{Recent developments in the PySCF program
  package}},\ }\href {https://doi.org/10.1063/5.0006074} {\bibfield  {journal}
  {\bibinfo  {journal} {The Journal of Chemical Physics}\ }\textbf {\bibinfo
  {volume} {153}},\ \bibinfo {pages} {024109} (\bibinfo {year}
  {2020})}\BibitemShut {NoStop}%
\bibitem [{\citenamefont {Markland}\ and\ \citenamefont
  {Manolopoulos}(2008{\natexlab{a}})}]{mark-mano08jcp}%
  \BibitemOpen
  \bibfield  {author} {\bibinfo {author} {\bibfnamefont {T.~E.}\ \bibnamefont
  {Markland}}\ and\ \bibinfo {author} {\bibfnamefont {D.~E.}\ \bibnamefont
  {Manolopoulos}},\ }\bibfield  {title} {\bibinfo {title} {{An efficient ring
  polymer contraction scheme for imaginary time path integral simulations}},\
  }\href {https://doi.org/10.1063/1.2953308} {\bibfield  {journal} {\bibinfo
  {journal} {The Journal of Chemical Physics}\ }\textbf {\bibinfo {volume}
  {129}},\ \bibinfo {pages} {024105} (\bibinfo {year}
  {2008}{\natexlab{a}})}\BibitemShut {NoStop}%
\bibitem [{\citenamefont {Markland}\ and\ \citenamefont
  {Manolopoulos}(2008{\natexlab{b}})}]{mark-mano08cpl}%
  \BibitemOpen
  \bibfield  {author} {\bibinfo {author} {\bibfnamefont {T.~E.}\ \bibnamefont
  {Markland}}\ and\ \bibinfo {author} {\bibfnamefont {D.~E.}\ \bibnamefont
  {Manolopoulos}},\ }\bibfield  {title} {\bibinfo {title} {{A refined ring
  polymer contraction scheme for systems with electrostatic interactions}},\
  }\href@noop {} {\bibfield  {journal} {\bibinfo  {journal} {Chem. Phys.
  Lett.}\ }\textbf {\bibinfo {volume} {464}},\ \bibinfo {pages} {256} (\bibinfo
  {year} {2008}{\natexlab{b}})}\BibitemShut {NoStop}%
\bibitem [{\citenamefont {Tisi}\ \emph {et~al.}(2024)\citenamefont {Tisi},
  \citenamefont {Grasselli}, \citenamefont {Gigli},\ and\ \citenamefont
  {Ceriotti}}]{tisi2024thermal}%
  \BibitemOpen
  \bibfield  {author} {\bibinfo {author} {\bibfnamefont {D.}~\bibnamefont
  {Tisi}}, \bibinfo {author} {\bibfnamefont {F.}~\bibnamefont {Grasselli}},
  \bibinfo {author} {\bibfnamefont {L.}~\bibnamefont {Gigli}},\ and\ \bibinfo
  {author} {\bibfnamefont {M.}~\bibnamefont {Ceriotti}},\ }\bibfield  {title}
  {\bibinfo {title} {Thermal conductivity of
  ${\mathrm{li}}_{3}{\mathrm{ps}}_{4}$ solid electrolytes with ab initio
  accuracy},\ }\href {https://doi.org/10.1103/PhysRevMaterials.8.065403}
  {\bibfield  {journal} {\bibinfo  {journal} {Phys. Rev. Mater.}\ }\textbf
  {\bibinfo {volume} {8}},\ \bibinfo {pages} {065403} (\bibinfo {year}
  {2024})}\BibitemShut {NoStop}%
\bibitem [{\citenamefont {Kapil}\ \emph
  {et~al.}(2016{\natexlab{a}})\citenamefont {Kapil}, \citenamefont
  {VandeVondele},\ and\ \citenamefont {Ceriotti}}]{kapi+16jcp}%
  \BibitemOpen
  \bibfield  {author} {\bibinfo {author} {\bibfnamefont {V.}~\bibnamefont
  {Kapil}}, \bibinfo {author} {\bibfnamefont {J.}~\bibnamefont
  {VandeVondele}},\ and\ \bibinfo {author} {\bibfnamefont {M.}~\bibnamefont
  {Ceriotti}},\ }\bibfield  {title} {\bibinfo {title} {{Accurate molecular
  dynamics and nuclear quantum effects at low cost by multiple steps in real
  and imaginary time: Using density functional theory to accelerate
  wavefunction methods}},\ }\href {https://doi.org/10.1063/1.4941091}
  {\bibfield  {journal} {\bibinfo  {journal} {The Journal of Chemical Physics}\
  }\textbf {\bibinfo {volume} {144}},\ \bibinfo {pages} {054111} (\bibinfo
  {year} {2016}{\natexlab{a}})}\BibitemShut {NoStop}%
\bibitem [{\citenamefont {Kapil}\ \emph
  {et~al.}(2019{\natexlab{b}})\citenamefont {Kapil}, \citenamefont {Wieme},
  \citenamefont {Vandenbrande}, \citenamefont {Lamaire}, \citenamefont
  {Van~Speybroeck},\ and\ \citenamefont {Ceriotti}}]{kapil_modeling_2019}%
  \BibitemOpen
  \bibfield  {author} {\bibinfo {author} {\bibfnamefont {V.}~\bibnamefont
  {Kapil}}, \bibinfo {author} {\bibfnamefont {J.}~\bibnamefont {Wieme}},
  \bibinfo {author} {\bibfnamefont {S.}~\bibnamefont {Vandenbrande}}, \bibinfo
  {author} {\bibfnamefont {A.}~\bibnamefont {Lamaire}}, \bibinfo {author}
  {\bibfnamefont {V.}~\bibnamefont {Van~Speybroeck}},\ and\ \bibinfo {author}
  {\bibfnamefont {M.}~\bibnamefont {Ceriotti}},\ }\bibfield  {title} {\bibinfo
  {title} {Modeling the {Structural} and {Thermal} {Properties} of {Loaded}
  {Metal}–{Organic} {Frameworks}. {An} {Interplay} of {Quantum} and
  {Anharmonic} {Fluctuations}},\ }\href
  {https://doi.org/10.1021/acs.jctc.8b01297} {\bibfield  {journal} {\bibinfo
  {journal} {Journal of Chemical Theory and Computation}\ }\textbf {\bibinfo
  {volume} {15}},\ \bibinfo {pages} {3237} (\bibinfo {year}
  {2019}{\natexlab{b}})}\BibitemShut {NoStop}%
\bibitem [{\citenamefont {Litman}\ \emph {et~al.}(2018)\citenamefont {Litman},
  \citenamefont {Donadio}, \citenamefont {Ceriotti},\ and\ \citenamefont
  {Rossi}}]{litm+17jcp}%
  \BibitemOpen
  \bibfield  {author} {\bibinfo {author} {\bibfnamefont {Y.}~\bibnamefont
  {Litman}}, \bibinfo {author} {\bibfnamefont {D.}~\bibnamefont {Donadio}},
  \bibinfo {author} {\bibfnamefont {M.}~\bibnamefont {Ceriotti}},\ and\
  \bibinfo {author} {\bibfnamefont {M.}~\bibnamefont {Rossi}},\ }\bibfield
  {title} {\bibinfo {title} {{Decisive role of nuclear quantum effects on
  surface mediated water dissociation at finite temperature}},\ }\href@noop {}
  {\bibfield  {journal} {\bibinfo  {journal} {The Journal of Chemical Physics}\
  }\textbf {\bibinfo {volume} {148}},\ \bibinfo {pages} {102320} (\bibinfo
  {year} {2018})}\BibitemShut {NoStop}%
\bibitem [{\citenamefont {Larsen}\ \emph {et~al.}(2017)\citenamefont {Larsen},
  \citenamefont {Mortensen}, \citenamefont {Blomqvist}, \citenamefont
  {Castelli}, \citenamefont {Christensen}, \citenamefont {Dułak},
  \citenamefont {Friis}, \citenamefont {Groves}, \citenamefont {Hammer},
  \citenamefont {Hargus}, \citenamefont {Hermes}, \citenamefont {Jennings},
  \citenamefont {Jensen}, \citenamefont {Kermode}, \citenamefont {Kitchin},
  \citenamefont {Kolsbjerg}, \citenamefont {Kubal}, \citenamefont {Kaasbjerg},
  \citenamefont {Lysgaard}, \citenamefont {Maronsson}, \citenamefont {Maxson},
  \citenamefont {Olsen}, \citenamefont {Pastewka}, \citenamefont {Peterson},
  \citenamefont {Rostgaard}, \citenamefont {Schiøtz}, \citenamefont {Schütt},
  \citenamefont {Strange}, \citenamefont {Thygesen}, \citenamefont {Vegge},
  \citenamefont {Vilhelmsen}, \citenamefont {Walter}, \citenamefont {Zeng},\
  and\ \citenamefont {Jacobsen}}]{ASE}%
  \BibitemOpen
  \bibfield  {author} {\bibinfo {author} {\bibfnamefont {A.~H.}\ \bibnamefont
  {Larsen}}, \bibinfo {author} {\bibfnamefont {J.~J.}\ \bibnamefont
  {Mortensen}}, \bibinfo {author} {\bibfnamefont {J.}~\bibnamefont
  {Blomqvist}}, \bibinfo {author} {\bibfnamefont {I.~E.}\ \bibnamefont
  {Castelli}}, \bibinfo {author} {\bibfnamefont {R.}~\bibnamefont
  {Christensen}}, \bibinfo {author} {\bibfnamefont {M.}~\bibnamefont {Dułak}},
  \bibinfo {author} {\bibfnamefont {J.}~\bibnamefont {Friis}}, \bibinfo
  {author} {\bibfnamefont {M.~N.}\ \bibnamefont {Groves}}, \bibinfo {author}
  {\bibfnamefont {B.}~\bibnamefont {Hammer}}, \bibinfo {author} {\bibfnamefont
  {C.}~\bibnamefont {Hargus}}, \bibinfo {author} {\bibfnamefont {E.~D.}\
  \bibnamefont {Hermes}}, \bibinfo {author} {\bibfnamefont {P.~C.}\
  \bibnamefont {Jennings}}, \bibinfo {author} {\bibfnamefont {P.~B.}\
  \bibnamefont {Jensen}}, \bibinfo {author} {\bibfnamefont {J.}~\bibnamefont
  {Kermode}}, \bibinfo {author} {\bibfnamefont {J.~R.}\ \bibnamefont
  {Kitchin}}, \bibinfo {author} {\bibfnamefont {E.~L.}\ \bibnamefont
  {Kolsbjerg}}, \bibinfo {author} {\bibfnamefont {J.}~\bibnamefont {Kubal}},
  \bibinfo {author} {\bibfnamefont {K.}~\bibnamefont {Kaasbjerg}}, \bibinfo
  {author} {\bibfnamefont {S.}~\bibnamefont {Lysgaard}}, \bibinfo {author}
  {\bibfnamefont {J.~B.}\ \bibnamefont {Maronsson}}, \bibinfo {author}
  {\bibfnamefont {T.}~\bibnamefont {Maxson}}, \bibinfo {author} {\bibfnamefont
  {T.}~\bibnamefont {Olsen}}, \bibinfo {author} {\bibfnamefont
  {L.}~\bibnamefont {Pastewka}}, \bibinfo {author} {\bibfnamefont
  {A.}~\bibnamefont {Peterson}}, \bibinfo {author} {\bibfnamefont
  {C.}~\bibnamefont {Rostgaard}}, \bibinfo {author} {\bibfnamefont
  {J.}~\bibnamefont {Schiøtz}}, \bibinfo {author} {\bibfnamefont
  {O.}~\bibnamefont {Schütt}}, \bibinfo {author} {\bibfnamefont
  {M.}~\bibnamefont {Strange}}, \bibinfo {author} {\bibfnamefont {K.~S.}\
  \bibnamefont {Thygesen}}, \bibinfo {author} {\bibfnamefont {T.}~\bibnamefont
  {Vegge}}, \bibinfo {author} {\bibfnamefont {L.}~\bibnamefont {Vilhelmsen}},
  \bibinfo {author} {\bibfnamefont {M.}~\bibnamefont {Walter}}, \bibinfo
  {author} {\bibfnamefont {Z.}~\bibnamefont {Zeng}},\ and\ \bibinfo {author}
  {\bibfnamefont {K.~W.}\ \bibnamefont {Jacobsen}},\ }\bibfield  {title}
  {\bibinfo {title} {The atomic simulation environment—a python library for
  working with atoms},\ }\href {https://doi.org/10.1088/1361-648X/aa680e}
  {\bibfield  {journal} {\bibinfo  {journal} {Journal of Physics: Condensed
  Matter}\ }\textbf {\bibinfo {volume} {29}},\ \bibinfo {pages} {273002}
  (\bibinfo {year} {2017})}\BibitemShut {NoStop}%
\bibitem [{\citenamefont {Thompson}\ \emph {et~al.}(2022)\citenamefont
  {Thompson}, \citenamefont {Aktulga}, \citenamefont {Berger}, \citenamefont
  {Bolintineanu}, \citenamefont {Brown}, \citenamefont {Crozier}, \citenamefont
  {in~'t Veld}, \citenamefont {Kohlmeyer}, \citenamefont {Moore}, \citenamefont
  {Nguyen}, \citenamefont {Shan}, \citenamefont {Stevens}, \citenamefont
  {Tranchida}, \citenamefont {Trott},\ and\ \citenamefont {Plimpton}}]{LAMMPS}%
  \BibitemOpen
  \bibfield  {author} {\bibinfo {author} {\bibfnamefont {A.~P.}\ \bibnamefont
  {Thompson}}, \bibinfo {author} {\bibfnamefont {H.~M.}\ \bibnamefont
  {Aktulga}}, \bibinfo {author} {\bibfnamefont {R.}~\bibnamefont {Berger}},
  \bibinfo {author} {\bibfnamefont {D.~S.}\ \bibnamefont {Bolintineanu}},
  \bibinfo {author} {\bibfnamefont {W.~M.}\ \bibnamefont {Brown}}, \bibinfo
  {author} {\bibfnamefont {P.~S.}\ \bibnamefont {Crozier}}, \bibinfo {author}
  {\bibfnamefont {P.~J.}\ \bibnamefont {in~'t Veld}}, \bibinfo {author}
  {\bibfnamefont {A.}~\bibnamefont {Kohlmeyer}}, \bibinfo {author}
  {\bibfnamefont {S.~G.}\ \bibnamefont {Moore}}, \bibinfo {author}
  {\bibfnamefont {T.~D.}\ \bibnamefont {Nguyen}}, \bibinfo {author}
  {\bibfnamefont {R.}~\bibnamefont {Shan}}, \bibinfo {author} {\bibfnamefont
  {M.~J.}\ \bibnamefont {Stevens}}, \bibinfo {author} {\bibfnamefont
  {J.}~\bibnamefont {Tranchida}}, \bibinfo {author} {\bibfnamefont
  {C.}~\bibnamefont {Trott}},\ and\ \bibinfo {author} {\bibfnamefont {S.~J.}\
  \bibnamefont {Plimpton}},\ }\bibfield  {title} {\bibinfo {title} {{LAMMPS} -
  a flexible simulation tool for particle-based materials modeling at the
  atomic, meso, and continuum scales},\ }\href
  {https://doi.org/10.1016/j.cpc.2021.108171} {\bibfield  {journal} {\bibinfo
  {journal} {Computer Physics Communications.}\ }\textbf {\bibinfo {volume}
  {271}},\ \bibinfo {pages} {108171} (\bibinfo {year} {2022})}\BibitemShut
  {NoStop}%
\bibitem [{Note1()}]{Note1}%
  \BibitemOpen
  \bibinfo {note} {{\protect \color {blue}{i-PI}\protect \xspace official
  documentation: \protect \url
  {https://ipi-code.org/i-pi/index.html}}}\BibitemShut {NoStop}%
\bibitem [{\citenamefont {Clark}\ \emph {et~al.}(2005)\citenamefont {Clark},
  \citenamefont {Segall}, \citenamefont {Pickard}, \citenamefont {Hasnip},
  \citenamefont {Probert}, \citenamefont {Refson},\ and\ \citenamefont
  {Payne}}]{CASTEP}%
  \BibitemOpen
  \bibfield  {author} {\bibinfo {author} {\bibfnamefont {S.~J.}\ \bibnamefont
  {Clark}}, \bibinfo {author} {\bibfnamefont {M.~D.}\ \bibnamefont {Segall}},
  \bibinfo {author} {\bibfnamefont {C.~J.}\ \bibnamefont {Pickard}}, \bibinfo
  {author} {\bibfnamefont {P.~J.}\ \bibnamefont {Hasnip}}, \bibinfo {author}
  {\bibfnamefont {M.~I.~J.}\ \bibnamefont {Probert}}, \bibinfo {author}
  {\bibfnamefont {K.}~\bibnamefont {Refson}},\ and\ \bibinfo {author}
  {\bibfnamefont {M.~C.}\ \bibnamefont {Payne}},\ }\bibfield  {title} {\bibinfo
  {title} {First principles methods using castep},\ }\href
  {https://doi.org/doi:10.1524/zkri.220.5.567.65075} {\bibfield  {journal}
  {\bibinfo  {journal} {Zeitschrift für Kristallographie - Crystalline
  Materials}\ }\textbf {\bibinfo {volume} {220}},\ \bibinfo {pages} {567}
  (\bibinfo {year} {2005})}\BibitemShut {NoStop}%
\bibitem [{\citenamefont {Kühne}\ \emph {et~al.}(2020)\citenamefont {Kühne},
  \citenamefont {Iannuzzi}, \citenamefont {Del~Ben}, \citenamefont {Rybkin},
  \citenamefont {Seewald}, \citenamefont {Stein}, \citenamefont {Laino},
  \citenamefont {Khaliullin}, \citenamefont {Schütt}, \citenamefont
  {Schiffmann}, \citenamefont {Golze}, \citenamefont {Wilhelm}, \citenamefont
  {Chulkov}, \citenamefont {Bani-Hashemian}, \citenamefont {Weber},
  \citenamefont {Borštnik}, \citenamefont {Taillefumier}, \citenamefont
  {Jakobovits}, \citenamefont {Lazzaro}, \citenamefont {Pabst}, \citenamefont
  {Müller}, \citenamefont {Schade}, \citenamefont {Guidon}, \citenamefont
  {Andermatt}, \citenamefont {Holmberg}, \citenamefont {Schenter},
  \citenamefont {Hehn}, \citenamefont {Bussy}, \citenamefont {Belleflamme},
  \citenamefont {Tabacchi}, \citenamefont {Glöß}, \citenamefont {Lass},
  \citenamefont {Bethune}, \citenamefont {Mundy}, \citenamefont {Plessl},
  \citenamefont {Watkins}, \citenamefont {VandeVondele}, \citenamefont
  {Krack},\ and\ \citenamefont {Hutter}}]{CP2K}%
  \BibitemOpen
  \bibfield  {author} {\bibinfo {author} {\bibfnamefont {T.~D.}\ \bibnamefont
  {Kühne}}, \bibinfo {author} {\bibfnamefont {M.}~\bibnamefont {Iannuzzi}},
  \bibinfo {author} {\bibfnamefont {M.}~\bibnamefont {Del~Ben}}, \bibinfo
  {author} {\bibfnamefont {V.~V.}\ \bibnamefont {Rybkin}}, \bibinfo {author}
  {\bibfnamefont {P.}~\bibnamefont {Seewald}}, \bibinfo {author} {\bibfnamefont
  {F.}~\bibnamefont {Stein}}, \bibinfo {author} {\bibfnamefont
  {T.}~\bibnamefont {Laino}}, \bibinfo {author} {\bibfnamefont {R.~Z.}\
  \bibnamefont {Khaliullin}}, \bibinfo {author} {\bibfnamefont
  {O.}~\bibnamefont {Schütt}}, \bibinfo {author} {\bibfnamefont
  {F.}~\bibnamefont {Schiffmann}}, \bibinfo {author} {\bibfnamefont
  {D.}~\bibnamefont {Golze}}, \bibinfo {author} {\bibfnamefont
  {J.}~\bibnamefont {Wilhelm}}, \bibinfo {author} {\bibfnamefont
  {S.}~\bibnamefont {Chulkov}}, \bibinfo {author} {\bibfnamefont {M.~H.}\
  \bibnamefont {Bani-Hashemian}}, \bibinfo {author} {\bibfnamefont
  {V.}~\bibnamefont {Weber}}, \bibinfo {author} {\bibfnamefont
  {U.}~\bibnamefont {Borštnik}}, \bibinfo {author} {\bibfnamefont
  {M.}~\bibnamefont {Taillefumier}}, \bibinfo {author} {\bibfnamefont {A.~S.}\
  \bibnamefont {Jakobovits}}, \bibinfo {author} {\bibfnamefont
  {A.}~\bibnamefont {Lazzaro}}, \bibinfo {author} {\bibfnamefont
  {H.}~\bibnamefont {Pabst}}, \bibinfo {author} {\bibfnamefont
  {T.}~\bibnamefont {Müller}}, \bibinfo {author} {\bibfnamefont
  {R.}~\bibnamefont {Schade}}, \bibinfo {author} {\bibfnamefont
  {M.}~\bibnamefont {Guidon}}, \bibinfo {author} {\bibfnamefont
  {S.}~\bibnamefont {Andermatt}}, \bibinfo {author} {\bibfnamefont
  {N.}~\bibnamefont {Holmberg}}, \bibinfo {author} {\bibfnamefont {G.~K.}\
  \bibnamefont {Schenter}}, \bibinfo {author} {\bibfnamefont {A.}~\bibnamefont
  {Hehn}}, \bibinfo {author} {\bibfnamefont {A.}~\bibnamefont {Bussy}},
  \bibinfo {author} {\bibfnamefont {F.}~\bibnamefont {Belleflamme}}, \bibinfo
  {author} {\bibfnamefont {G.}~\bibnamefont {Tabacchi}}, \bibinfo {author}
  {\bibfnamefont {A.}~\bibnamefont {Glöß}}, \bibinfo {author} {\bibfnamefont
  {M.}~\bibnamefont {Lass}}, \bibinfo {author} {\bibfnamefont {I.}~\bibnamefont
  {Bethune}}, \bibinfo {author} {\bibfnamefont {C.~J.}\ \bibnamefont {Mundy}},
  \bibinfo {author} {\bibfnamefont {C.}~\bibnamefont {Plessl}}, \bibinfo
  {author} {\bibfnamefont {M.}~\bibnamefont {Watkins}}, \bibinfo {author}
  {\bibfnamefont {J.}~\bibnamefont {VandeVondele}}, \bibinfo {author}
  {\bibfnamefont {M.}~\bibnamefont {Krack}},\ and\ \bibinfo {author}
  {\bibfnamefont {J.}~\bibnamefont {Hutter}},\ }\bibfield  {title} {\bibinfo
  {title} {{CP2K: An electronic structure and molecular dynamics software
  package - Quickstep: Efficient and accurate electronic structure
  calculations}},\ }\href {https://doi.org/10.1063/5.0007045} {\bibfield
  {journal} {\bibinfo  {journal} {The Journal of Chemical Physics}\ }\textbf
  {\bibinfo {volume} {152}},\ \bibinfo {pages} {194103} (\bibinfo {year}
  {2020})}\BibitemShut {NoStop}%
\bibitem [{\citenamefont {Hourahine}\ \emph {et~al.}(2020)\citenamefont
  {Hourahine}, \citenamefont {Aradi}, \citenamefont {Blum}, \citenamefont
  {Bonafé}, \citenamefont {Buccheri}, \citenamefont {Camacho}, \citenamefont
  {Cevallos}, \citenamefont {Deshaye}, \citenamefont {Dumitrică},
  \citenamefont {Dominguez}, \citenamefont {Ehlert}, \citenamefont {Elstner},
  \citenamefont {van~der Heide}, \citenamefont {Hermann}, \citenamefont {Irle},
  \citenamefont {Kranz}, \citenamefont {Köhler}, \citenamefont {Kowalczyk},
  \citenamefont {Kubař}, \citenamefont {Lee}, \citenamefont {Lutsker},
  \citenamefont {Maurer}, \citenamefont {Min}, \citenamefont {Mitchell},
  \citenamefont {Negre}, \citenamefont {Niehaus}, \citenamefont {Niklasson},
  \citenamefont {Page}, \citenamefont {Pecchia}, \citenamefont {Penazzi},
  \citenamefont {Persson}, \citenamefont {Řezáč}, \citenamefont {Sánchez},
  \citenamefont {Sternberg}, \citenamefont {Stöhr}, \citenamefont
  {Stuckenberg}, \citenamefont {Tkatchenko}, \citenamefont {Yu},\ and\
  \citenamefont {Frauenheim}}]{DFTB+}%
  \BibitemOpen
  \bibfield  {author} {\bibinfo {author} {\bibfnamefont {B.}~\bibnamefont
  {Hourahine}}, \bibinfo {author} {\bibfnamefont {B.}~\bibnamefont {Aradi}},
  \bibinfo {author} {\bibfnamefont {V.}~\bibnamefont {Blum}}, \bibinfo {author}
  {\bibfnamefont {F.}~\bibnamefont {Bonafé}}, \bibinfo {author} {\bibfnamefont
  {A.}~\bibnamefont {Buccheri}}, \bibinfo {author} {\bibfnamefont
  {C.}~\bibnamefont {Camacho}}, \bibinfo {author} {\bibfnamefont
  {C.}~\bibnamefont {Cevallos}}, \bibinfo {author} {\bibfnamefont {M.~Y.}\
  \bibnamefont {Deshaye}}, \bibinfo {author} {\bibfnamefont {T.}~\bibnamefont
  {Dumitrică}}, \bibinfo {author} {\bibfnamefont {A.}~\bibnamefont
  {Dominguez}}, \bibinfo {author} {\bibfnamefont {S.}~\bibnamefont {Ehlert}},
  \bibinfo {author} {\bibfnamefont {M.}~\bibnamefont {Elstner}}, \bibinfo
  {author} {\bibfnamefont {T.}~\bibnamefont {van~der Heide}}, \bibinfo {author}
  {\bibfnamefont {J.}~\bibnamefont {Hermann}}, \bibinfo {author} {\bibfnamefont
  {S.}~\bibnamefont {Irle}}, \bibinfo {author} {\bibfnamefont {J.~J.}\
  \bibnamefont {Kranz}}, \bibinfo {author} {\bibfnamefont {C.}~\bibnamefont
  {Köhler}}, \bibinfo {author} {\bibfnamefont {T.}~\bibnamefont {Kowalczyk}},
  \bibinfo {author} {\bibfnamefont {T.}~\bibnamefont {Kubař}}, \bibinfo
  {author} {\bibfnamefont {I.~S.}\ \bibnamefont {Lee}}, \bibinfo {author}
  {\bibfnamefont {V.}~\bibnamefont {Lutsker}}, \bibinfo {author} {\bibfnamefont
  {R.~J.}\ \bibnamefont {Maurer}}, \bibinfo {author} {\bibfnamefont {S.~K.}\
  \bibnamefont {Min}}, \bibinfo {author} {\bibfnamefont {I.}~\bibnamefont
  {Mitchell}}, \bibinfo {author} {\bibfnamefont {C.}~\bibnamefont {Negre}},
  \bibinfo {author} {\bibfnamefont {T.~A.}\ \bibnamefont {Niehaus}}, \bibinfo
  {author} {\bibfnamefont {A.~M.~N.}\ \bibnamefont {Niklasson}}, \bibinfo
  {author} {\bibfnamefont {A.~J.}\ \bibnamefont {Page}}, \bibinfo {author}
  {\bibfnamefont {A.}~\bibnamefont {Pecchia}}, \bibinfo {author} {\bibfnamefont
  {G.}~\bibnamefont {Penazzi}}, \bibinfo {author} {\bibfnamefont {M.~P.}\
  \bibnamefont {Persson}}, \bibinfo {author} {\bibfnamefont {J.}~\bibnamefont
  {Řezáč}}, \bibinfo {author} {\bibfnamefont {C.~G.}\ \bibnamefont
  {Sánchez}}, \bibinfo {author} {\bibfnamefont {M.}~\bibnamefont {Sternberg}},
  \bibinfo {author} {\bibfnamefont {M.}~\bibnamefont {Stöhr}}, \bibinfo
  {author} {\bibfnamefont {F.}~\bibnamefont {Stuckenberg}}, \bibinfo {author}
  {\bibfnamefont {A.}~\bibnamefont {Tkatchenko}}, \bibinfo {author}
  {\bibfnamefont {V.~W.-z.}\ \bibnamefont {Yu}},\ and\ \bibinfo {author}
  {\bibfnamefont {T.}~\bibnamefont {Frauenheim}},\ }\bibfield  {title}
  {\bibinfo {title} {{DFTB+, a software package for efficient approximate
  density functional theory based atomistic simulations}},\ }\href
  {https://doi.org/10.1063/1.5143190} {\bibfield  {journal} {\bibinfo
  {journal} {The Journal of Chemical Physics}\ }\textbf {\bibinfo {volume}
  {152}},\ \bibinfo {pages} {124101} (\bibinfo {year} {2020})}\BibitemShut
  {NoStop}%
\bibitem [{\citenamefont {Chmiela}\ \emph {et~al.}(2019)\citenamefont
  {Chmiela}, \citenamefont {Sauceda}, \citenamefont {Poltavsky}, \citenamefont
  {Müller},\ and\ \citenamefont {Tkatchenko}}]{ffsGDML}%
  \BibitemOpen
  \bibfield  {author} {\bibinfo {author} {\bibfnamefont {S.}~\bibnamefont
  {Chmiela}}, \bibinfo {author} {\bibfnamefont {H.~E.}\ \bibnamefont
  {Sauceda}}, \bibinfo {author} {\bibfnamefont {I.}~\bibnamefont {Poltavsky}},
  \bibinfo {author} {\bibfnamefont {K.-R.}\ \bibnamefont {Müller}},\ and\
  \bibinfo {author} {\bibfnamefont {A.}~\bibnamefont {Tkatchenko}},\ }\bibfield
   {title} {\bibinfo {title} {sgdml: Constructing accurate and data efficient
  molecular force fields using machine learning},\ }\href
  {https://doi.org/https://doi.org/10.1016/j.cpc.2019.02.007} {\bibfield
  {journal} {\bibinfo  {journal} {Computer Physics Communications}\ }\textbf
  {\bibinfo {volume} {240}},\ \bibinfo {pages} {38} (\bibinfo {year}
  {2019})}\BibitemShut {NoStop}%
\bibitem [{\citenamefont {Blum}\ \emph {et~al.}(2009)\citenamefont {Blum},
  \citenamefont {Gehrke}, \citenamefont {Hanke}, \citenamefont {Havu},
  \citenamefont {Havu}, \citenamefont {Ren}, \citenamefont {Reuter},\ and\
  \citenamefont {Scheffler}}]{FHI-AIMS}%
  \BibitemOpen
  \bibfield  {author} {\bibinfo {author} {\bibfnamefont {V.}~\bibnamefont
  {Blum}}, \bibinfo {author} {\bibfnamefont {R.}~\bibnamefont {Gehrke}},
  \bibinfo {author} {\bibfnamefont {F.}~\bibnamefont {Hanke}}, \bibinfo
  {author} {\bibfnamefont {P.}~\bibnamefont {Havu}}, \bibinfo {author}
  {\bibfnamefont {V.}~\bibnamefont {Havu}}, \bibinfo {author} {\bibfnamefont
  {X.}~\bibnamefont {Ren}}, \bibinfo {author} {\bibfnamefont {K.}~\bibnamefont
  {Reuter}},\ and\ \bibinfo {author} {\bibfnamefont {M.}~\bibnamefont
  {Scheffler}},\ }\bibfield  {title} {\bibinfo {title} {Ab initio molecular
  simulations with numeric atom-centered orbitals},\ }\href
  {https://doi.org/http://dx.doi.org/10.1016/j.cpc.2009.06.022} {\bibfield
  {journal} {\bibinfo  {journal} {Computer Physics Communications.}\ }\textbf
  {\bibinfo {volume} {180}},\ \bibinfo {pages} {2175 } (\bibinfo {year}
  {2009})}\BibitemShut {NoStop}%
\bibitem [{\citenamefont {Musil}\ \emph {et~al.}(2021)\citenamefont {Musil},
  \citenamefont {Stricker}, \citenamefont {Goscinski}, \citenamefont
  {F.Giberti}, \citenamefont {Veit}, \citenamefont {Junge}, \citenamefont
  {Fraux}, \citenamefont {Ceriotti}, \citenamefont {Cersonsky}, \citenamefont
  {Willatt},\ and\ \citenamefont {Grisafi}}]{librascal}%
  \BibitemOpen
  \bibfield  {author} {\bibinfo {author} {\bibfnamefont {F.}~\bibnamefont
  {Musil}}, \bibinfo {author} {\bibfnamefont {M.}~\bibnamefont {Stricker}},
  \bibinfo {author} {\bibfnamefont {A.}~\bibnamefont {Goscinski}}, \bibinfo
  {author} {\bibnamefont {F.Giberti}}, \bibinfo {author} {\bibfnamefont
  {M.}~\bibnamefont {Veit}}, \bibinfo {author} {\bibfnamefont {T.}~\bibnamefont
  {Junge}}, \bibinfo {author} {\bibfnamefont {G.}~\bibnamefont {Fraux}},
  \bibinfo {author} {\bibfnamefont {M.}~\bibnamefont {Ceriotti}}, \bibinfo
  {author} {\bibfnamefont {R.}~\bibnamefont {Cersonsky}}, \bibinfo {author}
  {\bibfnamefont {M.}~\bibnamefont {Willatt}},\ and\ \bibinfo {author}
  {\bibfnamefont {A.}~\bibnamefont {Grisafi}},\ }\href@noop {} {\bibinfo
  {title} {cosmo-epfl/librascal, zenodo.
  https://doi.org/10.5281/zenodo.4526063}} (\bibinfo {year} {2021})\BibitemShut
  {NoStop}%
\bibitem [{\citenamefont {Giannozzi}\ \emph {et~al.}(2009)\citenamefont
  {Giannozzi}, \citenamefont {Baroni}, \citenamefont {Bonini}, \citenamefont
  {Calandra}, \citenamefont {Car}, \citenamefont {Cavazzoni}, \citenamefont
  {Ceresoli}, \citenamefont {Chiarotti}, \citenamefont {Cococcioni},
  \citenamefont {Dabo}, \citenamefont {Corso}, \citenamefont {de~Gironcoli},
  \citenamefont {Fabris}, \citenamefont {Fratesi}, \citenamefont {Gebauer},
  \citenamefont {Gerstmann}, \citenamefont {Gougoussis}, \citenamefont
  {Kokalj}, \citenamefont {Lazzeri}, \citenamefont {Martin-Samos},
  \citenamefont {Marzari}, \citenamefont {Mauri}, \citenamefont {Mazzarello},
  \citenamefont {Paolini}, \citenamefont {Pasquarello}, \citenamefont
  {Paulatto}, \citenamefont {Sbraccia}, \citenamefont {Scandolo}, \citenamefont
  {Sclauzero}, \citenamefont {Seitsonen}, \citenamefont {Smogunov},
  \citenamefont {Umari},\ and\ \citenamefont {Wentzcovitch}}]{Qespresso}%
  \BibitemOpen
  \bibfield  {author} {\bibinfo {author} {\bibfnamefont {P.}~\bibnamefont
  {Giannozzi}}, \bibinfo {author} {\bibfnamefont {S.}~\bibnamefont {Baroni}},
  \bibinfo {author} {\bibfnamefont {N.}~\bibnamefont {Bonini}}, \bibinfo
  {author} {\bibfnamefont {M.}~\bibnamefont {Calandra}}, \bibinfo {author}
  {\bibfnamefont {R.}~\bibnamefont {Car}}, \bibinfo {author} {\bibfnamefont
  {C.}~\bibnamefont {Cavazzoni}}, \bibinfo {author} {\bibfnamefont
  {D.}~\bibnamefont {Ceresoli}}, \bibinfo {author} {\bibfnamefont {G.~L.}\
  \bibnamefont {Chiarotti}}, \bibinfo {author} {\bibfnamefont {M.}~\bibnamefont
  {Cococcioni}}, \bibinfo {author} {\bibfnamefont {I.}~\bibnamefont {Dabo}},
  \bibinfo {author} {\bibfnamefont {A.~D.}\ \bibnamefont {Corso}}, \bibinfo
  {author} {\bibfnamefont {S.}~\bibnamefont {de~Gironcoli}}, \bibinfo {author}
  {\bibfnamefont {S.}~\bibnamefont {Fabris}}, \bibinfo {author} {\bibfnamefont
  {G.}~\bibnamefont {Fratesi}}, \bibinfo {author} {\bibfnamefont
  {R.}~\bibnamefont {Gebauer}}, \bibinfo {author} {\bibfnamefont
  {U.}~\bibnamefont {Gerstmann}}, \bibinfo {author} {\bibfnamefont
  {C.}~\bibnamefont {Gougoussis}}, \bibinfo {author} {\bibfnamefont
  {A.}~\bibnamefont {Kokalj}}, \bibinfo {author} {\bibfnamefont
  {M.}~\bibnamefont {Lazzeri}}, \bibinfo {author} {\bibfnamefont
  {L.}~\bibnamefont {Martin-Samos}}, \bibinfo {author} {\bibfnamefont
  {N.}~\bibnamefont {Marzari}}, \bibinfo {author} {\bibfnamefont
  {F.}~\bibnamefont {Mauri}}, \bibinfo {author} {\bibfnamefont
  {R.}~\bibnamefont {Mazzarello}}, \bibinfo {author} {\bibfnamefont
  {S.}~\bibnamefont {Paolini}}, \bibinfo {author} {\bibfnamefont
  {A.}~\bibnamefont {Pasquarello}}, \bibinfo {author} {\bibfnamefont
  {L.}~\bibnamefont {Paulatto}}, \bibinfo {author} {\bibfnamefont
  {C.}~\bibnamefont {Sbraccia}}, \bibinfo {author} {\bibfnamefont
  {S.}~\bibnamefont {Scandolo}}, \bibinfo {author} {\bibfnamefont
  {G.}~\bibnamefont {Sclauzero}}, \bibinfo {author} {\bibfnamefont {A.~P.}\
  \bibnamefont {Seitsonen}}, \bibinfo {author} {\bibfnamefont {A.}~\bibnamefont
  {Smogunov}}, \bibinfo {author} {\bibfnamefont {P.}~\bibnamefont {Umari}},\
  and\ \bibinfo {author} {\bibfnamefont {R.~M.}\ \bibnamefont {Wentzcovitch}},\
  }\bibfield  {title} {\bibinfo {title} {Quantum espresso: a modular and
  open-source software project for quantum simulations of materials},\ }\href
  {https://doi.org/10.1088/0953-8984/21/39/395502} {\bibfield  {journal}
  {\bibinfo  {journal} {Journal of Physics: Condensed Matter}\ }\textbf
  {\bibinfo {volume} {21}},\ \bibinfo {pages} {395502} (\bibinfo {year}
  {2009})}\BibitemShut {NoStop}%
\bibitem [{\citenamefont {Giannozzi}\ \emph {et~al.}(2017)\citenamefont
  {Giannozzi}, \citenamefont {Andreussi}, \citenamefont {Brumme}, \citenamefont
  {Bunau}, \citenamefont {Nardelli}, \citenamefont {Calandra}, \citenamefont
  {Car}, \citenamefont {Cavazzoni}, \citenamefont {Ceresoli}, \citenamefont
  {Cococcioni}, \citenamefont {Colonna}, \citenamefont {Carnimeo},
  \citenamefont {Corso}, \citenamefont {de~Gironcoli}, \citenamefont {Delugas},
  \citenamefont {Jr}, \citenamefont {Ferretti}, \citenamefont {Floris},
  \citenamefont {Fratesi}, \citenamefont {Fugallo}, \citenamefont {Gebauer},
  \citenamefont {Gerstmann}, \citenamefont {Giustino}, \citenamefont {Gorni},
  \citenamefont {Jia}, \citenamefont {Kawamura}, \citenamefont {Ko},
  \citenamefont {Kokalj}, \citenamefont {Küçükbenli}, \citenamefont
  {Lazzeri}, \citenamefont {Marsili}, \citenamefont {Marzari}, \citenamefont
  {Mauri}, \citenamefont {Nguyen}, \citenamefont {Nguyen}, \citenamefont {de-la
  Roza}, \citenamefont {Paulatto}, \citenamefont {Poncé}, \citenamefont
  {Rocca}, \citenamefont {Sabatini}, \citenamefont {Santra}, \citenamefont
  {Schlipf}, \citenamefont {Seitsonen}, \citenamefont {Smogunov}, \citenamefont
  {Timrov}, \citenamefont {Thonhauser}, \citenamefont {Umari}, \citenamefont
  {Vast}, \citenamefont {Wu},\ and\ \citenamefont {Baroni}}]{QE-2017}%
  \BibitemOpen
  \bibfield  {author} {\bibinfo {author} {\bibfnamefont {P.}~\bibnamefont
  {Giannozzi}}, \bibinfo {author} {\bibfnamefont {O.}~\bibnamefont
  {Andreussi}}, \bibinfo {author} {\bibfnamefont {T.}~\bibnamefont {Brumme}},
  \bibinfo {author} {\bibfnamefont {O.}~\bibnamefont {Bunau}}, \bibinfo
  {author} {\bibfnamefont {M.~B.}\ \bibnamefont {Nardelli}}, \bibinfo {author}
  {\bibfnamefont {M.}~\bibnamefont {Calandra}}, \bibinfo {author}
  {\bibfnamefont {R.}~\bibnamefont {Car}}, \bibinfo {author} {\bibfnamefont
  {C.}~\bibnamefont {Cavazzoni}}, \bibinfo {author} {\bibfnamefont
  {D.}~\bibnamefont {Ceresoli}}, \bibinfo {author} {\bibfnamefont
  {M.}~\bibnamefont {Cococcioni}}, \bibinfo {author} {\bibfnamefont
  {N.}~\bibnamefont {Colonna}}, \bibinfo {author} {\bibfnamefont
  {I.}~\bibnamefont {Carnimeo}}, \bibinfo {author} {\bibfnamefont {A.~D.}\
  \bibnamefont {Corso}}, \bibinfo {author} {\bibfnamefont {S.}~\bibnamefont
  {de~Gironcoli}}, \bibinfo {author} {\bibfnamefont {P.}~\bibnamefont
  {Delugas}}, \bibinfo {author} {\bibfnamefont {R.~A.~D.}\ \bibnamefont {Jr}},
  \bibinfo {author} {\bibfnamefont {A.}~\bibnamefont {Ferretti}}, \bibinfo
  {author} {\bibfnamefont {A.}~\bibnamefont {Floris}}, \bibinfo {author}
  {\bibfnamefont {G.}~\bibnamefont {Fratesi}}, \bibinfo {author} {\bibfnamefont
  {G.}~\bibnamefont {Fugallo}}, \bibinfo {author} {\bibfnamefont
  {R.}~\bibnamefont {Gebauer}}, \bibinfo {author} {\bibfnamefont
  {U.}~\bibnamefont {Gerstmann}}, \bibinfo {author} {\bibfnamefont
  {F.}~\bibnamefont {Giustino}}, \bibinfo {author} {\bibfnamefont
  {T.}~\bibnamefont {Gorni}}, \bibinfo {author} {\bibfnamefont
  {J.}~\bibnamefont {Jia}}, \bibinfo {author} {\bibfnamefont {M.}~\bibnamefont
  {Kawamura}}, \bibinfo {author} {\bibfnamefont {H.-Y.}\ \bibnamefont {Ko}},
  \bibinfo {author} {\bibfnamefont {A.}~\bibnamefont {Kokalj}}, \bibinfo
  {author} {\bibfnamefont {E.}~\bibnamefont {Küçükbenli}}, \bibinfo {author}
  {\bibfnamefont {M.}~\bibnamefont {Lazzeri}}, \bibinfo {author} {\bibfnamefont
  {M.}~\bibnamefont {Marsili}}, \bibinfo {author} {\bibfnamefont
  {N.}~\bibnamefont {Marzari}}, \bibinfo {author} {\bibfnamefont
  {F.}~\bibnamefont {Mauri}}, \bibinfo {author} {\bibfnamefont {N.~L.}\
  \bibnamefont {Nguyen}}, \bibinfo {author} {\bibfnamefont {H.-V.}\
  \bibnamefont {Nguyen}}, \bibinfo {author} {\bibfnamefont {A.~O.}\
  \bibnamefont {de-la Roza}}, \bibinfo {author} {\bibfnamefont
  {L.}~\bibnamefont {Paulatto}}, \bibinfo {author} {\bibfnamefont
  {S.}~\bibnamefont {Poncé}}, \bibinfo {author} {\bibfnamefont
  {D.}~\bibnamefont {Rocca}}, \bibinfo {author} {\bibfnamefont
  {R.}~\bibnamefont {Sabatini}}, \bibinfo {author} {\bibfnamefont
  {B.}~\bibnamefont {Santra}}, \bibinfo {author} {\bibfnamefont
  {M.}~\bibnamefont {Schlipf}}, \bibinfo {author} {\bibfnamefont {A.~P.}\
  \bibnamefont {Seitsonen}}, \bibinfo {author} {\bibfnamefont {A.}~\bibnamefont
  {Smogunov}}, \bibinfo {author} {\bibfnamefont {I.}~\bibnamefont {Timrov}},
  \bibinfo {author} {\bibfnamefont {T.}~\bibnamefont {Thonhauser}}, \bibinfo
  {author} {\bibfnamefont {P.}~\bibnamefont {Umari}}, \bibinfo {author}
  {\bibfnamefont {N.}~\bibnamefont {Vast}}, \bibinfo {author} {\bibfnamefont
  {X.}~\bibnamefont {Wu}},\ and\ \bibinfo {author} {\bibfnamefont
  {S.}~\bibnamefont {Baroni}},\ }\bibfield  {title} {\bibinfo {title} {Advanced
  capabilities for materials modelling with quantum espresso},\ }\href
  {http://stacks.iop.org/0953-8984/29/i=46/a=465901} {\bibfield  {journal}
  {\bibinfo  {journal} {Journal of Physics: Condensed Matter}\ }\textbf
  {\bibinfo {volume} {29}},\ \bibinfo {pages} {465901} (\bibinfo {year}
  {2017})}\BibitemShut {NoStop}%
\bibitem [{\citenamefont {Giannozzi}\ \emph {et~al.}(2020)\citenamefont
  {Giannozzi}, \citenamefont {Baseggio}, \citenamefont {Bonfà}, \citenamefont
  {Brunato}, \citenamefont {Car}, \citenamefont {Carnimeo}, \citenamefont
  {Cavazzoni}, \citenamefont {de~Gironcoli}, \citenamefont {Delugas},
  \citenamefont {Ferrari~Ruffino}, \citenamefont {Ferretti}, \citenamefont
  {Marzari}, \citenamefont {Timrov}, \citenamefont {Urru},\ and\ \citenamefont
  {Baroni}}]{QE-2020}%
  \BibitemOpen
  \bibfield  {author} {\bibinfo {author} {\bibfnamefont {P.}~\bibnamefont
  {Giannozzi}}, \bibinfo {author} {\bibfnamefont {O.}~\bibnamefont {Baseggio}},
  \bibinfo {author} {\bibfnamefont {P.}~\bibnamefont {Bonfà}}, \bibinfo
  {author} {\bibfnamefont {D.}~\bibnamefont {Brunato}}, \bibinfo {author}
  {\bibfnamefont {R.}~\bibnamefont {Car}}, \bibinfo {author} {\bibfnamefont
  {I.}~\bibnamefont {Carnimeo}}, \bibinfo {author} {\bibfnamefont
  {C.}~\bibnamefont {Cavazzoni}}, \bibinfo {author} {\bibfnamefont
  {S.}~\bibnamefont {de~Gironcoli}}, \bibinfo {author} {\bibfnamefont
  {P.}~\bibnamefont {Delugas}}, \bibinfo {author} {\bibfnamefont
  {F.}~\bibnamefont {Ferrari~Ruffino}}, \bibinfo {author} {\bibfnamefont
  {A.}~\bibnamefont {Ferretti}}, \bibinfo {author} {\bibfnamefont
  {N.}~\bibnamefont {Marzari}}, \bibinfo {author} {\bibfnamefont
  {I.}~\bibnamefont {Timrov}}, \bibinfo {author} {\bibfnamefont
  {A.}~\bibnamefont {Urru}},\ and\ \bibinfo {author} {\bibfnamefont
  {S.}~\bibnamefont {Baroni}},\ }\bibfield  {title} {\bibinfo {title} {Quantum
  espresso toward the exascale},\ }\href {https://doi.org/10.1063/5.0005082}
  {\bibfield  {journal} {\bibinfo  {journal} {The Journal of Chemical Physics}\
  }\textbf {\bibinfo {volume} {152}},\ \bibinfo {pages} {154105} (\bibinfo
  {year} {2020})}\BibitemShut {NoStop}%
\bibitem [{\citenamefont {Kresse}\ and\ \citenamefont
  {Furthm{\"{u}}ller}(1996)}]{VASP}%
  \BibitemOpen
  \bibfield  {author} {\bibinfo {author} {\bibfnamefont {G.}~\bibnamefont
  {Kresse}}\ and\ \bibinfo {author} {\bibfnamefont {J.}~\bibnamefont
  {Furthm{\"{u}}ller}},\ }\bibfield  {title} {\bibinfo {title} {{Efficient
  iterative schemes for ab initio total-energy calculations using a plane-wave
  basis set}},\ }\href@noop {} {\bibfield  {journal} {\bibinfo  {journal}
  {Physiscal Review B}\ }\textbf {\bibinfo {volume} {54}},\ \bibinfo {pages}
  {11169} (\bibinfo {year} {1996})}\BibitemShut {NoStop}%
\bibitem [{\citenamefont {Berges}\ \emph {et~al.}(2017)\citenamefont {Berges},
  \citenamefont {Schobert}, \citenamefont {van Loon}, \citenamefont {Rösner},\
  and\ \citenamefont {Wehling}}]{elphmod}%
  \BibitemOpen
  \bibfield  {author} {\bibinfo {author} {\bibfnamefont {J.}~\bibnamefont
  {Berges}}, \bibinfo {author} {\bibfnamefont {A.}~\bibnamefont {Schobert}},
  \bibinfo {author} {\bibfnamefont {E.~G. C.~P.}\ \bibnamefont {van Loon}},
  \bibinfo {author} {\bibfnamefont {M.}~\bibnamefont {Rösner}},\ and\ \bibinfo
  {author} {\bibfnamefont {T.~O.}\ \bibnamefont {Wehling}},\ }\bibfield
  {title} {\bibinfo {title} {elphmod: Python modules for electron-phonon
  models},\ }\href@noop {} {\bibfield  {journal} {\bibinfo  {journal} {Zenodo}\
  } (\bibinfo {year} {2017})},\ \bibinfo {note}
  {doi:10.5281/zenodo.5919991}\BibitemShut {NoStop}%
\bibitem [{\citenamefont {T.~Verstraelen}\ and\ \citenamefont
  {Rogge}()}]{YAFF}%
  \BibitemOpen
  \bibfield  {author} {\bibinfo {author} {\bibfnamefont {S.~V.}\ \bibnamefont
  {T.~Verstraelen}, \bibfnamefont {L.~Vanduyfhuys}}\ and\ \bibinfo {author}
  {\bibfnamefont {S.~M.~J.}\ \bibnamefont {Rogge}},\ }\href@noop {} {\bibinfo
  {title} {Yaff, yet another force field,
  https://molmod.ugent.be/software/}}\BibitemShut {NoStop}%
\bibitem [{Note2()}]{Note2}%
  \BibitemOpen
  \bibinfo {note} {{\protect \color {blue}The \enquote {Path Integral Methods
  in Atomistic Modelling} is hosted at \protect \url
  {https://app.courseware.epfl.ch/learning/course/course-v1:EPFL+X+2022/home}
  and code tutorials at \protect \url
  {https://github.com/i-pi/piqm2023-tutorial/}}}\BibitemShut {NoStop}%
\bibitem [{Note3()}]{Note3}%
  \BibitemOpen
  \bibinfo {note} {{\protect \color {blue}See \protect \url
  {https://flake8.pycqa.org/}} for more information about Flake8}\BibitemShut
  {NoStop}%
\bibitem [{Note4()}]{Note4}%
  \BibitemOpen
  \bibinfo {note} {{\protect \color {blue}See \protect \url
  {https://github.com/psf/black} for more information about Black}}\BibitemShut
  {NoStop}%
\bibitem [{\citenamefont {Bussi}\ \emph
  {et~al.}(2007{\natexlab{a}})\citenamefont {Bussi}, \citenamefont {Donadio},\
  and\ \citenamefont {Parrinello}}]{buss+07jcp}%
  \BibitemOpen
  \bibfield  {author} {\bibinfo {author} {\bibfnamefont {G.}~\bibnamefont
  {Bussi}}, \bibinfo {author} {\bibfnamefont {D.}~\bibnamefont {Donadio}},\
  and\ \bibinfo {author} {\bibfnamefont {M.}~\bibnamefont {Parrinello}},\
  }\bibfield  {title} {\bibinfo {title} {{Canonical sampling through velocity
  rescaling}},\ }\href@noop {} {\bibfield  {journal} {\bibinfo  {journal} {The
  Journal of Chemical Physics}\ }\textbf {\bibinfo {volume} {126}},\ \bibinfo
  {pages} {14101} (\bibinfo {year} {2007}{\natexlab{a}})}\BibitemShut {NoStop}%
\bibitem [{\citenamefont {Ceriotti}\ \emph
  {et~al.}(2010{\natexlab{a}})\citenamefont {Ceriotti}, \citenamefont
  {Parrinello}, \citenamefont {Markland},\ and\ \citenamefont
  {Manolopoulos}}]{ceri+10jcp}%
  \BibitemOpen
  \bibfield  {author} {\bibinfo {author} {\bibfnamefont {M.}~\bibnamefont
  {Ceriotti}}, \bibinfo {author} {\bibfnamefont {M.}~\bibnamefont
  {Parrinello}}, \bibinfo {author} {\bibfnamefont {T.~E.}\ \bibnamefont
  {Markland}},\ and\ \bibinfo {author} {\bibfnamefont {D.~E.}\ \bibnamefont
  {Manolopoulos}},\ }\bibfield  {title} {\bibinfo {title} {{Efficient
  stochastic thermostatting of path integral molecular dynamics.}},\
  }\href@noop {} {\bibfield  {journal} {\bibinfo  {journal} {The Journal of
  Chemical Physics}\ }\textbf {\bibinfo {volume} {133}},\ \bibinfo {pages}
  {124104} (\bibinfo {year} {2010}{\natexlab{a}})}\BibitemShut {NoStop}%
\bibitem [{\citenamefont {Ceriotti}\ \emph
  {et~al.}(2009{\natexlab{a}})\citenamefont {Ceriotti}, \citenamefont {Bussi},\
  and\ \citenamefont {Parrinello}}]{ceri+09prl}%
  \BibitemOpen
  \bibfield  {author} {\bibinfo {author} {\bibfnamefont {M.}~\bibnamefont
  {Ceriotti}}, \bibinfo {author} {\bibfnamefont {G.}~\bibnamefont {Bussi}},\
  and\ \bibinfo {author} {\bibfnamefont {M.}~\bibnamefont {Parrinello}},\
  }\bibfield  {title} {\bibinfo {title} {{Langevin Equation with Colored Noise
  for Constant-Temperature Molecular Dynamics Simulations}},\ }\href@noop {}
  {\bibfield  {journal} {\bibinfo  {journal} {Physical Review Letters}\
  }\textbf {\bibinfo {volume} {102}},\ \bibinfo {pages} {020601} (\bibinfo
  {year} {2009}{\natexlab{a}})}\BibitemShut {NoStop}%
\bibitem [{\citenamefont {Ceriotti}\ \emph
  {et~al.}(2010{\natexlab{b}})\citenamefont {Ceriotti}, \citenamefont {Bussi},\
  and\ \citenamefont {Parrinello}}]{ceri+10jctc}%
  \BibitemOpen
  \bibfield  {author} {\bibinfo {author} {\bibfnamefont {M.}~\bibnamefont
  {Ceriotti}}, \bibinfo {author} {\bibfnamefont {G.}~\bibnamefont {Bussi}},\
  and\ \bibinfo {author} {\bibfnamefont {M.}~\bibnamefont {Parrinello}},\
  }\bibfield  {title} {\bibinfo {title} {{Colored-Noise Thermostats {\`{a}} la
  Carte}},\ }\href@noop {} {\bibfield  {journal} {\bibinfo  {journal} {Journal
  of Chemical Theory and Computation}\ }\textbf {\bibinfo {volume} {6}},\
  \bibinfo {pages} {1170} (\bibinfo {year} {2010}{\natexlab{b}})}\BibitemShut
  {NoStop}%
\bibitem [{\citenamefont {Ceriotti}\ \emph
  {et~al.}(2009{\natexlab{b}})\citenamefont {Ceriotti}, \citenamefont {Bussi},\
  and\ \citenamefont {Parrinello}}]{ceri+09prl2}%
  \BibitemOpen
  \bibfield  {author} {\bibinfo {author} {\bibfnamefont {M.}~\bibnamefont
  {Ceriotti}}, \bibinfo {author} {\bibfnamefont {G.}~\bibnamefont {Bussi}},\
  and\ \bibinfo {author} {\bibfnamefont {M.}~\bibnamefont {Parrinello}},\
  }\bibfield  {title} {\bibinfo {title} {{Nuclear Quantum Effects in Solids
  Using a Colored-Noise Thermostat}},\ }\href@noop {} {\bibfield  {journal}
  {\bibinfo  {journal} {Physical Review Letters}\ }\textbf {\bibinfo {volume}
  {103}},\ \bibinfo {pages} {30603} (\bibinfo {year}
  {2009}{\natexlab{b}})}\BibitemShut {NoStop}%
\bibitem [{\citenamefont {Ceriotti}\ and\ \citenamefont
  {Parrinello}(2010)}]{ceri-parr10pcs}%
  \BibitemOpen
  \bibfield  {author} {\bibinfo {author} {\bibfnamefont {M.}~\bibnamefont
  {Ceriotti}}\ and\ \bibinfo {author} {\bibfnamefont {M.}~\bibnamefont
  {Parrinello}},\ }\bibfield  {title} {\bibinfo {title} {{The
  $\delta$-thermostat: selective normal-modes excitation by colored-noise
  Langevin dynamics}},\ }\href@noop {} {\bibfield  {journal} {\bibinfo
  {journal} {Procedia Computer Science}\ }\textbf {\bibinfo {volume} {1}},\
  \bibinfo {pages} {1607} (\bibinfo {year} {2010})}\BibitemShut {NoStop}%
\bibitem [{\citenamefont {Dettori}\ \emph {et~al.}(2017)\citenamefont
  {Dettori}, \citenamefont {Ceriotti}, \citenamefont {Hunger}, \citenamefont
  {Melis}, \citenamefont {Colombo},\ and\ \citenamefont
  {Donadio}}]{dett+17jctc}%
  \BibitemOpen
  \bibfield  {author} {\bibinfo {author} {\bibfnamefont {R.}~\bibnamefont
  {Dettori}}, \bibinfo {author} {\bibfnamefont {M.}~\bibnamefont {Ceriotti}},
  \bibinfo {author} {\bibfnamefont {J.}~\bibnamefont {Hunger}}, \bibinfo
  {author} {\bibfnamefont {C.}~\bibnamefont {Melis}}, \bibinfo {author}
  {\bibfnamefont {L.}~\bibnamefont {Colombo}},\ and\ \bibinfo {author}
  {\bibfnamefont {D.}~\bibnamefont {Donadio}},\ }\bibfield  {title} {\bibinfo
  {title} {{Simulating Energy Relaxation in Pump–Probe Vibrational
  Spectroscopy of Hydrogen-Bonded Liquids}},\ }\href@noop {} {\bibfield
  {journal} {\bibinfo  {journal} {Journal of Chemical Theory and Computation}\
  }\textbf {\bibinfo {volume} {13}},\ \bibinfo {pages} {1284} (\bibinfo {year}
  {2017})}\BibitemShut {NoStop}%
\bibitem [{\citenamefont {Ceriotti}\ and\ \citenamefont
  {Manolopoulos}(2012)}]{ceri-mano12prl}%
  \BibitemOpen
  \bibfield  {author} {\bibinfo {author} {\bibfnamefont {M.}~\bibnamefont
  {Ceriotti}}\ and\ \bibinfo {author} {\bibfnamefont {D.~E.}\ \bibnamefont
  {Manolopoulos}},\ }\bibfield  {title} {\bibinfo {title} {{Efficient
  First-Principles Calculation of the Quantum Kinetic Energy and Momentum
  Distribution of Nuclei}},\ }\href@noop {} {\bibfield  {journal} {\bibinfo
  {journal} {Physical Review Letters}\ }\textbf {\bibinfo {volume} {109}},\
  \bibinfo {pages} {100604} (\bibinfo {year} {2012})}\BibitemShut {NoStop}%
\bibitem [{\citenamefont {Rossi}\ \emph {et~al.}(2018)\citenamefont {Rossi},
  \citenamefont {Kapil},\ and\ \citenamefont {Ceriotti}}]{ross+18jcp}%
  \BibitemOpen
  \bibfield  {author} {\bibinfo {author} {\bibfnamefont {M.}~\bibnamefont
  {Rossi}}, \bibinfo {author} {\bibfnamefont {V.}~\bibnamefont {Kapil}},\ and\
  \bibinfo {author} {\bibfnamefont {M.}~\bibnamefont {Ceriotti}},\ }\bibfield
  {title} {\bibinfo {title} {{Fine tuning classical and quantum molecular
  dynamics using a generalized Langevin equation}},\ }\href@noop {} {\bibfield
  {journal} {\bibinfo  {journal} {The Journal of Chemical Physics}\ }\textbf
  {\bibinfo {volume} {148}},\ \bibinfo {pages} {102301} (\bibinfo {year}
  {2018})}\BibitemShut {NoStop}%
\bibitem [{\citenamefont {Hijazi}\ \emph {et~al.}(2018)\citenamefont {Hijazi},
  \citenamefont {Wilkins},\ and\ \citenamefont {Ceriotti}}]{hija+18jcp}%
  \BibitemOpen
  \bibfield  {author} {\bibinfo {author} {\bibfnamefont {M.}~\bibnamefont
  {Hijazi}}, \bibinfo {author} {\bibfnamefont {D.~M.~D.}\ \bibnamefont
  {Wilkins}},\ and\ \bibinfo {author} {\bibfnamefont {M.}~\bibnamefont
  {Ceriotti}},\ }\bibfield  {title} {\bibinfo {title} {{Fast-forward Langevin
  dynamics with momentum flips}},\ }\href@noop {} {\bibfield  {journal}
  {\bibinfo  {journal} {The Journal of Chemical Physics}\ }\textbf {\bibinfo
  {volume} {148}},\ \bibinfo {pages} {184109} (\bibinfo {year}
  {2018})}\BibitemShut {NoStop}%
\bibitem [{\citenamefont {Krajewski}\ and\ \citenamefont
  {Parrinello}(2006)}]{PhysRevB.73.041105}%
  \BibitemOpen
  \bibfield  {author} {\bibinfo {author} {\bibfnamefont {F.~R.}\ \bibnamefont
  {Krajewski}}\ and\ \bibinfo {author} {\bibfnamefont {M.}~\bibnamefont
  {Parrinello}},\ }\bibfield  {title} {\bibinfo {title} {Linear scaling
  electronic structure calculations and accurate statistical mechanics sampling
  with noisy forces},\ }\href {https://doi.org/10.1103/PhysRevB.73.041105}
  {\bibfield  {journal} {\bibinfo  {journal} {Physical Review B}\ }\textbf
  {\bibinfo {volume} {73}},\ \bibinfo {pages} {041105} (\bibinfo {year}
  {2006})}\BibitemShut {NoStop}%
\bibitem [{\citenamefont {K\"uhne}\ \emph {et~al.}(2007)\citenamefont
  {K\"uhne}, \citenamefont {Krack}, \citenamefont {Mohamed},\ and\
  \citenamefont {Parrinello}}]{PhysRevLett.98.066401}%
  \BibitemOpen
  \bibfield  {author} {\bibinfo {author} {\bibfnamefont {T.~D.}\ \bibnamefont
  {K\"uhne}}, \bibinfo {author} {\bibfnamefont {M.}~\bibnamefont {Krack}},
  \bibinfo {author} {\bibfnamefont {F.~R.}\ \bibnamefont {Mohamed}},\ and\
  \bibinfo {author} {\bibfnamefont {M.}~\bibnamefont {Parrinello}},\ }\bibfield
   {title} {\bibinfo {title} {Efficient and accurate car-parrinello-like
  approach to born-oppenheimer molecular dynamics},\ }\href
  {https://doi.org/10.1103/PhysRevLett.98.066401} {\bibfield  {journal}
  {\bibinfo  {journal} {Physical Review Letters}\ }\textbf {\bibinfo {volume}
  {98}},\ \bibinfo {pages} {066401} (\bibinfo {year} {2007})}\BibitemShut
  {NoStop}%
\bibitem [{\citenamefont {Bussi}\ \emph {et~al.}(2009)\citenamefont {Bussi},
  \citenamefont {Zykova-Timan},\ and\ \citenamefont
  {Parrinello}}]{Bussi_JCP_2009}%
  \BibitemOpen
  \bibfield  {author} {\bibinfo {author} {\bibfnamefont {G.}~\bibnamefont
  {Bussi}}, \bibinfo {author} {\bibfnamefont {T.}~\bibnamefont
  {Zykova-Timan}},\ and\ \bibinfo {author} {\bibfnamefont {M.}~\bibnamefont
  {Parrinello}},\ }\bibfield  {title} {\bibinfo {title} {{Isothermal-isobaric
  molecular dynamics using stochastic velocity rescaling}},\ }\href
  {https://doi.org/10.1063/1.3073889} {\bibfield  {journal} {\bibinfo
  {journal} {The Journal of Chemical Physics}\ }\textbf {\bibinfo {volume}
  {130}},\ \bibinfo {pages} {074101} (\bibinfo {year} {2009})}\BibitemShut
  {NoStop}%
\bibitem [{\citenamefont {Raiteri}\ \emph {et~al.}(2011)\citenamefont
  {Raiteri}, \citenamefont {Gale},\ and\ \citenamefont
  {Bussi}}]{Raiteri_JPhys_2011}%
  \BibitemOpen
  \bibfield  {author} {\bibinfo {author} {\bibfnamefont {P.}~\bibnamefont
  {Raiteri}}, \bibinfo {author} {\bibfnamefont {J.~D.}\ \bibnamefont {Gale}},\
  and\ \bibinfo {author} {\bibfnamefont {G.}~\bibnamefont {Bussi}},\ }\bibfield
   {title} {\bibinfo {title} {Reactive force field simulation of proton
  diffusion in bazro3 using an empirical valence bond approach},\ }\href
  {https://doi.org/10.1088/0953-8984/23/33/334213} {\bibfield  {journal}
  {\bibinfo  {journal} {Journal of Physics: Condensed Matter}\ }\textbf
  {\bibinfo {volume} {23}},\ \bibinfo {pages} {334213} (\bibinfo {year}
  {2011})}\BibitemShut {NoStop}%
\bibitem [{\citenamefont {Martyna}\ \emph {et~al.}(1999)\citenamefont
  {Martyna}, \citenamefont {Hughes},\ and\ \citenamefont
  {Tuckerman}}]{Martyna_JCP_1999}%
  \BibitemOpen
  \bibfield  {author} {\bibinfo {author} {\bibfnamefont {G.~J.}\ \bibnamefont
  {Martyna}}, \bibinfo {author} {\bibfnamefont {A.}~\bibnamefont {Hughes}},\
  and\ \bibinfo {author} {\bibfnamefont {M.~E.}\ \bibnamefont {Tuckerman}},\
  }\bibfield  {title} {\bibinfo {title} {{Molecular dynamics algorithms for
  path integrals at constant pressure}},\ }\href
  {https://doi.org/10.1063/1.478193} {\bibfield  {journal} {\bibinfo  {journal}
  {The Journal of Chemical Physics}\ }\textbf {\bibinfo {volume} {110}},\
  \bibinfo {pages} {3275} (\bibinfo {year} {1999})}\BibitemShut {NoStop}%
\bibitem [{\citenamefont {Kapil}\ \emph
  {et~al.}(2019{\natexlab{c}})\citenamefont {Kapil}, \citenamefont {Wieme},
  \citenamefont {Vandenbrande}, \citenamefont {Lamaire}, \citenamefont
  {Van~Speybroeck},\ and\ \citenamefont {Ceriotti}}]{Kapil_JCTC_2019b}%
  \BibitemOpen
  \bibfield  {author} {\bibinfo {author} {\bibfnamefont {V.}~\bibnamefont
  {Kapil}}, \bibinfo {author} {\bibfnamefont {J.}~\bibnamefont {Wieme}},
  \bibinfo {author} {\bibfnamefont {S.}~\bibnamefont {Vandenbrande}}, \bibinfo
  {author} {\bibfnamefont {A.}~\bibnamefont {Lamaire}}, \bibinfo {author}
  {\bibfnamefont {V.}~\bibnamefont {Van~Speybroeck}},\ and\ \bibinfo {author}
  {\bibfnamefont {M.}~\bibnamefont {Ceriotti}},\ }\bibfield  {title} {\bibinfo
  {title} {Modeling the structural and thermal properties of loaded
  metal–organic frameworks. an interplay of quantum and anharmonic
  fluctuations},\ }\href {https://doi.org/10.1021/acs.jctc.8b01297} {\bibfield
  {journal} {\bibinfo  {journal} {Journal of Chemical Theory and Computation}\
  }\textbf {\bibinfo {volume} {15}},\ \bibinfo {pages} {3237} (\bibinfo {year}
  {2019}{\natexlab{c}})}\BibitemShut {NoStop}%
\bibitem [{\citenamefont {Henkelman}\ \emph {et~al.}(2000)\citenamefont
  {Henkelman}, \citenamefont {Uberuaga},\ and\ \citenamefont
  {Jónsson}}]{Henkelman_JCP_2000_I}%
  \BibitemOpen
  \bibfield  {author} {\bibinfo {author} {\bibfnamefont {G.}~\bibnamefont
  {Henkelman}}, \bibinfo {author} {\bibfnamefont {B.~P.}\ \bibnamefont
  {Uberuaga}},\ and\ \bibinfo {author} {\bibfnamefont {H.}~\bibnamefont
  {Jónsson}},\ }\bibfield  {title} {\bibinfo {title} {{A climbing image nudged
  elastic band method for finding saddle points and minimum energy paths}},\
  }\href {https://doi.org/10.1063/1.1329672} {\bibfield  {journal} {\bibinfo
  {journal} {The Journal of Chemical Physics}\ }\textbf {\bibinfo {volume}
  {113}},\ \bibinfo {pages} {9901} (\bibinfo {year} {2000})}\BibitemShut
  {NoStop}%
\bibitem [{\citenamefont {Henkelman}\ and\ \citenamefont
  {Jónsson}(2000)}]{Henkelman_JCP_2000_II}%
  \BibitemOpen
  \bibfield  {author} {\bibinfo {author} {\bibfnamefont {G.}~\bibnamefont
  {Henkelman}}\ and\ \bibinfo {author} {\bibfnamefont {H.}~\bibnamefont
  {Jónsson}},\ }\bibfield  {title} {\bibinfo {title} {{Improved tangent
  estimate in the nudged elastic band method for finding minimum energy paths
  and saddle points}},\ }\href {https://doi.org/10.1063/1.1323224} {\bibfield
  {journal} {\bibinfo  {journal} {The Journal of Chemical Physics}\ }\textbf
  {\bibinfo {volume} {113}},\ \bibinfo {pages} {9978} (\bibinfo {year}
  {2000})}\BibitemShut {NoStop}%
\bibitem [{\citenamefont {Nichols}\ \emph {et~al.}(1990)\citenamefont
  {Nichols}, \citenamefont {Taylor}, \citenamefont {Schmidt},\ and\
  \citenamefont {Simons}}]{Nichols_JCP_1990}%
  \BibitemOpen
  \bibfield  {author} {\bibinfo {author} {\bibfnamefont {J.}~\bibnamefont
  {Nichols}}, \bibinfo {author} {\bibfnamefont {H.}~\bibnamefont {Taylor}},
  \bibinfo {author} {\bibfnamefont {P.}~\bibnamefont {Schmidt}},\ and\ \bibinfo
  {author} {\bibfnamefont {J.}~\bibnamefont {Simons}},\ }\bibfield  {title}
  {\bibinfo {title} {{Walking on potential energy surfaces}},\ }\href
  {https://doi.org/10.1063/1.458435} {\bibfield  {journal} {\bibinfo  {journal}
  {The Journal of Chemical Physics}\ }\textbf {\bibinfo {volume} {92}},\
  \bibinfo {pages} {340} (\bibinfo {year} {1990})}\BibitemShut {NoStop}%
\bibitem [{\citenamefont {Rossi}\ \emph {et~al.}(2016)\citenamefont {Rossi},
  \citenamefont {Gasparotto},\ and\ \citenamefont {Ceriotti}}]{ross+16prl}%
  \BibitemOpen
  \bibfield  {author} {\bibinfo {author} {\bibfnamefont {M.}~\bibnamefont
  {Rossi}}, \bibinfo {author} {\bibfnamefont {P.}~\bibnamefont {Gasparotto}},\
  and\ \bibinfo {author} {\bibfnamefont {M.}~\bibnamefont {Ceriotti}},\
  }\bibfield  {title} {\bibinfo {title} {{Anharmonic and Quantum Fluctuations
  in Molecular Crystals: A First-Principles Study of the Stability of
  Paracetamol}},\ }\href@noop {} {\bibfield  {journal} {\bibinfo  {journal}
  {Physical Review Letters}\ }\textbf {\bibinfo {volume} {117}},\ \bibinfo
  {pages} {115702} (\bibinfo {year} {2016})}\BibitemShut {NoStop}%
\bibitem [{\citenamefont {Petraglia}\ \emph {et~al.}(2016)\citenamefont
  {Petraglia}, \citenamefont {Nicola{\"{i}}}, \citenamefont {Wodrich},
  \citenamefont {Ceriotti},\ and\ \citenamefont {Corminboeuf}}]{petr+15jcc}%
  \BibitemOpen
  \bibfield  {author} {\bibinfo {author} {\bibfnamefont {R.}~\bibnamefont
  {Petraglia}}, \bibinfo {author} {\bibfnamefont {A.}~\bibnamefont
  {Nicola{\"{i}}}}, \bibinfo {author} {\bibfnamefont {M.~M.~D.}\ \bibnamefont
  {Wodrich}}, \bibinfo {author} {\bibfnamefont {M.}~\bibnamefont {Ceriotti}},\
  and\ \bibinfo {author} {\bibfnamefont {C.}~\bibnamefont {Corminboeuf}},\
  }\bibfield  {title} {\bibinfo {title} {{Beyond static structures: Putting
  forth REMD as a tool to solve problems in computational organic chemistry}},\
  }\href@noop {} {\bibfield  {journal} {\bibinfo  {journal} {Journal of
  Computational Chemistry}\ }\textbf {\bibinfo {volume} {37}},\ \bibinfo
  {pages} {83} (\bibinfo {year} {2016})}\BibitemShut {NoStop}%
\bibitem [{\citenamefont {Liu}\ \emph {et~al.}(2013)\citenamefont {Liu},
  \citenamefont {Andino}, \citenamefont {Miller}, \citenamefont {Chen},
  \citenamefont {Wilkins}, \citenamefont {Ceriotti},\ and\ \citenamefont
  {Manolopoulos}}]{liu+13jpcc}%
  \BibitemOpen
  \bibfield  {author} {\bibinfo {author} {\bibfnamefont {J.}~\bibnamefont
  {Liu}}, \bibinfo {author} {\bibfnamefont {R.~S.}\ \bibnamefont {Andino}},
  \bibinfo {author} {\bibfnamefont {C.~M.}\ \bibnamefont {Miller}}, \bibinfo
  {author} {\bibfnamefont {X.}~\bibnamefont {Chen}}, \bibinfo {author}
  {\bibfnamefont {D.~M.}\ \bibnamefont {Wilkins}}, \bibinfo {author}
  {\bibfnamefont {M.}~\bibnamefont {Ceriotti}},\ and\ \bibinfo {author}
  {\bibfnamefont {D.~E.}\ \bibnamefont {Manolopoulos}},\ }\bibfield  {title}
  {\bibinfo {title} {{A Surface-Specific Isotope Effect in Mixtures of Light
  and Heavy Water}},\ }\href@noop {} {\bibfield  {journal} {\bibinfo  {journal}
  {The Journal of Physical Chemistry C}\ }\textbf {\bibinfo {volume} {117}},\
  \bibinfo {pages} {2944} (\bibinfo {year} {2013})}\BibitemShut {NoStop}%
\bibitem [{\citenamefont {Cheng}\ \emph {et~al.}(2016)\citenamefont {Cheng},
  \citenamefont {Behler},\ and\ \citenamefont {Ceriotti}}]{chen+16jpcl}%
  \BibitemOpen
  \bibfield  {author} {\bibinfo {author} {\bibfnamefont {B.}~\bibnamefont
  {Cheng}}, \bibinfo {author} {\bibfnamefont {J.}~\bibnamefont {Behler}},\ and\
  \bibinfo {author} {\bibfnamefont {M.}~\bibnamefont {Ceriotti}},\ }\bibfield
  {title} {\bibinfo {title} {{Nuclear Quantum Effects in Water at the Triple
  Point: Using Theory as a Link Between Experiments}},\ }\href@noop {}
  {\bibfield  {journal} {\bibinfo  {journal} {The Journal of Physical Chemistry
  Lettersrs}\ }\textbf {\bibinfo {volume} {7}},\ \bibinfo {pages} {2210}
  (\bibinfo {year} {2016})}\BibitemShut {NoStop}%
\bibitem [{\citenamefont {Craig}\ and\ \citenamefont
  {Manolopoulos}(2004)}]{crai-mano04jcp}%
  \BibitemOpen
  \bibfield  {author} {\bibinfo {author} {\bibfnamefont {I.~R.}\ \bibnamefont
  {Craig}}\ and\ \bibinfo {author} {\bibfnamefont {D.~E.}\ \bibnamefont
  {Manolopoulos}},\ }\bibfield  {title} {\bibinfo {title} {{Quantum statistics
  and classical mechanics: Real time correlation functions from ring polymer
  molecular dynamics}},\ }\href@noop {} {\bibfield  {journal} {\bibinfo
  {journal} {The Journal of Chemical Physics}\ }\textbf {\bibinfo {volume}
  {121}},\ \bibinfo {pages} {3368} (\bibinfo {year} {2004})}\BibitemShut
  {NoStop}%
\bibitem [{\citenamefont {Habershon}\ \emph {et~al.}(2013)\citenamefont
  {Habershon}, \citenamefont {Manolopoulos}, \citenamefont {Markland},\ and\
  \citenamefont {Miller}}]{habe+13arpc}%
  \BibitemOpen
  \bibfield  {author} {\bibinfo {author} {\bibfnamefont {S.}~\bibnamefont
  {Habershon}}, \bibinfo {author} {\bibfnamefont {D.~E.}\ \bibnamefont
  {Manolopoulos}}, \bibinfo {author} {\bibfnamefont {T.~E.}\ \bibnamefont
  {Markland}},\ and\ \bibinfo {author} {\bibfnamefont {T.~F.}\ \bibnamefont
  {Miller}},\ }\bibfield  {title} {\bibinfo {title} {{Ring-polymer molecular
  dynamics: quantum effects in chemical dynamics from classical trajectories in
  an extended phase space.}},\ }\href@noop {} {\bibfield  {journal} {\bibinfo
  {journal} {Annual review of physical chemistry}\ }\textbf {\bibinfo {volume}
  {64}},\ \bibinfo {pages} {387} (\bibinfo {year} {2013})}\BibitemShut
  {NoStop}%
\bibitem [{\citenamefont {Cao}\ and\ \citenamefont
  {Voth}(1993)}]{cao-voth93jcp}%
  \BibitemOpen
  \bibfield  {author} {\bibinfo {author} {\bibfnamefont {J.}~\bibnamefont
  {Cao}}\ and\ \bibinfo {author} {\bibfnamefont {G.~A.}\ \bibnamefont {Voth}},\
  }\bibfield  {title} {\bibinfo {title} {{A new perspective on quantum time
  correlation functions}},\ }\href@noop {} {\bibfield  {journal} {\bibinfo
  {journal} {The Journal of Chemical Physics}\ }\textbf {\bibinfo {volume}
  {99}},\ \bibinfo {pages} {10070} (\bibinfo {year} {1993})}\BibitemShut
  {NoStop}%
\bibitem [{\citenamefont {Cao}\ and\ \citenamefont {Voth}(1994)}]{cao_JCP_94}%
  \BibitemOpen
  \bibfield  {author} {\bibinfo {author} {\bibfnamefont {J.}~\bibnamefont
  {Cao}}\ and\ \bibinfo {author} {\bibfnamefont {G.~A.}\ \bibnamefont {Voth}},\
  }\bibfield  {title} {\bibinfo {title} {{The formulation of quantum
  statistical mechanics based on the Feynman path centroid density. IV.
  Algorithms for centroid molecular dynamics}},\ }\href
  {https://doi.org/10.1063/1.468399} {\bibfield  {journal} {\bibinfo  {journal}
  {The Journal of Chemical Physics}\ }\textbf {\bibinfo {volume} {101}},\
  \bibinfo {pages} {6168} (\bibinfo {year} {1994})}\BibitemShut {NoStop}%
\bibitem [{\citenamefont {Rossi}\ \emph {et~al.}(2014)\citenamefont {Rossi},
  \citenamefont {Ceriotti},\ and\ \citenamefont {Manolopoulos}}]{ross+14jcp}%
  \BibitemOpen
  \bibfield  {author} {\bibinfo {author} {\bibfnamefont {M.}~\bibnamefont
  {Rossi}}, \bibinfo {author} {\bibfnamefont {M.}~\bibnamefont {Ceriotti}},\
  and\ \bibinfo {author} {\bibfnamefont {D.~E.}\ \bibnamefont {Manolopoulos}},\
  }\bibfield  {title} {\bibinfo {title} {How to remove the spurious resonances
  from ring polymer molecular dynamics.},\ }\href@noop {} {\bibfield  {journal}
  {\bibinfo  {journal} {The Journal of Chemical Physics}\ }\textbf {\bibinfo
  {volume} {140}},\ \bibinfo {pages} {234116} (\bibinfo {year}
  {2014})}\BibitemShut {NoStop}%
\bibitem [{\citenamefont {Litman}\ \emph {et~al.}(2019)\citenamefont {Litman},
  \citenamefont {Richardson}, \citenamefont {Kumagai},\ and\ \citenamefont
  {Rossi}}]{Litman_JACS_2019}%
  \BibitemOpen
  \bibfield  {author} {\bibinfo {author} {\bibfnamefont {Y.}~\bibnamefont
  {Litman}}, \bibinfo {author} {\bibfnamefont {J.~O.}\ \bibnamefont
  {Richardson}}, \bibinfo {author} {\bibfnamefont {T.}~\bibnamefont
  {Kumagai}},\ and\ \bibinfo {author} {\bibfnamefont {M.}~\bibnamefont
  {Rossi}},\ }\bibfield  {title} {\bibinfo {title} {Elucidating the nuclear
  quantum dynamics of intramolecular double hydrogen transfer in porphycene},\
  }\href {https://doi.org/10.1021/jacs.8b12471} {\bibfield  {journal} {\bibinfo
   {journal} {Journal of the American Chemical Society}\ }\textbf {\bibinfo
  {volume} {141}},\ \bibinfo {pages} {2526} (\bibinfo {year}
  {2019})}\BibitemShut {NoStop}%
\bibitem [{\citenamefont {Kapil}\ \emph
  {et~al.}(2016{\natexlab{b}})\citenamefont {Kapil}, \citenamefont {Behler},\
  and\ \citenamefont {Ceriotti}}]{kapi+16jcp2}%
  \BibitemOpen
  \bibfield  {author} {\bibinfo {author} {\bibfnamefont {V.}~\bibnamefont
  {Kapil}}, \bibinfo {author} {\bibfnamefont {J.}~\bibnamefont {Behler}},\ and\
  \bibinfo {author} {\bibfnamefont {M.}~\bibnamefont {Ceriotti}},\ }\bibfield
  {title} {\bibinfo {title} {{High order path integrals made easy}},\
  }\href@noop {} {\bibfield  {journal} {\bibinfo  {journal} {The Journal of
  Chemical Physics}\ }\textbf {\bibinfo {volume} {145}},\ \bibinfo {pages}
  {234103} (\bibinfo {year} {2016}{\natexlab{b}})}\BibitemShut {NoStop}%
\bibitem [{\citenamefont {Poltavsky}\ and\ \citenamefont
  {Tkatchenko}(2016)}]{polt-tkat16cs}%
  \BibitemOpen
  \bibfield  {author} {\bibinfo {author} {\bibfnamefont {I.}~\bibnamefont
  {Poltavsky}}\ and\ \bibinfo {author} {\bibfnamefont {A.}~\bibnamefont
  {Tkatchenko}},\ }\bibfield  {title} {\bibinfo {title} {{Modeling quantum
  nuclei with perturbed path integral molecular dynamics}},\ }\href@noop {}
  {\bibfield  {journal} {\bibinfo  {journal} {Chemical Science}\ }\textbf
  {\bibinfo {volume} {7}},\ \bibinfo {pages} {1368} (\bibinfo {year}
  {2016})}\BibitemShut {NoStop}%
\bibitem [{\citenamefont {Yamamoto}(2005)}]{yama05jcp}%
  \BibitemOpen
  \bibfield  {author} {\bibinfo {author} {\bibfnamefont {T.~M.}\ \bibnamefont
  {Yamamoto}},\ }\bibfield  {title} {\bibinfo {title} {{Path-integral virial
  estimator based on the scaling of fluctuation coordinates: Application to
  quantum clusters with fourth-order propagators}},\ }\href@noop {} {\bibfield
  {journal} {\bibinfo  {journal} {The Journal of Chemical Physics}\ }\textbf
  {\bibinfo {volume} {123}},\ \bibinfo {pages} {104101} (\bibinfo {year}
  {2005})}\BibitemShut {NoStop}%
\bibitem [{\citenamefont {Ceriotti}\ and\ \citenamefont
  {Markland}(2013)}]{ceri-mark13jcp}%
  \BibitemOpen
  \bibfield  {author} {\bibinfo {author} {\bibfnamefont {M.}~\bibnamefont
  {Ceriotti}}\ and\ \bibinfo {author} {\bibfnamefont {T.~E.}\ \bibnamefont
  {Markland}},\ }\bibfield  {title} {\bibinfo {title} {{Efficient methods and
  practical guidelines for simulating isotope effects.}},\ }\href@noop {}
  {\bibfield  {journal} {\bibinfo  {journal} {The Journal of Chemical Physics}\
  }\textbf {\bibinfo {volume} {138}},\ \bibinfo {pages} {014112} (\bibinfo
  {year} {2013})}\BibitemShut {NoStop}%
\bibitem [{\citenamefont {Lin}\ \emph {et~al.}(2010)\citenamefont {Lin},
  \citenamefont {Morrone}, \citenamefont {Car},\ and\ \citenamefont
  {Parrinello}}]{lin+10prl}%
  \BibitemOpen
  \bibfield  {author} {\bibinfo {author} {\bibfnamefont {L.}~\bibnamefont
  {Lin}}, \bibinfo {author} {\bibfnamefont {J.~A.}\ \bibnamefont {Morrone}},
  \bibinfo {author} {\bibfnamefont {R.}~\bibnamefont {Car}},\ and\ \bibinfo
  {author} {\bibfnamefont {M.}~\bibnamefont {Parrinello}},\ }\bibfield  {title}
  {\bibinfo {title} {{Displaced Path Integral Formulation for the Momentum
  Distribution of Quantum Particles}},\ }\href@noop {} {\bibfield  {journal}
  {\bibinfo  {journal} {Physical Review Letters}\ }\textbf {\bibinfo {volume}
  {105}},\ \bibinfo {pages} {110602} (\bibinfo {year} {2010})}\BibitemShut
  {NoStop}%
\bibitem [{\citenamefont {Ceriotti}\ \emph {et~al.}(2011)\citenamefont
  {Ceriotti}, \citenamefont {Brain}, \citenamefont {Riordan},\ and\
  \citenamefont {Manolopoulos}}]{ceri+12prsa}%
  \BibitemOpen
  \bibfield  {author} {\bibinfo {author} {\bibfnamefont {M.}~\bibnamefont
  {Ceriotti}}, \bibinfo {author} {\bibfnamefont {G.~A.~R.}\ \bibnamefont
  {Brain}}, \bibinfo {author} {\bibfnamefont {O.}~\bibnamefont {Riordan}},\
  and\ \bibinfo {author} {\bibfnamefont {D.~E.}\ \bibnamefont {Manolopoulos}},\
  }\bibfield  {title} {\bibinfo {title} {{The inefficiency of re-weighted
  sampling and the curse of system size in high order path integration}},\
  }\href@noop {} {\bibfield  {journal} {\bibinfo  {journal} {Proceedings of the
  Royal Society A: Mathematical, Physical and Engineering Sciences}\ }\textbf
  {\bibinfo {volume} {468}},\ \bibinfo {pages} {2} (\bibinfo {year}
  {2011})}\BibitemShut {NoStop}%
\bibitem [{\citenamefont {Jang}\ and\ \citenamefont
  {Voth}(2001)}]{jang-voth01jcp}%
  \BibitemOpen
  \bibfield  {author} {\bibinfo {author} {\bibfnamefont {S.~S.}\ \bibnamefont
  {Jang}}\ and\ \bibinfo {author} {\bibfnamefont {G.~A.}\ \bibnamefont
  {Voth}},\ }\bibfield  {title} {\bibinfo {title} {{Applications of higher
  order composite factorization schemes in imaginary time path integral
  simulations}},\ }\href@noop {} {\bibfield  {journal} {\bibinfo  {journal}
  {The Journal of Chemical Physics}\ }\textbf {\bibinfo {volume} {115}},\
  \bibinfo {pages} {7832} (\bibinfo {year} {2001})}\BibitemShut {NoStop}%
\bibitem [{\citenamefont {Korol}\ \emph {et~al.}(2019)\citenamefont {Korol},
  \citenamefont {Bou-Rabee},\ and\ \citenamefont {Miller}}]{Korol2019}%
  \BibitemOpen
  \bibfield  {author} {\bibinfo {author} {\bibfnamefont {R.}~\bibnamefont
  {Korol}}, \bibinfo {author} {\bibfnamefont {N.}~\bibnamefont {Bou-Rabee}},\
  and\ \bibinfo {author} {\bibfnamefont {T.~F.}\ \bibnamefont {Miller}},\
  }\bibfield  {title} {\bibinfo {title} {Cayley modification for strongly
  stable path-integral and ring-polymer molecular dynamics},\ }\bibfield
  {journal} {\bibinfo  {journal} {The Journal of Chemical Physics}\ }\textbf
  {\bibinfo {volume} {151}},\ \href {https://doi.org/10.1063/1.5120282}
  {10.1063/1.5120282} (\bibinfo {year} {2019})\BibitemShut {NoStop}%
\bibitem [{\citenamefont {Martyna}\ \emph {et~al.}(1996)\citenamefont
  {Martyna}, \citenamefont {Tuckerman}, \citenamefont {Tobias},\ and\
  \citenamefont {Klein}}]{Martyna1996}%
  \BibitemOpen
  \bibfield  {author} {\bibinfo {author} {\bibfnamefont {G.~J.}\ \bibnamefont
  {Martyna}}, \bibinfo {author} {\bibfnamefont {M.~E.}\ \bibnamefont
  {Tuckerman}}, \bibinfo {author} {\bibfnamefont {D.~J.}\ \bibnamefont
  {Tobias}},\ and\ \bibinfo {author} {\bibfnamefont {M.~L.}\ \bibnamefont
  {Klein}},\ }\bibfield  {title} {\bibinfo {title} {Explicit reversible
  integrators for extended systems dynamics},\ }\href
  {https://doi.org/10.1080/00268979600100761} {\bibfield  {journal} {\bibinfo
  {journal} {Molecular Physics}\ }\textbf {\bibinfo {volume} {87}},\ \bibinfo
  {pages} {1117–1157} (\bibinfo {year} {1996})}\BibitemShut {NoStop}%
\bibitem [{\citenamefont {Kapil}\ \emph {et~al.}(2018)\citenamefont {Kapil},
  \citenamefont {Cuzzocrea},\ and\ \citenamefont {Ceriotti}}]{kapi+18jpcb}%
  \BibitemOpen
  \bibfield  {author} {\bibinfo {author} {\bibfnamefont {V.}~\bibnamefont
  {Kapil}}, \bibinfo {author} {\bibfnamefont {A.}~\bibnamefont {Cuzzocrea}},\
  and\ \bibinfo {author} {\bibfnamefont {M.}~\bibnamefont {Ceriotti}},\
  }\bibfield  {title} {\bibinfo {title} {{Anisotropy of the Proton Momentum
  Distribution in Water}},\ }\href@noop {} {\bibfield  {journal} {\bibinfo
  {journal} {The Journal of Physical Chemistry B}\ }\textbf {\bibinfo {volume}
  {122}},\ \bibinfo {pages} {6048} (\bibinfo {year} {2018})}\BibitemShut
  {NoStop}%
\bibitem [{\citenamefont {Kapil}\ \emph
  {et~al.}(2019{\natexlab{d}})\citenamefont {Kapil}, \citenamefont {Engel},
  \citenamefont {Rossi},\ and\ \citenamefont {Ceriotti}}]{Kapil_JCTC_2019}%
  \BibitemOpen
  \bibfield  {author} {\bibinfo {author} {\bibfnamefont {V.}~\bibnamefont
  {Kapil}}, \bibinfo {author} {\bibfnamefont {E.}~\bibnamefont {Engel}},
  \bibinfo {author} {\bibfnamefont {M.}~\bibnamefont {Rossi}},\ and\ \bibinfo
  {author} {\bibfnamefont {M.}~\bibnamefont {Ceriotti}},\ }\bibfield  {title}
  {\bibinfo {title} {Assessment of approximate methods for anharmonic free
  energies},\ }\href {https://doi.org/10.1021/acs.jctc.9b00596} {\bibfield
  {journal} {\bibinfo  {journal} {Journal of Chemical Theory and Computation}\
  }\textbf {\bibinfo {volume} {15}},\ \bibinfo {pages} {5845} (\bibinfo {year}
  {2019}{\natexlab{d}})}\BibitemShut {NoStop}%
\bibitem [{\citenamefont {Imbalzano}\ \emph {et~al.}(2021)\citenamefont
  {Imbalzano}, \citenamefont {Zhuang}, \citenamefont {Kapil}, \citenamefont
  {Rossi}, \citenamefont {Engel}, \citenamefont {Grasselli},\ and\
  \citenamefont {Ceriotti}}]{Imbalzano_JCP_2021}%
  \BibitemOpen
  \bibfield  {author} {\bibinfo {author} {\bibfnamefont {G.}~\bibnamefont
  {Imbalzano}}, \bibinfo {author} {\bibfnamefont {Y.}~\bibnamefont {Zhuang}},
  \bibinfo {author} {\bibfnamefont {V.}~\bibnamefont {Kapil}}, \bibinfo
  {author} {\bibfnamefont {K.}~\bibnamefont {Rossi}}, \bibinfo {author}
  {\bibfnamefont {E.~A.}\ \bibnamefont {Engel}}, \bibinfo {author}
  {\bibfnamefont {F.}~\bibnamefont {Grasselli}},\ and\ \bibinfo {author}
  {\bibfnamefont {M.}~\bibnamefont {Ceriotti}},\ }\bibfield  {title} {\bibinfo
  {title} {{Uncertainty estimation for molecular dynamics and sampling}},\
  }\href {https://doi.org/10.1063/5.0036522} {\bibfield  {journal} {\bibinfo
  {journal} {The Journal of Chemical Physics}\ }\textbf {\bibinfo {volume}
  {154}},\ \bibinfo {pages} {074102} (\bibinfo {year} {2021})}\BibitemShut
  {NoStop}%
\bibitem [{\citenamefont {Hirshberg}\ \emph {et~al.}(2019)\citenamefont
  {Hirshberg}, \citenamefont {Rizzi},\ and\ \citenamefont
  {Parrinello}}]{hirshberg2019path}%
  \BibitemOpen
  \bibfield  {author} {\bibinfo {author} {\bibfnamefont {B.}~\bibnamefont
  {Hirshberg}}, \bibinfo {author} {\bibfnamefont {V.}~\bibnamefont {Rizzi}},\
  and\ \bibinfo {author} {\bibfnamefont {M.}~\bibnamefont {Parrinello}},\
  }\bibfield  {title} {\bibinfo {title} {Path integral molecular dynamics for
  bosons},\ }\href@noop {} {\bibfield  {journal} {\bibinfo  {journal}
  {Proceedings of the National Academy of Sciences}\ }\textbf {\bibinfo
  {volume} {116}},\ \bibinfo {pages} {21445} (\bibinfo {year}
  {2019})}\BibitemShut {NoStop}%
\bibitem [{\citenamefont {Feldman}\ and\ \citenamefont
  {Hirshberg}(2023)}]{10.1063/5.0173749}%
  \BibitemOpen
  \bibfield  {author} {\bibinfo {author} {\bibfnamefont {Y.~M.~Y.}\
  \bibnamefont {Feldman}}\ and\ \bibinfo {author} {\bibfnamefont
  {B.}~\bibnamefont {Hirshberg}},\ }\bibfield  {title} {\bibinfo {title}
  {{Quadratic scaling bosonic path integral molecular dynamics}},\ }\href
  {https://doi.org/10.1063/5.0173749} {\bibfield  {journal} {\bibinfo
  {journal} {The Journal of Chemical Physics}\ }\textbf {\bibinfo {volume}
  {159}},\ \bibinfo {pages} {154107} (\bibinfo {year} {2023})}\BibitemShut
  {NoStop}%
\bibitem [{\citenamefont {Hirshberg}\ \emph {et~al.}(2020)\citenamefont
  {Hirshberg}, \citenamefont {Invernizzi},\ and\ \citenamefont
  {Parrinello}}]{10.1063/5.0008720}%
  \BibitemOpen
  \bibfield  {author} {\bibinfo {author} {\bibfnamefont {B.}~\bibnamefont
  {Hirshberg}}, \bibinfo {author} {\bibfnamefont {M.}~\bibnamefont
  {Invernizzi}},\ and\ \bibinfo {author} {\bibfnamefont {M.}~\bibnamefont
  {Parrinello}},\ }\bibfield  {title} {\bibinfo {title} {{Path integral
  molecular dynamics for fermions: Alleviating the sign problem with the
  Bogoliubov inequality}},\ }\href {https://doi.org/10.1063/5.0008720}
  {\bibfield  {journal} {\bibinfo  {journal} {The Journal of Chemical Physics}\
  }\textbf {\bibinfo {volume} {152}},\ \bibinfo {pages} {171102} (\bibinfo
  {year} {2020})}\BibitemShut {NoStop}%
\bibitem [{\citenamefont {Li}\ \emph {et~al.}(2020)\citenamefont {Li},
  \citenamefont {Subotnik},\ and\ \citenamefont {Nitzan}}]{Li2020Water}%
  \BibitemOpen
  \bibfield  {author} {\bibinfo {author} {\bibfnamefont {T.~E.}\ \bibnamefont
  {Li}}, \bibinfo {author} {\bibfnamefont {J.~E.}\ \bibnamefont {Subotnik}},\
  and\ \bibinfo {author} {\bibfnamefont {A.}~\bibnamefont {Nitzan}},\
  }\bibfield  {title} {\bibinfo {title} {{Cavity Molecular Dynamics Simulations
  of Liquid Water under Vibrational Ultrastrong Coupling}},\ }\href
  {https://doi.org/10.1073/pnas.2009272117} {\bibfield  {journal} {\bibinfo
  {journal} {Proc. Natl. Acad. Sci.}\ }\textbf {\bibinfo {volume} {117}},\
  \bibinfo {pages} {18324} (\bibinfo {year} {2020})}\BibitemShut {NoStop}%
\bibitem [{\citenamefont {Li}\ \emph {et~al.}(2022{\natexlab{a}})\citenamefont
  {Li}, \citenamefont {Nitzan}, \citenamefont {Hammes-Schiffer},\ and\
  \citenamefont {Subotnik}}]{Li2022RPMDCav}%
  \BibitemOpen
  \bibfield  {author} {\bibinfo {author} {\bibfnamefont {T.~E.}\ \bibnamefont
  {Li}}, \bibinfo {author} {\bibfnamefont {A.}~\bibnamefont {Nitzan}}, \bibinfo
  {author} {\bibfnamefont {S.}~\bibnamefont {Hammes-Schiffer}},\ and\ \bibinfo
  {author} {\bibfnamefont {J.~E.}\ \bibnamefont {Subotnik}},\ }\bibfield
  {title} {\bibinfo {title} {{Quantum Simulations of Vibrational Strong
  Coupling via Path Integrals}},\ }\href
  {https://doi.org/10.1021/acs.jpclett.2c00613} {\bibfield  {journal} {\bibinfo
   {journal} {The Journal of Physical Chemistry Letters}\ }\textbf {\bibinfo
  {volume} {13}},\ \bibinfo {pages} {3890} (\bibinfo {year}
  {2022}{\natexlab{a}})}\BibitemShut {NoStop}%
\bibitem [{\citenamefont {Litman}\ \emph
  {et~al.}(2022{\natexlab{a}})\citenamefont {Litman}, \citenamefont {Pós},
  \citenamefont {Box}, \citenamefont {Martinazzo}, \citenamefont {Maurer},\
  and\ \citenamefont {Rossi}}]{Litman_JCP_2022_I}%
  \BibitemOpen
  \bibfield  {author} {\bibinfo {author} {\bibfnamefont {Y.}~\bibnamefont
  {Litman}}, \bibinfo {author} {\bibfnamefont {E.~S.}\ \bibnamefont {Pós}},
  \bibinfo {author} {\bibfnamefont {C.~L.}\ \bibnamefont {Box}}, \bibinfo
  {author} {\bibfnamefont {R.}~\bibnamefont {Martinazzo}}, \bibinfo {author}
  {\bibfnamefont {R.~J.}\ \bibnamefont {Maurer}},\ and\ \bibinfo {author}
  {\bibfnamefont {M.}~\bibnamefont {Rossi}},\ }\bibfield  {title} {\bibinfo
  {title} {{Dissipative tunneling rates through the incorporation of
  first-principles electronic friction in instanton rate theory. I. Theory}},\
  }\href {https://doi.org/10.1063/5.0088399} {\bibfield  {journal} {\bibinfo
  {journal} {The Journal of Chemical Physics}\ }\textbf {\bibinfo {volume}
  {156}},\ \bibinfo {pages} {194106} (\bibinfo {year}
  {2022}{\natexlab{a}})}\BibitemShut {NoStop}%
\bibitem [{\citenamefont {Litman}\ \emph
  {et~al.}(2022{\natexlab{b}})\citenamefont {Litman}, \citenamefont {Pós},
  \citenamefont {Box}, \citenamefont {Martinazzo}, \citenamefont {Maurer},\
  and\ \citenamefont {Rossi}}]{Litman_JCP_2022_II}%
  \BibitemOpen
  \bibfield  {author} {\bibinfo {author} {\bibfnamefont {Y.}~\bibnamefont
  {Litman}}, \bibinfo {author} {\bibfnamefont {E.~S.}\ \bibnamefont {Pós}},
  \bibinfo {author} {\bibfnamefont {C.~L.}\ \bibnamefont {Box}}, \bibinfo
  {author} {\bibfnamefont {R.}~\bibnamefont {Martinazzo}}, \bibinfo {author}
  {\bibfnamefont {R.~J.}\ \bibnamefont {Maurer}},\ and\ \bibinfo {author}
  {\bibfnamefont {M.}~\bibnamefont {Rossi}},\ }\bibfield  {title} {\bibinfo
  {title} {{Dissipative tunneling rates through the incorporation of
  first-principles electronic friction in instanton rate theory. II. Benchmarks
  and applications}},\ }\href {https://doi.org/10.1063/5.0088400} {\bibfield
  {journal} {\bibinfo  {journal} {The Journal of Chemical Physics}\ }\textbf
  {\bibinfo {volume} {156}},\ \bibinfo {pages} {194107} (\bibinfo {year}
  {2022}{\natexlab{b}})}\BibitemShut {NoStop}%
\bibitem [{\citenamefont {Kellner}\ and\ \citenamefont
  {Ceriotti}(2024)}]{kellner2024uncertainty}%
  \BibitemOpen
  \bibfield  {author} {\bibinfo {author} {\bibfnamefont {M.}~\bibnamefont
  {Kellner}}\ and\ \bibinfo {author} {\bibfnamefont {M.}~\bibnamefont
  {Ceriotti}},\ }\href {https://doi.org/10.1088/2632-2153/ad594a} {\bibinfo
  {title} {Uncertainty quantification by direct propagation of shallow
  ensembles}} (\bibinfo {year} {2024})\BibitemShut {NoStop}%
\bibitem [{\citenamefont {Kapil}\ \emph {et~al.}(2020)\citenamefont {Kapil},
  \citenamefont {Wilkins}, \citenamefont {Lan},\ and\ \citenamefont
  {Ceriotti}}]{kapil_inexpensive_2020}%
  \BibitemOpen
  \bibfield  {author} {\bibinfo {author} {\bibfnamefont {V.}~\bibnamefont
  {Kapil}}, \bibinfo {author} {\bibfnamefont {D.~M.}\ \bibnamefont {Wilkins}},
  \bibinfo {author} {\bibfnamefont {J.}~\bibnamefont {Lan}},\ and\ \bibinfo
  {author} {\bibfnamefont {M.}~\bibnamefont {Ceriotti}},\ }\bibfield  {title}
  {\bibinfo {title} {Inexpensive modeling of quantum dynamics using path
  integral generalized {Langevin} equation thermostats},\ }\href
  {https://doi.org/10.1063/1.5141950} {\bibfield  {journal} {\bibinfo
  {journal} {The Journal of Chemical Physics}\ }\textbf {\bibinfo {volume}
  {152}},\ \bibinfo {pages} {124104} (\bibinfo {year} {2020})}\BibitemShut
  {NoStop}%
\bibitem [{\citenamefont {Hasegawa}\ and\ \citenamefont
  {Tanimura}(2006)}]{Hasegawa_JCP_2006}%
  \BibitemOpen
  \bibfield  {author} {\bibinfo {author} {\bibfnamefont {T.}~\bibnamefont
  {Hasegawa}}\ and\ \bibinfo {author} {\bibfnamefont {Y.}~\bibnamefont
  {Tanimura}},\ }\bibfield  {title} {\bibinfo {title} {{Calculating fifth-order
  Raman signals for various molecular liquids by equilibrium and nonequilibrium
  hybrid molecular dynamics simulation algorithms}},\ }\href
  {https://doi.org/10.1063/1.2217947} {\bibfield  {journal} {\bibinfo
  {journal} {The Journal of Chemical Physics}\ }\textbf {\bibinfo {volume}
  {125}},\ \bibinfo {pages} {074512} (\bibinfo {year} {2006})}\BibitemShut
  {NoStop}%
\bibitem [{\citenamefont {Begušić}\ \emph {et~al.}(2022)\citenamefont
  {Begušić}, \citenamefont {Tao}, \citenamefont {Blake},\ and\ \citenamefont
  {Miller}}]{Begusic_JCP_2022}%
  \BibitemOpen
  \bibfield  {author} {\bibinfo {author} {\bibfnamefont {T.}~\bibnamefont
  {Begušić}}, \bibinfo {author} {\bibfnamefont {X.}~\bibnamefont {Tao}},
  \bibinfo {author} {\bibfnamefont {G.~A.}\ \bibnamefont {Blake}},\ and\
  \bibinfo {author} {\bibfnamefont {I.}~\bibnamefont {Miller}, \bibfnamefont
  {Thomas~F.}},\ }\bibfield  {title} {\bibinfo {title}
  {{Equilibrium–nonequilibrium ring-polymer molecular dynamics for nonlinear
  spectroscopy}},\ }\href {https://doi.org/10.1063/5.0087156} {\bibfield
  {journal} {\bibinfo  {journal} {The Journal of Chemical Physics}\ }\textbf
  {\bibinfo {volume} {156}},\ \bibinfo {pages} {131102} (\bibinfo {year}
  {2022})}\BibitemShut {NoStop}%
\bibitem [{\citenamefont {Begu{\v{s}}i{\'{c}}}\ and\ \citenamefont
  {Blake}(2023)}]{Begusic_NatComm_2023}%
  \BibitemOpen
  \bibfield  {author} {\bibinfo {author} {\bibfnamefont {T.}~\bibnamefont
  {Begu{\v{s}}i{\'{c}}}}\ and\ \bibinfo {author} {\bibfnamefont {G.~A.}\
  \bibnamefont {Blake}},\ }\bibfield  {title} {\bibinfo {title}
  {Two-dimensional infrared-raman spectroscopy as a probe of water's
  tetrahedrality},\ }\href {https://doi.org/10.1038/s41467-023-37667-7}
  {\bibfield  {journal} {\bibinfo  {journal} {Nature Communications}\ }\textbf
  {\bibinfo {volume} {14}},\ \bibinfo {pages} {1950} (\bibinfo {year}
  {2023})}\BibitemShut {NoStop}%
\bibitem [{\citenamefont {Musil}\ \emph {et~al.}(2022)\citenamefont {Musil},
  \citenamefont {Zaporozhets}, \citenamefont {Noé}, \citenamefont {Clementi},\
  and\ \citenamefont {Kapil}}]{musil_quantum_2022}%
  \BibitemOpen
  \bibfield  {author} {\bibinfo {author} {\bibfnamefont {F.}~\bibnamefont
  {Musil}}, \bibinfo {author} {\bibfnamefont {I.}~\bibnamefont {Zaporozhets}},
  \bibinfo {author} {\bibfnamefont {F.}~\bibnamefont {Noé}}, \bibinfo {author}
  {\bibfnamefont {C.}~\bibnamefont {Clementi}},\ and\ \bibinfo {author}
  {\bibfnamefont {V.}~\bibnamefont {Kapil}},\ }\bibfield  {title} {\bibinfo
  {title} {Quantum dynamics using path integral coarse-graining},\ }\bibfield
  {journal} {\bibinfo  {journal} {The Journal of Chemical Physics}\ }\href
  {https://doi.org/10.1063/5.0120386} {10.1063/5.0120386} (\bibinfo {year}
  {2022})\BibitemShut {NoStop}%
\bibitem [{\citenamefont {Kapil}\ \emph {et~al.}(2024)\citenamefont {Kapil},
  \citenamefont {Kovács}, \citenamefont {Csányi},\ and\ \citenamefont
  {Michaelides}}]{kapil_first-principles_2023}%
  \BibitemOpen
  \bibfield  {author} {\bibinfo {author} {\bibfnamefont {V.}~\bibnamefont
  {Kapil}}, \bibinfo {author} {\bibfnamefont {D.~P.}\ \bibnamefont {Kovács}},
  \bibinfo {author} {\bibfnamefont {G.}~\bibnamefont {Csányi}},\ and\ \bibinfo
  {author} {\bibfnamefont {A.}~\bibnamefont {Michaelides}},\ }\bibfield
  {title} {\bibinfo {title} {First-principles spectroscopy of aqueous
  interfaces using machine-learned electronic and quantum nuclear effects},\
  }\href@noop {} {\bibfield  {journal} {\bibinfo  {journal} {Faraday Discuss.}\
  }\textbf {\bibinfo {volume} {249}},\ \bibinfo {pages} {50} (\bibinfo {year}
  {2024})}\BibitemShut {NoStop}%
\bibitem [{\citenamefont {Bussi}\ \emph
  {et~al.}(2007{\natexlab{b}})\citenamefont {Bussi}, \citenamefont {Donadio},\
  and\ \citenamefont {Parrinello}}]{bussi_canonical_2007}%
  \BibitemOpen
  \bibfield  {author} {\bibinfo {author} {\bibfnamefont {G.}~\bibnamefont
  {Bussi}}, \bibinfo {author} {\bibfnamefont {D.}~\bibnamefont {Donadio}},\
  and\ \bibinfo {author} {\bibfnamefont {M.}~\bibnamefont {Parrinello}},\
  }\bibfield  {title} {\bibinfo {title} {Canonical sampling through velocity
  rescaling},\ }\href {https://doi.org/10.1063/1.2408420} {\bibfield  {journal}
  {\bibinfo  {journal} {The Journal of Chemical Physics}\ }\textbf {\bibinfo
  {volume} {126}},\ \bibinfo {pages} {014101} (\bibinfo {year}
  {2007}{\natexlab{b}})}\BibitemShut {NoStop}%
\bibitem [{\citenamefont {Suzuki}(1995)}]{suzu95pla}%
  \BibitemOpen
  \bibfield  {author} {\bibinfo {author} {\bibfnamefont {M.}~\bibnamefont
  {Suzuki}},\ }\bibfield  {title} {\bibinfo {title} {{Hybrid exponential
  product formulas for unbounded operators with possible applications to Monte
  Carlo simulations}},\ }\href@noop {} {\bibfield  {journal} {\bibinfo
  {journal} {Physics Letters A}\ }\textbf {\bibinfo {volume} {201}},\ \bibinfo
  {pages} {425} (\bibinfo {year} {1995})}\BibitemShut {NoStop}%
\bibitem [{\citenamefont {Leimkuhler}\ and\ \citenamefont
  {Matthews}(2013)}]{leim+13jcp}%
  \BibitemOpen
  \bibfield  {author} {\bibinfo {author} {\bibfnamefont {B.}~\bibnamefont
  {Leimkuhler}}\ and\ \bibinfo {author} {\bibfnamefont {C.}~\bibnamefont
  {Matthews}},\ }\bibfield  {title} {\bibinfo {title} {{Robust and efficient
  configurational molecular sampling via Langevin dynamics}},\ }\href@noop {}
  {\bibfield  {journal} {\bibinfo  {journal} {The Journal of Chemical Physics}\
  }\textbf {\bibinfo {volume} {138}} (\bibinfo {year} {2013})}\BibitemShut
  {NoStop}%
\bibitem [{\citenamefont {Nagle}(1984)}]{Nagle1984}%
  \BibitemOpen
  \bibfield  {author} {\bibinfo {author} {\bibfnamefont {J.}~\bibnamefont
  {Nagle}},\ }\bibfield  {title} {\bibinfo {title} {Congestion control in
  ip/tcp internetworks},\ }\href {https://doi.org/10.1145/1024908.1024910}
  {\bibfield  {journal} {\bibinfo  {journal} {SIGCOMM Comput. Commun. Rev.}\
  }\textbf {\bibinfo {volume} {14}},\ \bibinfo {pages} {11–17} (\bibinfo
  {year} {1984})}\BibitemShut {NoStop}%
\bibitem [{\citenamefont {Han}\ \emph {et~al.}(2018)\citenamefont {Han},
  \citenamefont {Zhang}, \citenamefont {Car},\ and\ \citenamefont
  {E}}]{deepmd}%
  \BibitemOpen
  \bibfield  {author} {\bibinfo {author} {\bibfnamefont {J.}~\bibnamefont
  {Han}}, \bibinfo {author} {\bibfnamefont {L.}~\bibnamefont {Zhang}}, \bibinfo
  {author} {\bibfnamefont {R.}~\bibnamefont {Car}},\ and\ \bibinfo {author}
  {\bibfnamefont {W.}~\bibnamefont {E}},\ }\bibfield  {title} {\bibinfo {title}
  {Deep potential: A general representation of a many-body potential energy
  surface},\ }\href {https://doi.org/https://doi.org/10.4208/cicp.OA-2017-0213}
  {\bibfield  {journal} {\bibinfo  {journal} {Communications in Computational
  Physics}\ }\textbf {\bibinfo {volume} {23}},\ \bibinfo {pages} {629}
  (\bibinfo {year} {2018})}\BibitemShut {NoStop}%
\bibitem [{\citenamefont {Wang}\ \emph {et~al.}(2018)\citenamefont {Wang},
  \citenamefont {Zhang}, \citenamefont {Han},\ and\ \citenamefont
  {E}}]{DeePMD-kitv1}%
  \BibitemOpen
  \bibfield  {author} {\bibinfo {author} {\bibfnamefont {H.}~\bibnamefont
  {Wang}}, \bibinfo {author} {\bibfnamefont {L.}~\bibnamefont {Zhang}},
  \bibinfo {author} {\bibfnamefont {J.}~\bibnamefont {Han}},\ and\ \bibinfo
  {author} {\bibfnamefont {W.}~\bibnamefont {E}},\ }\bibfield  {title}
  {\bibinfo {title} {Deepmd-kit: A deep learning package for many-body
  potential energy representation and molecular dynamics},\ }\href
  {https://doi.org/https://doi.org/10.1016/j.cpc.2018.03.016} {\bibfield
  {journal} {\bibinfo  {journal} {Computer Physics Communications}\ }\textbf
  {\bibinfo {volume} {228}},\ \bibinfo {pages} {178} (\bibinfo {year}
  {2018})}\BibitemShut {NoStop}%
\bibitem [{\citenamefont {Zeng}\ \emph {et~al.}(2023)\citenamefont {Zeng},
  \citenamefont {Zhang}, \citenamefont {Lu}, \citenamefont {Mo}, \citenamefont
  {Li}, \citenamefont {Chen}, \citenamefont {Rynik}, \citenamefont {Huang},
  \citenamefont {Li}, \citenamefont {Shi}, \citenamefont {Wang}, \citenamefont
  {Ye}, \citenamefont {Tuo}, \citenamefont {Yang}, \citenamefont {Ding},
  \citenamefont {Li}, \citenamefont {Tisi}, \citenamefont {Zeng}, \citenamefont
  {Bao}, \citenamefont {Xia}, \citenamefont {Huang}, \citenamefont {Muraoka},
  \citenamefont {Wang}, \citenamefont {Chang}, \citenamefont {Yuan},
  \citenamefont {Bore}, \citenamefont {Cai}, \citenamefont {Lin}, \citenamefont
  {Wang}, \citenamefont {Xu}, \citenamefont {Zhu}, \citenamefont {Luo},
  \citenamefont {Zhang}, \citenamefont {Goodall}, \citenamefont {Liang},
  \citenamefont {Singh}, \citenamefont {Yao}, \citenamefont {Zhang},
  \citenamefont {Wentzcovitch}, \citenamefont {Han}, \citenamefont {Liu},
  \citenamefont {Jia}, \citenamefont {York}, \citenamefont {E}, \citenamefont
  {Car}, \citenamefont {Zhang},\ and\ \citenamefont {Wang}}]{DeePMD-kitv2}%
  \BibitemOpen
  \bibfield  {author} {\bibinfo {author} {\bibfnamefont {J.}~\bibnamefont
  {Zeng}}, \bibinfo {author} {\bibfnamefont {D.}~\bibnamefont {Zhang}},
  \bibinfo {author} {\bibfnamefont {D.}~\bibnamefont {Lu}}, \bibinfo {author}
  {\bibfnamefont {P.}~\bibnamefont {Mo}}, \bibinfo {author} {\bibfnamefont
  {Z.}~\bibnamefont {Li}}, \bibinfo {author} {\bibfnamefont {Y.}~\bibnamefont
  {Chen}}, \bibinfo {author} {\bibfnamefont {M.}~\bibnamefont {Rynik}},
  \bibinfo {author} {\bibfnamefont {L.}~\bibnamefont {Huang}}, \bibinfo
  {author} {\bibfnamefont {Z.}~\bibnamefont {Li}}, \bibinfo {author}
  {\bibfnamefont {S.}~\bibnamefont {Shi}}, \bibinfo {author} {\bibfnamefont
  {Y.}~\bibnamefont {Wang}}, \bibinfo {author} {\bibfnamefont {H.}~\bibnamefont
  {Ye}}, \bibinfo {author} {\bibfnamefont {P.}~\bibnamefont {Tuo}}, \bibinfo
  {author} {\bibfnamefont {J.}~\bibnamefont {Yang}}, \bibinfo {author}
  {\bibfnamefont {Y.}~\bibnamefont {Ding}}, \bibinfo {author} {\bibfnamefont
  {Y.}~\bibnamefont {Li}}, \bibinfo {author} {\bibfnamefont {D.}~\bibnamefont
  {Tisi}}, \bibinfo {author} {\bibfnamefont {Q.}~\bibnamefont {Zeng}}, \bibinfo
  {author} {\bibfnamefont {H.}~\bibnamefont {Bao}}, \bibinfo {author}
  {\bibfnamefont {Y.}~\bibnamefont {Xia}}, \bibinfo {author} {\bibfnamefont
  {J.}~\bibnamefont {Huang}}, \bibinfo {author} {\bibfnamefont
  {K.}~\bibnamefont {Muraoka}}, \bibinfo {author} {\bibfnamefont
  {Y.}~\bibnamefont {Wang}}, \bibinfo {author} {\bibfnamefont {J.}~\bibnamefont
  {Chang}}, \bibinfo {author} {\bibfnamefont {F.}~\bibnamefont {Yuan}},
  \bibinfo {author} {\bibfnamefont {S.~L.}\ \bibnamefont {Bore}}, \bibinfo
  {author} {\bibfnamefont {C.}~\bibnamefont {Cai}}, \bibinfo {author}
  {\bibfnamefont {Y.}~\bibnamefont {Lin}}, \bibinfo {author} {\bibfnamefont
  {B.}~\bibnamefont {Wang}}, \bibinfo {author} {\bibfnamefont {J.}~\bibnamefont
  {Xu}}, \bibinfo {author} {\bibfnamefont {J.-X.}\ \bibnamefont {Zhu}},
  \bibinfo {author} {\bibfnamefont {C.}~\bibnamefont {Luo}}, \bibinfo {author}
  {\bibfnamefont {Y.}~\bibnamefont {Zhang}}, \bibinfo {author} {\bibfnamefont
  {R.~E.~A.}\ \bibnamefont {Goodall}}, \bibinfo {author} {\bibfnamefont
  {W.}~\bibnamefont {Liang}}, \bibinfo {author} {\bibfnamefont {A.~K.}\
  \bibnamefont {Singh}}, \bibinfo {author} {\bibfnamefont {S.}~\bibnamefont
  {Yao}}, \bibinfo {author} {\bibfnamefont {J.}~\bibnamefont {Zhang}}, \bibinfo
  {author} {\bibfnamefont {R.}~\bibnamefont {Wentzcovitch}}, \bibinfo {author}
  {\bibfnamefont {J.}~\bibnamefont {Han}}, \bibinfo {author} {\bibfnamefont
  {J.}~\bibnamefont {Liu}}, \bibinfo {author} {\bibfnamefont {W.}~\bibnamefont
  {Jia}}, \bibinfo {author} {\bibfnamefont {D.~M.}\ \bibnamefont {York}},
  \bibinfo {author} {\bibfnamefont {W.}~\bibnamefont {E}}, \bibinfo {author}
  {\bibfnamefont {R.}~\bibnamefont {Car}}, \bibinfo {author} {\bibfnamefont
  {L.}~\bibnamefont {Zhang}},\ and\ \bibinfo {author} {\bibfnamefont
  {H.}~\bibnamefont {Wang}},\ }\bibfield  {title} {\bibinfo {title}
  {{DeePMD-kit v2: A software package for deep potential models}},\ }\href
  {https://doi.org/10.1063/5.0155600} {\bibfield  {journal} {\bibinfo
  {journal} {The Journal of Chemical Physics}\ }\textbf {\bibinfo {volume}
  {159}},\ \bibinfo {pages} {054801} (\bibinfo {year} {2023})}\BibitemShut
  {NoStop}%
\bibitem [{\citenamefont {Ravindra}\ \emph {et~al.}(2023)\citenamefont
  {Ravindra}, \citenamefont {Advincula}, \citenamefont {Schran}, \citenamefont
  {Michaelides},\ and\ \citenamefont
  {Kapil}}]{ravindra_quasi-one-dimensional_2023}%
  \BibitemOpen
  \bibfield  {author} {\bibinfo {author} {\bibfnamefont {P.}~\bibnamefont
  {Ravindra}}, \bibinfo {author} {\bibfnamefont {X.~R.}\ \bibnamefont
  {Advincula}}, \bibinfo {author} {\bibfnamefont {C.}~\bibnamefont {Schran}},
  \bibinfo {author} {\bibfnamefont {A.}~\bibnamefont {Michaelides}},\ and\
  \bibinfo {author} {\bibfnamefont {V.}~\bibnamefont {Kapil}},\ }\href
  {https://doi.org/10.48550/arXiv.2312.01340} {\bibinfo {title} {A
  quasi-one-dimensional hydrogen-bonded monolayer ice phase}} (\bibinfo {year}
  {2023}),\ \bibinfo {note} {arXiv:2312.01340 [cond-mat]}\BibitemShut {NoStop}%
\bibitem [{\citenamefont {Ceperley}(1995)}]{cepe95rmp}%
  \BibitemOpen
  \bibfield  {author} {\bibinfo {author} {\bibfnamefont {D.~M.}\ \bibnamefont
  {Ceperley}},\ }\bibfield  {title} {\bibinfo {title} {{Path integrals in the
  theory of condensed helium}},\ }\href@noop {} {\bibfield  {journal} {\bibinfo
   {journal} {Reviews of Modern Physics}\ }\textbf {\bibinfo {volume} {67}},\
  \bibinfo {pages} {279} (\bibinfo {year} {1995})}\BibitemShut {NoStop}%
\bibitem [{\citenamefont {Myung}\ \emph {et~al.}(2022)\citenamefont {Myung},
  \citenamefont {Hirshberg},\ and\ \citenamefont
  {Parrinello}}]{PhysRevLett.128.045301}%
  \BibitemOpen
  \bibfield  {author} {\bibinfo {author} {\bibfnamefont {C.~W.}\ \bibnamefont
  {Myung}}, \bibinfo {author} {\bibfnamefont {B.}~\bibnamefont {Hirshberg}},\
  and\ \bibinfo {author} {\bibfnamefont {M.}~\bibnamefont {Parrinello}},\
  }\bibfield  {title} {\bibinfo {title} {Prediction of a supersolid phase in
  high-pressure deuterium},\ }\href
  {https://doi.org/10.1103/PhysRevLett.128.045301} {\bibfield  {journal}
  {\bibinfo  {journal} {Physical Review Letters}\ }\textbf {\bibinfo {volume}
  {128}},\ \bibinfo {pages} {045301} (\bibinfo {year} {2022})}\BibitemShut
  {NoStop}%
\bibitem [{\citenamefont {Chandler}\ and\ \citenamefont
  {Wolynes}(1981)}]{chan-woly81jcp}%
  \BibitemOpen
  \bibfield  {author} {\bibinfo {author} {\bibfnamefont {D.}~\bibnamefont
  {Chandler}}\ and\ \bibinfo {author} {\bibfnamefont {P.~G.}\ \bibnamefont
  {Wolynes}},\ }\bibfield  {title} {\bibinfo {title} {{Exploiting the
  isomorphism between quantum theory and classical statistical mechanics of
  polyatomic fluids}},\ }\href@noop {} {\bibfield  {journal} {\bibinfo
  {journal} {The Journal of Chemical Physics}\ }\textbf {\bibinfo {volume}
  {74}},\ \bibinfo {pages} {4078} (\bibinfo {year} {1981})}\BibitemShut
  {NoStop}%
\bibitem [{\citenamefont {Tuckerman}\ \emph {et~al.}(1993)\citenamefont
  {Tuckerman}, \citenamefont {Berne}, \citenamefont {Martyna},\ and\
  \citenamefont {Klein}}]{Tuckerman1993}%
  \BibitemOpen
  \bibfield  {author} {\bibinfo {author} {\bibfnamefont {M.~E.}\ \bibnamefont
  {Tuckerman}}, \bibinfo {author} {\bibfnamefont {B.~J.}\ \bibnamefont
  {Berne}}, \bibinfo {author} {\bibfnamefont {G.~J.}\ \bibnamefont {Martyna}},\
  and\ \bibinfo {author} {\bibfnamefont {M.~L.}\ \bibnamefont {Klein}},\
  }\bibfield  {title} {\bibinfo {title} {{Efficient molecular dynamics and
  hybrid Monte Carlo algorithms for path integrals}},\ }\href
  {https://doi.org/10.1063/1.465188} {\bibfield  {journal} {\bibinfo  {journal}
  {The Journal of Chemical Physics}\ }\textbf {\bibinfo {volume} {99}},\
  \bibinfo {pages} {2796} (\bibinfo {year} {1993})},\ \Eprint
  {https://arxiv.org/abs/https://pubs.aip.org/aip/jcp/article-pdf/99/4/2796/19121187/2796\_1\_online.pdf}
  {https://pubs.aip.org/aip/jcp/article-pdf/99/4/2796/19121187/2796\_1\_online.pdf}
  \BibitemShut {NoStop}%
\bibitem [{\citenamefont {Dornheim}\ \emph {et~al.}(2020)\citenamefont
  {Dornheim}, \citenamefont {Invernizzi}, \citenamefont {Vorberger},\ and\
  \citenamefont {Hirshberg}}]{10.1063/5.0030760}%
  \BibitemOpen
  \bibfield  {author} {\bibinfo {author} {\bibfnamefont {T.}~\bibnamefont
  {Dornheim}}, \bibinfo {author} {\bibfnamefont {M.}~\bibnamefont
  {Invernizzi}}, \bibinfo {author} {\bibfnamefont {J.}~\bibnamefont
  {Vorberger}},\ and\ \bibinfo {author} {\bibfnamefont {B.}~\bibnamefont
  {Hirshberg}},\ }\bibfield  {title} {\bibinfo {title} {{Attenuating the
  fermion sign problem in path integral Monte Carlo simulations using the
  Bogoliubov inequality and thermodynamic integration}},\ }\href
  {https://doi.org/10.1063/5.0030760} {\bibfield  {journal} {\bibinfo
  {journal} {The Journal of Chemical Physics}\ }\textbf {\bibinfo {volume}
  {153}},\ \bibinfo {pages} {234104} (\bibinfo {year} {2020})}\BibitemShut
  {NoStop}%
\bibitem [{\citenamefont {Dornheim}\ \emph {et~al.}(2023)\citenamefont
  {Dornheim}, \citenamefont {Tolias}, \citenamefont {Groth}, \citenamefont
  {Moldabekov}, \citenamefont {Vorberger},\ and\ \citenamefont
  {Hirshberg}}]{10.1063/5.0171930}%
  \BibitemOpen
  \bibfield  {author} {\bibinfo {author} {\bibfnamefont {T.}~\bibnamefont
  {Dornheim}}, \bibinfo {author} {\bibfnamefont {P.}~\bibnamefont {Tolias}},
  \bibinfo {author} {\bibfnamefont {S.}~\bibnamefont {Groth}}, \bibinfo
  {author} {\bibfnamefont {Z.~A.}\ \bibnamefont {Moldabekov}}, \bibinfo
  {author} {\bibfnamefont {J.}~\bibnamefont {Vorberger}},\ and\ \bibinfo
  {author} {\bibfnamefont {B.}~\bibnamefont {Hirshberg}},\ }\bibfield  {title}
  {\bibinfo {title} {{Fermionic physics from ab initio path integral Monte
  Carlo simulations of fictitious identical particles}},\ }\href
  {https://doi.org/10.1063/5.0171930} {\bibfield  {journal} {\bibinfo
  {journal} {The Journal of Chemical Physics}\ }\textbf {\bibinfo {volume}
  {159}},\ \bibinfo {pages} {164113} (\bibinfo {year} {2023})}\BibitemShut
  {NoStop}%
\bibitem [{\citenamefont {Wodtke}(2016)}]{Wodtke_ChemSocRev_2016}%
  \BibitemOpen
  \bibfield  {author} {\bibinfo {author} {\bibfnamefont {A.~M.}\ \bibnamefont
  {Wodtke}},\ }\bibfield  {title} {\bibinfo {title} {Electronically
  non-adiabatic influences in surface chemistry and dynamics},\ }\href
  {https://doi.org/10.1039/C6CS00078A} {\bibfield  {journal} {\bibinfo
  {journal} {Chem. Soc. Rev.}\ }\textbf {\bibinfo {volume} {45}},\ \bibinfo
  {pages} {3641} (\bibinfo {year} {2016})}\BibitemShut {NoStop}%
\bibitem [{\citenamefont {Kavokine}\ \emph {et~al.}(2022)\citenamefont
  {Kavokine}, \citenamefont {Bocquet},\ and\ \citenamefont
  {Bocquet}}]{Kavokine2022}%
  \BibitemOpen
  \bibfield  {author} {\bibinfo {author} {\bibfnamefont {N.}~\bibnamefont
  {Kavokine}}, \bibinfo {author} {\bibfnamefont {M.-L.}\ \bibnamefont
  {Bocquet}},\ and\ \bibinfo {author} {\bibfnamefont {L.}~\bibnamefont
  {Bocquet}},\ }\bibfield  {title} {\bibinfo {title} {Fluctuation-induced
  quantum friction in nanoscale water flows},\ }\href
  {https://doi.org/10.1038/s41586-021-04284-7} {\bibfield  {journal} {\bibinfo
  {journal} {Nature}\ }\textbf {\bibinfo {volume} {602}},\ \bibinfo {pages}
  {84} (\bibinfo {year} {2022})}\BibitemShut {NoStop}%
\bibitem [{\citenamefont {Head‐Gordon}\ and\ \citenamefont
  {Tully}(1995)}]{HeadGordon_JCP_1995}%
  \BibitemOpen
  \bibfield  {author} {\bibinfo {author} {\bibfnamefont {M.}~\bibnamefont
  {Head‐Gordon}}\ and\ \bibinfo {author} {\bibfnamefont {J.~C.}\ \bibnamefont
  {Tully}},\ }\bibfield  {title} {\bibinfo {title} {{Molecular dynamics with
  electronic frictions}},\ }\href {https://doi.org/10.1063/1.469915} {\bibfield
   {journal} {\bibinfo  {journal} {The Journal of Chemical Physics}\ }\textbf
  {\bibinfo {volume} {103}},\ \bibinfo {pages} {10137} (\bibinfo {year}
  {1995})}\BibitemShut {NoStop}%
\bibitem [{\citenamefont {Dou}\ \emph {et~al.}(2017)\citenamefont {Dou},
  \citenamefont {Miao},\ and\ \citenamefont {Subotnik}}]{Dou_PRL_2017}%
  \BibitemOpen
  \bibfield  {author} {\bibinfo {author} {\bibfnamefont {W.}~\bibnamefont
  {Dou}}, \bibinfo {author} {\bibfnamefont {G.}~\bibnamefont {Miao}},\ and\
  \bibinfo {author} {\bibfnamefont {J.~E.}\ \bibnamefont {Subotnik}},\
  }\bibfield  {title} {\bibinfo {title} {Born-oppenheimer dynamics, electronic
  friction, and the inclusion of electron-electron interactions},\ }\href
  {https://doi.org/10.1103/PhysRevLett.119.046001} {\bibfield  {journal}
  {\bibinfo  {journal} {Physical Review Letters}\ }\textbf {\bibinfo {volume}
  {119}},\ \bibinfo {pages} {046001} (\bibinfo {year} {2017})}\BibitemShut
  {NoStop}%
\bibitem [{\citenamefont {Dou}\ and\ \citenamefont
  {Subotnik}(2018)}]{Dou_JCP_2018}%
  \BibitemOpen
  \bibfield  {author} {\bibinfo {author} {\bibfnamefont {W.}~\bibnamefont
  {Dou}}\ and\ \bibinfo {author} {\bibfnamefont {J.~E.}\ \bibnamefont
  {Subotnik}},\ }\bibfield  {title} {\bibinfo {title} {{Perspective: How to
  understand electronic friction}},\ }\href {https://doi.org/10.1063/1.5035412}
  {\bibfield  {journal} {\bibinfo  {journal} {The Journal of Chemical Physics}\
  }\textbf {\bibinfo {volume} {148}},\ \bibinfo {pages} {230901} (\bibinfo
  {year} {2018})}\BibitemShut {NoStop}%
\bibitem [{\citenamefont {Martinazzo}\ and\ \citenamefont
  {Burghardt}(2022{\natexlab{a}})}]{Martinazzo_PRA_2022}%
  \BibitemOpen
  \bibfield  {author} {\bibinfo {author} {\bibfnamefont {R.}~\bibnamefont
  {Martinazzo}}\ and\ \bibinfo {author} {\bibfnamefont {I.}~\bibnamefont
  {Burghardt}},\ }\bibfield  {title} {\bibinfo {title} {Quantum theory of
  electronic friction},\ }\href {https://doi.org/10.1103/PhysRevA.105.052215}
  {\bibfield  {journal} {\bibinfo  {journal} {Physical Review A}\ }\textbf
  {\bibinfo {volume} {105}},\ \bibinfo {pages} {052215} (\bibinfo {year}
  {2022}{\natexlab{a}})}\BibitemShut {NoStop}%
\bibitem [{\citenamefont {Martinazzo}\ and\ \citenamefont
  {Burghardt}(2022{\natexlab{b}})}]{Martinazzo_PRL_2022}%
  \BibitemOpen
  \bibfield  {author} {\bibinfo {author} {\bibfnamefont {R.}~\bibnamefont
  {Martinazzo}}\ and\ \bibinfo {author} {\bibfnamefont {I.}~\bibnamefont
  {Burghardt}},\ }\bibfield  {title} {\bibinfo {title} {Quantum dynamics with
  electronic friction},\ }\href
  {https://doi.org/10.1103/PhysRevLett.128.206002} {\bibfield  {journal}
  {\bibinfo  {journal} {Physical Review Letters}\ }\textbf {\bibinfo {volume}
  {128}},\ \bibinfo {pages} {206002} (\bibinfo {year}
  {2022}{\natexlab{b}})}\BibitemShut {NoStop}%
\bibitem [{\citenamefont {Martinazzo}\ and\ \citenamefont
  {Burghardt}(2023)}]{martinazzo2023quantum}%
  \BibitemOpen
  \bibfield  {author} {\bibinfo {author} {\bibfnamefont {R.}~\bibnamefont
  {Martinazzo}}\ and\ \bibinfo {author} {\bibfnamefont {I.}~\bibnamefont
  {Burghardt}},\ }\href@noop {} {\bibinfo {title} {Quantum hydrodynamics of
  coupled electron-nuclear systems}} (\bibinfo {year} {2023}),\ \Eprint
  {https://arxiv.org/abs/2310.08766} {arXiv:2310.08766} \BibitemShut {NoStop}%
\bibitem [{\citenamefont {Maurer}\ \emph {et~al.}(2016)\citenamefont {Maurer},
  \citenamefont {Askerka}, \citenamefont {Batista},\ and\ \citenamefont
  {Tully}}]{Maurer_PRB_2016}%
  \BibitemOpen
  \bibfield  {author} {\bibinfo {author} {\bibfnamefont {R.~J.}\ \bibnamefont
  {Maurer}}, \bibinfo {author} {\bibfnamefont {M.}~\bibnamefont {Askerka}},
  \bibinfo {author} {\bibfnamefont {V.~S.}\ \bibnamefont {Batista}},\ and\
  \bibinfo {author} {\bibfnamefont {J.~C.}\ \bibnamefont {Tully}},\ }\bibfield
  {title} {\bibinfo {title} {Ab initio tensorial electronic friction for
  molecules on metal surfaces: Nonadiabatic vibrational relaxation},\ }\href
  {https://doi.org/10.1103/PhysRevB.94.115432} {\bibfield  {journal} {\bibinfo
  {journal} {Physical Review B}\ }\textbf {\bibinfo {volume} {94}},\ \bibinfo
  {pages} {115432} (\bibinfo {year} {2016})}\BibitemShut {NoStop}%
\bibitem [{\citenamefont {Richardson}(2018)}]{Richardson_Review_2018}%
  \BibitemOpen
  \bibfield  {author} {\bibinfo {author} {\bibfnamefont {J.~O.}\ \bibnamefont
  {Richardson}},\ }\bibfield  {title} {\bibinfo {title} {Ring-polymer instanton
  theory},\ }\href {https://doi.org/10.1080/0144235X.2018.1472353} {\bibfield
  {journal} {\bibinfo  {journal} {International Reviews in Physical Chemistry}\
  }\textbf {\bibinfo {volume} {37}},\ \bibinfo {pages} {171} (\bibinfo {year}
  {2018})}\BibitemShut {NoStop}%
\bibitem [{\citenamefont {Litman}\ and\ \citenamefont
  {Rossi}(2020)}]{Litman_PRL_2020}%
  \BibitemOpen
  \bibfield  {author} {\bibinfo {author} {\bibfnamefont {Y.}~\bibnamefont
  {Litman}}\ and\ \bibinfo {author} {\bibfnamefont {M.}~\bibnamefont {Rossi}},\
  }\bibfield  {title} {\bibinfo {title} {Multidimensional hydrogen tunneling in
  supported molecular switches: The role of surface interactions},\ }\href
  {https://doi.org/10.1103/PhysRevLett.125.216001} {\bibfield  {journal}
  {\bibinfo  {journal} {Physical Review Letters}\ }\textbf {\bibinfo {volume}
  {125}},\ \bibinfo {pages} {216001} (\bibinfo {year} {2020})}\BibitemShut
  {NoStop}%
\bibitem [{\citenamefont {Box}\ \emph {et~al.}(2023)\citenamefont {Box},
  \citenamefont {Stark},\ and\ \citenamefont {Maurer}}]{Box_IOP_2023}%
  \BibitemOpen
  \bibfield  {author} {\bibinfo {author} {\bibfnamefont {C.~L.}\ \bibnamefont
  {Box}}, \bibinfo {author} {\bibfnamefont {W.~G.}\ \bibnamefont {Stark}},\
  and\ \bibinfo {author} {\bibfnamefont {R.~J.}\ \bibnamefont {Maurer}},\
  }\bibfield  {title} {\bibinfo {title} {Ab initio calculation of
  electron-phonon linewidths and molecular dynamics with electronic friction at
  metal surfaces with numeric atom-centred orbitals},\ }\href
  {https://doi.org/10.1088/2516-1075/acf3c4} {\bibfield  {journal} {\bibinfo
  {journal} {Electronic Structure}\ }\textbf {\bibinfo {volume} {5}},\ \bibinfo
  {pages} {035005} (\bibinfo {year} {2023})}\BibitemShut {NoStop}%
\bibitem [{\citenamefont {Richardson}\ and\ \citenamefont
  {Althorpe}(2009)}]{Richardson_JCP_2009}%
  \BibitemOpen
  \bibfield  {author} {\bibinfo {author} {\bibfnamefont {J.~O.}\ \bibnamefont
  {Richardson}}\ and\ \bibinfo {author} {\bibfnamefont {S.~C.}\ \bibnamefont
  {Althorpe}},\ }\bibfield  {title} {\bibinfo {title} {{Ring-polymer molecular
  dynamics rate-theory in the deep-tunneling regime: Connection with
  semiclassical instanton theory}},\ }\href@noop {} {\bibfield  {journal}
  {\bibinfo  {journal} {The Journal of Chemical Physics}\ }\textbf {\bibinfo
  {volume} {131}},\ \bibinfo {pages} {214106} (\bibinfo {year}
  {2009})}\BibitemShut {NoStop}%
\bibitem [{\citenamefont {Beyer}\ \emph {et~al.}(2016)\citenamefont {Beyer},
  \citenamefont {Richardson}, \citenamefont {Knowles}, \citenamefont {Rommel},\
  and\ \citenamefont {Althorpe}}]{Beyer_PCL_2016}%
  \BibitemOpen
  \bibfield  {author} {\bibinfo {author} {\bibfnamefont {A.~N.}\ \bibnamefont
  {Beyer}}, \bibinfo {author} {\bibfnamefont {J.~O.}\ \bibnamefont
  {Richardson}}, \bibinfo {author} {\bibfnamefont {P.~J.}\ \bibnamefont
  {Knowles}}, \bibinfo {author} {\bibfnamefont {J.}~\bibnamefont {Rommel}},\
  and\ \bibinfo {author} {\bibfnamefont {S.~C.}\ \bibnamefont {Althorpe}},\
  }\bibfield  {title} {\bibinfo {title} {Quantum tunneling rates of gas-phase
  reactions from on-the-fly instanton calculations},\ }\href
  {https://doi.org/10.1021/acs.jpclett.6b02115} {\bibfield  {journal} {\bibinfo
   {journal} {The Journal of Physical Chemistry Letters}\ }\textbf {\bibinfo
  {volume} {7}},\ \bibinfo {pages} {4374} (\bibinfo {year} {2016})}\BibitemShut
  {NoStop}%
\bibitem [{\citenamefont {Litman}(2020)}]{Litman_thesis}%
  \BibitemOpen
  \bibfield  {author} {\bibinfo {author} {\bibfnamefont {Y.}~\bibnamefont
  {Litman}},\ }\emph {\bibinfo {title} {Tunneling and Zero-Point Energy Effects
  in Multidimensional Hydrogen Transfer Reactions: From Gas Phase to Adsorption
  on Metal Surfaces}},\ \href@noop {} {Ph.D. thesis},\ \bibinfo  {school}
  {Freie Universit\"at Berlin} (\bibinfo {year} {2020})\BibitemShut {NoStop}%
\bibitem [{\citenamefont {Bridge}\ \emph {et~al.}(2024)\citenamefont {Bridge},
  \citenamefont {Lazzaroni}, \citenamefont {Martinazzo}, \citenamefont {Rossi},
  \citenamefont {Althorpe},\ and\ \citenamefont {Litman}}]{bridge2024}%
  \BibitemOpen
  \bibfield  {author} {\bibinfo {author} {\bibfnamefont {O.}~\bibnamefont
  {Bridge}}, \bibinfo {author} {\bibfnamefont {P.}~\bibnamefont {Lazzaroni}},
  \bibinfo {author} {\bibfnamefont {R.}~\bibnamefont {Martinazzo}}, \bibinfo
  {author} {\bibfnamefont {M.}~\bibnamefont {Rossi}}, \bibinfo {author}
  {\bibfnamefont {S.~C.}\ \bibnamefont {Althorpe}},\ and\ \bibinfo {author}
  {\bibfnamefont {Y.}~\bibnamefont {Litman}},\ }\href@noop {} {\bibinfo {title}
  {Quantum rates in dissipative systems with spatially varying friction}}
  (\bibinfo {year} {2024}),\ \Eprint {https://arxiv.org/abs/2405.00512}
  {arXiv:2405.00512} \BibitemShut {NoStop}%
\bibitem [{\citenamefont {Salén}\ \emph {et~al.}(2019)\citenamefont {Salén},
  \citenamefont {Basini}, \citenamefont {Bonetti}, \citenamefont {Hebling},
  \citenamefont {Krasilnikov}, \citenamefont {Nikitin}, \citenamefont
  {Shamuilov}, \citenamefont {Tibai}, \citenamefont {Zhaunerchyk},\ and\
  \citenamefont {Goryashko}}]{SALEN20191}%
  \BibitemOpen
  \bibfield  {author} {\bibinfo {author} {\bibfnamefont {P.}~\bibnamefont
  {Salén}}, \bibinfo {author} {\bibfnamefont {M.}~\bibnamefont {Basini}},
  \bibinfo {author} {\bibfnamefont {S.}~\bibnamefont {Bonetti}}, \bibinfo
  {author} {\bibfnamefont {J.}~\bibnamefont {Hebling}}, \bibinfo {author}
  {\bibfnamefont {M.}~\bibnamefont {Krasilnikov}}, \bibinfo {author}
  {\bibfnamefont {A.~Y.}\ \bibnamefont {Nikitin}}, \bibinfo {author}
  {\bibfnamefont {G.}~\bibnamefont {Shamuilov}}, \bibinfo {author}
  {\bibfnamefont {Z.}~\bibnamefont {Tibai}}, \bibinfo {author} {\bibfnamefont
  {V.}~\bibnamefont {Zhaunerchyk}},\ and\ \bibinfo {author} {\bibfnamefont
  {V.}~\bibnamefont {Goryashko}},\ }\bibfield  {title} {\bibinfo {title}
  {Matter manipulation with extreme terahertz light: Progress in the enabling
  thz technology},\ }\href
  {https://doi.org/https://doi.org/10.1016/j.physrep.2019.09.002} {\bibfield
  {journal} {\bibinfo  {journal} {Physics Reports}\ }\textbf {\bibinfo {volume}
  {836-837}},\ \bibinfo {pages} {1} (\bibinfo {year} {2019})}\BibitemShut
  {NoStop}%
\bibitem [{\citenamefont {de~la Torre}\ \emph {et~al.}(2021)\citenamefont
  {de~la Torre}, \citenamefont {Kennes}, \citenamefont {Claassen},
  \citenamefont {Gerber}, \citenamefont {McIver},\ and\ \citenamefont
  {Sentef}}]{RevModPhys.93.041002}%
  \BibitemOpen
  \bibfield  {author} {\bibinfo {author} {\bibfnamefont {A.}~\bibnamefont
  {de~la Torre}}, \bibinfo {author} {\bibfnamefont {D.~M.}\ \bibnamefont
  {Kennes}}, \bibinfo {author} {\bibfnamefont {M.}~\bibnamefont {Claassen}},
  \bibinfo {author} {\bibfnamefont {S.}~\bibnamefont {Gerber}}, \bibinfo
  {author} {\bibfnamefont {J.~W.}\ \bibnamefont {McIver}},\ and\ \bibinfo
  {author} {\bibfnamefont {M.~A.}\ \bibnamefont {Sentef}},\ }\bibfield  {title}
  {\bibinfo {title} {Colloquium: Nonthermal pathways to ultrafast control in
  quantum materials},\ }\href {https://doi.org/10.1103/RevModPhys.93.041002}
  {\bibfield  {journal} {\bibinfo  {journal} {Reviews of Modern Physics}\
  }\textbf {\bibinfo {volume} {93}},\ \bibinfo {pages} {041002} (\bibinfo
  {year} {2021})}\BibitemShut {NoStop}%
\bibitem [{\citenamefont {Disa}\ \emph {et~al.}(2021)\citenamefont {Disa},
  \citenamefont {Nova},\ and\ \citenamefont {Cavalleri}}]{Disa2021}%
  \BibitemOpen
  \bibfield  {author} {\bibinfo {author} {\bibfnamefont {A.~S.}\ \bibnamefont
  {Disa}}, \bibinfo {author} {\bibfnamefont {T.~F.}\ \bibnamefont {Nova}},\
  and\ \bibinfo {author} {\bibfnamefont {A.}~\bibnamefont {Cavalleri}},\
  }\bibfield  {title} {\bibinfo {title} {Engineering crystal structures with
  light},\ }\href {https://doi.org/10.1038/s41567-021-01366-1} {\bibfield
  {journal} {\bibinfo  {journal} {Nature Physics}\ }\textbf {\bibinfo {volume}
  {17}},\ \bibinfo {pages} {1087} (\bibinfo {year} {2021})}\BibitemShut
  {NoStop}%
\bibitem [{\citenamefont {Juraschek}\ \emph {et~al.}(2017)\citenamefont
  {Juraschek}, \citenamefont {Fechner},\ and\ \citenamefont
  {Spaldin}}]{PhysRevLett.118.054101}%
  \BibitemOpen
  \bibfield  {author} {\bibinfo {author} {\bibfnamefont {D.~M.}\ \bibnamefont
  {Juraschek}}, \bibinfo {author} {\bibfnamefont {M.}~\bibnamefont {Fechner}},\
  and\ \bibinfo {author} {\bibfnamefont {N.~A.}\ \bibnamefont {Spaldin}},\
  }\bibfield  {title} {\bibinfo {title} {Ultrafast structure switching through
  nonlinear phononics},\ }\href
  {https://doi.org/10.1103/PhysRevLett.118.054101} {\bibfield  {journal}
  {\bibinfo  {journal} {Physical Review Letters}\ }\textbf {\bibinfo {volume}
  {118}},\ \bibinfo {pages} {054101} (\bibinfo {year} {2017})}\BibitemShut
  {NoStop}%
\bibitem [{\citenamefont {Subedi}\ \emph {et~al.}(2014)\citenamefont {Subedi},
  \citenamefont {Cavalleri},\ and\ \citenamefont
  {Georges}}]{PhysRevB.89.220301}%
  \BibitemOpen
  \bibfield  {author} {\bibinfo {author} {\bibfnamefont {A.}~\bibnamefont
  {Subedi}}, \bibinfo {author} {\bibfnamefont {A.}~\bibnamefont {Cavalleri}},\
  and\ \bibinfo {author} {\bibfnamefont {A.}~\bibnamefont {Georges}},\
  }\bibfield  {title} {\bibinfo {title} {Theory of nonlinear phononics for
  coherent light control of solids},\ }\href
  {https://doi.org/10.1103/PhysRevB.89.220301} {\bibfield  {journal} {\bibinfo
  {journal} {Physical Review B}\ }\textbf {\bibinfo {volume} {89}},\ \bibinfo
  {pages} {220301} (\bibinfo {year} {2014})}\BibitemShut {NoStop}%
\bibitem [{\citenamefont {F{\"o}rst}\ \emph {et~al.}(2011)\citenamefont
  {F{\"o}rst}, \citenamefont {Manzoni}, \citenamefont {Kaiser}, \citenamefont
  {Tomioka}, \citenamefont {Tokura}, \citenamefont {Merlin},\ and\
  \citenamefont {Cavalleri}}]{Forst2011}%
  \BibitemOpen
  \bibfield  {author} {\bibinfo {author} {\bibfnamefont {M.}~\bibnamefont
  {F{\"o}rst}}, \bibinfo {author} {\bibfnamefont {C.}~\bibnamefont {Manzoni}},
  \bibinfo {author} {\bibfnamefont {S.}~\bibnamefont {Kaiser}}, \bibinfo
  {author} {\bibfnamefont {Y.}~\bibnamefont {Tomioka}}, \bibinfo {author}
  {\bibfnamefont {Y.}~\bibnamefont {Tokura}}, \bibinfo {author} {\bibfnamefont
  {R.}~\bibnamefont {Merlin}},\ and\ \bibinfo {author} {\bibfnamefont
  {A.}~\bibnamefont {Cavalleri}},\ }\bibfield  {title} {\bibinfo {title}
  {Nonlinear phononics as an ultrafast route to lattice control},\ }\href
  {https://doi.org/10.1038/nphys2055} {\bibfield  {journal} {\bibinfo
  {journal} {Nature Physics}\ }\textbf {\bibinfo {volume} {7}},\ \bibinfo
  {pages} {854} (\bibinfo {year} {2011})}\BibitemShut {NoStop}%
\bibitem [{\citenamefont {Spaldin}(2012)}]{SPALDIN20122}%
  \BibitemOpen
  \bibfield  {author} {\bibinfo {author} {\bibfnamefont {N.~A.}\ \bibnamefont
  {Spaldin}},\ }\bibfield  {title} {\bibinfo {title} {A beginner's guide to the
  modern theory of polarization},\ }\href
  {https://doi.org/https://doi.org/10.1016/j.jssc.2012.05.010} {\bibfield
  {journal} {\bibinfo  {journal} {Journal of Solid State Chemistry}\ }\textbf
  {\bibinfo {volume} {195}},\ \bibinfo {pages} {2} (\bibinfo {year} {2012})},\
  \bibinfo {note} {polar Inorganic Materials: Design Strategies and Functional
  Properties}\BibitemShut {NoStop}%
\bibitem [{\citenamefont {Baroni}\ \emph {et~al.}(2001)\citenamefont {Baroni},
  \citenamefont {de~Gironcoli}, \citenamefont {Dal~Corso},\ and\ \citenamefont
  {Giannozzi}}]{baroni2001phonons}%
  \BibitemOpen
  \bibfield  {author} {\bibinfo {author} {\bibfnamefont {S.}~\bibnamefont
  {Baroni}}, \bibinfo {author} {\bibfnamefont {S.}~\bibnamefont
  {de~Gironcoli}}, \bibinfo {author} {\bibfnamefont {A.}~\bibnamefont
  {Dal~Corso}},\ and\ \bibinfo {author} {\bibfnamefont {P.}~\bibnamefont
  {Giannozzi}},\ }\bibfield  {title} {\bibinfo {title} {Phonons and related
  crystal properties from density-functional perturbation theory},\ }\href
  {https://doi.org/10.1103/RevModPhys.73.515} {\bibfield  {journal} {\bibinfo
  {journal} {Reviews of Modern Physics}\ }\textbf {\bibinfo {volume} {73}},\
  \bibinfo {pages} {515} (\bibinfo {year} {2001})}\BibitemShut {NoStop}%
\bibitem [{\citenamefont {Shang}\ \emph {et~al.}(2018)\citenamefont {Shang},
  \citenamefont {Raimbault}, \citenamefont {Rinke}, \citenamefont {Scheffler},
  \citenamefont {Rossi},\ and\ \citenamefont {Carbogno}}]{Shang_2018}%
  \BibitemOpen
  \bibfield  {author} {\bibinfo {author} {\bibfnamefont {H.}~\bibnamefont
  {Shang}}, \bibinfo {author} {\bibfnamefont {N.}~\bibnamefont {Raimbault}},
  \bibinfo {author} {\bibfnamefont {P.}~\bibnamefont {Rinke}}, \bibinfo
  {author} {\bibfnamefont {M.}~\bibnamefont {Scheffler}}, \bibinfo {author}
  {\bibfnamefont {M.}~\bibnamefont {Rossi}},\ and\ \bibinfo {author}
  {\bibfnamefont {C.}~\bibnamefont {Carbogno}},\ }\bibfield  {title} {\bibinfo
  {title} {All-electron, real-space perturbation theory for homogeneous
  electric fields: theory, implementation, and application within dft},\ }\href
  {https://doi.org/10.1088/1367-2630/aace6d} {\bibfield  {journal} {\bibinfo
  {journal} {New Journal of Physics}\ }\textbf {\bibinfo {volume} {20}},\
  \bibinfo {pages} {073040} (\bibinfo {year} {2018})}\BibitemShut {NoStop}%
\bibitem [{\citenamefont {Partridge}\ and\ \citenamefont
  {Schwenke}(1997)}]{10.1063/1.473987}%
  \BibitemOpen
  \bibfield  {author} {\bibinfo {author} {\bibfnamefont {H.}~\bibnamefont
  {Partridge}}\ and\ \bibinfo {author} {\bibfnamefont {D.~W.}\ \bibnamefont
  {Schwenke}},\ }\bibfield  {title} {\bibinfo {title} {{The determination of an
  accurate isotope dependent potential energy surface for water from extensive
  ab initio calculations and experimental data}},\ }\href
  {https://doi.org/10.1063/1.473987} {\bibfield  {journal} {\bibinfo  {journal}
  {The Journal of Chemical Physics}\ }\textbf {\bibinfo {volume} {106}},\
  \bibinfo {pages} {4618} (\bibinfo {year} {1997})}\BibitemShut {NoStop}%
\bibitem [{\citenamefont {Musil}\ \emph {et~al.}(2019)\citenamefont {Musil},
  \citenamefont {Willatt}, \citenamefont {Langovoy},\ and\ \citenamefont
  {Ceriotti}}]{musi+19jctc}%
  \BibitemOpen
  \bibfield  {author} {\bibinfo {author} {\bibfnamefont {F.}~\bibnamefont
  {Musil}}, \bibinfo {author} {\bibfnamefont {M.~J.}\ \bibnamefont {Willatt}},
  \bibinfo {author} {\bibfnamefont {M.~A.}\ \bibnamefont {Langovoy}},\ and\
  \bibinfo {author} {\bibfnamefont {M.}~\bibnamefont {Ceriotti}},\ }\bibfield
  {title} {\bibinfo {title} {{Fast and Accurate Uncertainty Estimation in
  Chemical Machine Learning}},\ }\href@noop {} {\bibfield  {journal} {\bibinfo
  {journal} {Journal of Chemical Theory and Computation}\ }\textbf {\bibinfo
  {volume} {15}},\ \bibinfo {pages} {906} (\bibinfo {year} {2019})}\BibitemShut
  {NoStop}%
\bibitem [{\citenamefont {Lakshminarayanan}\ \emph {et~al.}(2017)\citenamefont
  {Lakshminarayanan}, \citenamefont {Pritzel},\ and\ \citenamefont
  {Blundell}}]{Lakshminarayanan2017}%
  \BibitemOpen
  \bibfield  {author} {\bibinfo {author} {\bibfnamefont {B.}~\bibnamefont
  {Lakshminarayanan}}, \bibinfo {author} {\bibfnamefont {A.}~\bibnamefont
  {Pritzel}},\ and\ \bibinfo {author} {\bibfnamefont {C.}~\bibnamefont
  {Blundell}},\ }\bibfield  {title} {\bibinfo {title} {Simple and scalable
  predictive uncertainty estimation using deep ensembles},\ }in\ \href
  {https://proceedings.neurips.cc/paper_files/paper/2017/file/9ef2ed4b7fd2c810847ffa5fa85bce38-Paper.pdf}
  {\emph {\bibinfo {booktitle} {Advances in Neural Information Processing
  Systems}}},\ Vol.~\bibinfo {volume} {30},\ \bibinfo {editor} {edited by\
  \bibinfo {editor} {\bibfnamefont {I.}~\bibnamefont {Guyon}}, \bibinfo
  {editor} {\bibfnamefont {U.~V.}\ \bibnamefont {Luxburg}}, \bibinfo {editor}
  {\bibfnamefont {S.}~\bibnamefont {Bengio}}, \bibinfo {editor} {\bibfnamefont
  {H.}~\bibnamefont {Wallach}}, \bibinfo {editor} {\bibfnamefont
  {R.}~\bibnamefont {Fergus}}, \bibinfo {editor} {\bibfnamefont
  {S.}~\bibnamefont {Vishwanathan}},\ and\ \bibinfo {editor} {\bibfnamefont
  {R.}~\bibnamefont {Garnett}}}\ (\bibinfo  {publisher} {Curran Associates,
  Inc.},\ \bibinfo {year} {2017})\BibitemShut {NoStop}%
\bibitem [{\citenamefont {Kunapuli}(2023)}]{kunapuli2023ensemble}%
  \BibitemOpen
  \bibfield  {author} {\bibinfo {author} {\bibfnamefont {G.}~\bibnamefont
  {Kunapuli}},\ }\href {https://books.google.ch/books?id=wXGazgEACAAJ} {\emph
  {\bibinfo {title} {Ensemble Methods for Machine Learning}}}\ (\bibinfo
  {publisher} {Manning},\ \bibinfo {year} {2023})\BibitemShut {NoStop}%
\bibitem [{\citenamefont {Ashukha}\ \emph {et~al.}(2021)\citenamefont
  {Ashukha}, \citenamefont {Lyzhov}, \citenamefont {Molchanov},\ and\
  \citenamefont {Vetrov}}]{ashukha2021pitfalls}%
  \BibitemOpen
  \bibfield  {author} {\bibinfo {author} {\bibfnamefont {A.}~\bibnamefont
  {Ashukha}}, \bibinfo {author} {\bibfnamefont {A.}~\bibnamefont {Lyzhov}},
  \bibinfo {author} {\bibfnamefont {D.}~\bibnamefont {Molchanov}},\ and\
  \bibinfo {author} {\bibfnamefont {D.}~\bibnamefont {Vetrov}},\ }\href@noop {}
  {\bibinfo {title} {Pitfalls of in-domain uncertainty estimation and
  ensembling in deep learning}} (\bibinfo {year} {2021}),\ \Eprint
  {https://arxiv.org/abs/2002.06470} {arXiv:2002.06470} \BibitemShut {NoStop}%
\bibitem [{\citenamefont {Rahaman}\ and\ \citenamefont
  {Thiery}(2021)}]{Rahaman2021}%
  \BibitemOpen
  \bibfield  {author} {\bibinfo {author} {\bibfnamefont {R.}~\bibnamefont
  {Rahaman}}\ and\ \bibinfo {author} {\bibfnamefont {A.}~\bibnamefont
  {Thiery}},\ }\bibfield  {title} {\bibinfo {title} {Uncertainty quantification
  and deep ensembles},\ }in\ \href
  {https://proceedings.neurips.cc/paper_files/paper/2021/file/a70dc40477bc2adceef4d2c90f47eb82-Paper.pdf}
  {\emph {\bibinfo {booktitle} {Advances in Neural Information Processing
  Systems}}},\ Vol.~\bibinfo {volume} {34},\ \bibinfo {editor} {edited by\
  \bibinfo {editor} {\bibfnamefont {M.}~\bibnamefont {Ranzato}}, \bibinfo
  {editor} {\bibfnamefont {A.}~\bibnamefont {Beygelzimer}}, \bibinfo {editor}
  {\bibfnamefont {Y.}~\bibnamefont {Dauphin}}, \bibinfo {editor} {\bibfnamefont
  {P.}~\bibnamefont {Liang}},\ and\ \bibinfo {editor} {\bibfnamefont {J.~W.}\
  \bibnamefont {Vaughan}}}\ (\bibinfo  {publisher} {Curran Associates, Inc.},\
  \bibinfo {year} {2021})\ pp.\ \bibinfo {pages} {20063--20075}\BibitemShut
  {NoStop}%
\bibitem [{\citenamefont {Bigi}\ \emph {et~al.}(2024)\citenamefont {Bigi},
  \citenamefont {Chong}, \citenamefont {Ceriotti},\ and\ \citenamefont
  {Grasselli}}]{bigi2024prediction}%
  \BibitemOpen
  \bibfield  {author} {\bibinfo {author} {\bibfnamefont {F.}~\bibnamefont
  {Bigi}}, \bibinfo {author} {\bibfnamefont {S.}~\bibnamefont {Chong}},
  \bibinfo {author} {\bibfnamefont {M.}~\bibnamefont {Ceriotti}},\ and\
  \bibinfo {author} {\bibfnamefont {F.}~\bibnamefont {Grasselli}},\ }\href@noop
  {} {\bibinfo {title} {A prediction rigidity formalism for low-cost
  uncertainties in trained neural networks}} (\bibinfo {year} {2024}),\ \Eprint
  {https://arxiv.org/abs/2403.02251} {arXiv:2403.02251} \BibitemShut {NoStop}%
\bibitem [{\citenamefont {Behler}\ and\ \citenamefont
  {Parrinello}(2007)}]{behl-parr07prl}%
  \BibitemOpen
  \bibfield  {author} {\bibinfo {author} {\bibfnamefont {J.}~\bibnamefont
  {Behler}}\ and\ \bibinfo {author} {\bibfnamefont {M.}~\bibnamefont
  {Parrinello}},\ }\bibfield  {title} {\bibinfo {title} {{Generalized
  Neural-Network Representation of High-Dimensional Potential-Energy
  Surfaces}},\ }\href@noop {} {\bibfield  {journal} {\bibinfo  {journal}
  {Physical Review Letters}\ }\textbf {\bibinfo {volume} {98}},\ \bibinfo
  {pages} {146401} (\bibinfo {year} {2007})}\BibitemShut {NoStop}%
\bibitem [{\citenamefont {Bart{\'{o}}k}\ \emph {et~al.}(2013)\citenamefont
  {Bart{\'{o}}k}, \citenamefont {Kondor},\ and\ \citenamefont
  {Cs{\'{a}}nyi}}]{bart+13prb}%
  \BibitemOpen
  \bibfield  {author} {\bibinfo {author} {\bibfnamefont {A.~P.}\ \bibnamefont
  {Bart{\'{o}}k}}, \bibinfo {author} {\bibfnamefont {R.}~\bibnamefont
  {Kondor}},\ and\ \bibinfo {author} {\bibfnamefont {G.}~\bibnamefont
  {Cs{\'{a}}nyi}},\ }\bibfield  {title} {\bibinfo {title} {{On representing
  chemical environments}},\ }\href@noop {} {\bibfield  {journal} {\bibinfo
  {journal} {Physical Review B}\ }\textbf {\bibinfo {volume} {87}},\ \bibinfo
  {pages} {184115} (\bibinfo {year} {2013})}\BibitemShut {NoStop}%
\bibitem [{\citenamefont {Pedersen}\ \emph {et~al.}(2015)\citenamefont
  {Pedersen}, \citenamefont {Hummel},\ and\ \citenamefont
  {Dellago}}]{pede+15jcp}%
  \BibitemOpen
  \bibfield  {author} {\bibinfo {author} {\bibfnamefont {U.~R.}\ \bibnamefont
  {Pedersen}}, \bibinfo {author} {\bibfnamefont {F.}~\bibnamefont {Hummel}},\
  and\ \bibinfo {author} {\bibfnamefont {C.}~\bibnamefont {Dellago}},\
  }\bibfield  {title} {\bibinfo {title} {{Computing the crystal growth rate by
  the interface pinning method.}},\ }\href@noop {} {\bibfield  {journal}
  {\bibinfo  {journal} {The Journal of Chemical Physics}\ }\textbf {\bibinfo
  {volume} {142}},\ \bibinfo {pages} {44104} (\bibinfo {year}
  {2015})}\BibitemShut {NoStop}%
\bibitem [{\citenamefont {Torrie}\ and\ \citenamefont
  {Valleau}(1977)}]{torr-vall99jcp}%
  \BibitemOpen
  \bibfield  {author} {\bibinfo {author} {\bibfnamefont {G.~M.}\ \bibnamefont
  {Torrie}}\ and\ \bibinfo {author} {\bibfnamefont {J.~P.}\ \bibnamefont
  {Valleau}},\ }\bibfield  {title} {\bibinfo {title} {{Nonphysical sampling
  distributions in Monte Carlo free-energy estimation: Umbrella sampling}},\
  }\href@noop {} {\bibfield  {journal} {\bibinfo  {journal} {Journal of
  Computational Physics}\ }\textbf {\bibinfo {volume} {23}},\ \bibinfo {pages}
  {187} (\bibinfo {year} {1977})}\BibitemShut {NoStop}%
\bibitem [{\citenamefont {Ribeiro}\ \emph {et~al.}(2018)\citenamefont
  {Ribeiro}, \citenamefont {Mart{\'{i}}nez-Mart{\'{i}}nez}, \citenamefont {Du},
  \citenamefont {Campos-Gonzalez-Angulo},\ and\ \citenamefont
  {Yuen-Zhou}}]{Ribeiro2018}%
  \BibitemOpen
  \bibfield  {author} {\bibinfo {author} {\bibfnamefont {R.~F.}\ \bibnamefont
  {Ribeiro}}, \bibinfo {author} {\bibfnamefont {L.~A.}\ \bibnamefont
  {Mart{\'{i}}nez-Mart{\'{i}}nez}}, \bibinfo {author} {\bibfnamefont
  {M.}~\bibnamefont {Du}}, \bibinfo {author} {\bibfnamefont {J.}~\bibnamefont
  {Campos-Gonzalez-Angulo}},\ and\ \bibinfo {author} {\bibfnamefont
  {J.}~\bibnamefont {Yuen-Zhou}},\ }\bibfield  {title} {\bibinfo {title}
  {{Polariton Chemistry: Controlling Molecular Dynamics with Optical
  Cavities}},\ }\href {https://doi.org/10.1039/C8SC01043A} {\bibfield
  {journal} {\bibinfo  {journal} {Chemical Science}\ }\textbf {\bibinfo
  {volume} {9}},\ \bibinfo {pages} {6325} (\bibinfo {year} {2018})}\BibitemShut
  {NoStop}%
\bibitem [{\citenamefont {Li}\ \emph {et~al.}(2022{\natexlab{b}})\citenamefont
  {Li}, \citenamefont {Cui}, \citenamefont {Subotnik},\ and\ \citenamefont
  {Nitzan}}]{Li2022Review}%
  \BibitemOpen
  \bibfield  {author} {\bibinfo {author} {\bibfnamefont {T.~E.}\ \bibnamefont
  {Li}}, \bibinfo {author} {\bibfnamefont {B.}~\bibnamefont {Cui}}, \bibinfo
  {author} {\bibfnamefont {J.~E.}\ \bibnamefont {Subotnik}},\ and\ \bibinfo
  {author} {\bibfnamefont {A.}~\bibnamefont {Nitzan}},\ }\bibfield  {title}
  {\bibinfo {title} {{Molecular Polaritonics: Chemical Dynamics Under Strong
  Light–Matter Coupling}},\ }\href
  {https://doi.org/10.1146/annurev-physchem-090519-042621} {\bibfield
  {journal} {\bibinfo  {journal} {Annu. Rev. Phys. Chem.}\ }\textbf {\bibinfo
  {volume} {73}},\ \bibinfo {pages} {43} (\bibinfo {year}
  {2022}{\natexlab{b}})}\BibitemShut {NoStop}%
\bibitem [{\citenamefont {Fregoni}\ \emph {et~al.}(2022)\citenamefont
  {Fregoni}, \citenamefont {Garcia-Vidal},\ and\ \citenamefont
  {Feist}}]{Fregoni2022}%
  \BibitemOpen
  \bibfield  {author} {\bibinfo {author} {\bibfnamefont {J.}~\bibnamefont
  {Fregoni}}, \bibinfo {author} {\bibfnamefont {F.~J.}\ \bibnamefont
  {Garcia-Vidal}},\ and\ \bibinfo {author} {\bibfnamefont {J.}~\bibnamefont
  {Feist}},\ }\bibfield  {title} {\bibinfo {title} {{Theoretical Challenges in
  Polaritonic Chemistry}},\ }\href
  {https://doi.org/10.1021/acsphotonics.1c01749} {\bibfield  {journal}
  {\bibinfo  {journal} {ACS Photonics}\ }\textbf {\bibinfo {volume} {9}},\
  \bibinfo {pages} {1096} (\bibinfo {year} {2022})}\BibitemShut {NoStop}%
\bibitem [{\citenamefont {Mandal}\ \emph {et~al.}(2023)\citenamefont {Mandal},
  \citenamefont {Taylor}, \citenamefont {Weight}, \citenamefont {Koessler},
  \citenamefont {Li},\ and\ \citenamefont {Huo}}]{Mandal2023ChemRev}%
  \BibitemOpen
  \bibfield  {author} {\bibinfo {author} {\bibfnamefont {A.}~\bibnamefont
  {Mandal}}, \bibinfo {author} {\bibfnamefont {M.~A.}\ \bibnamefont {Taylor}},
  \bibinfo {author} {\bibfnamefont {B.~M.}\ \bibnamefont {Weight}}, \bibinfo
  {author} {\bibfnamefont {E.~R.}\ \bibnamefont {Koessler}}, \bibinfo {author}
  {\bibfnamefont {X.}~\bibnamefont {Li}},\ and\ \bibinfo {author}
  {\bibfnamefont {P.}~\bibnamefont {Huo}},\ }\bibfield  {title} {\bibinfo
  {title} {{Theoretical Advances in Polariton Chemistry and Molecular Cavity
  Quantum Electrodynamics}},\ }\href
  {https://doi.org/10.1021/acs.chemrev.2c00855} {\bibfield  {journal} {\bibinfo
   {journal} {Chemical Reviews}\ }\textbf {\bibinfo {volume} {123}},\ \bibinfo
  {pages} {9786} (\bibinfo {year} {2023})}\BibitemShut {NoStop}%
\bibitem [{\citenamefont {Ruggenthaler}\ \emph {et~al.}(2023)\citenamefont
  {Ruggenthaler}, \citenamefont {Sidler},\ and\ \citenamefont
  {Rubio}}]{Ruggenthaler2023}%
  \BibitemOpen
  \bibfield  {author} {\bibinfo {author} {\bibfnamefont {M.}~\bibnamefont
  {Ruggenthaler}}, \bibinfo {author} {\bibfnamefont {D.}~\bibnamefont
  {Sidler}},\ and\ \bibinfo {author} {\bibfnamefont {A.}~\bibnamefont
  {Rubio}},\ }\bibfield  {title} {\bibinfo {title} {{Understanding Polaritonic
  Chemistry from Ab Initio Quantum Electrodynamics}},\ }\href
  {https://doi.org/10.1021/acs.chemrev.2c00788} {\bibfield  {journal} {\bibinfo
   {journal} {Chemical Reviews}\ }\textbf {\bibinfo {volume} {123}},\ \bibinfo
  {pages} {11191} (\bibinfo {year} {2023})}\BibitemShut {NoStop}%
\bibitem [{\citenamefont {Shalabney}\ \emph {et~al.}(2015)\citenamefont
  {Shalabney}, \citenamefont {George}, \citenamefont {Hutchison}, \citenamefont
  {Pupillo}, \citenamefont {Genet},\ and\ \citenamefont
  {Ebbesen}}]{Shalabney2015}%
  \BibitemOpen
  \bibfield  {author} {\bibinfo {author} {\bibfnamefont {A.}~\bibnamefont
  {Shalabney}}, \bibinfo {author} {\bibfnamefont {J.}~\bibnamefont {George}},
  \bibinfo {author} {\bibfnamefont {J.}~\bibnamefont {Hutchison}}, \bibinfo
  {author} {\bibfnamefont {G.}~\bibnamefont {Pupillo}}, \bibinfo {author}
  {\bibfnamefont {C.}~\bibnamefont {Genet}},\ and\ \bibinfo {author}
  {\bibfnamefont {T.~W.}\ \bibnamefont {Ebbesen}},\ }\bibfield  {title}
  {\bibinfo {title} {{Coherent Coupling of Molecular Resonators with a
  Microcavity Mode}},\ }\href {https://doi.org/10.1038/ncomms6981} {\bibfield
  {journal} {\bibinfo  {journal} {Nature Communications}\ }\textbf {\bibinfo
  {volume} {6}},\ \bibinfo {pages} {5981} (\bibinfo {year} {2015})}\BibitemShut
  {NoStop}%
\bibitem [{\citenamefont {Long}\ and\ \citenamefont
  {Simpkins}(2015)}]{Long2015}%
  \BibitemOpen
  \bibfield  {author} {\bibinfo {author} {\bibfnamefont {J.~P.}\ \bibnamefont
  {Long}}\ and\ \bibinfo {author} {\bibfnamefont {B.~S.}\ \bibnamefont
  {Simpkins}},\ }\bibfield  {title} {\bibinfo {title} {{Coherent Coupling
  between a Molecular Vibration and Fabry–Perot Optical Cavity to Give
  Hybridized States in the Strong Coupling Limit}},\ }\href
  {https://doi.org/10.1021/ph5003347} {\bibfield  {journal} {\bibinfo
  {journal} {ACS Photonics}\ }\textbf {\bibinfo {volume} {2}},\ \bibinfo
  {pages} {130} (\bibinfo {year} {2015})}\BibitemShut {NoStop}%
\bibitem [{\citenamefont {Thomas}\ \emph {et~al.}(2016)\citenamefont {Thomas},
  \citenamefont {George}, \citenamefont {Shalabney}, \citenamefont {Dryzhakov},
  \citenamefont {Varma}, \citenamefont {Moran}, \citenamefont {Chervy},
  \citenamefont {Zhong}, \citenamefont {Devaux}, \citenamefont {Genet},
  \citenamefont {Hutchison},\ and\ \citenamefont {Ebbesen}}]{Thomas2016}%
  \BibitemOpen
  \bibfield  {author} {\bibinfo {author} {\bibfnamefont {A.}~\bibnamefont
  {Thomas}}, \bibinfo {author} {\bibfnamefont {J.}~\bibnamefont {George}},
  \bibinfo {author} {\bibfnamefont {A.}~\bibnamefont {Shalabney}}, \bibinfo
  {author} {\bibfnamefont {M.}~\bibnamefont {Dryzhakov}}, \bibinfo {author}
  {\bibfnamefont {S.~J.}\ \bibnamefont {Varma}}, \bibinfo {author}
  {\bibfnamefont {J.}~\bibnamefont {Moran}}, \bibinfo {author} {\bibfnamefont
  {T.}~\bibnamefont {Chervy}}, \bibinfo {author} {\bibfnamefont
  {X.}~\bibnamefont {Zhong}}, \bibinfo {author} {\bibfnamefont
  {E.}~\bibnamefont {Devaux}}, \bibinfo {author} {\bibfnamefont
  {C.}~\bibnamefont {Genet}}, \bibinfo {author} {\bibfnamefont {J.~A.}\
  \bibnamefont {Hutchison}},\ and\ \bibinfo {author} {\bibfnamefont {T.~W.}\
  \bibnamefont {Ebbesen}},\ }\bibfield  {title} {\bibinfo {title}
  {{Ground-State Chemical Reactivity under Vibrational Coupling to the Vacuum
  Electromagnetic Field}},\ }\href {https://doi.org/10.1002/anie.201605504}
  {\bibfield  {journal} {\bibinfo  {journal} {Angew. Chemie Int. Ed.}\ }\textbf
  {\bibinfo {volume} {55}},\ \bibinfo {pages} {11462} (\bibinfo {year}
  {2016})}\BibitemShut {NoStop}%
\bibitem [{\citenamefont {Thomas}\ \emph {et~al.}(2019)\citenamefont {Thomas},
  \citenamefont {Lethuillier-Karl}, \citenamefont {Nagarajan}, \citenamefont
  {Vergauwe}, \citenamefont {George}, \citenamefont {Chervy}, \citenamefont
  {Shalabney}, \citenamefont {Devaux}, \citenamefont {Genet}, \citenamefont
  {Moran},\ and\ \citenamefont {Ebbesen}}]{Thomas2019_science}%
  \BibitemOpen
  \bibfield  {author} {\bibinfo {author} {\bibfnamefont {A.}~\bibnamefont
  {Thomas}}, \bibinfo {author} {\bibfnamefont {L.}~\bibnamefont
  {Lethuillier-Karl}}, \bibinfo {author} {\bibfnamefont {K.}~\bibnamefont
  {Nagarajan}}, \bibinfo {author} {\bibfnamefont {R.~M.~A.}\ \bibnamefont
  {Vergauwe}}, \bibinfo {author} {\bibfnamefont {J.}~\bibnamefont {George}},
  \bibinfo {author} {\bibfnamefont {T.}~\bibnamefont {Chervy}}, \bibinfo
  {author} {\bibfnamefont {A.}~\bibnamefont {Shalabney}}, \bibinfo {author}
  {\bibfnamefont {E.}~\bibnamefont {Devaux}}, \bibinfo {author} {\bibfnamefont
  {C.}~\bibnamefont {Genet}}, \bibinfo {author} {\bibfnamefont
  {J.}~\bibnamefont {Moran}},\ and\ \bibinfo {author} {\bibfnamefont {T.~W.}\
  \bibnamefont {Ebbesen}},\ }\bibfield  {title} {\bibinfo {title} {{Tilting a
  Ground-State Reactivity Landscape by Vibrational Strong Coupling}},\ }\href
  {https://doi.org/10.1126/science.aau7742} {\bibfield  {journal} {\bibinfo
  {journal} {Science}\ }\textbf {\bibinfo {volume} {363}},\ \bibinfo {pages}
  {615} (\bibinfo {year} {2019})}\BibitemShut {NoStop}%
\bibitem [{\citenamefont {Xiang}\ \emph {et~al.}(2020)\citenamefont {Xiang},
  \citenamefont {Ribeiro}, \citenamefont {Du}, \citenamefont {Chen},
  \citenamefont {Yang}, \citenamefont {Wang}, \citenamefont {Yuen-Zhou},\ and\
  \citenamefont {Xiong}}]{Xiang2020Science}%
  \BibitemOpen
  \bibfield  {author} {\bibinfo {author} {\bibfnamefont {B.}~\bibnamefont
  {Xiang}}, \bibinfo {author} {\bibfnamefont {R.~F.}\ \bibnamefont {Ribeiro}},
  \bibinfo {author} {\bibfnamefont {M.}~\bibnamefont {Du}}, \bibinfo {author}
  {\bibfnamefont {L.}~\bibnamefont {Chen}}, \bibinfo {author} {\bibfnamefont
  {Z.}~\bibnamefont {Yang}}, \bibinfo {author} {\bibfnamefont {J.}~\bibnamefont
  {Wang}}, \bibinfo {author} {\bibfnamefont {J.}~\bibnamefont {Yuen-Zhou}},\
  and\ \bibinfo {author} {\bibfnamefont {W.}~\bibnamefont {Xiong}},\ }\bibfield
   {title} {\bibinfo {title} {{Intermolecular Vibrational Energy Transfer
  Enabled by Microcavity Strong Light--Matter Coupling}},\ }\href
  {https://doi.org/10.1126/science.aba3544} {\bibfield  {journal} {\bibinfo
  {journal} {Science}\ }\textbf {\bibinfo {volume} {368}},\ \bibinfo {pages}
  {665} (\bibinfo {year} {2020})}\BibitemShut {NoStop}%
\bibitem [{\citenamefont {Flick}\ \emph {et~al.}(2017)\citenamefont {Flick},
  \citenamefont {Ruggenthaler}, \citenamefont {Appel},\ and\ \citenamefont
  {Rubio}}]{Flick2017}%
  \BibitemOpen
  \bibfield  {author} {\bibinfo {author} {\bibfnamefont {J.}~\bibnamefont
  {Flick}}, \bibinfo {author} {\bibfnamefont {M.}~\bibnamefont {Ruggenthaler}},
  \bibinfo {author} {\bibfnamefont {H.}~\bibnamefont {Appel}},\ and\ \bibinfo
  {author} {\bibfnamefont {A.}~\bibnamefont {Rubio}},\ }\bibfield  {title}
  {\bibinfo {title} {{Atoms and Molecules in Cavities, from Weak to Strong
  Coupling in Quantum-Electrodynamics (QED) Chemistry}},\ }\href
  {https://doi.org/10.1073/pnas.1615509114} {\bibfield  {journal} {\bibinfo
  {journal} {Proc. Natl. Acad. Sci.}\ }\textbf {\bibinfo {volume} {114}},\
  \bibinfo {pages} {3026} (\bibinfo {year} {2017})}\BibitemShut {NoStop}%
\bibitem [{\citenamefont {Li}\ \emph {et~al.}(2021)\citenamefont {Li},
  \citenamefont {Nitzan},\ and\ \citenamefont {Subotnik}}]{Li2020Nonlinear}%
  \BibitemOpen
  \bibfield  {author} {\bibinfo {author} {\bibfnamefont {T.~E.}\ \bibnamefont
  {Li}}, \bibinfo {author} {\bibfnamefont {A.}~\bibnamefont {Nitzan}},\ and\
  \bibinfo {author} {\bibfnamefont {J.~E.}\ \bibnamefont {Subotnik}},\
  }\bibfield  {title} {\bibinfo {title} {{Cavity Molecular Dynamics Simulations
  of Vibrational Polariton-Enhanced Molecular Nonlinear Absorption}},\ }\href
  {https://doi.org/10.1063/5.0037623} {\bibfield  {journal} {\bibinfo
  {journal} {The Journal of Chemical Physics}\ }\textbf {\bibinfo {volume}
  {154}},\ \bibinfo {pages} {094124} (\bibinfo {year} {2021})}\BibitemShut
  {NoStop}%
\bibitem [{\citenamefont {Li}\ \emph {et~al.}(2022{\natexlab{c}})\citenamefont
  {Li}, \citenamefont {Nitzan},\ and\ \citenamefont {Subotnik}}]{Li2021Solute}%
  \BibitemOpen
  \bibfield  {author} {\bibinfo {author} {\bibfnamefont {T.~E.}\ \bibnamefont
  {Li}}, \bibinfo {author} {\bibfnamefont {A.}~\bibnamefont {Nitzan}},\ and\
  \bibinfo {author} {\bibfnamefont {J.~E.}\ \bibnamefont {Subotnik}},\
  }\bibfield  {title} {\bibinfo {title} {{Energy-Efficient Pathway for
  Selectively Exciting Solute Molecules to High Vibrational States via Solvent
  Vibration-Polariton Pumping}},\ }\href
  {https://doi.org/10.1038/s41467-022-31703-8} {\bibfield  {journal} {\bibinfo
  {journal} {Nature Communications}\ }\textbf {\bibinfo {volume} {13}},\
  \bibinfo {pages} {4203} (\bibinfo {year} {2022}{\natexlab{c}})}\BibitemShut
  {NoStop}%
\bibitem [{\citenamefont {Li}\ and\ \citenamefont
  {Hammes-Schiffer}(2023)}]{Li2023QMMM}%
  \BibitemOpen
  \bibfield  {author} {\bibinfo {author} {\bibfnamefont {T.~E.}\ \bibnamefont
  {Li}}\ and\ \bibinfo {author} {\bibfnamefont {S.}~\bibnamefont
  {Hammes-Schiffer}},\ }\bibfield  {title} {\bibinfo {title} {{QM/MM Modeling
  of Vibrational Polariton Induced Energy Transfer and Chemical Dynamics}},\
  }\href {https://doi.org/10.1021/JACS.2C10170/SUPPL_FILE/JA2C10170_SI_001.PDF}
  {\bibfield  {journal} {\bibinfo  {journal} {Journal of American Chemical
  Society}\ }\textbf {\bibinfo {volume} {145}},\ \bibinfo {pages} {377}
  (\bibinfo {year} {2023})}\BibitemShut {NoStop}%
\bibitem [{\citenamefont {Althorpe}(2021)}]{althorpe_path-integral_2021}%
  \BibitemOpen
  \bibfield  {author} {\bibinfo {author} {\bibfnamefont {S.~C.}\ \bibnamefont
  {Althorpe}},\ }\bibfield  {title} {\bibinfo {title} {Path-integral
  approximations to quantum dynamics},\ }\href
  {https://doi.org/10.1140/epjb/s10051-021-00155-2} {\bibfield  {journal}
  {\bibinfo  {journal} {The European Physical Journal B}\ }\textbf {\bibinfo
  {volume} {94}},\ \bibinfo {pages} {155} (\bibinfo {year} {2021})}\BibitemShut
  {NoStop}%
\bibitem [{\citenamefont {Willatt}\ \emph {et~al.}(2018)\citenamefont
  {Willatt}, \citenamefont {Ceriotti},\ and\ \citenamefont
  {Althorpe}}]{willatt_approximating_2018}%
  \BibitemOpen
  \bibfield  {author} {\bibinfo {author} {\bibfnamefont {M.~J.}\ \bibnamefont
  {Willatt}}, \bibinfo {author} {\bibfnamefont {M.}~\bibnamefont {Ceriotti}},\
  and\ \bibinfo {author} {\bibfnamefont {S.~C.}\ \bibnamefont {Althorpe}},\
  }\bibfield  {title} {\bibinfo {title} {Approximating {Matsubara} dynamics
  using the planetary model: {Tests} on liquid water and ice},\ }\href
  {https://doi.org/10.1063/1.5004808} {\bibfield  {journal} {\bibinfo
  {journal} {The Journal of Chemical Physics}\ }\textbf {\bibinfo {volume}
  {148}},\ \bibinfo {pages} {102336} (\bibinfo {year} {2018})}\BibitemShut
  {NoStop}%
\bibitem [{\citenamefont {Trenins}\ \emph {et~al.}(2019)\citenamefont
  {Trenins}, \citenamefont {Willatt},\ and\ \citenamefont
  {Althorpe}}]{trenins_path-integral_2019}%
  \BibitemOpen
  \bibfield  {author} {\bibinfo {author} {\bibfnamefont {G.}~\bibnamefont
  {Trenins}}, \bibinfo {author} {\bibfnamefont {M.~J.}\ \bibnamefont
  {Willatt}},\ and\ \bibinfo {author} {\bibfnamefont {S.~C.}\ \bibnamefont
  {Althorpe}},\ }\bibfield  {title} {\bibinfo {title} {Path-integral dynamics
  of water using curvilinear centroids},\ }\href
  {https://doi.org/10.1063/1.5100587} {\bibfield  {journal} {\bibinfo
  {journal} {The Journal of Chemical Physics}\ }\textbf {\bibinfo {volume}
  {151}},\ \bibinfo {pages} {054109} (\bibinfo {year} {2019})}\BibitemShut
  {NoStop}%
\bibitem [{\citenamefont {Haggard}\ \emph {et~al.}(2021)\citenamefont
  {Haggard}, \citenamefont {Sadhasivam}, \citenamefont {Trenins},\ and\
  \citenamefont {Althorpe}}]{haggard_testing_2021}%
  \BibitemOpen
  \bibfield  {author} {\bibinfo {author} {\bibfnamefont {C.}~\bibnamefont
  {Haggard}}, \bibinfo {author} {\bibfnamefont {V.~G.}\ \bibnamefont
  {Sadhasivam}}, \bibinfo {author} {\bibfnamefont {G.}~\bibnamefont
  {Trenins}},\ and\ \bibinfo {author} {\bibfnamefont {S.~C.}\ \bibnamefont
  {Althorpe}},\ }\bibfield  {title} {\bibinfo {title} {Testing the
  quasicentroid molecular dynamics method on gas-phase ammonia},\ }\href
  {https://doi.org/10.1063/5.0068250} {\bibfield  {journal} {\bibinfo
  {journal} {The Journal of Chemical Physics}\ }\textbf {\bibinfo {volume}
  {155}},\ \bibinfo {pages} {174120} (\bibinfo {year} {2021})}\BibitemShut
  {NoStop}%
\bibitem [{\citenamefont {Witt}\ \emph {et~al.}(2009)\citenamefont {Witt},
  \citenamefont {Ivanov}, \citenamefont {Shiga}, \citenamefont {Forbert},\ and\
  \citenamefont {Marx}}]{witt+09jcp}%
  \BibitemOpen
  \bibfield  {author} {\bibinfo {author} {\bibfnamefont {A.}~\bibnamefont
  {Witt}}, \bibinfo {author} {\bibfnamefont {S.~D.}\ \bibnamefont {Ivanov}},
  \bibinfo {author} {\bibfnamefont {M.}~\bibnamefont {Shiga}}, \bibinfo
  {author} {\bibfnamefont {H.}~\bibnamefont {Forbert}},\ and\ \bibinfo {author}
  {\bibfnamefont {D.}~\bibnamefont {Marx}},\ }\bibfield  {title} {\bibinfo
  {title} {{On the applicability of centroid and ring polymer path integral
  molecular dynamics for vibrational spectroscopy.}},\ }\href@noop {}
  {\bibfield  {journal} {\bibinfo  {journal} {The Journal of Chemical Physics}\
  }\textbf {\bibinfo {volume} {130}},\ \bibinfo {pages} {194510} (\bibinfo
  {year} {2009})}\BibitemShut {NoStop}%
\bibitem [{\citenamefont {Singraber}\ \emph {et~al.}(2019)\citenamefont
  {Singraber}, \citenamefont {Behler},\ and\ \citenamefont
  {Dellago}}]{singraber_library-based_2019}%
  \BibitemOpen
  \bibfield  {author} {\bibinfo {author} {\bibfnamefont {A.}~\bibnamefont
  {Singraber}}, \bibinfo {author} {\bibfnamefont {J.}~\bibnamefont {Behler}},\
  and\ \bibinfo {author} {\bibfnamefont {C.}~\bibnamefont {Dellago}},\
  }\bibfield  {title} {\bibinfo {title} {Library-{Based} {LAMMPS}
  {Implementation} of {High}-{Dimensional} {Neural} {Network} {Potentials}},\
  }\href {https://doi.org/10.1021/acs.jctc.8b00770} {\bibfield  {journal}
  {\bibinfo  {journal} {Journal of Chemical Theory and Computation}\ }\textbf
  {\bibinfo {volume} {15}},\ \bibinfo {pages} {1827} (\bibinfo {year}
  {2019})}\BibitemShut {NoStop}%
\bibitem [{\citenamefont {Batzner}\ \emph {et~al.}(2022)\citenamefont
  {Batzner}, \citenamefont {Musaelian}, \citenamefont {Sun}, \citenamefont
  {Geiger}, \citenamefont {Mailoa}, \citenamefont {Kornbluth}, \citenamefont
  {Molinari}, \citenamefont {Smidt},\ and\ \citenamefont
  {Kozinsky}}]{batzner_e3-equivariant_2022}%
  \BibitemOpen
  \bibfield  {author} {\bibinfo {author} {\bibfnamefont {S.}~\bibnamefont
  {Batzner}}, \bibinfo {author} {\bibfnamefont {A.}~\bibnamefont {Musaelian}},
  \bibinfo {author} {\bibfnamefont {L.}~\bibnamefont {Sun}}, \bibinfo {author}
  {\bibfnamefont {M.}~\bibnamefont {Geiger}}, \bibinfo {author} {\bibfnamefont
  {J.~P.}\ \bibnamefont {Mailoa}}, \bibinfo {author} {\bibfnamefont
  {M.}~\bibnamefont {Kornbluth}}, \bibinfo {author} {\bibfnamefont
  {N.}~\bibnamefont {Molinari}}, \bibinfo {author} {\bibfnamefont {T.~E.}\
  \bibnamefont {Smidt}},\ and\ \bibinfo {author} {\bibfnamefont
  {B.}~\bibnamefont {Kozinsky}},\ }\bibfield  {title} {\bibinfo {title}
  {E(3)-equivariant graph neural networks for data-efficient and accurate
  interatomic potentials},\ }\href {https://doi.org/10.1038/s41467-022-29939-5}
  {\bibfield  {journal} {\bibinfo  {journal} {Nature Communications}\ }\textbf
  {\bibinfo {volume} {13}},\ \bibinfo {pages} {2453} (\bibinfo {year}
  {2022})}\BibitemShut {NoStop}%
\bibitem [{\citenamefont {Batatia}\ \emph {et~al.}(2022)\citenamefont
  {Batatia}, \citenamefont {Kovacs}, \citenamefont {Simm}, \citenamefont
  {Ortner},\ and\ \citenamefont {Csanyi}}]{batatia_mace_2022}%
  \BibitemOpen
  \bibfield  {author} {\bibinfo {author} {\bibfnamefont {I.}~\bibnamefont
  {Batatia}}, \bibinfo {author} {\bibfnamefont {D.~P.}\ \bibnamefont {Kovacs}},
  \bibinfo {author} {\bibfnamefont {G.}~\bibnamefont {Simm}}, \bibinfo {author}
  {\bibfnamefont {C.}~\bibnamefont {Ortner}},\ and\ \bibinfo {author}
  {\bibfnamefont {G.}~\bibnamefont {Csanyi}},\ }\bibfield  {title} {\bibinfo
  {title} {{MACE}: {Higher} {Order} {Equivariant} {Message} {Passing} {Neural}
  {Networks} for {Fast} and {Accurate} {Force} {Fields}},\ }\href
  {https://proceedings.neurips.cc/paper\_files/paper/2022/hash/4a36c3c51af11ed9f34615b81edb5bbc-Abstract-Conference.html}
  {\bibfield  {journal} {\bibinfo  {journal} {Advances in Neural Information
  Processing Systems}\ }\textbf {\bibinfo {volume} {35}},\ \bibinfo {pages}
  {11423} (\bibinfo {year} {2022})}\BibitemShut {NoStop}%
\bibitem [{\citenamefont {Grechko}\ \emph {et~al.}(2018)\citenamefont
  {Grechko}, \citenamefont {Hasegawa}, \citenamefont {D'Angelo}, \citenamefont
  {Ito}, \citenamefont {Turchinovich}, \citenamefont {Nagata},\ and\
  \citenamefont {Bonn}}]{Grechko_NatComm_2018}%
  \BibitemOpen
  \bibfield  {author} {\bibinfo {author} {\bibfnamefont {M.}~\bibnamefont
  {Grechko}}, \bibinfo {author} {\bibfnamefont {T.}~\bibnamefont {Hasegawa}},
  \bibinfo {author} {\bibfnamefont {F.}~\bibnamefont {D'Angelo}}, \bibinfo
  {author} {\bibfnamefont {H.}~\bibnamefont {Ito}}, \bibinfo {author}
  {\bibfnamefont {D.}~\bibnamefont {Turchinovich}}, \bibinfo {author}
  {\bibfnamefont {Y.}~\bibnamefont {Nagata}},\ and\ \bibinfo {author}
  {\bibfnamefont {M.}~\bibnamefont {Bonn}},\ }\bibfield  {title} {\bibinfo
  {title} {{Coupling between intra- and intermolecular motions in liquid water
  revealed by two-dimensional terahertz-infrared-visible spectroscopy}},\
  }\href {https://doi.org/10.1038/s41467-018-03303-y} {\bibfield  {journal}
  {\bibinfo  {journal} {Nature Communications}\ }\textbf {\bibinfo {volume}
  {9}},\ \bibinfo {pages} {885} (\bibinfo {year} {2018})}\BibitemShut {NoStop}%
\bibitem [{\citenamefont {Mukamel}(1999)}]{Mukamel}%
  \BibitemOpen
  \bibfield  {author} {\bibinfo {author} {\bibfnamefont {S.}~\bibnamefont
  {Mukamel}},\ }\href@noop {} {\emph {\bibinfo {title} {Principles of Nonlinear
  Optical Spectroscopy}}}\ (\bibinfo  {publisher} {Oxford University Press},\
  \bibinfo {year} {1999})\BibitemShut {NoStop}%
\bibitem [{\citenamefont {Seliya}\ \emph {et~al.}(2024)\citenamefont {Seliya},
  \citenamefont {Bonn},\ and\ \citenamefont {Grechko}}]{Seliya_JCP_2024}%
  \BibitemOpen
  \bibfield  {author} {\bibinfo {author} {\bibfnamefont {P.}~\bibnamefont
  {Seliya}}, \bibinfo {author} {\bibfnamefont {M.}~\bibnamefont {Bonn}},\ and\
  \bibinfo {author} {\bibfnamefont {M.}~\bibnamefont {Grechko}},\ }\bibfield
  {title} {\bibinfo {title} {{On selection rules in two-dimensional
  terahertz–infrared–visible spectroscopy}},\ }\href
  {https://doi.org/10.1063/5.0179041} {\bibfield  {journal} {\bibinfo
  {journal} {The Journal of Chemical Physics}\ }\textbf {\bibinfo {volume}
  {160}},\ \bibinfo {pages} {034201} (\bibinfo {year} {2024})}\BibitemShut
  {NoStop}%
\bibitem [{\citenamefont {Jung}\ \emph {et~al.}(2020)\citenamefont {Jung},
  \citenamefont {Videla},\ and\ \citenamefont {Batista}}]{Jung_JCP_2020}%
  \BibitemOpen
  \bibfield  {author} {\bibinfo {author} {\bibfnamefont {K.~A.}\ \bibnamefont
  {Jung}}, \bibinfo {author} {\bibfnamefont {P.~E.}\ \bibnamefont {Videla}},\
  and\ \bibinfo {author} {\bibfnamefont {V.~S.}\ \bibnamefont {Batista}},\
  }\bibfield  {title} {\bibinfo {title} {{Ring-polymer, centroid, and
  mean-field approximations to multi-time Matsubara dynamics}},\ }\href
  {https://doi.org/10.1063/5.0021843} {\bibfield  {journal} {\bibinfo
  {journal} {The Journal of Chemical Physics}\ }\textbf {\bibinfo {volume}
  {153}},\ \bibinfo {pages} {124112} (\bibinfo {year} {2020})}\BibitemShut
  {NoStop}%
\bibitem [{\citenamefont {Jung}\ \emph {et~al.}(2018)\citenamefont {Jung},
  \citenamefont {Videla},\ and\ \citenamefont {Batista}}]{Jung_JCP_2018}%
  \BibitemOpen
  \bibfield  {author} {\bibinfo {author} {\bibfnamefont {K.~A.}\ \bibnamefont
  {Jung}}, \bibinfo {author} {\bibfnamefont {P.~E.}\ \bibnamefont {Videla}},\
  and\ \bibinfo {author} {\bibfnamefont {V.~S.}\ \bibnamefont {Batista}},\
  }\bibfield  {title} {\bibinfo {title} {{Inclusion of nuclear quantum effects
  for simulations of nonlinear spectroscopy}},\ }\href
  {https://doi.org/10.1063/1.5036768} {\bibfield  {journal} {\bibinfo
  {journal} {The Journal of Chemical Physics}\ }\textbf {\bibinfo {volume}
  {148}},\ \bibinfo {pages} {244105} (\bibinfo {year} {2018})}\BibitemShut
  {NoStop}%
\bibitem [{\citenamefont {Tong}\ \emph {et~al.}(2020)\citenamefont {Tong},
  \citenamefont {Videla}, \citenamefont {Jung}, \citenamefont {Batista},\ and\
  \citenamefont {Sun}}]{Tong_JCP_2020}%
  \BibitemOpen
  \bibfield  {author} {\bibinfo {author} {\bibfnamefont {Z.}~\bibnamefont
  {Tong}}, \bibinfo {author} {\bibfnamefont {P.~E.}\ \bibnamefont {Videla}},
  \bibinfo {author} {\bibfnamefont {K.~A.}\ \bibnamefont {Jung}}, \bibinfo
  {author} {\bibfnamefont {V.~S.}\ \bibnamefont {Batista}},\ and\ \bibinfo
  {author} {\bibfnamefont {X.}~\bibnamefont {Sun}},\ }\bibfield  {title}
  {\bibinfo {title} {{Two-dimensional Raman spectroscopy of Lennard-Jones
  liquids via ring-polymer molecular dynamics}},\ }\href
  {https://doi.org/10.1063/5.0015436} {\bibfield  {journal} {\bibinfo
  {journal} {The Journal of Chemical Physics}\ }\textbf {\bibinfo {volume}
  {153}},\ \bibinfo {pages} {034117} (\bibinfo {year} {2020})}\BibitemShut
  {NoStop}%
\bibitem [{\citenamefont {Aryasetiawan}\ and\ \citenamefont
  {Nilsson}(2022)}]{aryas+book}%
  \BibitemOpen
  \bibfield  {author} {\bibinfo {author} {\bibfnamefont {F.}~\bibnamefont
  {Aryasetiawan}}\ and\ \bibinfo {author} {\bibfnamefont {F.}~\bibnamefont
  {Nilsson}},\ }\href {https://doi.org/10.1063/9780735422490} {\emph {\bibinfo
  {title} {{Downfolding Methods in Many-Electron Theory}}}}\ (\bibinfo
  {publisher} {AIP Publishing LLC},\ \bibinfo {year} {2022})\ \Eprint
  {https://arxiv.org/abs/https://pubs.aip.org/book-pdf/12252618/9780735422490.pdf}
  {https://pubs.aip.org/book-pdf/12252618/9780735422490.pdf} \BibitemShut
  {NoStop}%
\bibitem [{\citenamefont {Berges}\ \emph {et~al.}(2023)\citenamefont {Berges},
  \citenamefont {Girotto}, \citenamefont {Wehling}, \citenamefont {Marzari},\
  and\ \citenamefont {Ponc\'e}}]{berge+23prx}%
  \BibitemOpen
  \bibfield  {author} {\bibinfo {author} {\bibfnamefont {J.}~\bibnamefont
  {Berges}}, \bibinfo {author} {\bibfnamefont {N.}~\bibnamefont {Girotto}},
  \bibinfo {author} {\bibfnamefont {T.}~\bibnamefont {Wehling}}, \bibinfo
  {author} {\bibfnamefont {N.}~\bibnamefont {Marzari}},\ and\ \bibinfo {author}
  {\bibfnamefont {S.}~\bibnamefont {Ponc\'e}},\ }\bibfield  {title} {\bibinfo
  {title} {Phonon self-energy corrections: To screen, or not to screen},\
  }\href {https://doi.org/10.1103/PhysRevX.13.041009} {\bibfield  {journal}
  {\bibinfo  {journal} {Physical Review X}\ }\textbf {\bibinfo {volume} {13}},\
  \bibinfo {pages} {041009} (\bibinfo {year} {2023})}\BibitemShut {NoStop}%
\bibitem [{\citenamefont {Zhao}\ \emph {et~al.}(2020)\citenamefont {Zhao},
  \citenamefont {Zhang}, \citenamefont {Cao}, \citenamefont {Zhao},
  \citenamefont {Struzhkin}, \citenamefont {Goncharov}, \citenamefont {Wang},
  \citenamefont {Gavriliuk}, \citenamefont {Mao},\ and\ \citenamefont
  {Chen}}]{zhao+20prb}%
  \BibitemOpen
  \bibfield  {author} {\bibinfo {author} {\bibfnamefont {X.-M.}\ \bibnamefont
  {Zhao}}, \bibinfo {author} {\bibfnamefont {K.}~\bibnamefont {Zhang}},
  \bibinfo {author} {\bibfnamefont {Z.-Y.}\ \bibnamefont {Cao}}, \bibinfo
  {author} {\bibfnamefont {Z.-W.}\ \bibnamefont {Zhao}}, \bibinfo {author}
  {\bibfnamefont {V.~V.}\ \bibnamefont {Struzhkin}}, \bibinfo {author}
  {\bibfnamefont {A.~F.}\ \bibnamefont {Goncharov}}, \bibinfo {author}
  {\bibfnamefont {H.-K.}\ \bibnamefont {Wang}}, \bibinfo {author}
  {\bibfnamefont {A.~G.}\ \bibnamefont {Gavriliuk}}, \bibinfo {author}
  {\bibfnamefont {H.-K.}\ \bibnamefont {Mao}},\ and\ \bibinfo {author}
  {\bibfnamefont {X.-J.}\ \bibnamefont {Chen}},\ }\bibfield  {title} {\bibinfo
  {title} {Pressure tuning of the charge density wave and superconductivity in
  $2h\ensuremath{-}{\mathrm{tas}}_{2}$},\ }\href
  {https://doi.org/10.1103/PhysRevB.101.134506} {\bibfield  {journal} {\bibinfo
   {journal} {Physical Review B}\ }\textbf {\bibinfo {volume} {101}},\ \bibinfo
  {pages} {134506} (\bibinfo {year} {2020})}\BibitemShut {NoStop}%
\end{thebibliography}
\end{document}